\documentclass[11pt,twoside,epsfig]{mybook}

\usepackage{amsmath,amssymb,graphicx,rotating}
\usepackage{natbib}
\renewcommand{\baselinestretch} {1.5}

\begin{document}

\frontmatter

\pagestyle{empty}

\begin{center}
\vspace*{-20mm}
\Large{\bf UNIVERSIT\`A DEGLI STUDI DI TRIESTE\\}
\vspace*{5mm}
\normalsize{\bf Sede Amministrativa del Dottorato di Ricerca}
\vspace*{2mm}

\Large{\bf XIX ciclo del Dottorato di Ricerca in Fisica}
\vspace*{15mm}

%\begin{center}
\Huge{\bf Chemical evolution of neutron capture elements in our Galaxy 
and in the dwarf spheroidal galaxies of the Local Group }

\vspace*{8mm}
\normalsize{Settore scientifico-disciplinare: FIS/05 ASTRONOMIA E ASTROFISICA}
\vspace*{25mm}

%\large{by}
%

\leftline{\normalsize {DOTTORANDO        \hspace*{48mm}   COORDINATORE DEL COLLEGIO DEI DOCENTI}}
\vspace*{-.5mm}
\leftline{\normalsize{\bf GABRIELE CESCUTTI \hspace*{29mm} \bf  PROF. GAETANO SENATORE}}               
\vspace*{12mm}
\leftline{\normalsize{ \hspace*{76.5mm}   TUTORE}}		   
\vspace*{-.5mm}
\leftline{\normalsize {\hspace*{78.5mm}\bf PROF. FRANCESCA MATTEUCCI}}
\vspace*{12mm}
\leftline{\normalsize{ \hspace*{77mm}   RELATORE}}		   
\vspace*{-.5mm}
\leftline{\normalsize {\hspace*{78.5mm} \bf PROF. FRANCESCA MATTEUCCI}}

\normalsize{}
\vspace*{15mm}

\large{\bf Anno Accademico 2005/2006\\}
\end{center}
\cleardoublepage

\renewcommand{\baselinestretch} {1.0}

\newpage
\vspace{40cm}
\rightline{\emph{I dedicate this work to}}
\rightline{\emph{my mother }}
\rightline{\emph{my father}}
\rightline{\emph{and}}
\rightline{\emph{my brother}}

\chapter{Abstract}

We model the evolution of the abundances of several neutron capture elements
 (Ba, Eu, La, Sr, Y and Zr) in the Milky Way and then we extend our predictions
 to some dwarf spheroidal galaxies of the Local Group.

 Two major neutron capture mechanisms on iron seeds are generally invoked:
 the slow process (s-process) and the rapid process (r-process),
 where the slow and the rapid are defined relative to the timescale of the $\beta$-decay.
 Nucleosynthesis calculations for r-process are very few, owing to the difficulties
 in modelling the physics the r-process and the lack of knowledge about the sites 
of productions of these elements. For s-process elements instead some calculations 
are available but the sites of production are also uncertain.

By adopting a chemical evolution model for the Milky Way already reproducing
 the evolution of several chemical elements (H, He, C, N, O, $\alpha$-elements
 and iron peak elements), we compare our theoretical results with accurate and new stellar
 data of neutron capture elements and we are able to impose strong constraints on
 the nucleosynthesis of the studied elements. We can suggest the stellar sites of
 production for each  element. In particular, the r-process component of each element
 (if any) is produced in the mass range from 10 to 30 $M_{\odot}$, whereas 
the s-process component arises from stars in the range from 1 to 3 $M_{\odot}$.

Using the same chemical evolution model, extended to different galactocentric
 distances, we obtain results on the radial gradients of the Milky Way. We compare 
the results of the model not only for the neutron capture elements but also for 
$\alpha$-elements and iron peak elements with new data of Cepheids stars. For the first time
 with these data, it is possible to verify the predictions for the gradients of very
 heavy elements. We conclude that the model, with an inside-out scenario for the
 building up of the disc and a constant density distribution of the gas for the halo 
phase, can be considered successful; in fact, for almost all the considered elements
 with  our nucleosynthesis prescriptions, the model well reproduces the observed 
abundance gradients.

We give a possible explanation to the considerable scatter 
of neutron capture elements observed in low metallicity stars 
 in the solar vicinity, compared to the small star to star scatter 
observed for the $\alpha$-elements. In fact, we have developed a stochastic
 chemical evolution model, in which the main assumption is a random formation
 of new stars, subject to the condition that the cumulative mass distribution
 follows a given initial mass function. With our model we are able to reproduce 
the different features of neutron capture elements and $\alpha$-elements. The reason
 for this resides in the random birth of stars coupled with different stellar mass
 ranges from where $\alpha$-elements and neutron capture elements originate. In particular,
 the site of production of $\alpha$-elements is the whole range of the massive stars,
 whereas the mass range of production for neutron capture elements has an upper limit
 of 30$M_{\odot}$.

Finally, we test the prescriptions for neutron capture elements also for the dwarf
 spheroidal galaxies of the Local Group. We use a chemical evolution model already
 able to reproduce the  abundances for $\alpha$-elements in these systems. We conclude
 that the same prescriptions used for the Milky Way well reproduce the main features
 of neutron capture elements also in the dwarf spheroidal galaxies for which we have
 observational data. In dwarf spheroidal galaxies for which we do not have
 observational data we only give predictions.
 We predict that the chemical evolution of these elements in dwarf spheroidal galaxies
  is different from the evolution in the solar vicinity. This is due to their different histories 
of star formation relative to our Galaxy and indicates that dwarf spheroidal
 galaxies (we see nowadays) cannot be the building blocks of our Galaxy.

\renewcommand{\baselinestretch} {1.5}

\normalsize{

\pagestyle{headings}

\tableofcontents

\mainmatter

\chapter{Introduction}

\rightline{\emph{``It is written in the stars above'' by Depeche Mode}}
\rightline{\emph{``The stars above us, govern our conditions'' by William Shakespeare}}

\vspace{1cm}

\section{The galactic chemical evolution}
Galactic chemical evolution is the study of the evolution in time and in space
of the abundances of the chemical elements in the interstellar gas in galaxies.
This process is influenced by many parameters such as the initial conditions,
the star formation and evolution, the nucleosynthesis and possible gas flows.
So, in order to build a chemical evolution model one needs to specify 
the initial conditions, namely whether the system is closed or open
and whether the gas is primordial (no metals \footnote{In astrophysics all the
 chemical elements heavier than $^{4}$He}) or already chemically enriched. 
Then, it is necessary to define the stellar birthrate function, 
which is generally expressed as the product of two independent functions, the star
formation rate (SFR) and the initial mass function (IMF), namely:
\begin{equation}
B(m,t)=\psi(t)\varphi(m)\ dt\ dm
\end{equation}
where $\varphi(t)$ is the  SFR  and $\psi(m)$ is the IMF.
The SFR is assumed to be only a function of the time and the IMF only
a function of the mass. This  oversimplification is due to
the absence of a clear knowledge of the star formation process.
Moreover, it is necessary to know the stellar evolution ad the nuclear burnings 
which take place in the stellar interiors during the stellar 
lifetime and produce new chemical elements, in particular metals. 
These metals, together with the pristine stellar material are restored into the interstellar medium
 (ISM) at the star death. This process clearly affects crucially the chemical
 evolution of the ISM. In order to take in account the elemental production
by stars we define the ``yields'', in particular the stellar yields, as 
the amount of elements produced by a single star.
Finally, the supplementary parameters are the infall of extragalactic gas,
radial flows and the galactic winds, which are important ingredients in
building galactic chemical evolution models.

The SFR is one of the most important drivers of galactic chemical evolution: it
describes the rate at which the gas is turned into stars in galaxies. Since the physics of the
star formation process is still not well known, several parameterizations are used to describe
the SFR. A common aspect to the different formulations of the SFR is that they include a
dependence upon the gas density. Here we recall the most commonly used parameterizations
for the SFR adopted so far in the literature.
An exponentially decreasing SFR provides an easy to handle formula:
\begin{equation}
\psi(t)=\nu e^{-t/\tau_{\star}}
\end{equation}
with $\tau_{\star}$ = 5-15 Gyr in order to obtain a good fit to the properties of the solar neighborhood
(Tosi, 1988) and  $\nu$= 1 - 2 Gyr$^{-1}$, being $\nu$ the efficiency of star formation which is
expressed as the inverse of the timescale of star formation.
However, the most famous formulation and most widely adopted for the SFR is the
Schmidt (1959) law:
\begin{equation}
\psi(t)=\nu \sigma^{k}_{gas}
\end{equation}
which assumes that the SFR is proportional to some power of the volume or surface gas
density ($\sigma_{gas}$). The exponent suggested by Schmidt was k = 2 but Kennicutt (1998) suggested that
the best fit to the observational data on spiral disks and starburst galaxies is obtained with an
exponent k = 1.4$\pm $ 0.15.
A more complex formulation, including a dependence also on the total surface mass
density ($\sigma_{tot}$), which is induced by the SN feedback, was suggested by the observations of Dopita
\& Ryder (1994) who proposed the following formulation:
\begin{equation}
\psi(t)=\nu \sigma^{k1}_{tot} \sigma^{k2}_{gas}
\end{equation}
Kennicutt suggested also an alternative law to the Schmidt-like one discussed above:
\begin{equation}
\psi(t)=0.017 \Omega_{gas} \sigma_{gas} 
\end{equation}
being $ \Omega_{gas}$ the angular rotation speed of the gas.

The IMF is a probability distribution function and 
the most common parameterization for the IMF is that proposed by Salpeter (1955),
which assumes a one-slope power law with x = 1.35, in particular:
\begin{equation}
\varphi(m)=A m^{-(1+x)}
\end{equation}
$\varphi(m)$ is the number of stars with masses 
in the interval M, M+dM, and A is a normalization constant.
The IMF is generally normalized as:
\begin{equation}
\int_{0}^{\infty} m \varphi(m)dm = 1
\end{equation}
More recently, multi-slope expressions of the IMF have been adopted since they
better describe the luminosity function of the main sequence stars in the solar vicinity
(Scalo 1986, 1998; Kroupa et al. 1993). Generally, the IMF is assumed to be constant in
space and time, with some exceptions such as the one suggested by Larson (1998), which
adopts a variable slope:
\begin{equation}
x= 1.35(1+m/m_{1})^{-1}
\end{equation}
where $m_{1}$ is variable with time and associated to the Jeans mass,
(the typical mass at which the internal pressure is no longer strong enough to prevent 
gravitational collapse). The effects of a variable
IMF on the galactic disk properties have been studied by Chiappini et al. (2000), who
concluded that only a very ``ad hoc'' variation of the IMF can reproduce the majority of
observational constraints, thus favoring chemical evolution models with IMF constant in
space and time.

The stellar yields are fundamental ingredients in galactic chemical evolution.
In the past ten years a large number of calculations 
 has become available for stars of all masses and
metallicities. However, uncertainties in stellar yields are still present.
  This is due to uncertainties in the nuclear reaction rates, treatment of
convection, mass cut, explosion energies, neutron fluxes and possible fall-back of matter onto
the proto-neutron star. Moreover, the $^{14}$N nucleosynthesis and its primary and/or secondary
nature are still under debate. The most recent calculations are summarized below:
\begin {itemize}

\item Low and Intermediate mass stars (0.8 $M_{\odot}<$M$<$ 8.0 $M_{\odot}$) 
(van den Hoeck \& Groenewegen 1997; Meynet \& Maeder 2002; Ventura et al. 2002; 
Siess et al. 2002).
These stars produce $^{4}$He, C, N and heavy  elements (A $>$ 90).

\item Massive stars (M $>$  10 $M_{\odot}$) (Woosley \& Weaver 1995;
 Langer \& Henkel 1995; Thielemann et al. 1996; 
Nomoto et al. 1997; Rauscher et al. 2002; Limongi \& Chieffi 2003).
These stars produce mainly $\alpha$-elements (O, Ne, Mg, Si, S, Ca), some Fe-peak elements,
heavy elements.

\item Type Ia SNe (Nomoto et al. 1997; Iwamoto et al. 1999) produce mainly Fe-peak elements. 

\item Very massive objects (M $>$ 100$M_{\odot}$), if they exist, should produce mostly oxygen 
(Portinari et al. 1998; Umeda \& Nomoto 2001; Nakamura \& Umemura 2001).
\end{itemize}

Depending on the galactic system, the infall rate (IR) can be assumed to be constant in
space and time, or more realistically the infall rate can be variable in space and time:
\begin{equation}
IR(r,t) = A(r)e^{-t/\tau(r)} 
\end{equation}
with $\tau(r)$ constant or varying along the disk. 
The parameter A(r) is derived by fitting the
present time total surface mass density in the disk of the Galaxy, $\sigma_{tot}(t_{now})$.
Otherwise, for the formation of the Galaxy one can assume two independent episodes of infall during which
the halo and perhaps part of the thick-disk formed first and then the thin-disk, 
as in the two-infall model of Chiappini et al. (1997).
For the rate of gas outflow or galactic wind there are no specific prescriptions 
but generally one simply assumes that the wind rate (W) is proportional to the
 star formation rate (Hartwick 1976, Matteucci \& Chiosi 1983):
\begin{equation}
W(t) = -\lambda \psi(t) 
\end{equation}
with  $\lambda$ being a free parameter

A good model of chemical evolution should be able to reproduce a minimum number
of observational constraints and the number of indipendent observational constraints should be larger
than the number of free parameters which are: $\tau$, $k_{1}$, $k_{2}$, the $\varphi(m)$ slope(s) 
and the parameter describing the wind, $\lambda$ if adopted.

%\begin{figure}
%\begin{center}     
%\includegraphics[width=0.5\textwidth]{milkyway} 
%\caption{An image of the Milky Way in an artist's rendering. Understanding 
%its formation and evolution is fundamental to improve the knowledge of the formation of 
%spiral galaxies and in general of all the galaxies.}
%\label{Milky}
%\end{center}
%\end{figure}

The main observational constraints in the Milky Way,
% (see an image in Fig\ref{Milky}) 
in particular in the solar vicinity, 
that a good model should reproduce (see Chiappini et al. 2001) are:

\begin{itemize}
\item The present time surface gas density: $\Sigma_{gas}$ = 13$\pm$3  M$_{\odot}$pc$^{-2}$
\item The present time surface star density $\Sigma_{\star}$= 35$\pm$5 M$_{\odot}$pc$^{-2}$
\item The present time total surface mass density: $\Sigma_{tot}$ = 48$\pm$9 M$_{\odot}$pc$^{-2}$
\item The present time SFR: $\psi_{0}$  = 2 - 5M$_{\odot}$ pc$^{-2}$Gyr$^{-1}$
\item The present time infall rate:  0.3 - 1.5M$_{\odot}$ pc$^{-2}$ Gyr$^{-1}$
\item The present day mass function (PDMF).
\item The solar abundances, namely the chemical abundances of the ISM at the time of birth of
the solar system 4.5 Gyr ago. 
\item The age-metallicity relation, namely the relation between the ages of the stars
and the metal abundances of their photospheres, assumed to be equivalent
to the stellar [Fe/H]\footnote{We adopt the usual spectroscopic notations that
[A/B]= $log_{10}(N_{A}/N_{B})_{\star}-log_{10}(N_{A}/N_{B})_{\odot}$ 
and that $log\epsilon(A)=log_{10}(N_{A}/N_{H})+12.0$, for elements A and B}.
\item The G-dwarf metallicity distribution, namely the number of stars with a lifetime equal
or larger than the age of the Galaxy as a function of their metallicities.
\item The distributions of gas and stars formation rate along the disk.
\item The average SNII and Ia rates along the disk 
(SNII=1.2$\pm$0.8 cen$^{-1}$ and SNIa=0.3$\pm$0.2 cen$^{-1}$).
\item  The observed abundances in the stars and the [A/Fe]  vs. [Fe/H] relations.
\end{itemize}

The chemical compositions of stars of all Galactic generations contains very 
important information about the cumulative nucleosynthesis history of the Galaxy. 
The difference in the timescales for the occurrence of SNII and Ia produces a timedelay
in the Fe production relative to the $\alpha$-elements (Tinsley 1979; Greggio \& Renzini
1983b; Matteucci 1986). In fact, in the single degenerate 
scenario for a SNIa,  originally proposed by Whelan and Iben (1973), 
the SNIa explodes due to a C-deflagration in a C-O white dwarf (WD) reaching the Chandrasekhar
mass limit, M$_{Ch}$= 1.44M$_{\odot}$, after accreting material from a red giant companion.
The progenitors of C-O WDs lie in the range 0.8 - 8.0M$_{\odot}$,
therefore, the most massive binary system of two C-O WDs is the 8M$_{\odot}$ + 8M$_{\odot}$ one. 
The clock of the system in this scenario is provided by the lifetime of the secondary star 
(i.e. the less massive one in the binary system). This implies
that the minimum timescale for the appearence of the first type Ia SNe is the lifetime
of the most massive secondary star. In this case the time is t$_{SNIamin}$=0.03 Gyr (Greggio \&
Renzini 1983a; Matteucci \& Greggio, 1986; Matteucci \& Recchi, 2001).
On this basis we can interpret all the observed abundance ratios
plotted as functions of metallicity. In particular, this interpretation is known as time-delay
model and only a model including both contributions in the percentages of
30\% (SNII) and 70\% (SNIa) can reproduce the data.
Moreover, the stars formed near the beginning of our Galaxy have a very low metallicity because
their chemical compositions were produced by few previous generations of massive stars.
So, studies of elemental abundances in very old and metal poor stars serve as tests of nucleosynthesis
theories and galactic chemical evolution models.

\section{The neutron capture elements}\label{intro_nc}
In this thesis we will mainly deal with very heavy elements.
Early work by Gilroy et al. (1988) first proposed that the stellar
abundances of very heavy elements with respect to iron, particularly
[Eu/Fe], showed a large scatter at low metallicities. 
Their work suggested that this scatter appeared to diminish with increasing
metallicity. This was confirmed by the 
 large spread observed in the [Ba/Fe] and [Eu/Fe] ratios in halo 
stars (e.g. McWilliam et al. 1995; Ryan et al. 1996).
A more extensive study by Burris et al.
(2000) confirmed the very large star-to-star scatter in the early
Galaxy, while studies of stars at higher metallicities, involving
mostly disk stars (Edvardsson et al. 1993; Woolf, Tomkin \&
Lambert 1995), show little scatter.
In the last few years a great deal of observational work on galactic stars 
appeared: Fulbright (2000), Mashonkina \& Gehren (2000, 2001),
Koch \& Edvardsson (2002), Honda et al. (2004), Ishimaru et al. (2004).
All these works confirmed the presence of the spread for these elements.
It is worth noting that this spread is not found for the  
[$\alpha$/Fe] ratios in very metal poor stars (down to $[Fe/H]=-4.0$, 
Cayrel et al. 2004). 
This fact suggests that the spread, if real, is a characteristic of  these heavy elements
and not only due to an inhomogeneous mixing in the early halo phases,
as suggested by several authors (Tsujimoto et al. 1999; Ishimaru \& Wanajo 1999).

To have an insight on this peculiar behaviour of these elements, 
it is important to understand how they are formed.
The neutron capture is the main
mechanism which forms elements heavier than iron; the other
mechanism, the p-process, is required for proton-rich isotopes, whose abundances in the
solar system is less than 1\%.
Two major neutron capture mechanisms are generally invoked: the slow process (s-process) 
and the rapid process(r-process), where the slow and rapid are defined relative to the
timescale of $\beta$-decay.

The s-process requires a relatively low neutron density and moves along the valley of
$\beta$ stability. The s-process feeds in particular the elements Sr-Y-Zr, 
Ba-La-Ce-Pr-Nd and Pb, the three major s-peaks.
The reason for the existence of these peaks is that:
the neutron capture process imposes certain features on the "spectrum" 
of the heavy element abundances.
For certain neutron numbers N =  50, 82, 126 the neutron capture cross-sections 
are much
smaller than for neighbouring neutron numbers. This means that once one of 
these "magic" numbers
is reached, it becomes significantly less likely for the nucleus to capture 
more neutrons.
These numbers represent  a quantum mechanical effect of closed shells, in 
precisely the same way
that closed electron shells produce high chemical stability in the noble 
gases.  
Therefore, if the neutron capture process operates in some environment for 
some finite length of time and then shuts off, we expect a fair number of 
nuclei to be "stuck" at these "magic" numbers. 
Elements that correspond to these "magic" numbers of neutrons will thus 
be especially abundant. The relevant properties necessary for describing 
the s-process chain are the neutron capture 
cross section and, in addition, the $\beta$-decay rates of those unstable isotopes,
which are long-lived enough to allow neutron captures to compete with the $\beta$-decay.
 K\"appeler et al. (1989) have calculated the s-process abundances in the solar system, 
scaling the abundances for the all the isotopes to the  s-only isotope $^{150}Sm$. 
They obtained their results by means of the classical analysis.
The classical analysis is a phenomenological model and 
in first approximation any time dependence of the physical parameters is neglected. 
\begin{figure}
\begin{center}     
\includegraphics[width=0.99\textwidth]{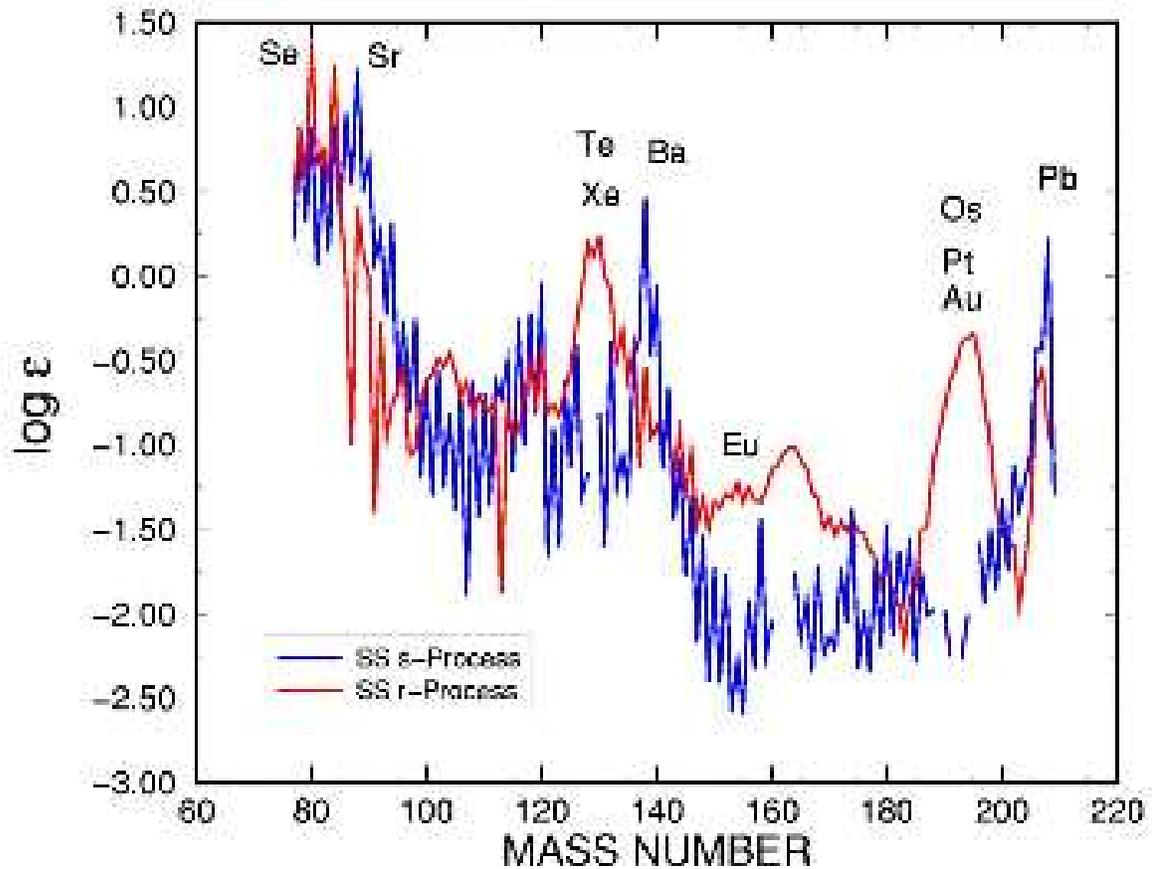} 
\caption{s-process (blue line) and r-process (red line) abundances in solar system matter, 
based upon the work by Ka\"ppeler et al. (1989).
 Note the distinctive s-process signatures at masses $\sim$ 138, and 208.
 The total solar system abundances for the heavy elements are those compiled by Anders 
\& Grevesse (1989).}
\label{r-s}
\end{center}
\end{figure}
The classical s-process pattern is shown in Figure \ref{r-s}. 
The r-process pattern, also shown in the figure, is obtained by 
subtracting the predicted abundance of s-process fraction of each element
to the solar system abundance of the same element.
As we said, the classical analysis is a phenomenological model but 
nevertheless, more detailed considerations (see K\"appeler et  al. 1989) 
lead to a physical environment characteristic of helium shell-burning zones.

The site of production of the main component of the  s-process, 
accounting for the s-process in the atomic mass number range $90<A<208$, was shown to
occur in the low-mass (1.5-3.0 $M_{\odot}$) asymptotic giant branch (AGB) stars during recurrent 
thermal pulses by stellar models (Gallino et al. 1998; Busso et al. 1999). 
A dredge down of protons can occur during these thermal pulses , as first 
suggested by Iben and Renzini (1982a, b). These protons move from the hydrogen-rich
envelope into the helium zone in low mass AGB stars, due to the operation of  
a semiconvective mixing. Subsequently, these protons are captured by $^{12}$C in the 
carbon-rich layer forming a $^{13}$C pocket.
This pocket is then engulfed by the growing convective region of the next pulse, releasing neutrons 
via the $^{13}C(\alpha,n)^{16}O$ reaction.
The s-process mechanism operating in the AGB model depends on the 
initial stellar metallicity. 
In fact, although the $^{13}C$ pocket, which acts as a 
neutron producer with the reaction $^{13}C(\alpha,n)^{16}O$, is of 
primary origin\footnote{We define a primary element an element 
synthesized from the original H and He, independently of metals. 
On the other hand, a secondary element is an element which is produced
proportionally to the abundance of metals already present in the stars and not made in situ.}
 in the work of Gallino et al. (1998) and  Busso et al. (1999, 2001),
the ensuing s-process production is dependent on the initial abundance 
of the Fe-group seeds, i.e. on the stellar metallicity.
The neutron exposure (the neutron flux per nucleus seed) is roughly proportional 
to the number of available neutron sources (the $^{13}C$ nuclei) 
per seed (the iron nuclei), hence inversely proportional to the stellar metallicity.
The strong s-component introduced by Clayton \& Rassbach (1967) in order to reproduce more than 50\%
 of solar $^{208}Pb$, is not necessary in the these stellar models, being 
naturally obtained in very metal poor stars by metallicity effects, where the iron seeds
are very rare. Moreover this scenario provide  to feed also chemical elements in the mass
number $\sim$ 90 such Sr, Y and Zr. 
The  $^{22}Ne(\alpha,n)^{25}Mg$ reaction plays also a marginal role in the production
of neutron capture elements in low mass ABG stars.
In fact, this reaction is triggered when the temperature T
exceeds $3\cdot 10^{8}$ K and in low mass stars  the maximum temperature at the bottom
of thermal pulses, although gradually increasing with the pulse
number, barely reaches the  mentioned value. 

The weak s-component is responsible for the s-process nuclides up to  $A\simeq90$
and it is recognized as the result of neutron capture during the helium burning core of
massive stars. In this environment, the $^{22}Ne(\alpha,n)^{25}Mg$ neutron producer 
reaction  can operate and first studies by Peters (1968) and Lamb et al. (1977) 
suggested that the lightest s-process nuclei can be provided by this source.
Recent studies (Raiteri et al. 1992, Baraffe et al. 1992) discovered a decrease of the production
at low metallicity. In fact, the elevated levels of nuclei from Ne to Ca in the
He-burning core of a massive star prevent the neutrons to be captured by the relative rare iron seeds.
 For these reasons, the solar system contribution from weak s-process is less than 
10\% for Sr and negligible for Y and Zr (Travaglio et al. 2004).

The r-process takes place in extremely neutron-rich environments in which 
the mean time between two successive neutron captures is very short, compared with the time 
necessary for the $\beta$-decay. 
For some time (cfr. Truran et al. 1981, Mathews \& Cowan 1990) the similar abundance 
distributions for Z$>$56 elements in metal deficient stars have been interpreted 
as evidence for a universal r-process abundance distribution in the early Galaxy.  
In particular, more recent observations (Sneden et al. 2002, 1998, 2000;
Johnson \& Bolte 2001) all seem to confirm this feature,  
that is generally referred as the ``universality'' of r-process.
Hence, it is generally believed that, at least for Z$>$56 elements, the astrophysical site
and associated yields of r-process nucleosynthesis are unique.

There are, however, some reasons to question the assumption of a universal
r-process abundance curve. The material out of which these metal poor stars
were formed is likely to have experienced the enrichment by only one or two supernovae before
incorporation into the stars. Depending upon which particular progenitor supernova
was involved, there might be substantially different abundance
distribution curves for these stars, compared to the one represented
in Solar-System material, which is a average on many episodes of chemical enrichment
 (Ishimaru \& Wanajo 1999). 
%Moreover, quite recently, Cayrel et al. (2001) have reported the observation of peculiar r-process
%elements in the metal deficient star CS31082-001. 
Recent work by Otsuki et al. (2003) indicates that the coincidence of the observed
abundance distribution for 56 $<$ Z $<$ 75 elements with the Solar r-process abundances 
does not necessarily mean that all r-process events occur in the
same universal environment. Moreover, different abundance distributions for
Z $>$ 75 and Z $<$ 56 elements are produced even when the universal 56 $<$ Z $<$
75 abundances are reproduced.
 Several scenarios have been proposed for the origin of r-process elements: neutrino winds in 
core-collapse supernovae (Woosley et al. 1994), the collapse of ONeMg cores resulting from stars
with initial masses in the range 8-10$M_{\odot}$ (Wanajo et al. 2003) and neutron star mergers 
(Freiburghaus et al. 1999),  even if this last scenario seems to be ruled out from the recent
 work of Argast et al. (2004) at least as the major one responsible for r-process enrichment in our Galaxy.
So if the r-process is generally accepted to take place in SNe II 	
explosions (Hill et al. 2002; Cowan et al. 2002),
no clear consensus has been achieved and r-process nucleosynthesis 
still remains uncertain. Theoretical predictions for r-process
 production still do not exist,  with the exception of the results of Wanajo et al. (2003) and
Woosley and Hoffmann (1992). 
However, the results of the model of Wanajo et al.  cannot be used in galactic chemical evolution models
 because they do not take into account fallback (after the SN explosion 
some material can fall back onto central collapsing neutron star) and so the amount
 of neutron capture elements produced 
is probably too high (about 2 orders of magnitude higher than the chemical evolution predictions).
Furthermore, Woosley and Hoffmann (1992) have given prescriptions for r-processes only until $^{107}Ru$.
In order to shed light on the nature (s- and/or r- processes)
of heavy elements such as Ba and Eu one should examine the abundances of 
these elements in Galactic stars of all metallicities. These abundances can
 give us clues to interpret their nucleosynthetic origin when compared with detailed chemical
evolution models.

\section{Chemical evolution of s- and r-process elements}
Previous studies of the evolution of the abundances of s- and r-process elements 
in the Galaxy are from Wheeler, Sneden \& Truran (1989), Mathews et al. (1992), 
Pagel \& Tautvaisiene (1997), Travaglio et al. (1999). 
In the Mathews et al. (1992) paper it was suggested that the observed 
apparent decrease of the abundance of Eu for [Fe/H] $<-2.5$ could be due to 
the fact that Eu originates mainly in low mass core-collapse SNe 
(7-8 $M_{\odot}$). 
Wheeler, Sneden \& Truran (1989) and
Pagel \& Tautvaisiene (1997) suggested that to reproduce the observed 
behaviour of Ba it is necessary to assume that at early stages Ba is 
produced as an r-process element by a not well identified range of massive stars.
A similar conclusion was reached by Travaglio et al. (1999) who 
showed that the evolution of Ba cannot be explained by assuming that 
this element is only an s-process element mainly formed in stars with initial 
masses 2-4$M_{\odot}$, but an r-process origin for it should be considered. 
In fact, in the hypothesis of a production of Ba only by s-process,
 a very late appearance of Ba should be expected, at variance 
with the observations indicating that Ba is already 
produced at [Fe/H]=-4.0. They suggested that low mass SNII 
(from 8 to 10$M_{\odot}$) could be responsible for the r-component of Ba.
Travaglio et al. (2004) compared their theoretical predictions with the abundance
pattern observed in the very r-process rich star CS
22892052 (Sneden et al. 2003). They considered this star  as having  
a pure r-process signature. They extracted from
this star the r-fraction of Sr, Y, and Zr (10\% of the solar
value). In the light of their nucleosynthesis calculations in
AGB stars at different metallicities, they concluded that the
s-process from AGB stars contributes to the solar abundances
of Sr, Y, and Zr by 71\%, 69\%, and 65\%, respectively. Concerning
the solar Sr abundance, they  also added a small contribution
(10\%) from the  weak s-component from massive stars.
As a consequence of the above results, they concluded that a
primary component from massive stars is needed to explain
8\% of the solar abundance of Sr and 18\% of solar Y and Zr.
 This process is of primary nature, unrelated to the classical
metallicity-dependent weak s-component.
 As already said, another important aspect of the [Ba/Fe] and [Eu/Fe] vs [Fe/H]
relations is the observed spread. 
An attempt to explain  the observed spread in s- and r-elements 
can be found in Tsujimoto et 
al. (1999) and Ishimaru \& Wanajo (1999),
who claim an inefficient mixing in 
the early galactic phases and attribute the spread to the fact that 
we observe the pollution due to single supernovae.
Ishimaru \& Wanajo (1999) also concluded that the Eu should originate as an 
r-process element in stars with masses in the range 8-10$M_{\odot}$.
This latter suggestion was confirmed by Ishimaru et al. (2004) by comparing model
 predictions with new data from Subaru indicating subsolar [Eu/Fe] ratios in three 
very metal poor stars ([Fe/H]$<-3.0$).

%Recently, many chemical evolution models have been developed to explain the
%chemical composition of the solar vicinity and the [el/Fe] vs. [Fe/H] patterns (e.g. Henry et al. 2000;
%Liang et al. 2001; Chiappini et al. 2003a, 2003b; Akerman et al. 2004, Fran\c cois 
%et al. 2004). 
%The paper is organized as follows: in Sect. 2 we present the observational 
%data, in Sect. 3 the chemical evolution model and in Sect. 
%4 the adopted nucleosynthesis prescriptions  are described.
%In Sect. 5  we present the results and in Sect. 6 some conclusions are 
%drawn.

Recently, several studies have attempted to follow the
enrichment history of the Galactic halo with special emphasis given
to the gas dynamical processes occurring in the early
Galaxy: Tsujimoto, Shigeyama, \& Yoshii (1999) provided
an explanation for the spread of Eu observed in the oldest
halo stars in the context of a model of supernova-induced
star formation; Ikuta \& Arimoto (1999) and McWilliam \&
Searle (1999) studied the metal enrichment of the Galactic
halo with the help of a stochastic model aimed at reproducing
the observed Sr abundances; Raiteri et al. (1999)
followed the Galactic evolution of Ba by means of a hydrodynamical
N-body/SPH code;  Argast et al. (2000)
concentrated on the effects of local inhomogeneities in the
Galactic halo produced by individual supernova events,
accounting in this way for the observed scatter of some (but
not all) elements typically produced by type II SNe. They predicted, however,
a too large spread for Mg and other $\alpha$-elements which is not observed.
Finally, Travaglio et al. (2004) investigated whether incomplete mixing of
the gas in the Galactic halo can lead to local chemical inhomogeneities
in the ISM of the heavy elements, in particular Eu, Ba, and Sr. However
what has still to be explained is why the spread is present only
for neutron capture elements whereas it is very small for the other 
elements (for example $\alpha$-elements).

Other important constraints, which are connected  to the evolution
of the Galaxy disk, are the abundance gradients of the elements along the disk
of the Milky Way.
Abundance gradients are a feature commonly observed in many
galaxies with their metallicities decreasing outward from the galactic centers.
The study of the gradients provides strong constraints to the mechanism of galaxy formation;
in fact, star formation and the accretion history as function of galactocentric distance
in the galactic disk strongly influence the formation and the development of the abundance gradients
(see Matteucci \& Fran\c cois 1989, Boisser \& Prantzos 1999, Chiappini et al. 2001).
Many models have been computed to explain the behaviour of abundances and abundance ratios as functions
of galactocentric radius (e.g  Hou et al. 2000; Chang et al. 1999;
 Chiappini et al. 2003b; Alib\'es et al. 2001) but they restrict
 their predictions to a small number of chemical elements and do not 
consider very heavy elements.

\section{S- and r-process elements in dwarf spheroidal galaxies}
The proximity and the relative simplicity of the Local Group (LG) 
dwarf spheroidal (dSph) galaxies make these systems excellent 
laboratories to test assumptions regarding the nucleosynthesis of
chemical elements and theories of galaxy evolution.
The Local Group is the group of galaxies that includes the Milky Way.
 The group comprises about 40 galaxies, 
with its gravitational center located between 
the Milky Way and the Andromeda Galaxy (see Fig. \ref{LG}).
\begin{figure}
\includegraphics[width=0.99\textwidth]{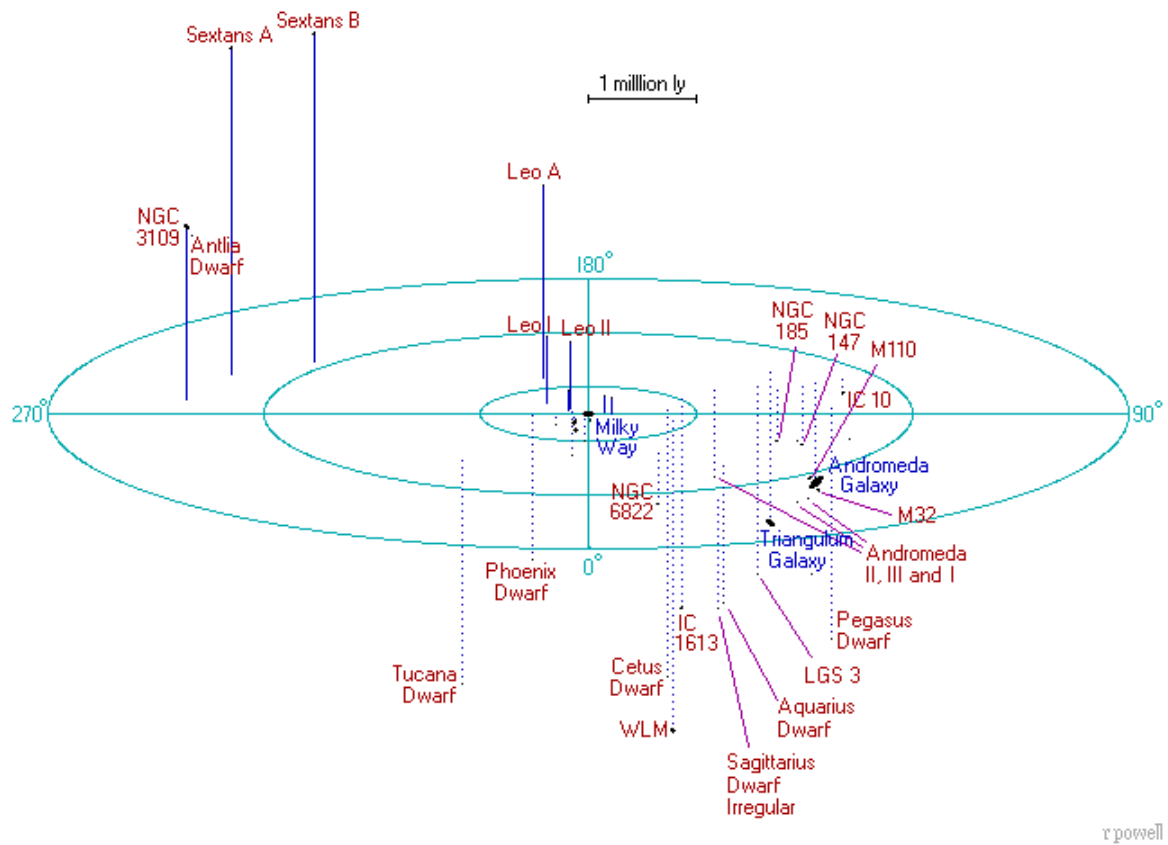} 
\caption{The map of the Local Group.The Milky Way is one of three
 large galaxies belonging to the group
 of galaxies called the Local Group which also contains several dozen
 dwarf galaxies. Most of these galaxies are depicted on the map.}
\label{LG}
\end{figure}
At the present time, the Milky Way  has 12 identified dwarf spheroidal (dSph)
galaxy companions and three of them  have been discovered very recently
(see table \ref{dsphtable}). 
\begin{table*}
\caption{The dSph galaxies of the Milky Way and their characteristic}\label{dsphtable}
\vspace{1.cm}

\begin{center}
\begin{tabular}{|c|c|c|c|c|c|c|}
\hline

\hline
Name         &  distance    & Visual luminosity      &  Absolute Visual& Total virial  mass & [Fe/H]& Year \\
             &   (kpc)      & L(V)/$10^{6}M_{\odot}$ &  Magnitude      & M/$10^{6}M_{\odot}$&       &      \\
\hline			      	          		   
			                       		   
Sagittarius  &    24 $\pm$2  & 18.1  &  -13.4&  --&   -1.0$\pm$0.2& 1994 \\
Bootes       &    60 $\pm$?  &  --   &   -5.7&  --&  	--        & 2006 \\ 
Ursa Minor   &    66 $\pm$3  &  0.3  &   -8.9&  23&   -2.2$\pm$0.1& 1954 \\
Sculptor     &    79 $\pm$4  &  2.2  &  -11.1&   6&   -1.8$\pm$0.1& 1938 \\
Draco        &    82 $\pm$6  &  0.3  &   -8.8&  22&   -2.0$\pm$0.1& 1954 \\
Sextans      &    86 $\pm$4  &  0.5  &   -9.5&  19&   -1.7$\pm$0.2& 1990 \\
Ursa Major   &   100 $\pm$?  &  --   &   -6.8&  --&     --        & 2005 \\
Carina       &   101 $\pm$5  &  0.4  &   -9.3&  13&   -2.0$\pm$0.2& 1977 \\
Fornax       &   138 $\pm$8  & 15.5  &  -13.2&  68&   -1.3$\pm$0.2& 1938 \\
Leo II       &   205 $\pm$12 &  0.6  &   -9.6&  10&   -1.9$\pm$0.1& 1950 \\
Canes Venatici&  220 $\pm$?  &  --   &   -7.9&  --&  	--        & 2006 \\
Leo I        &   250 $\pm$30 &  4.8  &  -11.9&  22&   -1.5$\pm$0.4& 1950 \\

\hline\hline                                                    

\end{tabular}

\end{center}

\end{table*}
DSph galaxies can be defined as  a low luminosity ($M_{V}>$ -14), non-nucleated dwarf elliptical 
galaxy with low surface brightness.
Several studies addressing the observation of red giant stars in these 
dSph galaxies with high resolution spectroscopy allow one to 
infer accurately the abundances of several elements including 
$\alpha$-, iron-peak and very heavy elements, 
such as barium and europium (Smecker-Hane $\&$ McWilliam 
1999; Bonifacio et al. 2000; Shetrone, Cot\'e $\&$ Sargent 
2001; Shetrone et al. 2003; Bonifacio et al. 2004; Sadakane 
et al. 2004; Geisler et al. 2005). These abundances and abundance 
ratios are not only central ingredients in galactic chemical 
evolution studies but are also very important in the attempt 
to clarify some aspects of the processes responsible for the
formation of chemical elements.

%\begin{figure}
%\includegraphics[width=0.99\textwidth]{draco} 
%\caption{The dSph galaxy Draco.}
%\label{Draco}
%\end{figure}

Shetrone, Cot\'e $\&$ Sargent (2001) argued that Draco 
%(see Fig \ref{Draco})
and Ursa Minor 
%(see Fig. \ref{Ursamin}) 
stars exhibit an abundance pattern consistent with one 
dominated by the r-process, i.e. 
[Ba/Eu] ranges from solar values at high metallicities to [Ba/Eu] 
$\sim$ -0.5 at [Fe/H] $\leq$ -1 dex. The pattern of [Ba/Fe] and [Eu/Fe] 
also resembles the one observed in the halo field stars according to
these authors. The same conclusion was reached by Shetrone et al. (2003), 
who analysed these abundance ratios in Sculptor, Fornax and Carina. 
%\begin{figure}
%\includegraphics[width=0.99\textwidth]{umidwarf} 
%\caption{The dSph galaxy Ursa minor.}
%\label{Ursamin}
%\end{figure}
Shetrone et al. (2003) claimed also that in Sculptor, Fornax and LeoI 
the pattern of [Eu/Fe] is consistent with the production of Eu in SNe II.
On the other hand, Venn et al. (2004), pointed out that, despite the 
general similarity, the dSph stars span a larger range in [Ba/Fe] 
and [Eu/Fe] ratios at intermediate metallicities than 
the Galactic stars and, more important, that about half of the dSph
stars exhibit lower [Y/Eu] and 2/3 higher [Ba/Y] than the Galactic stars
at the same metallicity, thus suggesting a clear difference between the
chemical evolution of our Galaxy and the one of dSph galaxies. 
The [$\alpha$/Fe] ratios observed in dSphs also are different from the same 
ratios in the Milky Way showing in general lower [$\alpha$/Fe] ratios than 
the Galactic stars with the same [Fe/H] (Smecker-Hane $\&$ McWilliam 
1999; Bonifacio et al. 2000; Shetrone, Cot\'e $\&$ Sargent 
2001; Shetrone et al. 2003; Bonifacio et al. 2004; Sadakane et al. 2004;
Geisler et al. 2005).

These observations not only shed some light on
the chemical evolution history of these galaxies but allowed also
the construction of chemical evolution models aimed at reproducing
important observational constraints, such as the elemental abundance 
ratios, the present gas mass and total mass (Carraro et al. 2001; 
Carigi, Hernandez $\&$ Gilmore 2002; Ikuta $\&$ Arimoto 2002; 
Lanfranchi $\&$ Matteucci 2003 (LM03); Lanfranchi $\&$ Matteucci
2004 (LM04)). 

Among these models the one proposed by LM03 and LM04
for 5 local dSph galaxies (namely Draco, Carina, Sculptor,
Ursa Minor and Sagittarius) succeeded  in reproducing the observed [$\alpha$/Fe] 
ratios, the present gas mass and final total mass by adopting a very low 
star formation rates proceedings in relatively long bursts as indicated by
the color-magnitude diagrams of these galaxies.
%$\nu$ $\sim$ 0.01 to 0.5 Gyr$^{-1}$ (with lower values for Draco 
%and higher ones for Sagittarius) and a high wind efficiency 
%(6-13 times the star formation rate).  
%Besides that, LM04 predicted the stellar metallicity
%distribution of these galaxies which were later on compared to 
%observational data for Carina with a reasonably good 
%agreement (Koch et al. 2004).

This thesis is organized as follows: 

in chapter 2, we present the results of a chemical evolution model
based on the model developed by Chiappini et al. (2003a) for the Milky Way;
we compare our theoretical results relative to the evolution of neutron capture elements
(Sr, Y, Zr, Ba, La and Eu) with the newest data  of Fran\c cois et al. (2006)  
and we impose constraints on the nucleosynthesis of the studied elements. 

In chapter 3, we calculate the abundance gradients of the largest  number of heavy elements
(O, Mg, Si, S, Ca, Sc, Ti, Co, V, Fe, Ni, Zn, Cu, Mn, Cr, Sr, Y, Zr, Ba, La and Eu) 
ever considered in a chemical evolution model; therefore, we are able to test nucleosynthesis 
prescriptions obtained in the previous chapter as well as the recent 
nucleosynthesis by  Fran\c cois et al. (2004) for the $\alpha$- and iron peak elements.
Chemical evolution models adopting the above nucleosynthesis prescriptions have been shown
to reproduce the evolution of the abundances in the solar neighborhood. Here we extend our 
predictions to the whole disk and we compare our model predictions  to new  observational data collected 
by Andrievsky et al. (2002abc, 2004) and Luck et al. (2003) (hereafter 4AL). They measured the abundances of all the selected elements (except Ba)
in a  sample of 130 galactic Cepheids found in the galactocentric distance range from 5 to 17 kpc.

In chapter 4, we  show the results of a stochastic chemical evolution model that we develop
with the same nucleosynthesis of the models of the previous chapters. 
We test if this model  is able to reproduce
 the large spread of the abundances of 
neutron capture elements observed in low metallicity stars 
in the solar vicinity and, at the same time, the small star to star scatter 
observed for the $\alpha$-elements.  

In chapter 5, we adopt  the nucleosynthesis prescriptions for Ba and Eu  which are able to
reproduce the most recent observed data for our Galaxy, as shown in chapter 2,
 and we compare the predictions of the models with observational data in 5 dSph galaxies.
In this way, it is possible to verify if the assumptions made regarding the 
nucleosynthesis of Ba and Eu can also fit also the data of local dSph galaxies.

In chapter 6, we use the results of 
chapter 2 and chapter 5, to compare the predictions of the Milky Way
to those of the dSph galaxies. We choose, as typical dwarf spheroidal
galaxy, Sculptor. We do not show all the data for all the dwarf spheroidal galaxies
because as will be shown in chapter 5, the star formation histories
are different and also the chemical evolution is different among
the dSph galxies, even if they share common behaviors.

Finally, in chapter 7, the main conclusions of our work are drawn.

\chapter [Chemical evolution in the solar vicinity]  {Chemical evolution of neutron capture elements in the solar vicinity}

\rightline{\emph{``"Man," I cried, "how ignorant art thou in thy pride of wisdom!" }}
\rightline{\emph{ by Mary Shelley}}

\vspace{1cm}

\section{Barium and europium}

We present the results of a chemical evolution model
based on the original two-infall model of Chiappini et al. (1997) for the Milky Way 
in the latest version developed by Chiappini et al. (2003a) and adopted in Fran\c cois et al. (2004).
We compare our theoretical results relative to the evolution of Ba and Eu 
with the newest and very accurate data  of Fran\c cois et al. (2007)  
and we impose constraints on the nucleosynthesis of the studied 
elements. 

\subsection{Observational data}\label{data}

We preferentially used the most recent available data based on high quality spectra collected
with efficient spectrographs and 8-10 m class telescopes.
In particular, for the extremely  metal poor stars ([Fe/H] between $-4$ and $-3$), we adopted the recent
 results from UVES Large Program
"First Star'' (Cayrel  et al. 2004, Fran\c cois et al. 2007). This sample consists of 31 extremely metal-poor
halo stars selected in the HK survey (Beers et al. 1992, 1999). 
We can deduce from the kinematics of these stars that they were born at very different places in the 
Galactic halo.
This overcomes the possibility of a selection bias.
The analysis is made in a systematic and homogeneous way, from very high 
quality data, giving  abundance ratios of unprecedented  accuracy in this metallicity range.
For the abundances in the remaining range of [Fe/H], we took published high quality data in the
literature from various sources: Burris et al. (2000), Fulbright (2000),
 Mashonkina \& Gehren (2000, 2001),
Koch \& Edvardsson (2002), Honda et al. (2004), Ishimaru et al. (2004).
 All of these data are relative to solar abundances of Grevesse \& Sauval (1998).

\subsection{Chemical evolution model for the solar vicinity}
We model the formation of the Galaxy assuming two main infall episodes: the first forms the halo
 and the thick disk, the second the thin disk. The timescale for the formation of 
the halo-thick disk is $\sim1Gyr$. The timescale 
for the thin disk is much longer, implying that the infalling gas
 forming the thin disk comes mainly from the
 intergalactic medium and not only from the halo (Chiappini et al. 1997). 
Moreover, the formation of the thin disk is assumed to be a function of the galactocentric distance,
 leading to an inside out scenario for the Galaxy  disk build up (Matteucci \& Fran\c cois 1989).
In this chapter, all the results shown are for the assumed solar  galactocentric distance: 8 kpc.
 The main characteristic  of the two-infall model is an almost independent evolution
 between the halo and the thin disk
(see also Pagel \& Tautvaisiene 1995). A threshold in the star formation process (Kennicutt 1989, 1998,
 Martin \& Kennicutt 2001) is also adopted.
 The model well reproduces an extended set of observational constraints both for the solar
neighborhood and for the whole disc. One of the most important observational constraints
is represented by the various relations between the abundances of metals (C,N,O,$\alpha$-elements,
iron peak elements) as functions of the [Fe/H] abundance  (see Chiappini et al. 2003)
Although this model is probably not unique,
it reproduces the majority of the observed features of the Milky Way.
Many of the assumptions of the model  are shared by other authors (e.g. Prantzos \& Boissier
2000, Alib\'es  et al. 2001, Chang et al. 1999). 
The equation below describes the time evolution  of $G_{i}$, namely the mass fraction of
 the element $i$ in the gas:

\begin{displaymath}
\dot{G_{i}}(t)=-\psi(r,t)X_{i}(r,t)
\end{displaymath}
\smallskip
\begin{displaymath}
+\int\limits^{M_{Bm}}_{M_{L}}\psi(t-\tau_{m})Q_{mi}(t-\tau_{m})\phi(m)dm
\end{displaymath}
\begin{displaymath}
+A\int\limits^{M_{BM}}_{M_{Bm}}\phi(M_{B})\cdot\left[\int\limits_{\mu_{m}}^{0.5}f(\mu)\psi(t-\tau_{m2})Q^{SNIa}_{mi}(t-\tau_{m2})d\mu\right]dM_{B}
\end{displaymath}
\begin{displaymath}
+(1-A)\int\limits^{M_{BM}}_{M_{Bm}}\psi(t-\tau_{m})Q_{mi}(t-\tau_{m})\phi(m)dm
\end{displaymath}
\begin{displaymath}
+\int\limits^{M_{U}}_{M_{BM}}\psi(t-\tau_{m})Q_{mi}(t-\tau_{m})\phi(m)dm
\end{displaymath}
\smallskip
\begin{equation}
+X_{A_{i}}A(r,t).
\end{equation}
where $X_{i}(r,t)$ is the abundance by mass of the element $i$ and $Q_{mi}$ indicates
the fraction of mass restored by a star of mass $m$ in the form of the element $i$, the so-called
``production matrix'' as originally defined by Talbot and Arnett (1973). We indicate with $M_{L}$ the lightest
mass that contributes to the chemical enrichment and it is set at $0.8M_{\odot}$; the upper mass limit, $M_{U}$, 
is set at $100M_{\odot}$. 

The star formation rate (SFR) $\psi(r,t)$ is defined:
\begin{equation}
\psi(r,t)=\nu\left(\frac{\Sigma(r,t)}{\Sigma(r_{\odot},t)}\right)^{2(k-1)}
\left(\frac{\Sigma(r,t_{Gal})}{\Sigma(r,t)}\right)^{k-1}G^{k}_{gas}(r,t).
\end{equation}
$\nu$ is the efficiency of the star formation process and is set to be $2Gyr^{-1}$
for the Galactic halo ($t<1Gyr$) and $1Gyr^{-1}$ for the disk ($t\ge1Gyr$).
$\Sigma(r,t)$ is the total
surface mass density, $\Sigma(r_{\odot},t)$ the total surface mass density at the 
solar position, $G_{gas}(r,t)$ the surface density normalized to the present time
total surface mass density in the disk $\Sigma_{D}(r,t_{Gal})$, where $t_{Gal}=14Gyr$ is the age 
assumed for the Milky Way and $r_{\odot}=8$ kpc the solar galactocentric distance 
(Reid 1993). The gas surface exponent, $k$, is set equal to 1.5.
With these values for the parameters the observational constraints, in particular in the solar vicinity,
are well fitted.
Below a critical threshold of the gas surface density ($7M_{\odot}pc^{-2}$) we assume no star formation.
This naturally produces a hiatus in the SFR between the halo-thick disk phase and the thin disk phase.
In Fig. (\ref{SFR}) we show the predicted star formation rate for the halo-thick disc phase
and the thin disc phase, respectively.

\begin{figure}[h!]
\begin{center}
\includegraphics[width=0.49\textwidth]{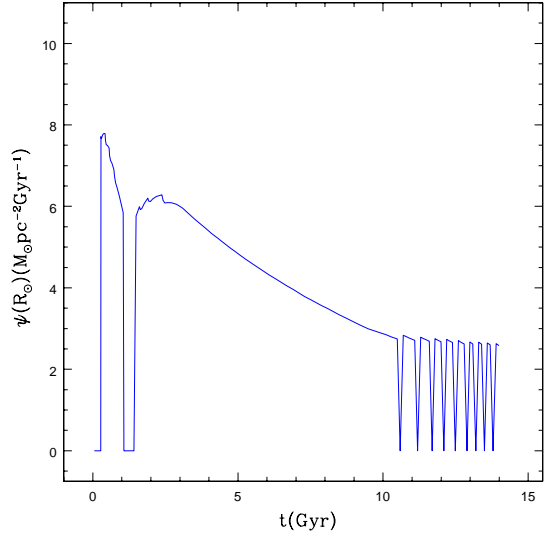}
\caption{ The SFR expressed  in $M_{\odot}pc^{-2}Gyr^{-1}$ as predicted by the two infall model. The 
gap in the SFR at the end of the halo-thick disc phase is evident. The oscillations are due to the fact
that at late times in the galactic disc the surface gas density is always close to the threshold density.}
\label{SFR}
\end{center}
\end{figure}

For $\phi$, the initial mass function (IMF), we use the Scalo (1986) one, constant in time and space,
 while $\tau_{m}$ is the evolutionary lifetime of stars as a function of their mass ``m''.

The SNIa rate has been computed following Greggio \& Renzini (1983a) and Matteucci \& Greggio (1986) and it is expressed as:
\begin{equation}
R_{SNeIa}=A\int\limits^{M_{BM}}_{M_{Bm}}\phi(M_{B})(\int\limits^{0.5}_{\mu_{m}}f(\mu)\psi(t-\tau_{M_{2}})d\mu) dM_{B}.
\end{equation}
where $M_{2}$ is the mass of the secondary, $M_{B}$ is the total mass of the binary
system, $\mu=M_{2}/M_{B}$, $\mu_{m}=max\left[M(t)_{2}/M_{B},(M_{B}-0.5M_{BM})/M_{B}\right]$, 
$M_{Bm}= 3 M_{\odot}$, $M_{BM}= 16 M_{\odot}$. The IMF is represented by $\phi(M_{B})$
and refers to the total mass of the binary system for the computation of the SNeIa rate,
$f(\mu)$ is the distribution function for the mass fraction of the secondary:
\begin{equation}
f(\mu)=2^{1+\gamma}(1+\gamma)\mu^{\gamma}.  
\end{equation}
with $\gamma=2$; A is the fraction of systems in the appropriate mass range that can give rise
to SNeIa events. This quantity is fixed to 0.05 by reproducing the observed SNeIa rate at
the present time (Cappellaro et al. 1999). Note that  in the case of SNIa the``production matrix''
is indicated by  $Q^{SNIa}_{mi}$ because of its different nucleosynthesis contribution
(for details see Matteucci and Greggio 1986).
In Fig. \ref{SNII} we show the predicted type II and Ia SN rates. The type II SN rate follows
the SFR, as expected, whereas the type Ia SN rate does not have this feature due to the nature
of type Ia SN progenitors, which are assumed to be low-intermediate mass stars with  long 
evolutionary time scales.

\begin{figure}[h!]
\begin{center}
\includegraphics[width=0.49\textwidth]{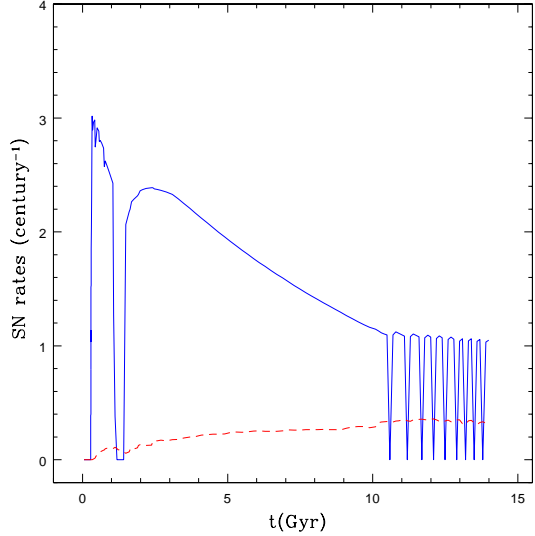}
\caption{Predicted SN II (continuous line) and Ia (dashed line) rates by the two infall model.}
\label{SNII}
\end{center}
\end{figure}

The last term in equation 1 represents the gas accretion and it is defined as:
\begin{equation}
A(r,t)=a(r)e^{-t/\tau_{H}}+b(r)e^{(t-t_{max})/\tau_{D}(r)}.
\end{equation}
where $X_{A_{i}}$ are the abundances of the infalling material, assumed to have a primordial
chemical composition , $t_{max}=1Gyr$
is the time for the maximum infall rate on the thin disk, $\tau_{H}=2.0Gyr$ is the time scale
for the formation of the halo thick-disk and $\tau_{D}$ is the timescale of the thin disk at the 
solar galactocentric distance ($\tau_{D}= 7Gyr$). The timescale $\tau_{D}$ increases with
the galactocentric distance as we will see in chapter 3.
The coefficients $a(r)$ and $b(r)$ are constrained by the present day
 total surface mass density.
In particular, $b(r)$ is assumed to be different from zero for $t>t_{max}$
(see Chiappini et al. 2003, for details).

\subsection{Nucleosynthesis prescriptions for Ba and Eu}{\label{NP}}
\subsubsection{S-process}{\label{NP_BaS}}

 We have adopted the yields of Busso et al. (2001) in the mass
 range 1.5-3$M_{\odot}$ for the s-main component. 
In this process, the dependence on the metallicity is very important. The
 s-process elements are made by accretion of neutrons on seed elements
(in particular iron) already present in the star. Therefore, this Ba 
component behaves like a secondary element.
The neutron flux is due to the reaction $^{13}C(\alpha,n)^{16}O$ which
can easily be activated at the low temperature of these stars
(see Busso et al. 1999).
The yields are shown in Table \ref{SBa}
and Fig. \ref{sBa2} as functions of the initial metallicity of the stars.
The theoretical results by Busso et al. (2001) suggest 
negligible europium production  
in the s-process and therefore we neglected this component in our work.
We have added for models 1 and 2 (see Table \ref{model}) an extension to the
theoretical result of Busso et al. (2001)  in the mass range $1-1.5M_{\odot}$
by simply scaling with the mass the values obtained for stars of $1.5 M_{\odot}$.
We have extended the prescription in order to better fit the data with a [Fe/H]
supersolar and verified that it does not change the results of the model for $[Fe/H]<0$.

\begin{table*}

\caption{The stellar yields in the range $1.5-3M_{\odot}$ from the paper of  Busso et al. (2001).}\label{SBa}
\vspace{1.5cm}

\begin{center}
\begin{tabular}{|c|c|c|}
\hline

$ Metallicity $ & $X^{new}_{Ba}$ for $1.5M_{\odot}$& $X^{new}_{Ba}$ for $3M_{\odot}$  \\

\hline\hline

 0.20$\cdot10^{-3}$ &  0.69$\cdot10^{-8}$  & 0.13$\cdot10^{-7}$ \\
 0.10$\cdot10^{-2}$ &  0.38$\cdot10^{-7}$  & 0.46$\cdot10^{-7}$ \\
 0.20$\cdot10^{-2}$ &  0.63$\cdot10^{-7}$  & 0.87$\cdot10^{-7}$ \\
 0.30$\cdot10^{-2}$ &  0.72$\cdot10^{-7}$  & 0.11$\cdot10^{-6}$ \\
 0.40$\cdot10^{-2}$ &  0.73$\cdot10^{-7}$  & 0.12$\cdot10^{-6}$ \\
 0.50$\cdot10^{-2}$ &  0.68$\cdot10^{-7}$  & 0.13$\cdot10^{-6}$ \\
 0.60$\cdot10^{-2}$ &  0.58$\cdot10^{-7}$  & 0.13$\cdot10^{-6}$ \\ 
 0.70$\cdot10^{-2}$ &  0.47$\cdot10^{-7}$  & 0.12$\cdot10^{-6}$ \\
 0.80$\cdot10^{-2}$ &  0.39$\cdot10^{-7}$  & 0.11$\cdot10^{-6}$ \\
 0.90$\cdot10^{-2}$ &  0.34$\cdot10^{-7}$  & 0.98$\cdot10^{-7}$ \\
 0.10$\cdot10^{-1}$ &  0.16$\cdot10^{-7}$  & 0.43$\cdot10^{-7}$ \\ 
 0.11$\cdot10^{-1}$ &  0.14$\cdot10^{-7}$  & 0.39$\cdot10^{-7}$ \\ 
 0.12$\cdot10^{-1}$ &  0.13$\cdot10^{-7}$  & 0.34$\cdot10^{-7}$  \\
 0.13$\cdot10^{-1}$ &  0.12$\cdot10^{-7}$  & 0.32$\cdot10^{-7}$  \\
 0.14$\cdot10^{-1}$ &  0.11$\cdot10^{-7}$  & 0.29$\cdot10^{-7}$  \\
 0.15$\cdot10^{-1}$ &  0.99$\cdot10^{-8}$  & 0.27$\cdot10^{-7}$  \\
 0.16$\cdot10^{-1}$ &  0.90$\cdot10^{-8}$  & 0.25$\cdot10^{-7}$  \\
 0.17$\cdot10^{-1}$ &  0.81$\cdot10^{-8}$  & 0.23$\cdot10^{-7}$  \\
 0.18$\cdot10^{-1}$ &  0.73$\cdot10^{-8}$  & 0.22$\cdot10^{-7}$  \\
 0.19$\cdot10^{-1}$ &  0.66$\cdot10^{-8}$  & 0.20$\cdot10^{-7}$  \\
 0.20$\cdot10^{-1}$ &  0.59$\cdot10^{-8}$  & 0.19$\cdot10^{-7}$  \\   
 0.30$\cdot10^{-1}$ &  0.24$\cdot10^{-8}$  & 0.94$\cdot10^{-8}$  \\   
 0.40$\cdot10^{-1}$ &  0.12$\cdot10^{-8}$  & 0.50$\cdot10^{-8}$  \\

\hline\hline                                                    

\end{tabular}

\end{center}

\end{table*}

\begin{figure}[h!]
\begin{center}
\includegraphics[width=0.49\textwidth]{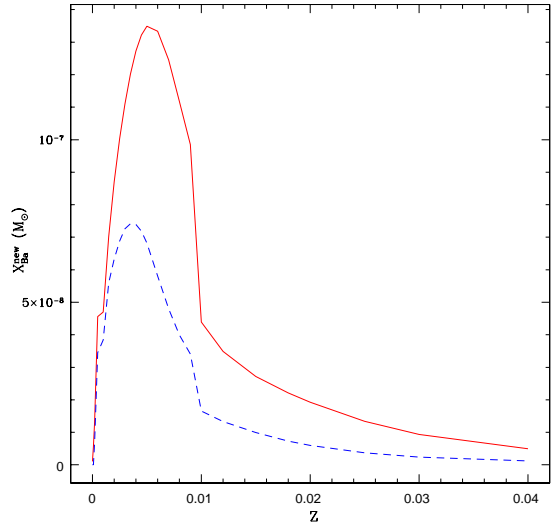}
\caption{The stellar yields  $X^{new}_{Ba}$ from  Busso et al. (2001) plotted versus metallicity.
 Dashed line: the prescriptions for stars of $1.5M_{\odot}$, solid line for stars of $3M_{\odot}$.} \label{sBa2}
\end{center}
\end{figure}

\subsubsection{R-process}{\label{NP_BaR}}
The production of r-process elements is still a challenge
for astrophysics and even for nuclear physics, due to the fact that 
the nuclear properties of thousands of nuclei located between 
the valley of $\beta$ stability and the neutron drip line, necessary to
correctly  compute this process, are ignored.
In our models we 
have tested 6 different nucleosynthesis prescriptions for the r-process Ba and Eu, 
as shown in Tables  \ref{model}, \ref{rBa} and \ref{rBa2}.
Some of the prescriptions refer to models by Travaglio et al. (2001) (model 3)
and Ishimaru et al. (2004) (models 4, 5 and 6), whereas the others contain yields chosen 
``ad hoc'', namely in order to reproduce the observational data.

\begin{table*}

\centering

\caption{Model parameters. The yields $X^{new}_{Ba}$ are expressed as mass fractions. The subscript  
``ext'' stands for extended (the yields have been extrapolated down to $1M_{\odot}$) and $M_{*}$ for
 the mass of the star.}\label{model}
\vspace{1.5cm}
\begin{tabular}{|c|c|c|c|c|}
\hline\hline

Mod             &  s-process           &  r-process    & s-process & r-process  \\
                &  Ba                  &  Ba           & Eu        & Eu        \\

\hline\hline
1               & $1.-3M_{\odot}$        & $12-30M_{\odot}$ &   none            & $12-30M_{\odot}$ \\  
                & Busso et al. (2001)ext.    & yields Table \ref{rBa}   &                   &  yields Table \ref{rBa}  \\
\hline\hline
2               &  $1.-3M_{\odot}$        & $10-25M_{\odot}$ &   none            & $10-25M_{\odot}$ \\  
                &  Busso et al. (2001)ext. & yields Table \ref{rBa2}   &                   &  yields Table \ref{rBa2}  \\
\hline 
3               &  $1.5-3M_{\odot}$      & $8-10M_{\odot}$  &   none            & $12-30M_{\odot}$ \\  
                &  Busso et al. (2001)    & $X^{new}_{Ba}=5.7\cdot10^{-6}/M_{*}$& &  yields Table \ref{rBa} \\
                &                        & (Travaglio et al. 2001)             & &  \\

\hline\hline
4               &  $1.5-3M_{\odot}$      & $10-30M_{\odot}$ &   none            &  $8-10M_{\odot}$\\  
                &  Busso et al. (2001)    & yields Table \ref{rBa}   &                   &  $X^{new}_{Eu}=3.1\cdot10^{-7}/M_{*}$ \\
                &                        &                  &                   &  (Ishimaru et al.2004 Mod.A)\\
\hline
5               &  $1.5-3M_{\odot}$      & $10-30M_{\odot}$ &   none            &  $20-25M_{\odot}$\\  
                &  Busso et al. (2001)    & yields Table  \ref{rBa}   &                   &  $X^{new}_{Eu}=1.1\cdot10^{-6}/M_{*}$ \\
                &                        &                  &                   &  (Ishimaru et al.2004 Mod.B)\\
\hline
6               &  $1.5-3M_{\odot}$      & $10-30M_{\odot}$ &   none            &  $>30M_{\odot}$\\  
                &  Busso et al. (2001)    & yields Table  \ref{rBa}   &                   &  $X^{new}_{Eu}=7.8\cdot10^{-7}/M_{*}$ \\
                &                        &                  &                   &  (Ishimaru et al.2004 Mod.C)\\

\hline\hline
\end{tabular}
\end{table*}

\begin{table*}

\centering
\caption{The stellar yields for barium and europium in massive stars (r-process)
in the case of a primary origin.} \label{rBa}

\begin{minipage}{90mm}

\begin{tabular}{|c|c|c|}
\hline

$M_{star}$  & $ X_{Ba}^{new}$  & $ X_{Eu}^{new}$\\

\hline\hline

12.   & 9.00$\cdot10^{-7}$ &  4.50$\cdot10^{-8}$  \\ 
15.   & 3.00$\cdot10^{-8}$ &  3.00$\cdot10^{-9}$ \\   
30.   & 1.00$\cdot10^{-9}$ &  5.00$\cdot10^{-10}$ \\

\hline\hline

\end{tabular}

\end{minipage}

\end{table*}

\begin{table*}

\centering

\caption{The stellar yields for Ba and Eu in massive stars (r-process)
in the case of secondary origin. The mass fraction does not change 
as a function of the stellar mass.} \label{rBa2}
\vspace{1.5cm}

\begin{minipage}{90mm}

\begin{tabular}{|c|c|c|c|}
\hline

$Z_{star}$  & $X_{Ba}^{new}$   & $ X_{Eu}^{new}$ \\
            & $10-25M_{\odot}$ & $10-25M_{\odot}$ \\
\hline\hline

           $Z<5\cdot10^{-7}$.   &  1.00$\cdot10^{-8}$ &  5.00$\cdot10^{-10}$  \\ 
$5\cdot10^{-7}<Z<1\cdot10^{-5}$ &  1.00$\cdot10^{-6}$ &  5.00$\cdot10^{-8}$  \\   
           $Z>1\cdot10^{-5}$    &  1.60$\cdot10^{-7}$ &  8.00$\cdot10^{-9}$  \\

\hline\hline

\end{tabular}

\end{minipage}

\end{table*}

In the case of Ba we have included an r-process component,
produced in massive stars in the range 12-30$M_{\odot}$ in  model 1
and in the range 10-25$M_{\odot}$ in model 2.
In Fig. \ref{MORTE} we show the lightest stellar mass dying as a function of
the metallicity of the ISM ([Fe/H]) in our chemical evolution model; it is clear 
from this plot that  it is impossible to explain the 
observed abundances of [Ba/Fe] in stars with $[Fe/H]<-2$ without the Ba component produced
in massive stars. The first stars,
which produce s-processed Ba (see Sect. \ref{NP_BaS}), have a mass of $3M_{\odot}$ 
and they start to enrich the ISM only for $[Fe/H] \ge -2$.

\begin{figure}[h!]
\begin{center}
\includegraphics[width=0.49\textwidth]{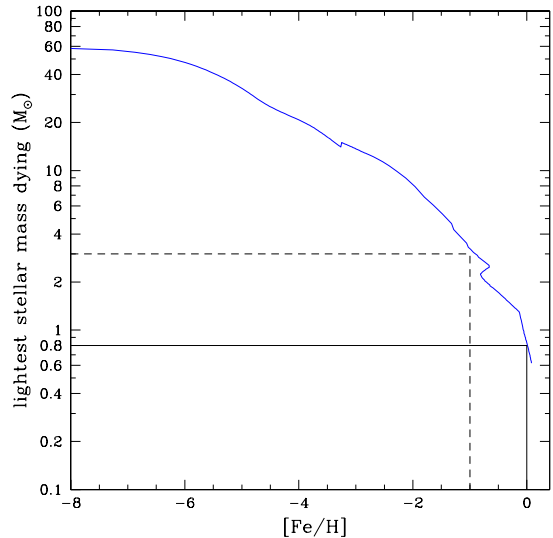}
\caption{In the plot we show the lightest stellar mass dying at the time corresponding to a given [Fe/H].
The solid line indicates the solar abundance ([Fe/H]=0), corresponding to a lightest dying mass star of $0.8M_{\odot}$,
the dashed line indicates the [Fe/H]=-1 corresponding to a lightest dying star mass of $3M_{\odot}$.
The adopted stellar lifetimes are from Maeder \& Meynet (1989).}
\label{MORTE}
\end{center}
\end{figure}

We stress that Travaglio et al. (2001) predicted r-process Ba only from 
stars in the range 8-10$M_{\odot}$, but their conclusions were based on 
an older set of observational data.

Moreover, we considered another independent indication for the r-process 
production of barium; 
Mazzali and Chugai (1995) observed Ba lines 
in SN 1987A, which  had a progenitor star of 20$M_{\odot}$.
These lines of Ba are well reproduced with a overabundance
factor $f=X_{obs}/X_{i}=5$ (typical metal abundance for LMC $X_{i}=(1/2.75)\times$ solar).
 From this observational data we can derive
a $X_{Ba}^{new}\sim 2 \cdot 10^{-8}$for a 20 $M_{\odot}$ star,
 which is in agreement with our prescriptions.

For Eu  we assumed that it is completely due to the r-process
and that the yields originate from massive stars  in the range 
12-30$M_{\odot}$ in model 1 and 10-25$M_{\odot}$ in model 2,
 as shown in Table \ref{model}.

In particular, our  choice is made to best fit
 the plots [Ba/Fe] vs.[Fe/H], 
[Eu/Fe] vs.[Fe/H] and [Ba/Eu] vs [Fe/H] as well as the Ba and Eu
 solar abundances (taking into account the contribution of the
 low-intermediate mass star in case of the Ba). 

We have tested prescriptions for Ba and Eu both for a primary production
 and a secondary production (with a dependence on the metallicity).
In the first case the main feature of the  yields
is a strong enhancement in the mass range $12-15M_{\odot}$
(model 1) with no dependence on the metallicity and so
the elements are considered as primary elements.
In the case of metallicity dependence (model 2), the yield behavior
is chosen to have a strong enhancement in the range of metallicity $5\cdot10^{-7}<Z<1\cdot10^{-5}$
 with almost constant yields for Eu and Ba in the whole mass range for
a given metallicity.

\subsubsection{Iron}
For the nucleosynthesis prescriptions of Fe, we adopted those suggested in Fran\c cois 
et al. (2004), in particular  the yields of Woosley \& Weaver (1995) (hereafter WW95)
 for a solar chemical composition. The yields for several 
elements suggested by Fran\c cois et al. (2004) are those best reproducing the observed 
[X/Fe] vs. [Fe/H] at all metallicities in the solar vicinity.

\subsection{Results}

\subsubsection{Trends}\label{trends}

We investigate how the different models fit the 
the trends of the abundance ratios for [Ba/Fe],
[Eu/Fe] and [Ba/Eu] versus [Fe/H] and even for [Ba/Eu] versus [Ba/H].

\begin{table*}

\centering
\caption{Results after the computation of the mean for the data inside bins
 along the [Fe/H] axis for the values of [Ba/Fe].} \label{meanBa}
\vspace{1.5cm}

\begin{tabular}{|c|c|c|c|c|}
\hline

bin center [Fe/H]& bin dim.[Fe/H]  & mean [Ba/Fe] & SD [Ba/Fe] &  N. of data in the bin \\
\hline\hline

 -3.82 & 0.75 & -1.25 &  0.30 &   6\\
 -3.32 & 0.25 & -0.96 &  0.50 &   7\\
 -3.07 & 0.25 & -0.65 &  0.65 &  11\\
 -2.82 & 0.25 & -0.37 &  0.60 &  17\\
 -2.57 & 0.25 & -0.15 &  0.40 &  11\\
 -2.32 & 0.25 &  0.09 &  0.58 &  13\\
 -2.07 & 0.25 &  0.23 &  0.50 &  15\\
 -1.82 & 0.25 &  0.10 &  0.20 &  20\\
 -1.58 & 0.25 &  0.08 &  0.15 &  27\\
 -1.33 & 0.25 &  0.20 &  0.22 &  16\\
 -1.08 & 0.25 &  0.07 &  0.19 &  20\\
 -0.83 & 0.25 & -0.03 &  0.08 &  30\\
 -0.58 & 0.25 & -0.04 &  0.14 &  59\\
 -0.33 & 0.25 &  0.05 &  0.20 &  46\\
 -0.08 & 0.25 &  0.03 &  0.13 &  53\\
  0.17 & 0.25 & -0.01 &  0.11 &  26\\

\hline\hline

\end{tabular}

\end{table*}

\begin{table}

\centering
\caption{Results after the computation of the mean for the data inside bins
 along the [Fe/H] axis for the values of [Eu/Fe] and [Ba/Eu].} \label{meanEu}
\vspace{1.5cm}

\begin{tabular}{|c|c|c|c|c|c|c|}
\hline

bin            & bin dim.& mean  & SD    & mean  & SD    & N.of data  \\ 
center [Fe/H]  & [Fe/H]   &[Eu/Fe]&[Eu/Fe]&[Ba/Eu]&[Ba/Eu]&in the bin  \\

\hline\hline
 -3.22 & 0.24 &  -0.10 &  0.21 & -0.71  & 0.25  &  5\\
 -2.98 & 0.24 &   0.08 &  0.60 & -0.57  & 0.13  & 12\\
 -2.74 & 0.24 &   0.46 &  0.60 & -0.64  & 0.11  & 14\\
 -2.49 & 0.24 &   0.45 &  0.28 & -0.52  & 0.17  &  7\\
 -2.25 & 0.24 &   0.38 &  0.36 & -0.38  & 0.33  & 11\\
 -2.01 & 0.24 &   0.51 &  0.34 & -0.36  & 0.26  & 10\\
 -1.77 & 0.24 &   0.29 &  0.22 & -0.20  & 0.19  & 19\\
 -1.53 & 0.24 &   0.44 &  0.15 & -0.39  & 0.22  & 21\\
 -1.28 & 0.24 &   0.42 &  0.20 & -0.26  & 0.31  & 18\\
 -1.04 & 0.24 &   0.39 &  0.13 & -0.38  & 0.15  & 16\\
 -0.80 & 0.24 &   0.32 &  0.12 & -0.35  & 0.14  & 36\\
 -0.56 & 0.24 &   0.23 &  0.14 & -0.27  & 0.20  & 55\\
 -0.32 & 0.24 &   0.18 &  0.10 & -0.13  & 0.23  & 44\\
 -0.07 & 0.24 &   0.04 &  0.07 & -0.02  & 0.14  & 51\\
  0.17 & 0.24 &  -0.02 &  0.07 &  0.00  & 0.12  & 26\\

\hline\hline

\end{tabular}

\end{table}

\begin{table*}
\centering
\caption{Results after the computation of the mean for the data inside bins
 along the [Ba/H] axis for the values of [Ba/Eu].} \label{meanBaH}
\vspace{1.5cm}
\begin{tabular}{|c|c|c|c|c|}
\hline

bin center [Ba/H]& bin dim.[Ba/H]  & mean [Ba/Eu] & SD [Ba/Eu] &  N of data in the bin \\
\hline\hline

 -4.35 & 0.58 &  -0.75 & 0.26 &  4\\
 -3.76 & 0.58 &  -0.60 & 0.14 & 12\\
 -3.32 & 0.29 &  -0.55 & 0.14 &  3\\
 -3.02 & 0.29 &  -0.62 & 0.13 &  4\\
 -2.73 & 0.29 &  -0.58 & 0.24 & 13\\
 -2.43 & 0.29 &  -0.58 & 0.21 &  4\\
 -2.14 & 0.29 &  -0.44 & 0.13 &  7\\
 -1.84 & 0.29 &  -0.33 & 0.28 & 20\\
 -1.54 & 0.29 &  -0.33 & 0.20 & 25\\
 -1.25 & 0.29 &  -0.39 & 0.19 & 21\\
 -0.95 & 0.29 &  -0.31 & 0.20 & 36\\
 -0.66 & 0.29 &  -0.33 & 0.18 & 64\\
 -0.36 & 0.29 &  -0.13 & 0.14 & 43\\
 -0.07 & 0.29 &  -0.03 & 0.09 & 68\\

\hline\hline

\end{tabular}

\end{table*}

To better investigate the trends of the data we divide
in several bins the [Fe/H] axis and the [Ba/H] axis and compute
the mean and the standard deviation from the mean of the ratios 
[Ba/Fe], [Eu/Fe] and [Ba/Eu] for  all the data inside each bin.
In Table \ref{meanBa} we show the results of this computation
for [Ba/Fe] versus [Fe/H], in Table \ref{meanEu}
for [Eu/Fe] and [Ba/Eu] versus [Fe/H] and in Table \ref{meanBaH}
for [Ba/Eu] versus [Ba/H].
Since  the ranges of [Ba/H] and [Fe/H] are different, we have bins
of different width. 
We have divided in a different way the [Fe/H] for [Ba/Fe] 
ratio and the [Fe/H] for [Eu/Fe] and [Ba/Eu] ratios, because
the [Eu/Fe] ratio for 12 stars at very low
metallicity is only an upper limit and therefore the data for these stars
have not been considered in the computation 
of the mean and the standard deviation for [Eu/Fe] and [Ba/Eu] ratios.
In the case of [Ba/Eu] and [Eu/Fe] we have simply divided the
[Fe/H] axis in 15 bins of equal dimension (see Table \ref{meanEu}); for [Ba/Fe]
 we have divided  the [Fe/H] in 18 bins but 
we have merged the first three bins (starting from the lowest value in [Fe/H])
 into a single bin in order to have enough data in the first bin  (see Table \ref{meanBa}).
For [Ba/Eu] versus [Ba/H] we have split the data into 16 equal bins but
again we have merged the first two pairs in two bins for the same reason 
 (see Table \ref{meanBaH}).

In Fig. \ref{trava} we show the results for the model 3 (with
the yields used in Travaglio et al. 2001) for [Ba/Fe] versus [Fe/H].
As evident from Fig. \ref{trava}, this model does not fit the most recent 
data.

\begin{figure}[h!]
\begin{center}
\includegraphics[width=0.99\textwidth]{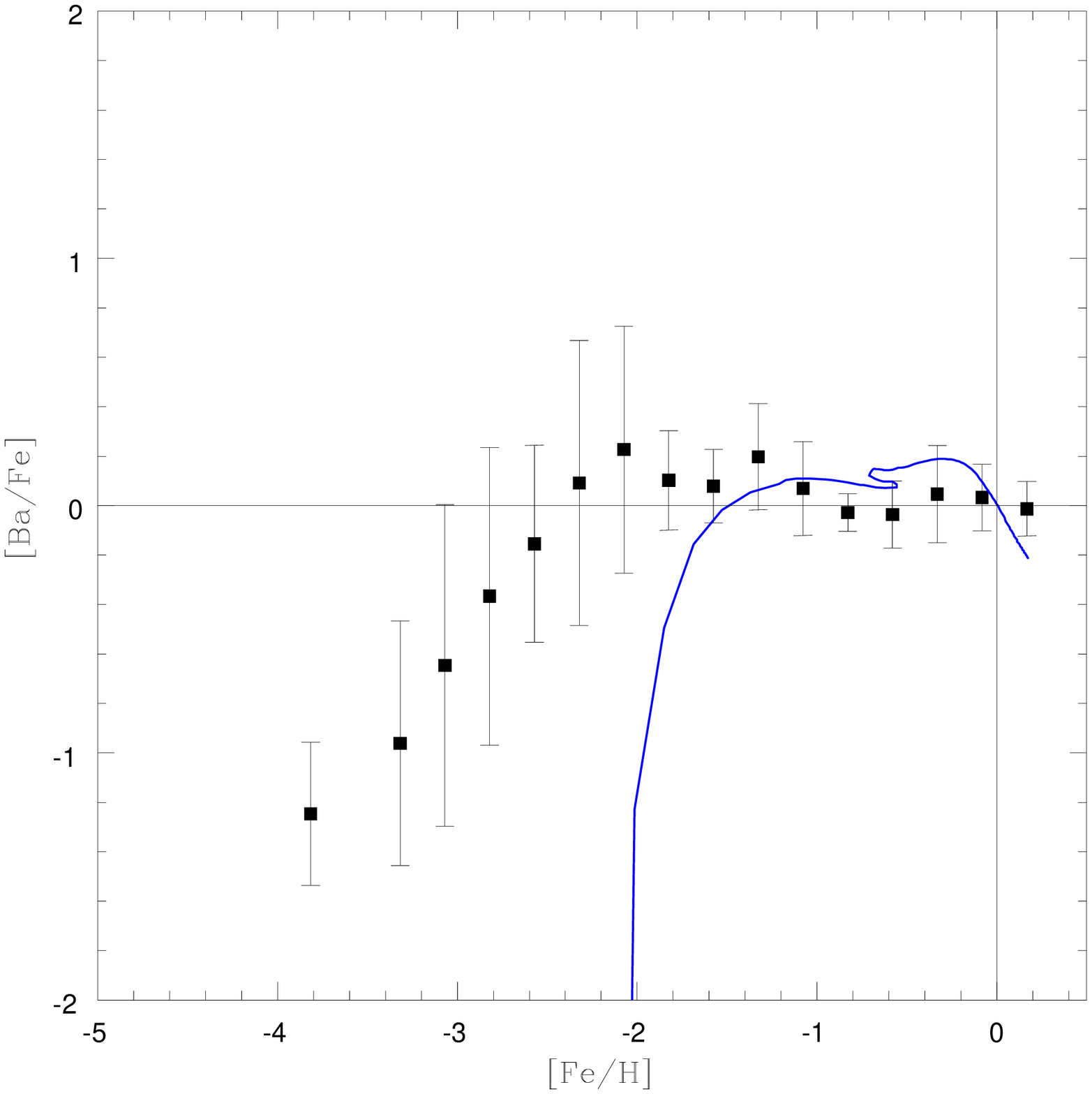}
\caption{The  ratio [Ba/Fe] versus [Fe/H]. The squares are
  the mean values of the data bins described in Table \ref{meanBa}. 
For error bars we use the standard deviation (see Table \ref{meanBa}).
Solid line: the results of model 3 (Models are described in Table \ref{model}).}\label{trava}
\end{center}
\end{figure}

Moreover, the model in Fig. \ref{trava} is different from the similar
model computed by Travaglio et al. (1999). 
We are using a different chemical
evolution model and this gives rise to different results.
The main difference between the two chemical evolution models
(the one of Travaglio and the present one) is the age-[Fe/H] relation
which grows more slowly in the model of Travaglio.
The cause for this difference is probably the different adopted
stellar lifetimes, the different $M_{up}$ (i.e. the most massive star ending  its life
as C-O white dwarf) and to the yield  prescriptions for iron  which
are probably the WW95 metallicity-dependent ones in the model of Travaglio et al. (1999),
whereas we use the WW95 yields for the solar chemical composition,
 which produce a faster rise of iron and generally a better agreement with 
the [X/Fe] vs [Fe/H] plots. 

To better fit the new data
we have to extend the mass range for the production of the r-processed
barium toward higher mass in order to reproduce [Ba/Fe] at lower metallicity.

In Fig. \ref{best1}, where we have plotted the 
predictions of model 1 and model 2  for [Ba/Fe] versus [Fe/H];
 it is clear that these models better fit the trend of the data.
\begin{figure}[h!]
\begin{center}
\includegraphics[width=0.99\textwidth]{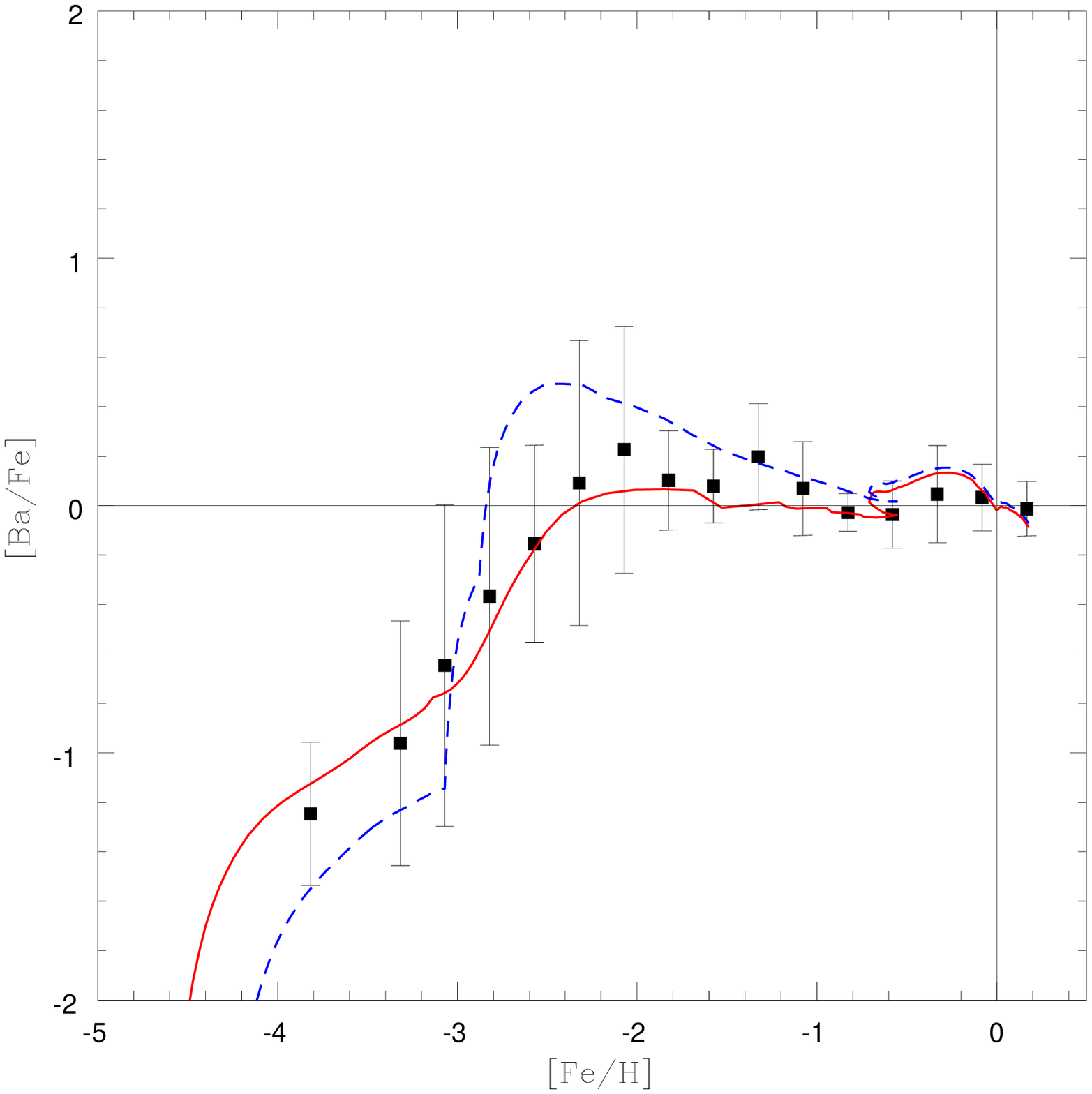}
\caption{The data are the same as in  Fig. \ref{trava}. Solid line:
 the model 1; dashed line: the model 2
 (Models  are described in Table \ref{model}).}\label{best1}
\end{center}
\end{figure}
In these models the upper mass limit for the production
of the r-processed Ba is 30$M_{\odot}$ in the case of model 1, and
25$M_{\odot}$ in the case of model 2.
However, model 2 does not fit the trend of the data as well as model 1.
In model 2 there is no dependence on stellar mass for a given metallicity
in the yields of Ba and Eu. This prescription is clearly an oversimplification
but shows how a model with yields only dependent on metallicity
works, allowing us to estimate whether it is appropriate or not.

We have obtained similar results comparing the trend of the abundances
of [Eu/Fe] versus [Fe/H] with the three models of 
Ishimaru et al. (2004) (Model 4, 5 and 6 in Table \ref{model}).
The chemical evolution of this r-process element
is shown in Fig. \ref{ISHI}. Note that that these authors 
 used a different chemical model.
Again model 4 does not explain the low metallicity abundances
and  model 5 and 6 do not fit the trend of the data well.

\begin{figure}[h!]
\begin{center}
\includegraphics[width=0.99\textwidth]{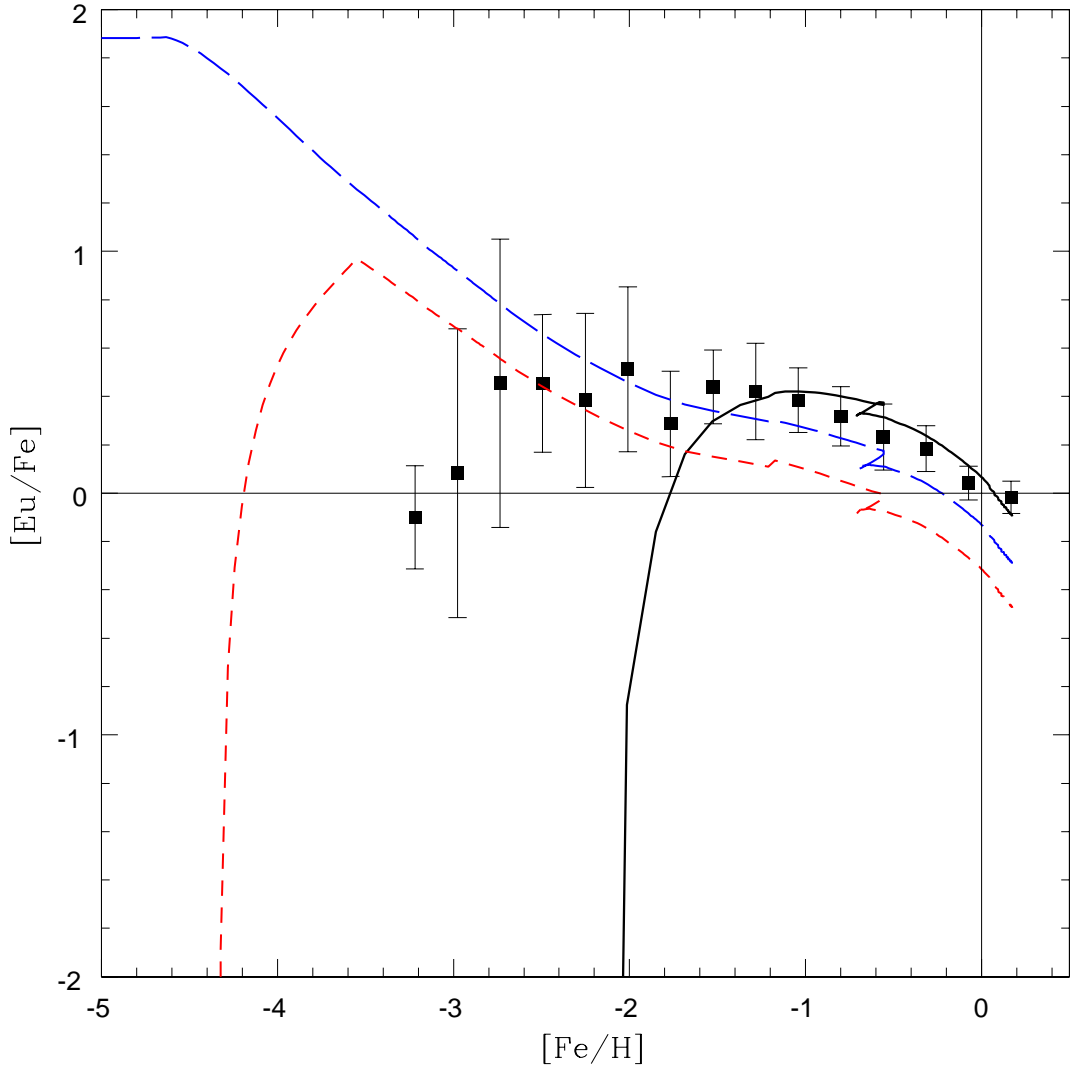}
\caption{[Eu/Fe] versus [Fe/H]. The squares are
  the mean values of the data bins described 
in the Table \ref{meanEu}. For error bars we use the standard deviation (see Table \ref{meanEu}).
Solid line: the results of model 4, short dashed line the results of model 5,
 long dashed line the results of model 6 (Models  are described in Table \ref{model}).}\label{ISHI}
\end{center}
\end{figure}

In Fig. \ref{best2} we show the results of models 1 and
2 in this case for [Eu/Fe] versus [Fe/H]. The trend of the data
is  followed well by both models from low metallicity
to solar metallicity.

\begin{figure}[h!]
\begin{center}
\includegraphics[width=0.99\textwidth]{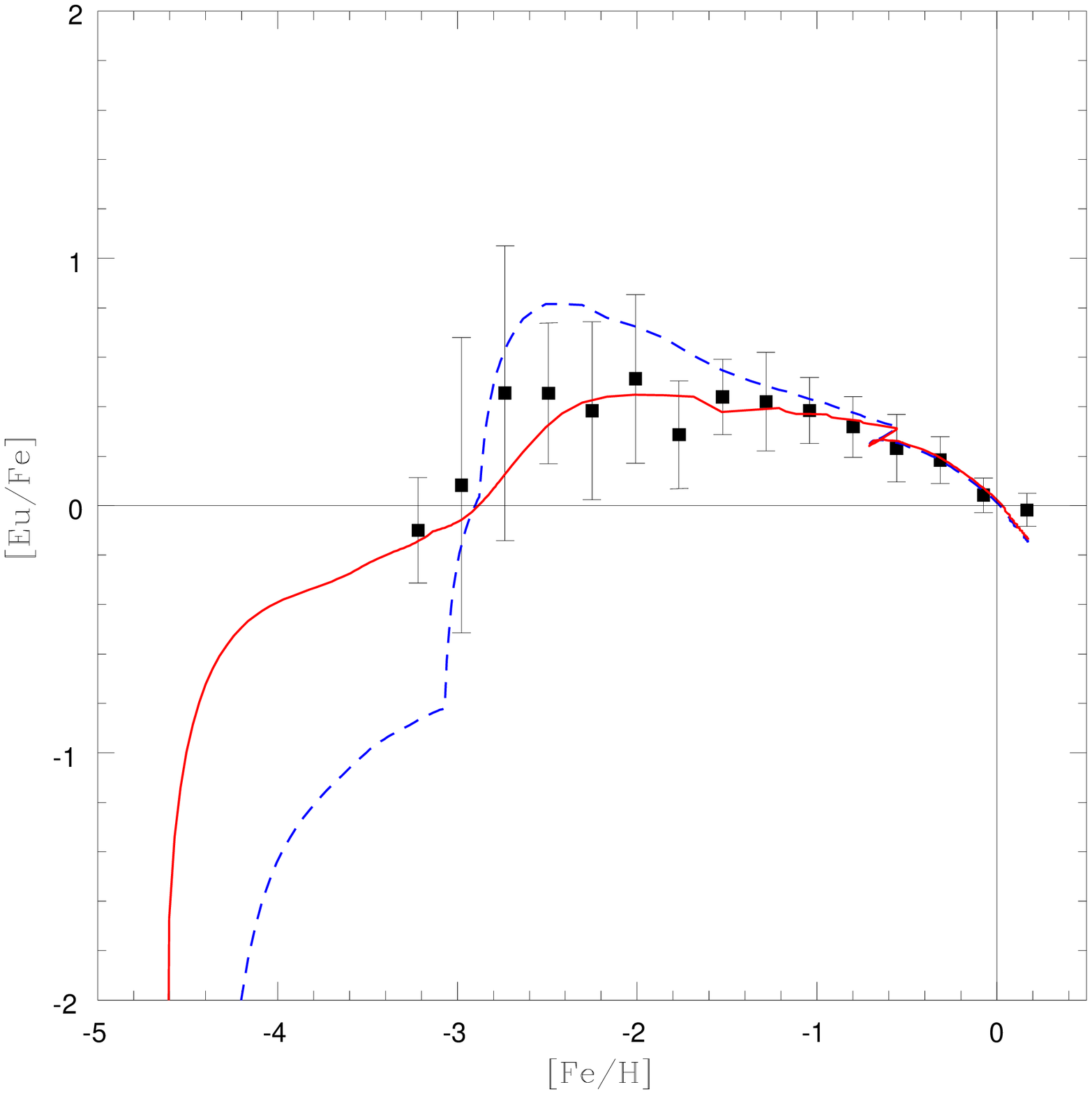}
\caption{Data as in Fig. \ref{ISHI}. Solid line: the results of model 1, dashed line the 
results of model 2 (Models  are described in Table \ref{model}).}\label{best2}
\end{center}
\end{figure}

In Table  \ref{bariumsol}  we show the predicted 
solar abundances 
of Eu and Ba for all our  models compared to the solar abundances
 by  Grevesse \& Sauval (1998). We also give
 the predicted s-process fraction in the barium solar abundance.
The results of almost all our models are in good agreement with the 
solar abundances with the exception of model 5 which underpredicts 
the Eu abundance by a factor of $\sim$ 2.
Note that we predict a different s-process fraction in the solar mixture
(nearly 60\% instead of 80\%) as compared to the s-process fraction
obtained by other authors with different chemical evolution codes as
Travaglio et al. (1999) and  Raiteri et al. (1999).
In fact, although we use the same yields as Travaglio et al. (1999) for
the production of Ba in low-intermediate mass stars,
we obtain different results. This is  due to the adopted
chemical evolution model, which produces a different
age-[Fe/H] relation which in turn affects the Ba production.
This fraction of s-process Ba is also lower than the results obtained
by means of stellar evolution
models  (e.g. Arlandini et al. 1999), although different s-process Ba 
fractions are possible in these models.

\begin{table*}
\centering
\caption{Solar abundances of Ba and Eu, as predicted by our models, 
compared to the observed ones from  Grevesse \& Sauval (1998).}\label{bariumsol}
\vspace{1.5cm}

\begin{tabular}{|c|c|c|c|c|c|}
\hline\hline

Mod               & $(X_{Ba})_{pr}$        & \%$Ba_{s}/Ba$ & $X_{Ba_{\odot}}$  &$(X_{Eu})_{pr}$       & $X_{Eu_{\odot}}$      \\        
\hline\hline                                                                                                                            
1                 & $1.55\cdot 10^{-8}$    &54\%           &$1.62\cdot 10^{-8}$&$4.06\cdot 10^{-10}$  & $3.84\cdot 10^{-10}$  \\         
\hline                                                                                                                          
2                 & $1.62\cdot 10^{-8}$    &51\%           &                   & $3.96\cdot 10^{-10}$ &                       \\         
\hline                                                                                                                          
3                 & $1.64\cdot 10^{-8}$    &44\%           &                   & As model 1           &                       \\        
\hline                                                                                                                          
4                 & As model 1             &As model 1     &                   & $4.48.\cdot 10^{-10}$&                       \\        
\hline                                                                                                                          
5                 & As model 1             &As model 1     &                   & $1.86\cdot 10^{-10}$ &                       \\        
\hline                                                                                                                          
6                 & As model 1             &As model 1     &                   & $2.84\cdot 10^{-10}$ &                       \\        
\hline\hline

\end{tabular}

\end{table*}

Fig. \ref{BaEu}, where we have plotted
the abundances of [Ba/Eu] versus [Fe/H], and Fig. \ref{BaEu2}, where we plot
 [Ba/Eu] versus [Ba/H],
have  important features.
\begin{figure}[h!]
\begin{center}
\includegraphics[width=0.99\textwidth]{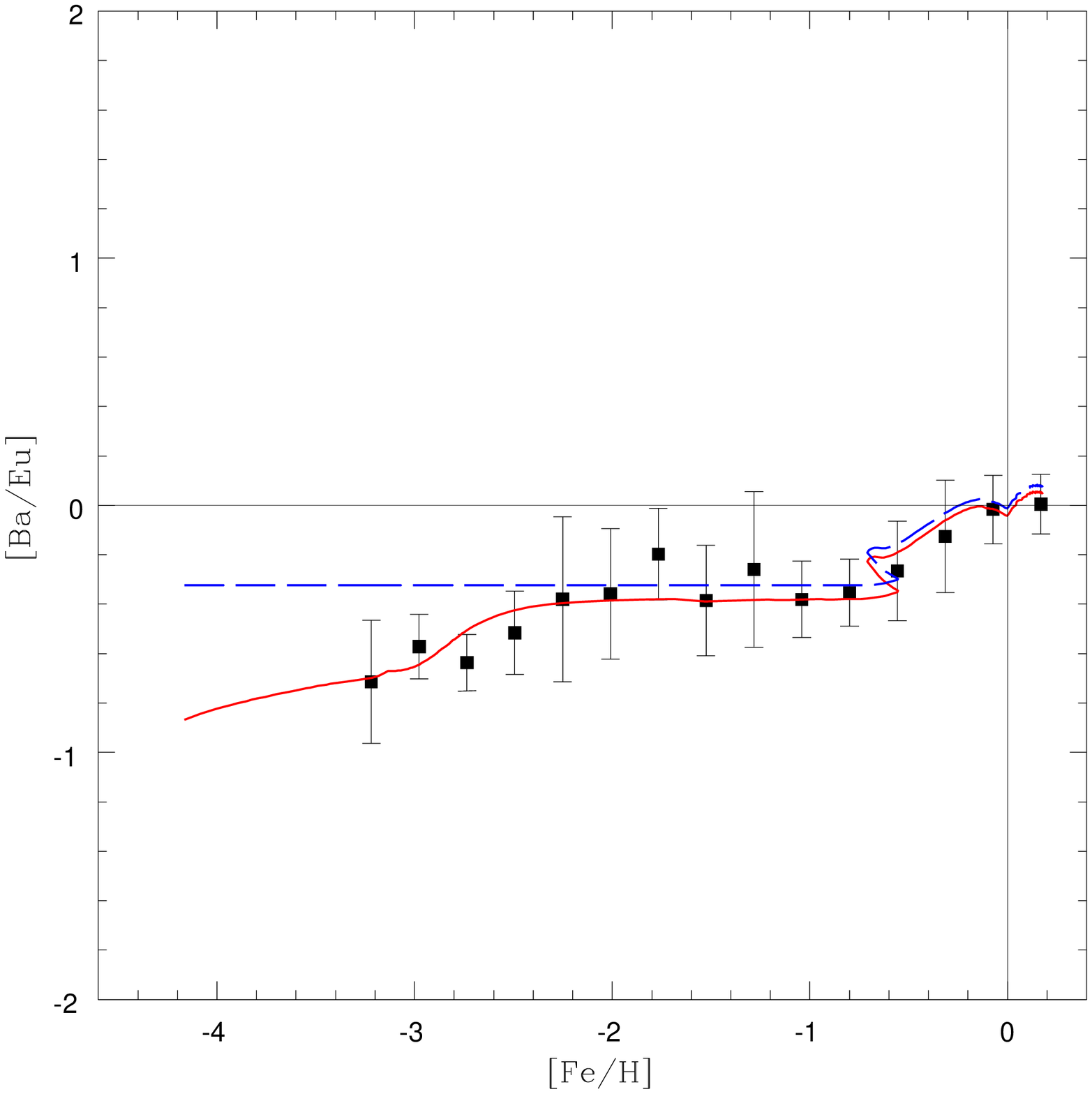}
\caption{The ratio of [Ba/Eu] versus [Fe/H]. The squares are
  the mean values of the data bins described 
in Table \ref{meanEu}. For error bars we use the standard deviation (see Table \ref{meanEu}).
Model 1: solid line, model 2: long dashed line
 (Models  are described in Table \ref{model}).}\label{BaEu}
\end{center}
\end{figure}
\begin{figure}[h!]
\begin{center}
\includegraphics[width=0.49\textwidth]{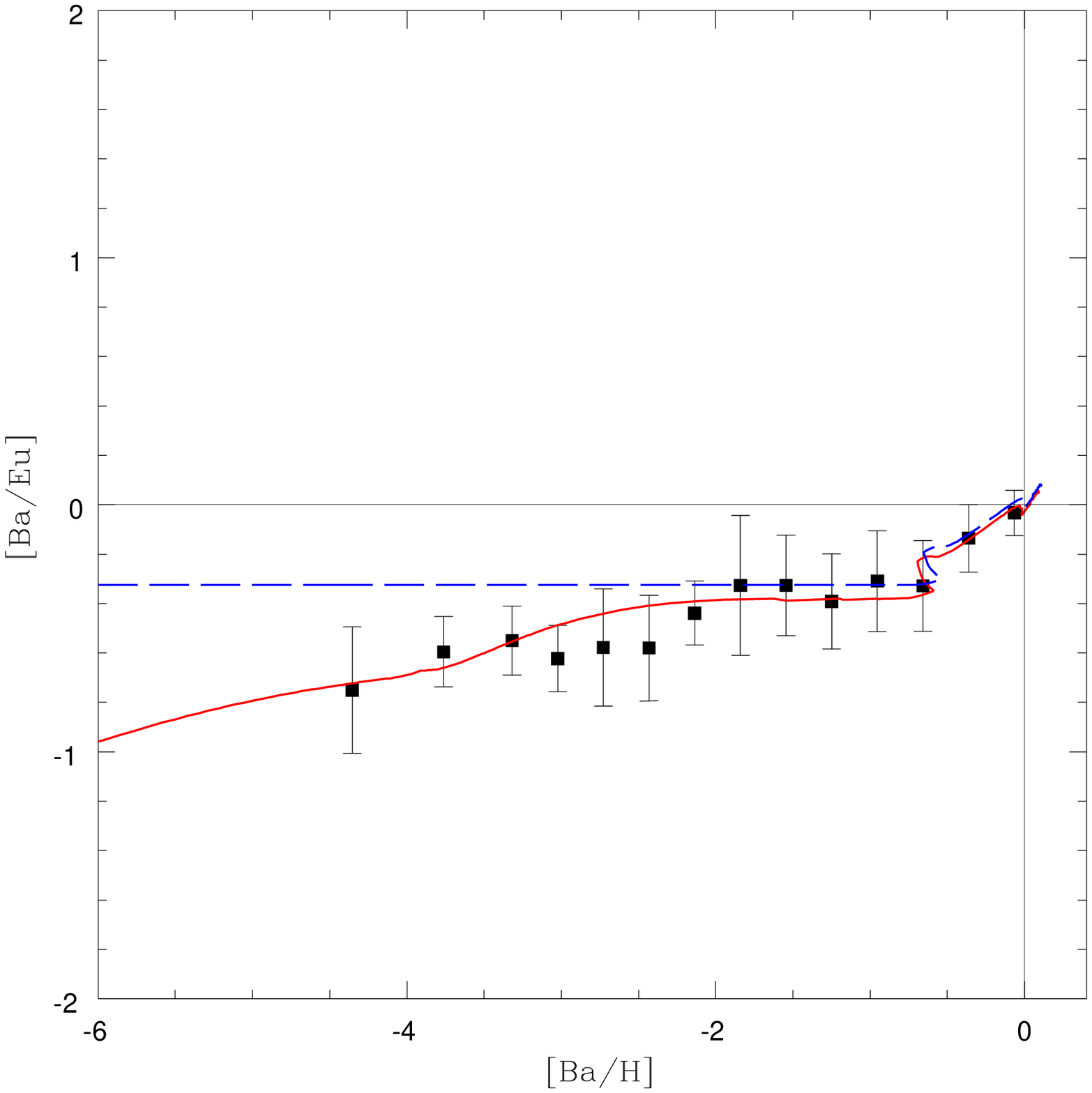}
\caption{[Ba/Eu] versus [Ba/H]. The squares are
  the mean values of the data bins described 
in Table \ref{meanBaH}. For error bars we use the standard deviation (see Table \ref{meanBaH}).
Model 1: solid line, model 2: long dashed line (Models  are described in Table \ref{model}). }\label{BaEu2}
\end{center}
\end{figure}
The first is that the spread, that we can infer in these 
plots from the standard deviation of each bin, is smaller 
if we use the [Ba/H] ratio on the x axis; 
the second feature is that it is evident from the data
 that there is a plateau in the [Ba/Eu] ratio
that is seen before the production of s-process Ba by the low-intermediate
mass stars starts to be non negligible, at $[Fe/H]\sim-1$ and $[Ba/H]\sim-0.8$;
 finally, the timescale of the rise of the [Ba/Eu] value, due to the production of Ba 
by low-intermediate mass stars, is very well reproduced by our model.

%%% The value of [Ba/Eu] at low metallicity is important  to understand
%%% the fraction of slow processed Ba in the solar abundance
%%% In fact, if the ratio  $\frac{Ba_{rapid}}{Eu}$
%%%  has a constant value over cosmic time,
%%%  then it must have the same
%%% value at the solar system formation time
%%% (if we do not want to add some peculiar effect during
%%% the last part of the Galaxy evolution).
%%% In this case the Ba s-process fraction is simply:
%%% \begin{displaymath}
%%% \frac{Ba_{slow}}{Ba_{total}}=1-10^{[\frac{Ba_{rapid}}{Eu}]}.
%%% \end{displaymath}
%%% Since we have a mean value for [Ba/Eu] versus [Fe/H] in the
%%% range $-3<[Fe/H]<-1$ of -0.44 (taken from the mean
%%% value in the bins which fall in that range)
%%% and a similar value for [Ba/Eu] versus [Ba/H] in the range
%%%  $-4<[Ba/H]<-0.8$ of -0.41 (computed in the same way as above), 
%%% then it turns out that  the s-process fraction for 
%%% slow processed barium has to be less than the claimed 80\%, with
%%% a value of $\sim 60\%$.
%%% 

The spread in the ratio of
[Ba/Eu] both versus [Fe/H] and [Ba/H] is 
lower than the spread
in [Ba/Fe] and [Eu/Fe], in particular when using as an
 evolutionary tracer the [Ba/H].
Considering the computed standard deviations  as spread tracers,
 where the spread for [Ba/Fe] and [Eu/Fe] is higher ($[Fe/H]\sim -3$),
their standard deviations are larger than 0.6 dex whereas  the standard deviations
 for [Ba/Eu] is less than 0.15 dex.

For this reason we believe that the mechanism which
produces the observational spread does not affect
the ratio of these two elements.
We propose that the explanation of the smaller spread in the ratio
 of [Ba/Eu] is that the site of production of these
two elements is the same: the neutronized shell close to the
mass cut in a SNII (see Woosley et al. 1994). What changes
could be the amount of the neutronized material that each
massive star expels during the SNII explosion.  
The mass cut and also the ejected neutronized material
are still uncertain quantities and usually they are considered as parameters
in the nucleosynthesis codes for massive stars (see Rauscher et al. 2002, 
Woosley \& Weaver 1995, Woosley et al. 1994).

\subsubsection{Upper and lower limit to the r-process production}\label{limits}

The purpose of this section is to give upper and lower limits
to the yields to reproduce the observed spread at low metallicities 
for Ba and Eu.
An inhomogeneous model would provide better 
predictions about the dispersion in the [r-process/Fe] ratios if due to 
yield variations, but it is still useful to study the 
effect of the yield variations by means of our model.

First we explore the range of variations of the yields as functions
of the  stellar mass. To do this we have used model 1:  in particular,
we  have modified
the yields of model 1 for both elements (Ba and Eu), leaving untouched 
the s-process yields and changing only the yields of the r-process.
Models 1Max and 1min and their characteristics are summarized in Table \ref{Mm}, where
are indicated the adopted yields and the factors by which they have been modified
relative to Model 1.
\begin{table*}

\caption{The stellar yields for model 1Max and 1Min for barium and europium in massive stars (r-process)
in the case of a primary origin.} \label{Mm}
\vspace{1.5cm}

\centering
\begin{minipage}{90mm}

\begin{tabular}{|c|c|c|c|c|}
\hline

             & Model 1Max         &    & Model 1Min         &    \\     
\hline\hline
$M_{star}$   & $ X_{Ba}^{new}$    & Factor & $ X_{Ba}^{new}$     & Factor\\

12.          & 1.35$\cdot10^{-6}$ & 1.5    & 4.50$\cdot10^{-7}$  & 0.5\\ 
$ < 15$.     & 4.50$\cdot10^{-8}$ & 1.5    & 1.50$\cdot10^{-8}$  & 0.5\\   
$ \ge 15$    & 3.00$\cdot10^{-7}$ & 10.    & 1.50$\cdot10^{-9}$  & 0.05\\
30.          & 1.00$\cdot10^{-8}$ & 10.    & 5.00$\cdot10^{-11}$ & 0.05\\ 

\hline\hline         
                                            
$M_{star}$   & $ X_{Eu}^{new}$    & Factor& $ X_{Eu}^{new}$     & Factor\\
12.          & 4.50$\cdot10^{-8}$        & 1.    & 2.25$\cdot10^{-8}$         & 0.5 \\ 
$ < 15$.     & 3.00$\cdot10^{-9}$        & 1.    & 1.50$\cdot10^{-9}$         & 0.5 \\   
$ \ge 15$    & 3.00$\cdot10^{-8}$        & 10.   & 1.50$\cdot10^{-10}$        & 0.05\\
30.          & 5.00$\cdot10^{-9}$        & 10.   & 2.50$\cdot10^{-11}$ & 0.05\\

\hline\hline

\end{tabular}

\end{minipage}

\end{table*}
In Fig. \ref{Mm1Ba} and \ref{Mm1Eu}  we plot ratios [Ba/Fe] vs [Fe/H] and [Eu/Fe] vs [Fe/H]
for the new models 1Max and  1min compared to the observational data;
 we show the same plot  for the ratios [Ba/Eu] vs [Fe/H] in  Fig.\ref{Mm1BaEu} and
and for [Ba/Eu] versus [Ba/H]  in Fig. \ref{Mm1BaEu2}.

\begin{figure}[h!]
\begin{center}
\includegraphics[width=0.99\textwidth]{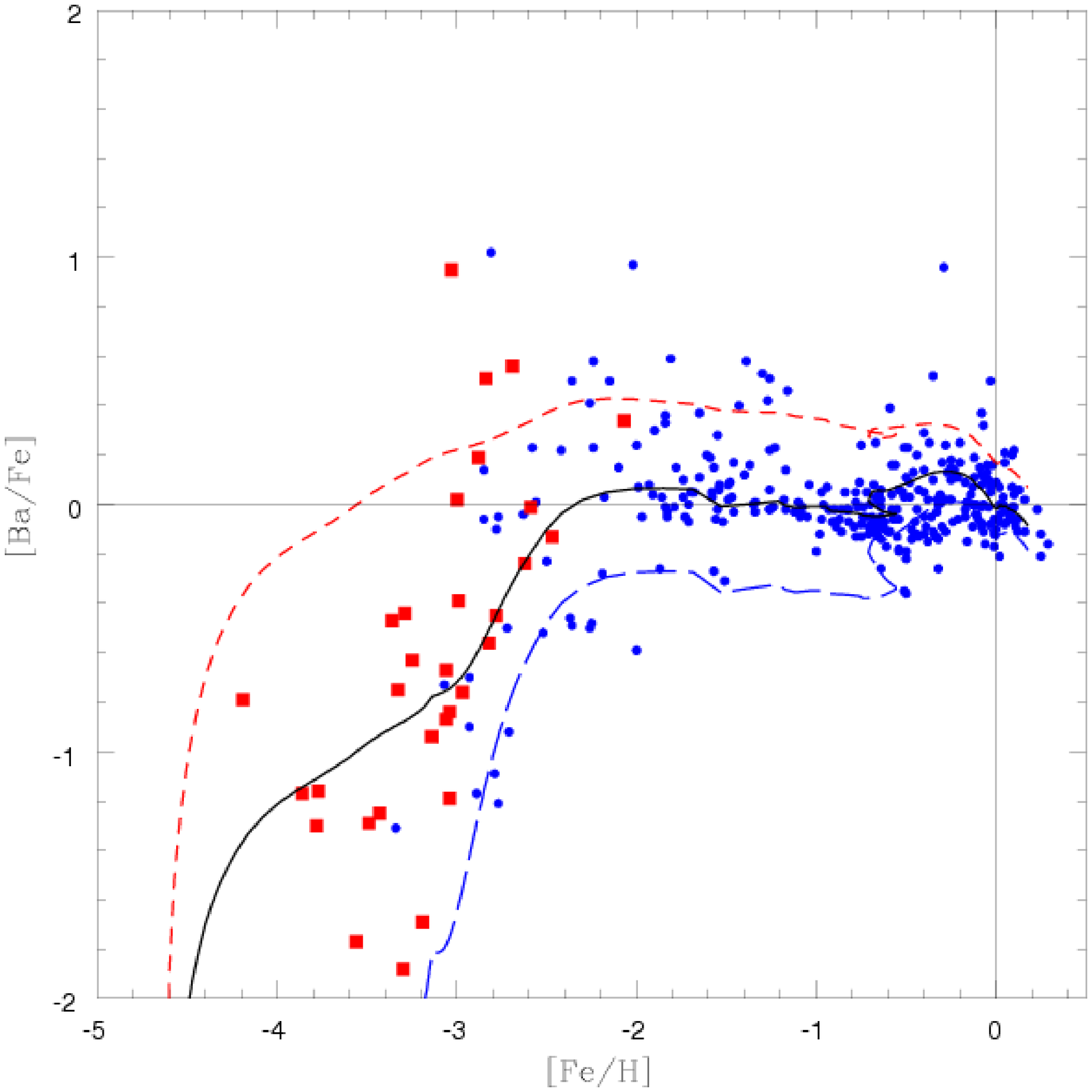}
\caption{[Ba/Fe] versus [Fe/H] for  the data by 
Fran\c cois et al. (2007) (filled squares)  and
 for the other observational data (see Sect. 2 in the text, the filled circles).
The solid line is the prediction of model 1, the short dashed line the prediction of model 1Max
and  the long dashed line the prediction of model 1min.}
\label{Mm1Ba}
\end{center}
\end{figure}

\begin{figure}[h!]
\begin{center}
\includegraphics[width=0.99\textwidth]{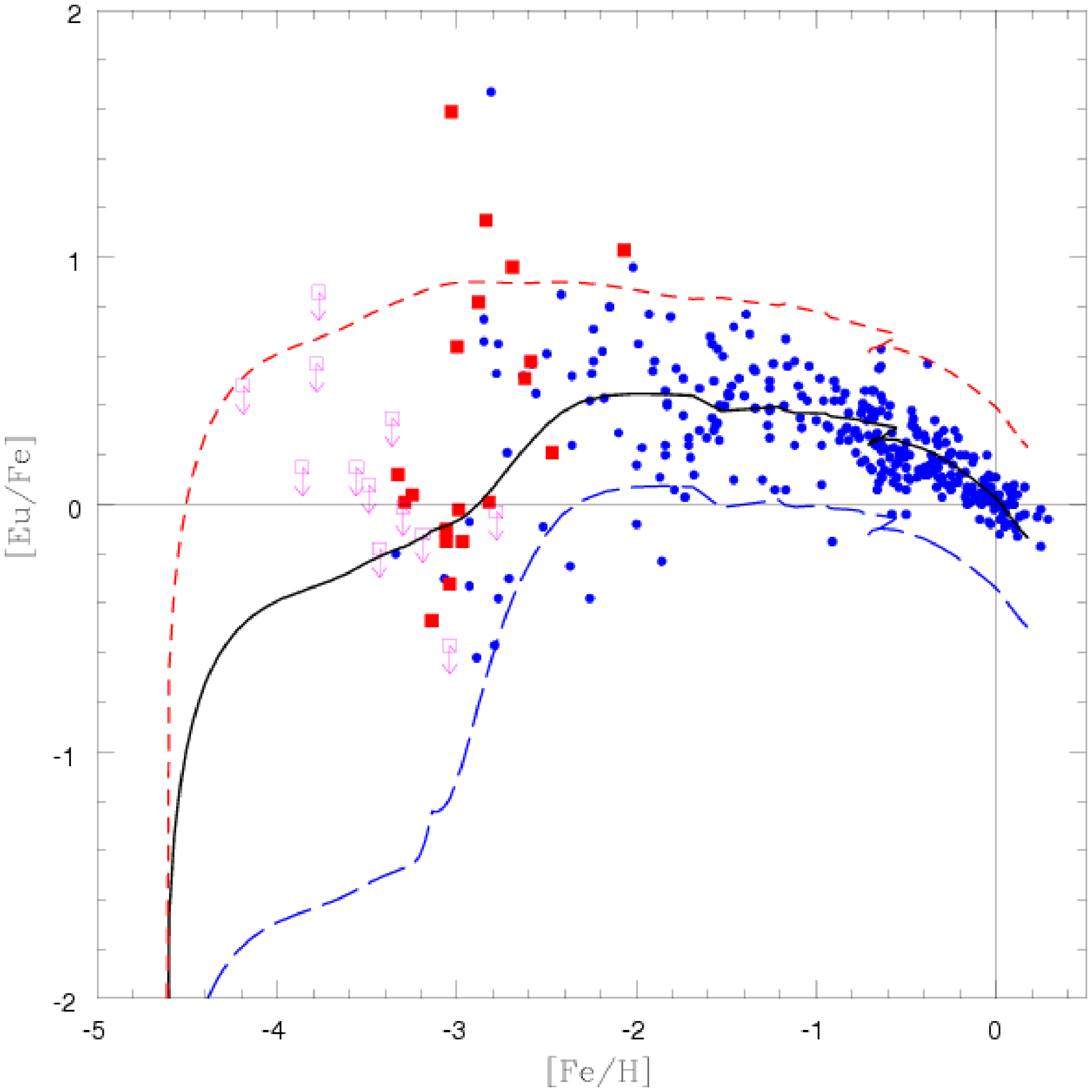}
\caption{[Eu/Fe] versus [Fe/H]. The data by 
Fran\c cois et al. (2007) are filled squares, the open squares
are upper limits (Fran\c cois et al. 2007). The filled circles are data 
by other observational works (see Sect. 2 in the text).
The solid line is the prediction of model 1, the short dashed line the prediction of model 1Max
and  the long dashed line the prediction of model 1min.}
\label{Mm1Eu}
\end{center}
\end{figure}

\begin{figure}[h!]
\begin{center}
\includegraphics[width=0.49\textwidth]{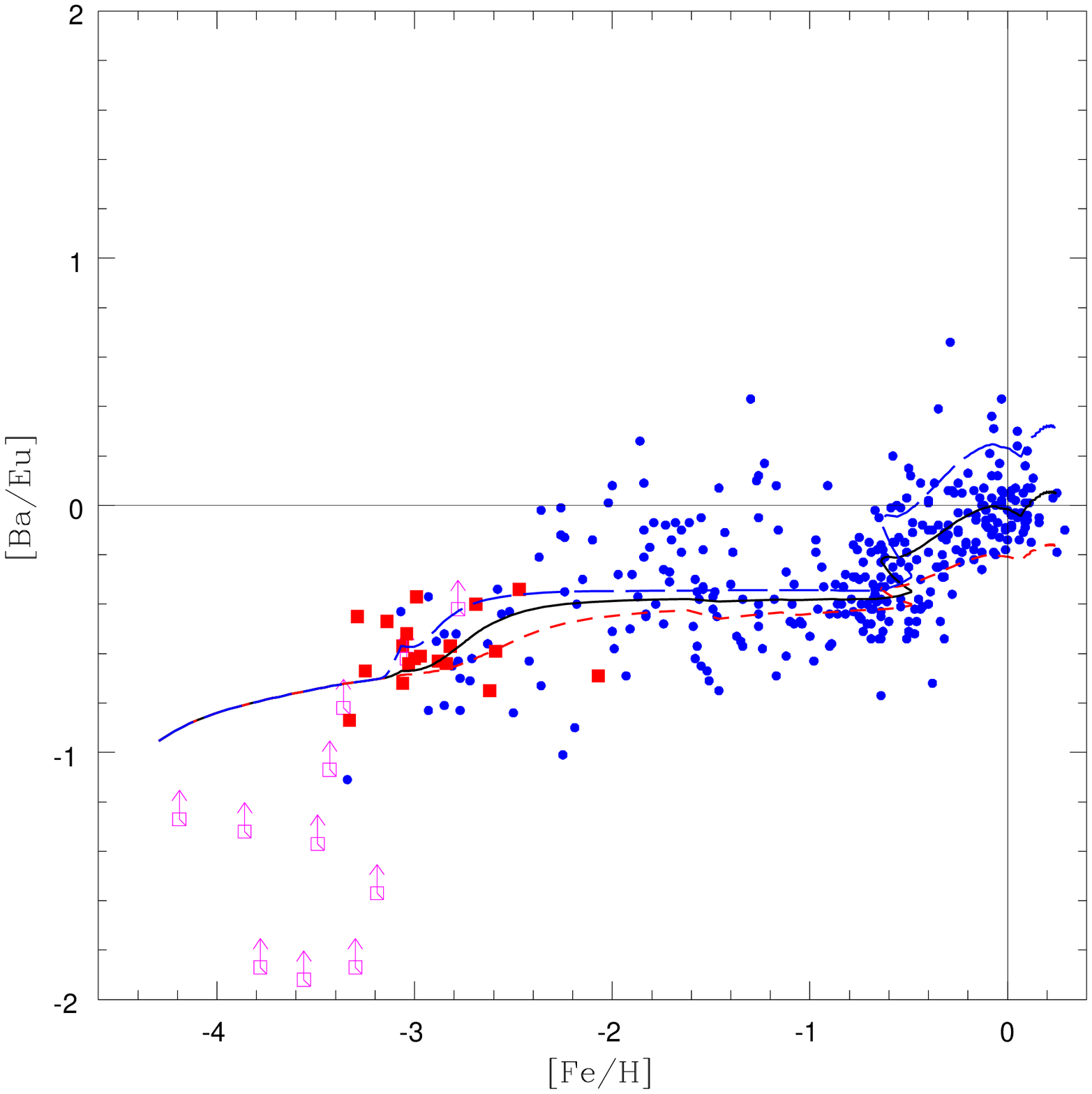}
\caption{[Ba/Eu] versus [Fe/H]. The data by 
Fran\c cois et al. (2007) are filled squares, the open squares
are lower limits (Fran\c cois et al. 2007). The filled circles are data 
by other observational works (see Sect. 2 in the text).
The solid line is the prediction of model 1, the short dashed line the prediction of model 1Max
and  the long dashed line the prediction of model 1min.}
\label{Mm1BaEu}
\end{center}
\end{figure}

\begin{figure}[h!]
\begin{center}
\includegraphics[width=0.49\textwidth]{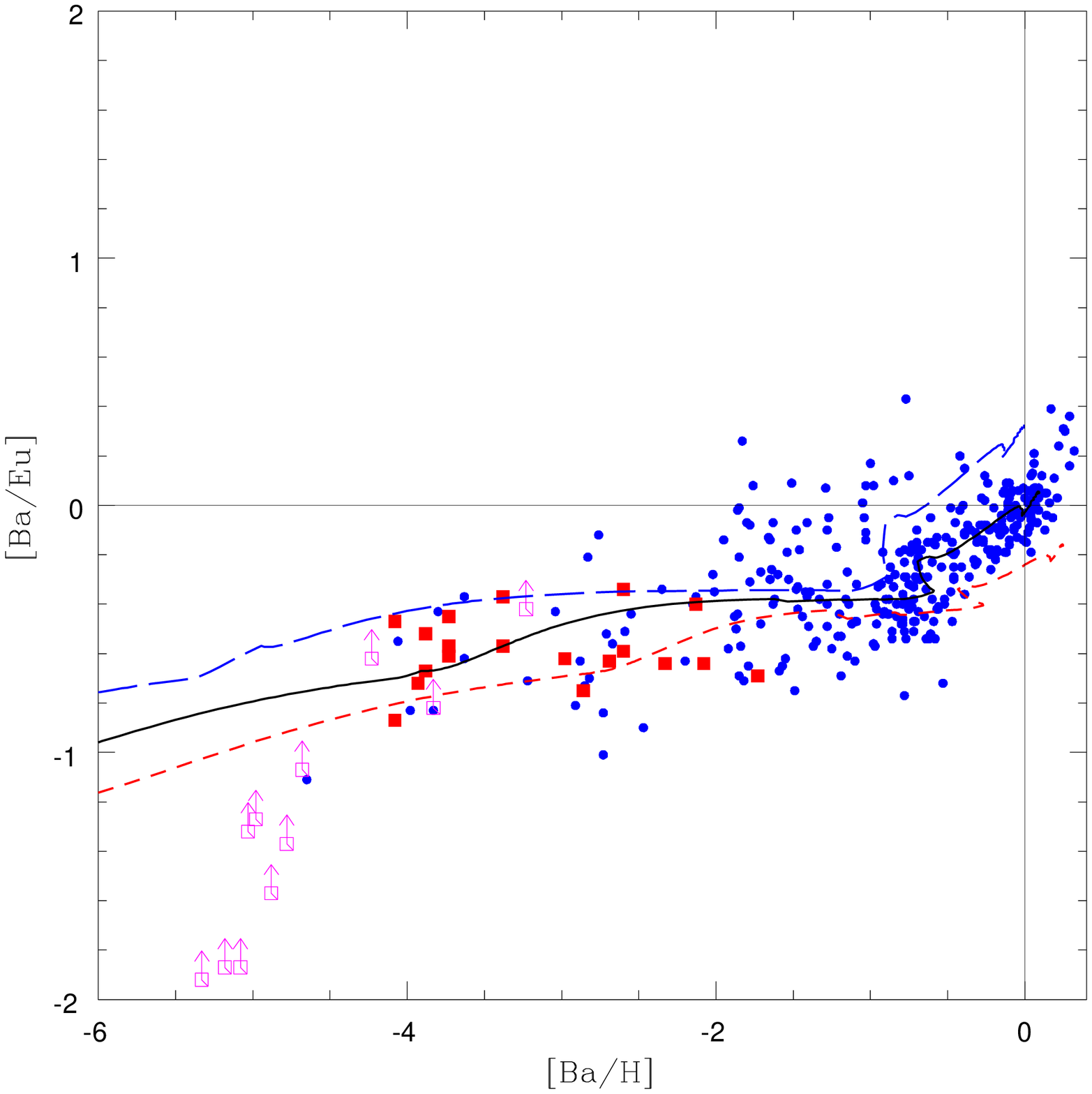}
\caption{As in Fig. \ref{Mm1BaEu} but  for [Ba/Eu] versus [Ba/H].}
\label{Mm1BaEu2}
\end{center}
\end{figure}

We can deduce from  these upper and lower limit models that the large observed spread 
could also be  due to a  different production of heavy elements among massive stars ($>15M_{\odot}$).
This type of stars could produce different amounts of these elements
independently of the mass.
As we have introduced in the previous subsect. (\ref{trends}), it is possible to link this fact 
with the problems of mass cut and the fallback during the explosion of
a SNII. If these elements are produced in a shell close to the iron core of the star, differences in the 
explosion behaviour can give rise to a different quantity of r-process elements  expelled by each star.
In this way we are able to explain the presence of the spread for the
heavy elements and the absence of the same spread, for example, in the $\alpha$-elements.
 The $\alpha$-elements are produced mostly
during the hydrostatic burning of massive stars and then  ejected by the explosion.

Another approach can be followed to derive
 upper and lower limits for the model 
by changing the yields as functions of metallicity.
 Model 2, which is the model with yields independent of the mass and depending only on metallicity, 
will be our test model. 
In particular, model 2 assumes for the massive
stars different yields for Ba and Eu in three ranges of  metallicity
(see Table \ref{model}).
The new prescriptions  for both Ba and Eu are summarized in Table \ref{rBa3}, where
are indicated the new limits for the ranges of metallicities for the model 2min
and 2Max. Note that for the model 2Max we assume only two regimes of production
of r-process elements as a function of the metallicities and so we distinguish
only two ranges of metallicities.

\begin{table*}

\caption{The stellar  yields of model 2Max and model 2min for Ba and Eu 
in massive stars (r-process).} \label{rBa3}

\centering
\vspace{1.5cm}

\begin{minipage}{90mm}

\begin{tabular}{|c|c|c|c|c|}

\hline\hline

 $Z_{star}$                &  $Z_{star}$                     & $X_{Ba}^{new}$     & $ X_{Eu}^{new}$ \\
 Model 2Max                &  Model 2min                     & $10-25M_{\odot}$ & $10-25M_{\odot}$ \\
                                                         
\hline                           
         --                &            $Z<8\cdot10^{-6}$.   & 1.00$\cdot10^{-8}$ &  5.00$\cdot10^{-10}$  \\ 
      $Z<1\cdot10^{-5}$    & $8\cdot10^{-6}<Z<1\cdot10^{-5}$ & 1.00$\cdot10^{-6}$ &  5.00$\cdot10^{-8}$  \\   
      $Z>1\cdot10^{-5}$    &            $Z>1\cdot10^{-5}$    & 1.60$\cdot10^{-7}$ &  8.00$\cdot10^{-9}$  \\ 

\hline\hline

\end{tabular}

\end{minipage}

\end{table*}

In Fig. \ref{Mm2Ba} and \ref{Mm2Eu}  we show the results of these two models (2Max and 2min) 
and of the original model 2 for [Ba/Fe] vs [Fe/H] and  for [Eu/Fe] versus [Fe/H]
 compared to the observational data.

\begin{figure}[h!]
\begin{center}
\includegraphics[width=0.49\textwidth]{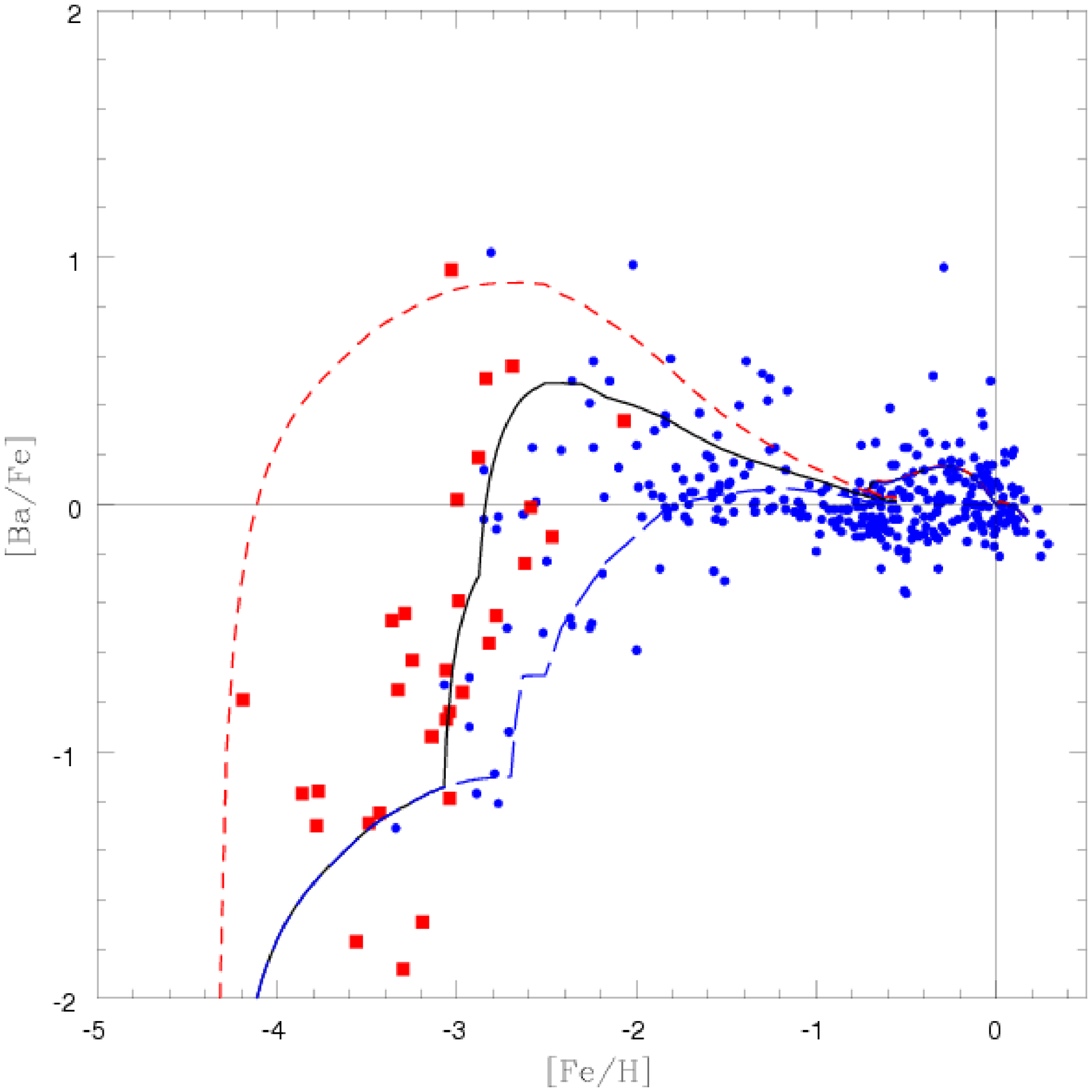}
\caption{[Ba/Fe] versus [Fe/H]. The data are as in Fig. \ref{Mm1Ba}.
The solid line is the prediction of model 2, the short dashed line the prediction of model 2Max
and  the long dashed line the prediction of model 2min.}\label{Mm2Ba}
\end{center}
\end{figure}

\begin{figure}[h!]
\begin{center}
\includegraphics[width=0.49\textwidth]{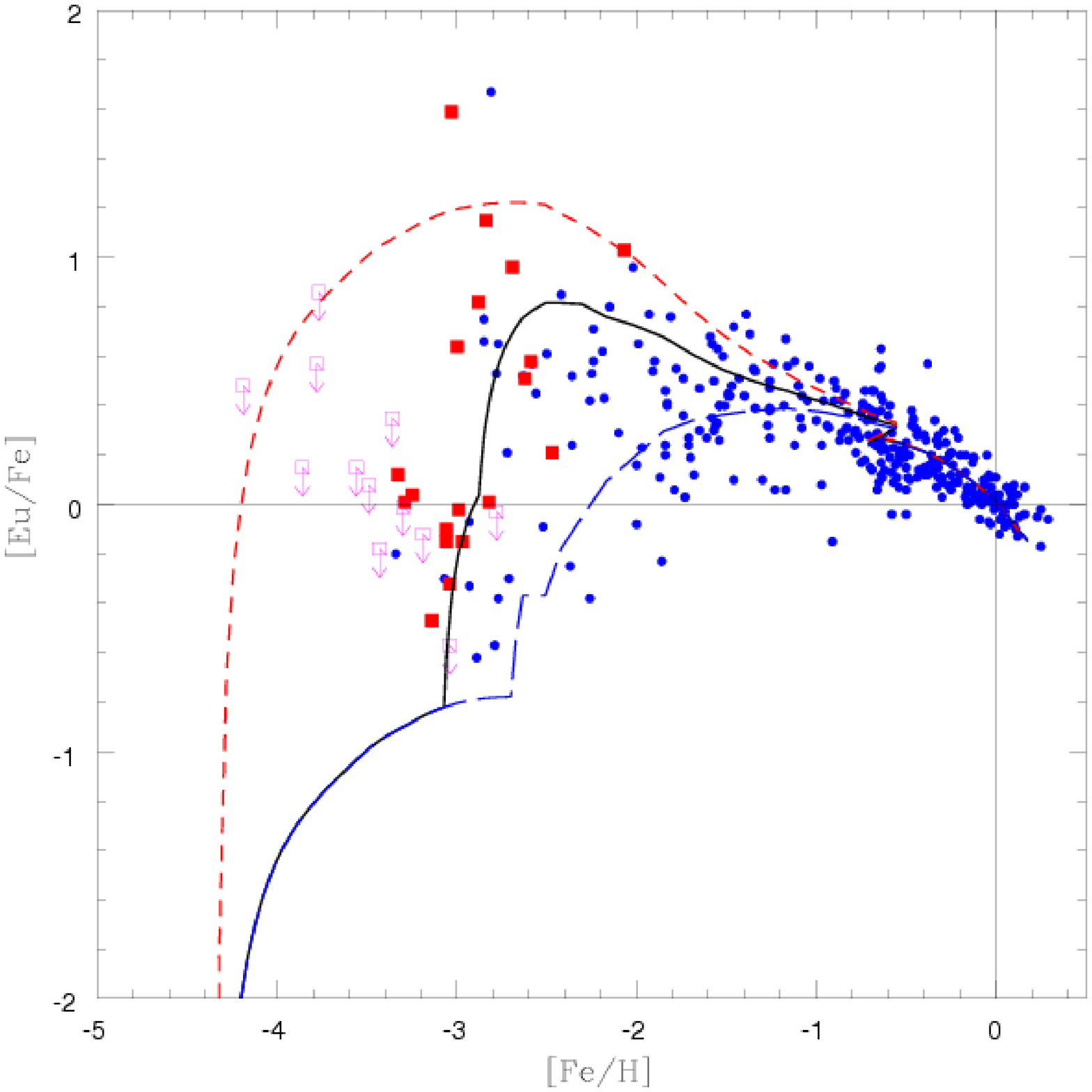}
\caption{[Eu/Fe] versus [Fe/H]. The data are as in Fig. \ref{Mm1Eu}.
The solid line is the prediction of model 2, the short dashed line the prediction of model 2Max
and  the long dashed line the prediction of model 2min.}\label{Mm2Eu}
\end{center}
\end{figure}

Changing the central range of metallicity, in which there is an enhancement 
of the production of Ba and Eu, it is possible to produce the upper and 
lower limits.
These two new models envelope the majority of the data at low metallicities.
At higher metallicity the two models overlap the best model and so they do not
contain all the spread in this part of the plot but most of them could 
be explained inside the typical observational error of 0.1 dex.

\section{Lanthanum}

In this Section we study the evolution of La. This element has a nucleosynthesis origin
very similar to that of Ba. Therefore we will follow the same method of study that we
adopted for Ba.

\subsection{Observational data}

We adopted the very recent results by Fran\c cois (2007).
For the abundances in the remaining range of [Fe/H],
 we took published high quality data in the
literature from various sources: Cowan et al. (2005),
 Burris (2000), Johnson (2002), Pompeia et al. (2003)
  and McWilliam \& Rich (1994).
 All of these data are relative to
 solar abundances of Grevesse \& Sauval (1998).

\subsection{Nucleosynthesis prescriptions for La}{\label{NP_BaS2}}
For the nucleosynthesis prescriptions of the s-process component of La
 we have adopted the yields of Busso et al. (2001) in the mass
range 1.5-3$M_{\odot}$.
We have extended the theoretical results of Busso 
et al. (2001)  in the mass range $1.5-1M_{\odot}$,
by simply scaling  the values obtained for stars of $1.5 M_{\odot}$ by the mass.
We have extended the prescription to better fit the data with a [Fe/H]
higher than solar. This hypothesis does not change the results of the model at [Fe/H]$<$0.

\begin{figure}
\begin{center}
\includegraphics[width=0.49\textwidth]{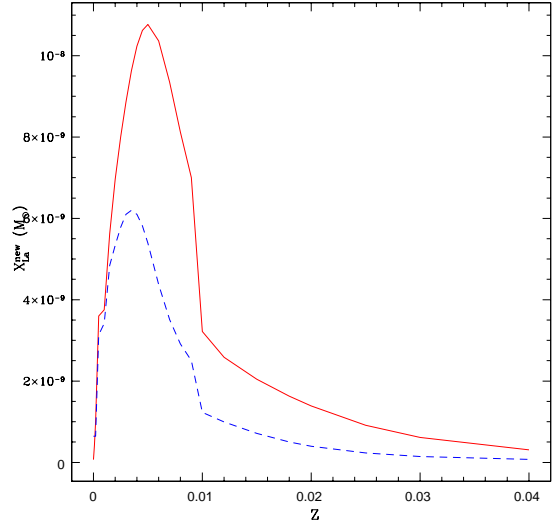}
\caption{The stellar yields  $X^{new}_{La}$ from the paper of  Busso et al. (2001) plotted versus 
stellar metallicity. The dashed
line represents the prescriptions for stars of $1.5M_{\odot}$, 
the solid line those for stars of $3M_{\odot}$.} \label{sLa}
\end{center}
\end{figure}

For the r-process contribution of La we give new prescriptions  following the same method 
as for Ba in model 1: we assume an r-process contribution in massive stars (12 - 30 $M_{\odot}$), besides
the s-process contribution from low mass stars.
The yields of this r-process contribution are summarized in Table \ref{rLa},
in which the mass fraction of newly produced La is given as function of the
mass.

\begin{table*}

\caption{The stellar yields for La in massive stars (r-process)
in the case of primary origin.} \label{rLa}
\vspace{1.5cm}
\centering
\begin{minipage}{90mm}

\begin{tabular}{|c|c|c|}
\hline

$M_{star}$  & $ X_{La}^{new}$\\

\hline\hline

12.   & 9.00$\cdot10^{-8}$ \\ 
15.   & 3.00$\cdot10^{-9}$ \\   
30.   & 1.00$\cdot10^{-10}$ \\

\hline\hline

\end{tabular}

\end{minipage}

\end{table*}

\begin{table*}

\caption{The mean and the standard deviations for the abundance of [La/Fe] for the stars 
inside each  bin along the [Fe/H] axis.} \label{meanLa}
\vspace{1.5cm}

\begin{tabular}{|c|c|c|c|c|}
\hline

bin center [Fe/H]& bin dim.[Fe/H]  & mean [La/Fe] & SD [La/Fe] &  N. of data in the bin \\
\hline\hline

 -2.97 & 1.20 & -0.13&  0.48& 29\\
 -2.17 & 0.40 &  0.04&  0.26& 15\\
 -1.78 & 0.40 &  0.06&  0.17& 7 \\
 -1.38 & 0.40 &  0.18&  0.17& 5 \\
 -0.99 & 0.40 & -0.04&  0.29& 4 \\
 -0.59 & 0.40 &  0.19&  0.32& 5 \\
 -0.19 & 0.40 &  0.09&  0.14& 13\\
  0.20 & 0.40 & -0.08&  0.09& 7 \\

\hline \hline
       
\end{tabular}

\end{table*}

\subsection{Results}

We divide the [Fe/H] axis in several bins  and we compute the mean and the 
standard deviations from the mean of the ratios
[La/Fe] for all the data inside each bin. These results are shown in Table \ref{meanLa},
where we also summarize the center and the dimension of each bin and the number of data points
contained in each of them.
In Fig.\ref{Laresult} we show  the predictions of the chemical 
evolution  model  for La in the solar neighborhood using our prescriptions for the yields in
 massive stars and the prescriptions of Busso et al. (2001) for low mass stars, 
as described in the previous section.
These results are new and the model well reproduces the trend 
of the stellar abundances at different [Fe/H] as well as the solar abundance 
of lanthanum. We obtain a La solar mass fraction of $1.35 \cdot 10^{-9}$
very close to the solar value of Asplund et al. (2005) of $1.38 \cdot 10^{-9}$.  

\begin{figure}
\begin{center}
\includegraphics[width=0.99\textwidth]{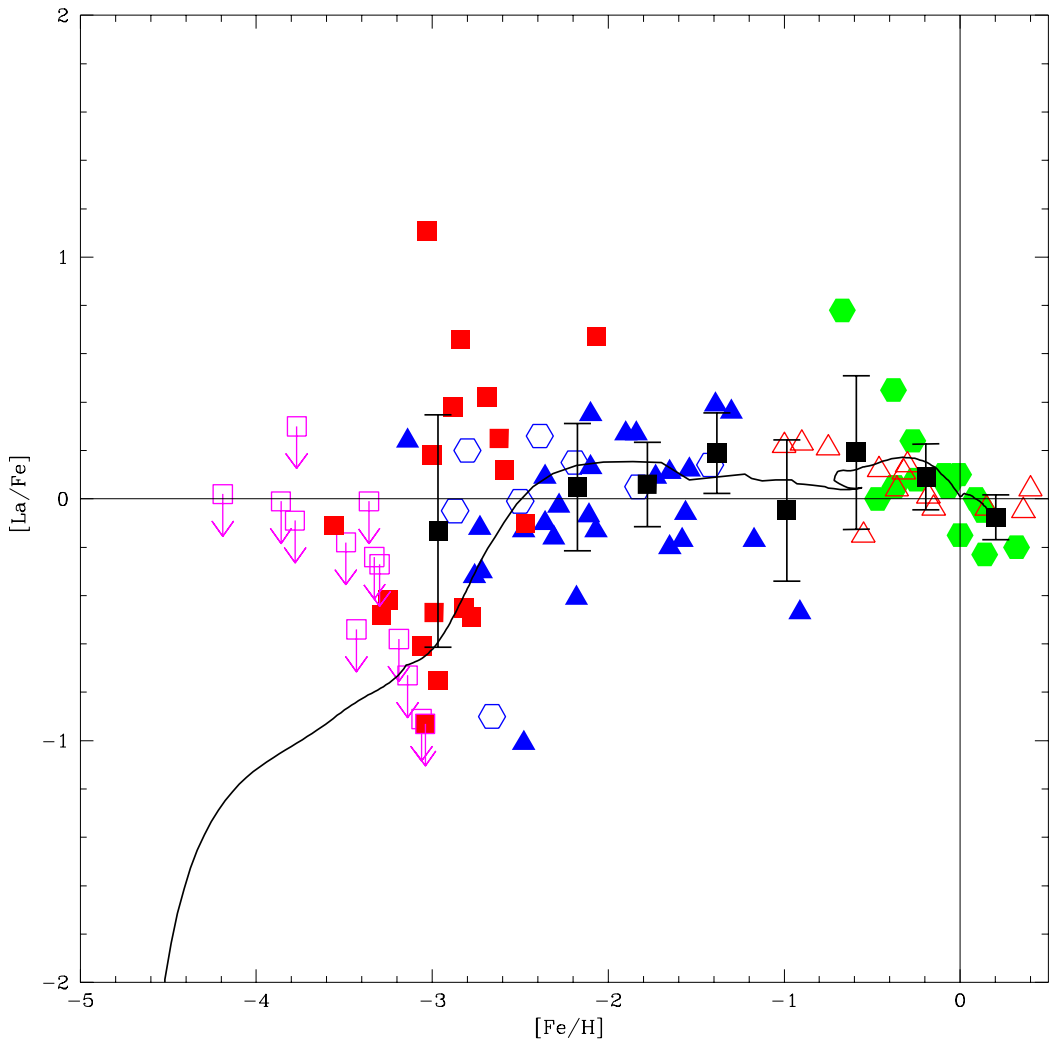}
\caption{[La/Fe] versus [Fe/H]. The data are taken
 from Fran\c cois et al. (2007), (filled red squares, whereas the pink open
 squares are only upper limits), Cowan et al. (2005)
(blue open hexagons), Johnson (2002) and Burris (2000) (blue solid triangles), Pompeia et al.
(2003) (green filled hexagons) and McWilliam \& Rich (1994)  (open red triangles).
 The black squares are  the mean values of the data bins described in the Table \ref{meanLa}.
 As error bars we consider the standard deviation (see Table \ref{meanLa}).
The solid line is the result of our model for La (see Table \ref{rLa}), normalized 
to the solar abundance as measured by Asplund et al. (2005).}
\label{Laresult}
\end{center}
\end{figure}

Unfortunately, for this element there are not so many data. Only recently, 
even if its solar abundance is about a tenth of the Ba solar abundance,
it has been measured because with only one stable isotope, La has more favorable
atomic properties than Ba does, making the abundance analysis
more straightforward. 
The predicted trend of La, with the new adopted yields  well fits 
the observational data in the solar neighborhood.

\section{Strontium, zirconium and yttrium}

In this Section we use the same  methodology used for heavy neutron capture elements
to constrain the yields in massive stars for light neutron capture elements, Sr, Zr and Y.
Light elements are produced also by weak s-process in massive stars (see \ref{intro_nc}).
In this work we do not distinguish between 
r-process production and weak s-process production and we
 rather set a nucleosynthesis related to the whole production 
in the massive stars. Nevertheless,  as Travaglio et al. (2004) have shown,
the fraction of solar Zr and Y, produced by the weak s-process,
 is marginal  and the fraction of solar Sr is less than 10\%. 
Moreover, we do not distinguish, as Travaglio
et al. (2004) do, between a r-process contribution and another primary source,
called in their work LEPP (lighter element primary process).
We do not distinguish these two contributions because, from a theoretical point of view,
it is not clear the difference between the two process.

\subsection{Observational data}

We adopted the new and accurate results by Fran\c cois et al. (2007).
For the abundances in the remaining range of [Fe/H],
 we took published high quality data in the
literature from various sources: Burris et al. (2000), McWilliam et al. (1995),
Fulbright (2000, 2002),  Mashonkina \& Geheren (2001), Johnson (2002),  
Nissen \& Schuster (1997),  Prochaska et al. (2000)  
Gratton \& Sneden (1994), Edvardsson et al. (1993), Stephens \& Boesgaard (2002)
Honda et al. (2004).
All of these data are relative to  solar abundances of Grevesse \& Sauval (1998).

\subsection{Nucleosynthesis prescriptions for Sr, Y and Zr}{\label{NP_SrS}}
For the nucleosynthesis prescriptions of s-process we have adopted
the yields by Travaglio et al. (2004) (see Fig.\ref{sSrYZr}.
We have extended the theoretical results in the mass range $1-3M_{\odot}$,
by simply scaling  the values obtained for stars of $1.5 M_{\odot}$ by the mass.

\begin{figure*}
\begin{center}
\includegraphics[width=0.99\textwidth]{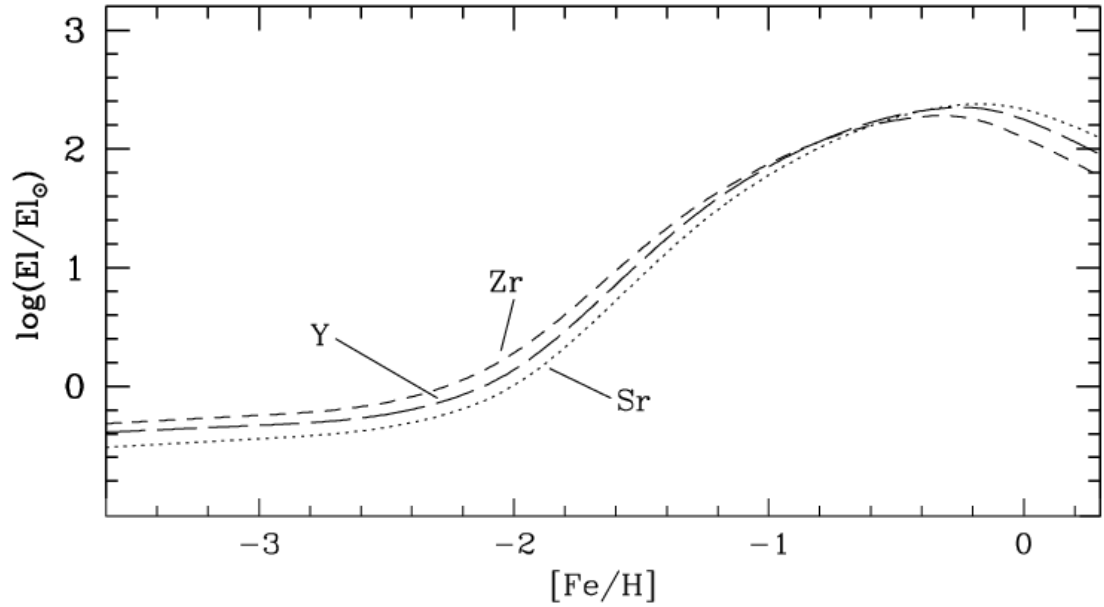}
\caption{Production factors of Sr, Y, Zr 
 respect to the solar value for 1.5 $M_{\odot}$ star
 at different metallicities.} \label{sSrYZr}
\end{center}
\end{figure*}

For the r-process, we give new prescriptions  following the same method 
adopted for Ba in model 1:
 we assume an r-process contribution in massive stars (12 - 30 $M_{\odot}$), besides
the s-process contribution from low mass stars.
The yields are summarized in Table \ref{rSrYZr},
in which the mass fraction of newly produced of Sr, Y, Zr
 is given as function of the mass.

\begin{table*}

\caption{The stellar yields for Sr, Y and Zr in massive stars.} \label{rSrYZr}

\vspace{1.5cm}
\centering
\begin{minipage}{90mm}

\begin{tabular}{|c|c|c|c|}
\hline

$M_{star}$  & $ X_{Sr}^{new}$ & $ X_{Y}^{new}$ & $ X_{Zr}^{new}$\\

\hline\hline

12.   & 1.80$\cdot10^{-6}$& 3.60$\cdot10^{-7}$&1.80 $\cdot10^{-6}$ \\ 
15.   & 7.50$\cdot10^{-8}$& 2.10$\cdot10^{-8}$&1.65 $\cdot10^{-7}$ \\   
30.   & 3.25$\cdot10^{-9}$& 1.00$\cdot10^{-9}$&5.00 $\cdot10^{-9}$ \\

\hline\hline

\end{tabular}

\end{minipage}

\end{table*}

\begin{table*}

\caption{The mean and the standard deviations for the abundance of [Sr/Fe] for the stars 
inside each  bin along the [Fe/H] axis.} \label{meanSrYZr1}

\vspace{1.5cm}

\begin{tabular}{|c|c|c|c|c|}
\hline

bin center [Fe/H]& bin dim.[Fe/H]  & mean [Sr/Fe] & SD [Sr/Fe] &  N. of data in the bin \\
\hline\hline

 -3.78  & 0.80 & -0.86 &   0.68 &    9      \\
 -3.18  & 0.40 & -0.41 &   0.66 &   28      \\
 -2.77  & 0.40 & -0.14 &   0.38 &   29      \\
 -2.37  & 0.40 & -0.43 &   0.42 &   12      \\
 -1.97  & 0.40 & -0.28 &   0.35 &   17      \\
 -1.57  & 0.40 & -0.14 &   0.23 &    8      \\
 -1.16  & 0.40 & -0.06 &   0.15 &    9      \\
 -0.76  & 0.40 &  0.00 &   0.06 &   11      \\
 -0.36  & 0.40 & -0.10 &   0.09 &   21      \\ 
  0.05  & 0.40 & -0.04 &   0.10 &   11      \\

\hline \hline

\end{tabular}

\end{table*}

\begin{table*}
\caption{The mean and the standard deviations for the abundance of [Y/Fe] for the stars 
inside each  bin along the [Fe/H] axis.} \label{meanSrYZr2}

\vspace{1.5cm}

\begin{tabular}{|c|c|c|c|c|}
\hline

bin center [Fe/H]& bin dim.[Fe/H]  & mean [Y/Fe] & SD [Y/Fe] &  N. of data in the bin \\
\hline \hline

 -3.82 &0.74&-0.75&  0.22 &  3\\
 -3.27 &0.37&-0.31&  0.36 & 13\\
 -2.90 &0.37&-0.18&  0.34 & 47\\
 -2.53 &0.37&-0.17&  0.27 & 35\\
 -2.16 &0.37&-0.17&  0.16 & 26\\
 -1.79 &0.37&-0.11&  0.22 & 44\\
 -1.42 &0.37&-0.05&  0.18 & 35\\
 -1.05 &0.37&-0.04&  0.15 & 32\\
 -0.68 &0.37&-0.03&  0.10 & 81\\
 -0.31 &0.37& 0.01&  0.17 & 74\\
  0.06 &0.37&-0.01&  0.14 & 61\\
\hline \hline

\end{tabular}

\end{table*}

\begin{table*}

\caption{The mean and the standard deviations for the abundance of [Zr/Fe] for the stars 
inside each  bin along the [Fe/H] axis.} \label{meanSrYZr3}

\vspace{1.5cm}

\begin{tabular}{|c|c|c|c|c|}
\hline

bin center [Fe/H]& bin dim.[Fe/H]  & mean [Zr/Fe] & SD [Zr/Fe] &  N. of data in the bin \\

\hline \hline

 -3.44& 0.66 &-0.05&  0.34&  10\\
 -2.94& 0.34 & 0.17&  0.31&  30\\
 -2.60& 0.34 & 0.22&  0.27&  27\\
 -2.27& 0.34 & 0.20&  0.21&  23\\
 -1.94& 0.34 & 0.26&  0.26&  23\\
 -1.60& 0.34 & 0.19&  0.18&  29\\
 -1.27& 0.34 & 0.29&  0.18&  19\\
 -0.93& 0.34 & 0.13&  0.15&  25\\
 -0.60& 0.34 & 0.07&  0.15&  44\\
 -0.26& 0.34 & 0.14&  0.20&  41\\
  0.07& 0.34 & 0.02&  0.09&  36\\
\hline \hline

\end{tabular}

\end{table*}

\subsection{Results}

We divide the [Fe/H] axis in several bins  and we compute the mean and the 
standard deviations from the mean of the ratios
[Sr/Fe],[Y/Fe] and [Zr/Fe] for all the data inside each bin.
 These results are shown in Tables \ref{meanSrYZr1},\ref{meanSrYZr2} and \ref{meanSrYZr3},
where we also summarize the center and the dimension of each bin and the number of data points
contained in each of them.
In Fig.\ref{Srresult}, \ref{Zrresult} and \ref{Yresult}  we show  the predictions of the chemical 
evolution  model  for Sr, Y and Zr in the solar neighborhood using our prescriptions for the yields in
 massive stars and the prescriptions by Travaglio et al. (2004) for low mass stars, 
as described in the previous section.
The model well reproduces the trend of the stellar abundances at different [Fe/H] 
as well as the solar abundances. In Table \ref{solarSrYZr} we show the solar mass fractions 
for Sr, Y and Zr  compared to the solar values by Asplund et al. (2005) and Grevesse \& Sauval (1998),
measured in meteorites and the resulting fraction produced by the main s-process 
in low mass stars, for our model. The Zr has a larger contribution by r-process than
Sr and Y. These have almost the same contribution by s-process.
These results are quite different from other chemical evolution models (see Travaglio et al. 2004),
and this is probably due to the different treatment of the coupling 
between halo phase and disc phase.

\begin{table*}
\centering
\caption{Solar abundances of Sr, Y and Zr, as predicted by our model, 
compared to the observed ones from  Grevesse \& Sauval (1998)
and the ones by Asplund et al. (2005). In the last column, the 
resulting fraction of each element  produced by the main s-process 
in the solar abundance}\label{solarSrYZr}

\vspace{1.5cm}

\begin{tabular}{|c|c|c|c|c|c|}
\hline
 & Elem.  &  Meteorites       &   Meteorites     & Model & s-process \\
 & 	  & Asplund et al.    & Grevesse \& Sauval&       &solar fraction\\	
\hline\hline
38 & Sr   &   $2.88 \pm 0.04$ &$2.92 \pm 0.05$   &  2.84 &  68\% \\
39 & Y    &   $2.17 \pm 0.04$ &$2.23 \pm 0.02$   &  2.12 &  63\% \\
40 & Zr   &   $2.57 \pm 0.02$ &$2.61 \pm 0.04$   &  2.58 &  33\% \\

\hline\hline
\end{tabular}
\end{table*}

\begin{figure}
\begin{center}
\includegraphics[width=0.99\textwidth]{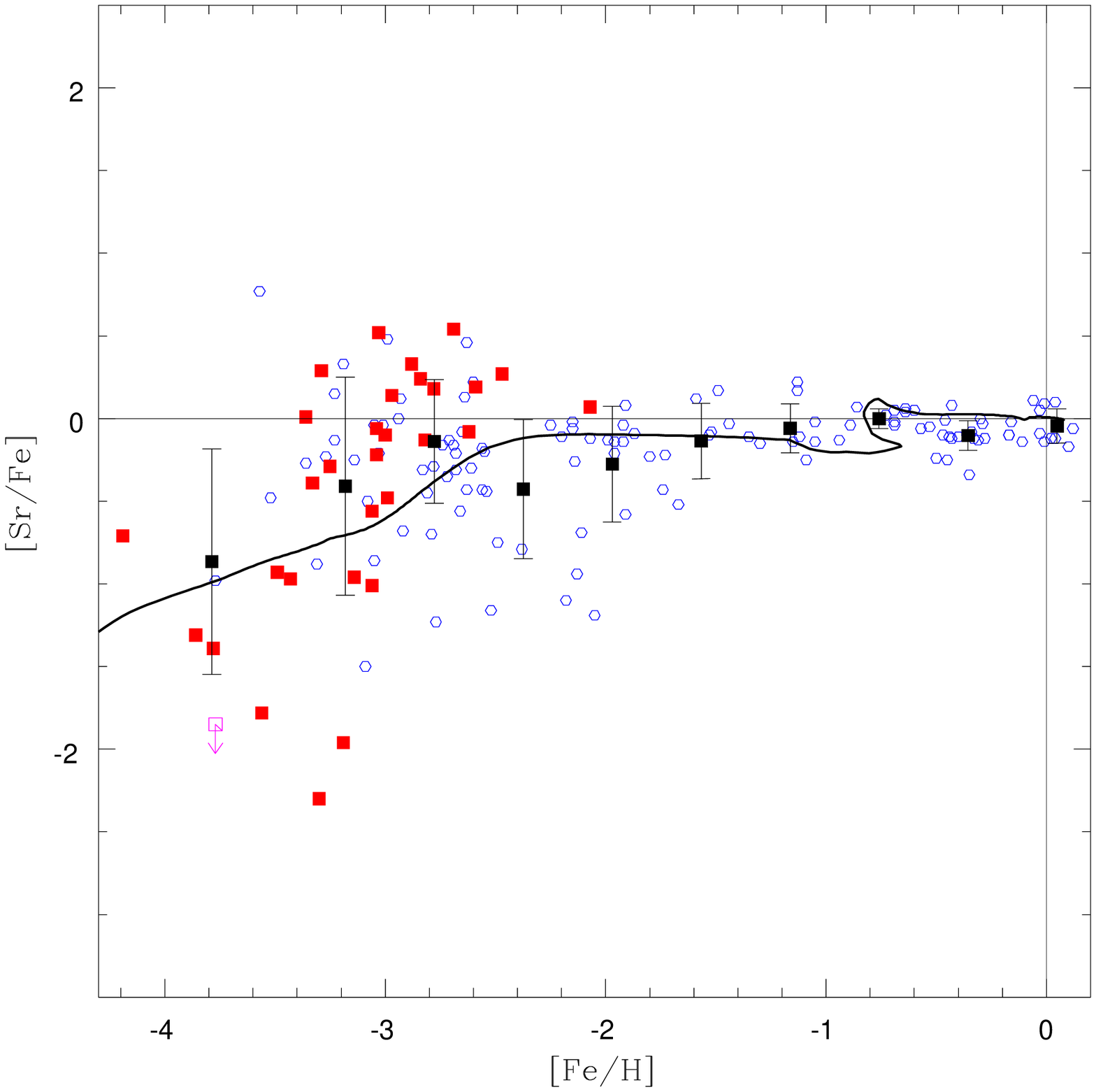}
\caption{ [Sr/Fe] versus [Fe/H]. The data are taken
 from Fran\c cois et al. (2007), (filled red squares, whereas the pink open
 squares are only upper limits) and other authors (blue dots).
 The black squares are  the mean values of the data bins described in the Table \ref{meanSrYZr1}.
 As error bars we consider the standard deviation (see Table \ref{meanSrYZr1}).
The solid line is the result of our model for Sr (see Table \ref{rSrYZr}), normalized 
to the solar abundance as measured by Asplund et al. (2005).}
\label{Srresult}
\end{center}
\end{figure}

\begin{figure}
\begin{center}
\includegraphics[width=0.99\textwidth]{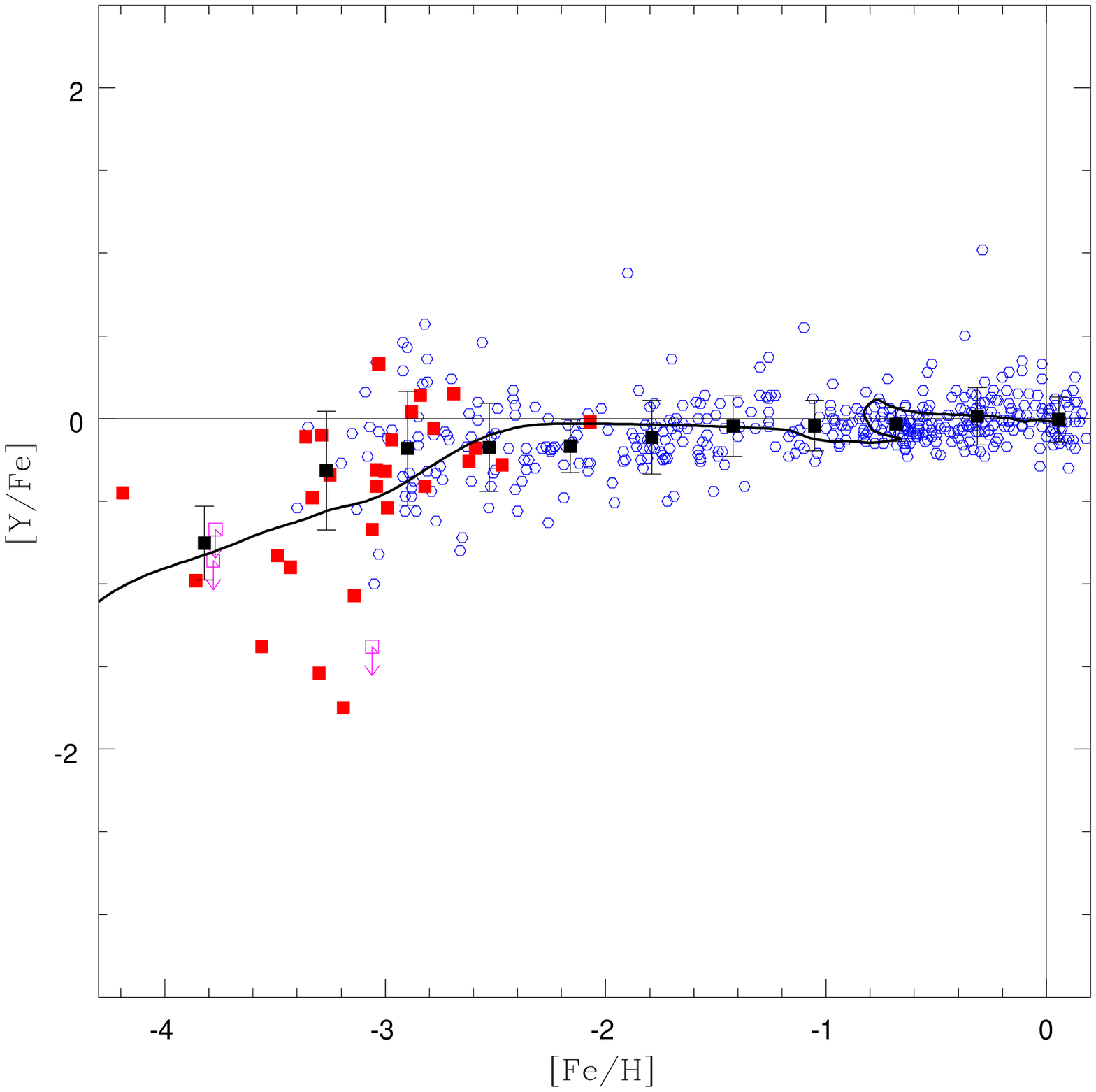}
\caption{[Y/Fe] versus [Fe/H]. The data are taken
 from Fran\c cois et al. (2007), (filled red squares, whereas the pink open
 squares are only upper limits) and other authors (blue dots).
 The black squares are  the mean values of the data bins described in the Table \ref{meanSrYZr2}.
 As error bars we consider the standard deviation (see Table \ref{meanSrYZr2}).
The solid line is the result of our model for Y (see Table \ref{rSrYZr}), normalized 
to the solar abundance as measured by Asplund et al. (2005).}
\label{Yresult}
\end{center}
\end{figure}

\begin{figure}
\begin{center}
\includegraphics[width=0.99\textwidth]{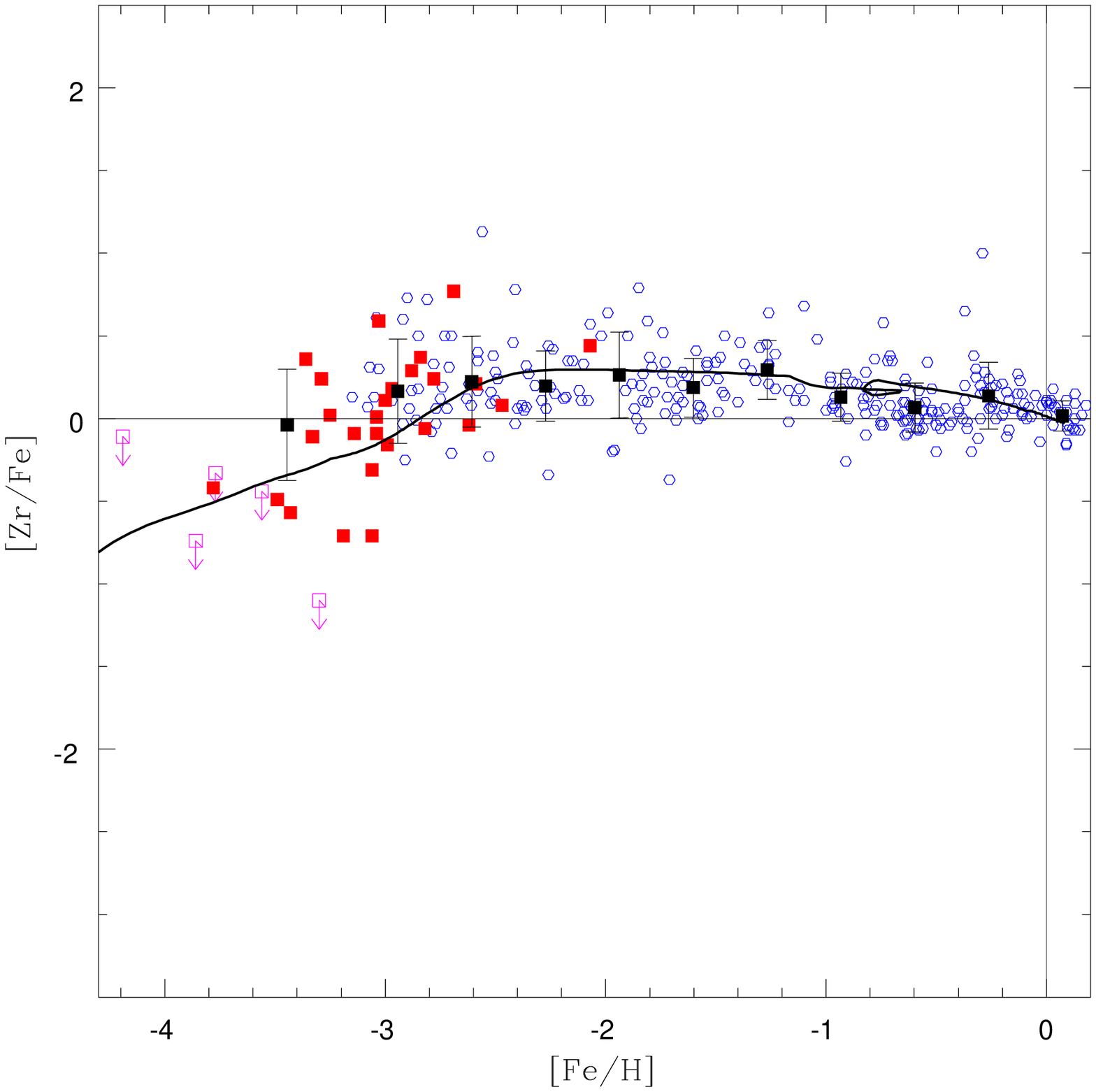}
\caption{[Zr/Fe] versus [Fe/H]. The data are taken
 from Fran\c cois et al. (2007), (filled red squares, whereas the pink open
 squares are only upper limits) and other authors (blue dots).
 The black squares are  the mean values of the data bins described in the Table \ref{meanSrYZr3}.
 As error bars we consider the standard deviation (see Table \ref{meanSrYZr3}).
The solid line is the result of our model for Zr (see Table \ref{rSrYZr}), normalized 
to the solar abundance as measured by Asplund et al. (2005).}
\label{Zrresult}
\end{center}
\end{figure}

Sr is the element which presents the larger spread in the
data at low metallicities, whereas Y and Zr seem to have rather smaller and similar spread.
The mean ratio of [Zr/Fe] is super-solar in the range $-2.5<[Fe/H]<0$, whereas the 
[Sr/Fe] and [Y/Fe] have  mean abundances slightly smaller than solar in the same range.
This difference is due to different contributions of the s-process production
to the solar abundances.  For values of $[Fe/H]< - 2.5$ 
the  ratios of [Sr/Fe], [Y/Fe] and [Zr/Fe] start to decrease.
The yields adopted in this work for Sr, Y and Zr show to well fit the trends of
the observational data in the solar neighborhood over the whole range of metallicities.

\chapter [Abundance gradients in the MW]
{Abundance gradients in the Milky Way for $\alpha$-elements, iron peak elements,
 barium, lanthanum and europium}

\rightline{\emph{``The sky was clear -- remarkably clear --}}
\rightline{\emph{ and the twinkling of all the stars seemed}}
\rightline{\emph{ to be but throbs of one body, timed by a common pulse.''}}
\rightline{\emph{ by Thomas Hardy}}

\vspace{1cm}

We calculate here the behaviour of the abundance gradients of the largest  number of heavy elements
(O, Mg, Si, S, Ca, Sc, Ti, Co, V, Fe, Ni, Zn, Cu, Mn, Cr, Sr, Y, Zr, Ba, La and Eu) 
ever considered in this kind of models.
 We are also able to test the the recent  nucleosynthesis 
prescriptions described in Fran\c cois et al. (2004) for the $\alpha$- and iron peak elements
 and in the previous chapter for neutron capture elements; 

Chemical evolution models adopting the above nucleosynthesis prescriptions have been shown
to reproduce the evolution of the abundances in the solar neighborhood. Here we extend our 
predictions to the whole disk to check whether these prescriptions can also reproduce the abundance
 gradients.
We compare our model predictions  to new  observational data collected 
by Andrievsky et al. (2002abc, 2004) and Luck et al. (2003)
(hereafter 4AL). They measured the abundances of all the selected elements (except Ba)
in a  sample of 130 galactic Cepheids found in the galactocentric distance range from 5 to 17 kpc.
In addition to the data by 4AL we compare our theoretical predictions
with abundance measurements in giants and open cluster located at even larger 
galactocentric distances.

\section{Observational data}\label{data2}

In this work we use  the data by 4AL for all the studied elemental gradients.
 These accurate data have been derived for a large sample of galactic Cepheids.
Cepheids variables have a distinct role in the determination of radial abundance
gradients for a number of reasons. First, they are usually bright enough that they can 
be observed at large distances, providing accurate abundances; second, their distances are generally 
well determined, as these objects are often used as distance calibrators (see Feast \& Walker 1987);
 third, their ages are also well determined, on the basis of relations involving their periods, 
luminosities, masses and ages (see Bono et al. 2005). 
They generally have ages close to a few hundred million years.
We can thus safely assume that they are representative of the present day gradients.
The 4AL sample contains abundance measurements for 130 Cepheid stars located 
between  5 and 17 kpc  from the Galactic center, for all the elements we  want to study
but Ba. 
The advantage of this data is that it constitutes homogeneous sample for a large number
of stars and measured elements. Therefore, the abundance gradients can be more closely traced
with better statistics. Moreover, only for Cepheids it is possible to obtain
abundances for so many elements, as well as a good estimate of the distance, necessary to compute
the gradients.
4AL obtained multiphase observations for the majority of stars.
For the distant Cepheids they used 3-4 spectra to 
derive the abundances, while for nearby stars 2-3 spectra were used.
We also adopt the data by Yong et al. (2006), who
computed the chemical abundances of Fe, Mg, Si, Ca, Ti, La and Eu for 30 Cepheids stars. 
Among these 30 stars, we choose only the 20 which are not in common with the sample
of 4AL. We apply the off-set found by Yong et al. (2006) with respect to 
the work of 4AL, in order to homogenize the two samples.

We use  the data of Daflon \& Cuhna  (2004) to compare the results 
on Cepheids with another class of young objects. Their database contains abundances
of C, N, O, Mg, Al, Si, S for 69 late O- to early B-type star members of 25 OB associations,
open clusters and HII regions. They determine the mean abundances of the different
clusters or associations of young objects. These objects all have ages less than 50 Myr.
Therefore, we assume that they also represent the present day gradients.

To extend the comparison between our model and the 
observational data toward the outer disk,
we also include the datasets of Carraro et al. (2004)
 and Yong et al. (2005); these authors observed distant open clusters up to 22 kpc.
We also show the average values of individual stars belonging to a cluster.
These stars are red giants with an estimated age ranging  from  2 Gyr to 5 Gyr.
We compare these data with the results of our model at the sun formation epoch,
 i.e. 4.5 Gyr before the present time.
Yong et al. (2005) measured the surface abundances of O, Mg, Si, Ca, Ti, Mn, Co,
Ni, Fe La, Eu and Ba for 5 clusters, whereas Carraro et al. (2004) computed the surface abundances 
of O, Mg, Si,Ca, Ti, Ni, Fe in 2 clusters. One of the clusters of Carraro et al. (2004), 
Berkeley 29, is in common with the sample of Yong et al. (2005) and we show  both measurements.
The galactocentric distance of this  object is 22 kpc and hence is the most distant open cluster
 ever observed.

We show the abundances of three field red giants,  which have been identified
 in the direction of the southern warp of the Galaxy by Carney et al. (2005).
In their work, they measure the abundances of O, Mg, Si, Ca, Ti, Mn, Co, Ni, Fe, La, Eu and Ba 
for the three red giants. The galactocentric distance of these object ranges from 10 kpc to 15 kpc.
The age of these three stars is unknown but it is likely that it is similar to the age of the
 red giants measured in the old open clusters. Therefore, we may compare them with the abundances at the 
solar system formation time.

\section{Chemical evolution model for the Milky Way}\label{chemma}

In our model, the Galaxy is assumed to have formed by means of two main infall episodes:
the first forms the halo and the thick disk, the second the thin disk.
The galactic thin disk is approximated by several independent rings, 2 kpc wide, without 
exchange of matter between them.
We assume that the thin disk forms inside-out as first suggested by Matteucci \& Fran\c cois (1989)
and then by Chiappini et al. (2001). In particular, the time scale $\tau_{D}$, 
expressed in Gyr, for the formation of the thin disk at different galactocentric 
distances is:
\begin{equation}
\tau_{D}(r) = 1.033 r - 1.267.
\end{equation}
where  $r$ is expressed in kpc.
Chiappini et al. (2001) have shown that the chemical evolution of the halo can
have an impact on the abundance gradients in the outer parts of the disk.
They  analyzed the influence of the halo surface mass density on
the formation of the abundance gradients  of  O, S, Fe and N at large galactocentric distances.
In their model A, the halo surface mass density is assumed to be constant and equal to  $17M_{\odot}pc^{-2}$
for $R\le8 kpc$ and decreases as $R^{-1}$ outwards. A threshold gas density 
is assumed also for the halo and set to $4M_{\odot}pc^{-2}$.
Then model B has a constant surface mass density  equal to  $17M_{\odot}pc^{-2}$ 
 for all galactocentric distances and the threshold in the halo phase is the same as in 
 model A. 
In their model C the halo surface mass density is assumed as in model A 
but it does not have a threshold in the halo phase.
In their model D, both the halo surface mass density and the threshold are as in model A
but the time scale for the halo formation at galactocentric distances greater than 10 kpc
is set to 2 Gyr and to 0.8 Gyr for distances less than 10 kpc. In all the other models,
the halo timescale is constant for all the galactocentric distances and equal to 0.8 Gyr.
 Here we will show our model predictions for model B.
The differences among model A,B,C and D arise primarily in the predicted steepness
 of the gradients for the outermost disc regions of the galactic disc. 
In this zone, the model B predicts the flattest gradients among the models 
of Chiappini et al. (2001), and provides the best fit according to observed 
flatness in the recent data by 4AL and in the distant open clusters.
Model A is also in good agreement with the abundance gradients traced by Cepheids up to
$\sim$ 12 kpc, whereas for larger galactocentric distances this model tends to be systematically
below observations. We will show the predictions of this model only for the $\alpha$-elements.
Model C shows a trend similar to model A, whereas model D tends to be below the observations 
for galactocentric distances greater than $10kpc$, so we chose to not show their 
predictions.
We recall that below a critical threshold for the gas surface density 
($7M_{\odot}pc^{-2}$ for the thin disk and $4M_{\odot}pc^{-2}$ for the halo phase)
we assume no star formation.
In Fig. (\ref{SFR}) we  show the predicted star formation rate for three different 
galactocentric distances: 4, 8 and 12 kpc;
the SFR is the same for all galactocentric distances during the halo phase,
 due to the fact that the assumed halo mass density in the selected model B
 is not a function of galactocentric distance; the critical threshold of the gas surface 
density  naturally produces  a bursting star formation history in the outer part of the disk,
  whereas at the solar neighborhood, it happens only toward the end of the evolution.
We note that at the solar galactocentric distance, which is assumed to be 8 kpc,
 the threshold also produces a hiatus between the halo phase and the thin disk phase.

\begin{figure}
\begin{center}
\includegraphics[width=0.99\textwidth]{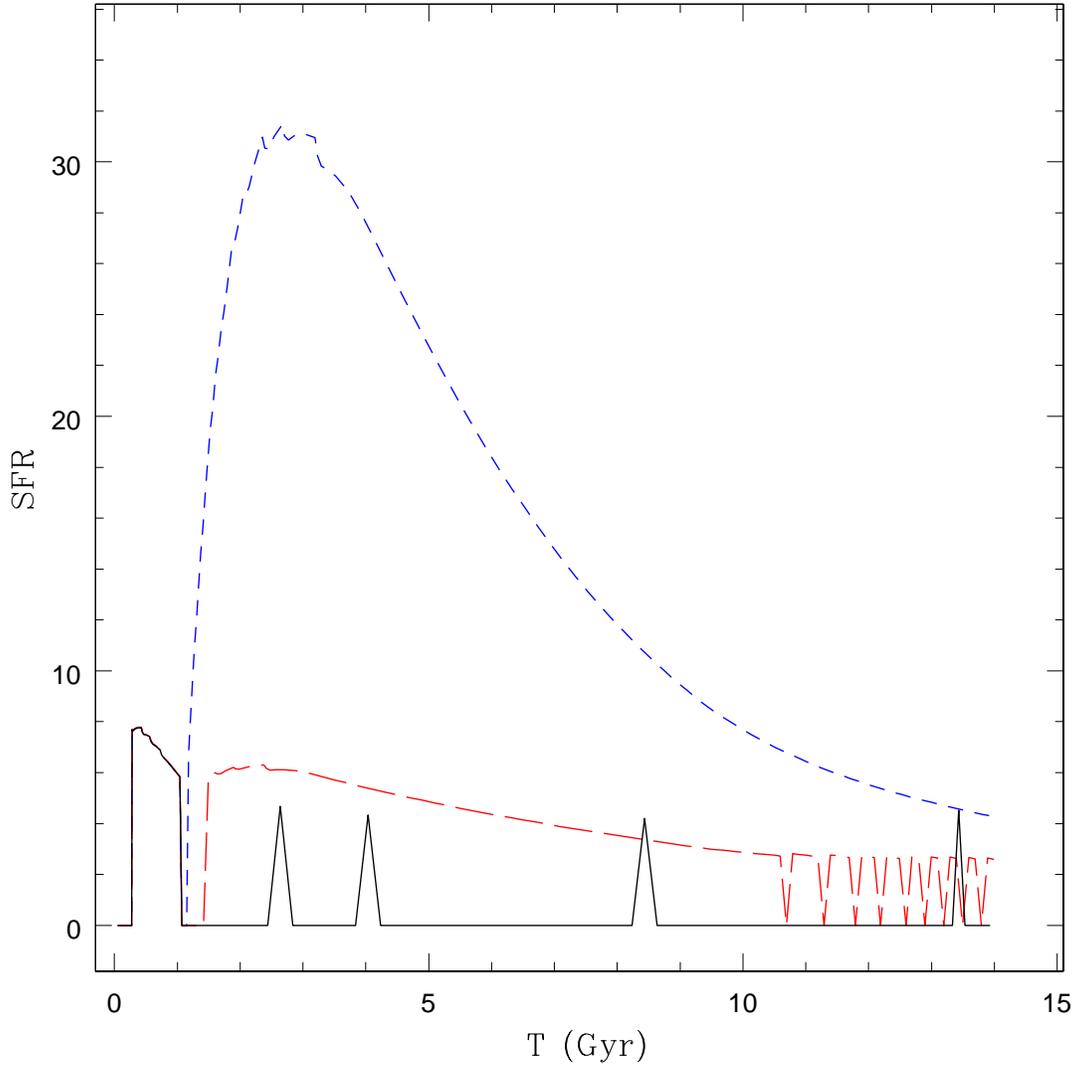}
\caption{ The SFR expressed  in $M_{\odot}pc^{-2}Gyr^{-1}$, as predicted by the two infall model for 
different galactocentric distances: 4 kpc (short dashed line), 8 kpc (long dashed line) and 12 kpc 
(solid line).
The SFR in the halo phase  (indicated by the solid line) up to 0.8 Gyr,
 is the same for all the galactocentric distances, 
whereas in the disk the SFR changes according to the different infall rates. 
Note that at 4 kpc distance the SFR in the disk is much higher 
than at larger galactocentric distances.
The gap in the SFR at the end of the halo-thick disk phase is evident in the solar neighborhood.
The oscillations are due to to the threshold density.}

\label{SFR2}
\end{center}
\end{figure}

\section{Nucleosynthesis prescriptions}{\label{NP2}}

For the nucleosynthesis prescriptions of the Fe and the others elements (O, S, Si, Ca, Mg, Sc, Ti,
 V, Cr, Zn, Cu, Ni, Co and Mn), we adopted those suggested in Fran\c cois et al. (2004). 
They compared theoretical predictions of the [el/Fe] vs. [Fe/H] trends in 
the solar neighborhood for the mentioned above elements and they selected the best sets
of yields to fit the data.
For the yields of SNe II they found that the Woosley \& Weaver (1995) ones
 provide the best fit. No modifications are required
for the yields of Ca, Fe, Zn and Ni as computed for solar chemical composition. For oxygen the best results 
are given by the Woosley \& Weaver (1995) yields computed as functions of the metallicity.
For the other elements, variations in the predicted yields are required to best fit the data
(see Fran\c cois et al. 2004 for details).
For  yields from type Ia SNe, revisions in the theoretical yields by Iwamoto et al. (1999) 
are needed for Mg, Ti, Sc, K, Co, Ni and Zn to best fit the data.
The prescriptions for single low-intermediate mass stars are by van den Hoek \& Groenewegen (1997),
for the case of the mass loss parameter which varies with metallicity (see Chiappini et al. 2003a, model5).
For neutron capture elements we adopt  the nucleosynthesis prescriptions described 
in the previous chapter. For Ba and Eu  the reference model used 
is model 1.
In Table \ref{t:sun} the predicted and observed solar abundances
are compared.  
In the next sections we check whether with the same  
nucleosynthesis prescriptions  our model can explain the data in the other parts
 of the galactic disk.

\begin{table*}
\caption{Element abundances by Asplund et al. (2005) 
in the present-day solar photosphere and in meteorites (C1 chondrites) 
 compared to the results of our model at the solar formation epoch.}
\vspace{1.5cm}

\center

\label{t:sun}
\smallskip
\begin{tabular}{llccc}
\hline\hline

 & Elem.  &  Photosphere  &   Meteorites  & Model \\

\hline

8  & O   &  $8.66 \pm 0.05$   &  $8.39 \pm 0.02$ &  8.67   \\
12 & Mg  &  $7.53 \pm 0.09$   &  $7.53 \pm 0.03$ &  7.57   \\
14 & Si  &  $7.51 \pm 0.04$   &  $7.51 \pm 0.02$ &  7.58   \\
16 & S   &  $7.14 \pm 0.05$   &  $7.16 \pm 0.04$ &  7.20   \\
20 & Ca  &  $6.31 \pm 0.04$   &  $6.29 \pm 0.03$ &  6.25   \\
21 & Sc  &  $3.05 \pm 0.08$   &  $3.04 \pm 0.04$ &  3.05   \\
22 & Ti  &  $4.90 \pm 0.06$   &  $4.89 \pm 0.03$ &  4.90   \\
23 & V   &  $4.00 \pm 0.02$   &  $3.97 \pm 0.03$ &  3.59   \\
24 & Cr  &  $5.64 \pm 0.10$   &  $5.63 \pm 0.05$ &  5.59   \\
25 & Mn  &  $5.39 \pm 0.03$   &  $5.47 \pm 0.03$ &  5.44   \\
26 & Fe  &  $7.45 \pm 0.05$   &  $7.45 \pm 0.03$ &  7.41   \\
27 & Co  &  $4.92 \pm 0.08$   &  $4.86 \pm 0.03$ &  4.88   \\
28 & Ni  &  $6.23 \pm 0.04$   &  $6.19 \pm 0.03$ &  6.23   \\
29 & Cu  &  $4.21 \pm 0.04$   &  $4.23 \pm 0.06$ &  4.13   \\
30 & Zn  &  $4.60 \pm 0.03$   &  $4.61 \pm 0.04$ &  4.53   \\
38 & Sr  &  $2.92 \pm 0.05$   &  $2.88 \pm 0.04$ &  2.84   \\
39 & Y  &   $2.21 \pm 0.02$   &  $2.17 \pm 0.04$ &  2.12   \\
40 & Zr  &  $2.59 \pm 0.04$   &  $2.57 \pm 0.02$ &  2.58   \\
56 & Ba  &  $2.17 \pm 0.07$   &  $2.16 \pm 0.03$ &  2.19   \\
57 & La  &  $1.13 \pm 0.05$   &  $1.15 \pm 0.06$ &  1.11   \\
63 & Eu  &  $0.52 \pm 0.06$   &  $0.49 \pm 0.04$ &  0.56   \\

\hline\hline
\end{tabular}

\end{table*}

\section{Abundance gradients compared with the 4AL data}

We used the model described in Sect.\ref{chemma} to predict the variation of the 
abundances of the studied elements along the galactic disk in the galactocentric range 
5 - 17 kpc, at the present time.
We then compared the abundances predicted by our model at the present time 
for all the elements with the observational data.
To better understand the trend of the data, we divide the data in 6 bins as functions 
of the galactocentric distance. In each bin we compute the mean value and the standard deviations for
all the elements. The results are shown in Table \ref{meanAbunda}:
in the first column we show the galactocentric distance range chosen for each bin,
 in the second column the mean galactocentric distance for the stars inside the considered bin,
 in the  other columns the mean and the standard deviation 
computed for the abundances of every chemical elements, inside the considered bin.
For some stars it has not been possible to measure all the abundances.
We plot the results of our model at the present day normalized to both
the solar observed abundances by Asplund et al. (2005) and to the mean 
value of the abundance data by 4AL at the solar distance. 
We adopt the solar abundances by Asplund et al. (2005) 
because they are more recent. These solar abundances have important differences 
for what concern elements as O and S, compared to the ones by Grevesse \& Sauval (1998),
whereas the differences are negligible for the neutron capture elements.

\begin{table*}

\caption{The mean value and the standard deviation inside each bin for 
 all the elements.}
 
\label{meanAbunda}
\vspace{1.5cm}

\begin{tabular}{|c|c|c|c|c|c|c|c|c|c|c|c||c|c|c|c|c|}
\hline

 galactocentric    & mean GC      &   mean  &   SD    &  mean   &  SD    & mean   &  SD    & mean &  SD       \\
 distance range    & distance(kpc)&  [O/H]  &  [O/H]  &  [Mg/H] & [Mg/H] & [Si/H] & [Si/H] & [S/H]& [S/H]    \\  
\hline	  
         $<$6.5kpc &   5.76       &   0.16  &    0.17 &  -0.19  &  0.17  &  0.21  &  0.13  &  0.37 & 0.19     \\ 
 6.5$<$--$<$7.5kpc &   7.10       &  -0.08  &    0.13 &  -0.19  &  0.10  &  0.07  &  0.06  &  0.17 & 0.08     \\
 7.5$<$--$<$8.5kpc &   8.00       &  -0.06  &    0.13 &  -0.19  &  0.13  &  0.06  &  0.06  &  0.09 & 0.10     \\ 
 8.5$<$--$<$9.5kpc &   8.96       &  -0.12  &    0.16 &  -0.21  &  0.11  &  0.04  &  0.04  &  0.08 & 0.17     \\ 
 9.5$<$--$<$11 kpc &  10.09       &  -0.16  &    0.19 &  -0.22  &  0.18  & -0.07  &  0.07  & -0.11 & 0.20     \\
         $>$11 kpc &  12.33       &  -0.19  &    0.21 &  -0.32  &  0.13  & -0.16  &  0.08  & -0.23 & 0.15     \\
\hline\hline

 galactocentric    & mean GC      & mean   & SD   & mean   &  SD     & mean   &  SD     &   mean   &  SD      \\
 distance range    & distance(kpc)& [Ca/H] &[Ca/H]& [Sc/H] & [Sc/H]  & [Ti/H] & [Ti/H]  &   [V/H]  & [V/H]    \\  
 \hline	  	                   	  	 
 $<$6.5kpc &   5.76       &  0.11  & 0.19&   0.15  &   0.18  &  0.19  &   0.13  &    0.14 &   0.14     \\ 
 6.5$<$--$<$7.5kpc &   7.10       &  0.00  & 0.10&  -0.05  &   0.18  &  0.05  &   0.08  &    0.04 &   0.05     \\
 7.5$<$--$<$8.5kpc &   8.00       & -0.04  & 0.07&  -0.06  &   0.13  &  0.05  &   0.07  &    0.03 &   0.09     \\ 
 8.5$<$--$<$9.5kpc &   8.96       & -0.04  & 0.11&  -0.09  &   0.13  &  0.04  &   0.06  &   -0.01 &   0.09     \\ 
 9.5$<$--$<$11 kpc &  10.09       & -0.13  & 0.09&  -0.12  &   0.09  & -0.05  &   0.14  &   -0.08 &   0.15     \\
 $>$11 kpc &  12.33       & -0.19  & 0.11&  -0.21  &   0.11  & -0.15  &   0.08  &   -0.21 &   0.11     \\
 \hline\hline

 galactocentric    & mean GC      & mean   & SD     &  mean   & SD    & mean   & SD     & mean   & SD     \\
 distance range    & distance(kpc)& [Cr/H]  & [Cr/H]&  [Mn/H] & [Mn/H]& [Fe/H] & [Fe/H] & [Co/H] & [Co/H] \\
 \hline	  	                                     	  	                    
 $<$6.5kpc &   5.76       &  0.11  &   0.11 &  0.09   &  0.14 &  0.17  &   0.13 &  0.06  &   0.13 \\
 6.5$<$--$<$7.5kpc &   7.10       &  0.05  &   0.08 &  0.05   &  0.12 &  0.05  &   0.07 & -0.10  &   0.07 \\
 7.5$<$--$<$8.5kpc &   8.00       &  0.04  &   0.11 &  0.01   &  0.11 &  0.01  &   0.06 & -0.10  &   0.11 \\
 8.5$<$--$<$9.5kpc &   8.96       &  0.03  &   0.12 &  0.00   &  0.09 & -0.01  &   0.08 & -0.06  &   0.11 \\
 9.5$<$--$<$11 kpc &  10.09       & -0.06  &   0.12 & -0.18   &  0.13 & -0.09  &   0.09 & -0.17  &   0.18 \\
 $>$11 kpc &  12.33       & -0.20  &   0.10 & -0.31   &  0.18 & -0.22  &   0.09 & -0.14  &   0.17 \\
 
 \hline

 \end{tabular}
\end{table*}

\begin{table*}

 \begin{tabular}{|c|c|c|c|c|c|c|c|c|c|}
 
 \hline
 galactocentric & mean GC     &  mean  & SD    & mean  & SD    & mean    & SD    &  mean  & SD      \\
 distance range &distance(kpc)& [Ni/H] & [Ni/H]&[Cu/H] &[Cu/H] & [Zn/H]  & [Zn/H]& [Y/H] & [Y/H]    \\
 \hline  	                                                   	  	                  	
 $<$6.5kpc 	   &  5.76 &  0.18 &  0.16 & 0.15  &  0.18 &  0.51 &  0.20   	 &   0.22 &  0.15   \\
 6.5$<$--$<$7.5kpc &  7.10 &  0.02 &  0.07 & 0.07  &  0.12 &  0.28   &  0.10 	 &   0.20 &  0.13   \\
 7.5$<$--$<$8.5kpc &  8.00 & -0.02 &  0.08 & 0.08  &  0.30 &  0.26   &  0.12 	 &   0.18 &  0.10   \\
 8.5$<$--$<$9.5kpc &  8.96 & -0.04 &  0.09 & 0.12  &  0.19 &  0.28   &  0.14 	 &   0.17 &  0.09   \\
 9.5$<$--$<$11 kpc & 10.09 & -0.14 &  0.10 &-0.35  &  0.29 &  0.16   &  0.09 	 &   0.05 &  0.13   \\
 $>$11 kpc         & 12.33 & -0.23 &  0.12 &-0.09  &  0.17 &  0.10 &  0.11   	 &  -0.05 &  0.11   \\

 \hline\hline
 
\end{tabular}

\begin{tabular}{|c|c|c|c|c|c|c|c|}
\hline

 galactocentric & mean GC     & mean   & SD    &  mean   & SD     & mean   & SD    \\
 distance range &distance(kpc)& [Zr/H] & [Zr/H] & [La/H] & [La/H] & [Eu/H] & [Eu/H]   \\
 \hline  	              	    	                   		            
 $<$6.5kpc 	   &  5.76    & -0.01  &   0.13 &   0.21 &  0.10  & 0.17  &   0.14  \\
 6.5$<$--$<$7.5kpc &  7.10    & -0.05  &   0.09 &   0.19  & 0.07  & 0.05 &   0.06   \\
 7.5$<$--$<$8.5kpc &  8.00    & -0.08  &   0.10 &   0.22  & 0.06  & 0.08 &   0.08   \\
 8.5$<$--$<$9.5kpc &  8.96    & -0.04  &   0.10 &   0.26  & 0.08  & 0.08 &   0.10   \\
 9.5$<$--$<$11 kpc & 10.09    & -0.10  &   0.12 &   0.23  & 0.09  & 0.04 &   0.13   \\
 $>$11 kpc         & 12.33    & -0.23  &   0.12 &   0.12 &  0.11  & 0.00  &   0.14  \\
\hline\hline

\end{tabular}

\end{table*}

\begin{center}
\begin{table*}

\caption{Model results for present time  gradients for each element. We show the gradients computed as 
a single slope, for all the range of galactocentric distance considered,  and as two slopes: 
from 4 to 14 kpc and from 16 to 22 kpc.}\label{gradients}
\vspace{1.5cm}
\begin{minipage}{90mm}
\smallskip
\begin{tabular}{l|ccccccccc}
\hline\hline

\hline

 							   &         Fe &     O &    Mg &   Si &     S &    Ca  &   Cu  \\
\hline
 $\frac{\Delta [el/H]}{\Delta R}(dex/kpc)$from 4 to 22 kpc &	-0.036	&-0.028	&-0.031	&-0.033	&-0.034	&-0.034	&-0.050	\\
 $\frac{\Delta [el/H]}{\Delta R}(dex/kpc)$from 4 to 14 kpc &   	-0.052	&-0.035	&-0.039	&-0.045	&-0.047	&-0.047	&-0.070	 \\
 $\frac{\Delta [el/H]}{\Delta R}(dex/kpc)$from 16 to 22 kpc&  	-0.012	&-0.011	&-0.012	&-0.012	&-0.012	&-0.012	&-0.014	\\

\hline\hline
							   &    Zn &   Ni 	&Sc     &Ti	& V	& Cr	& Mn	\\
\hline							                   
 $\frac{\Delta [el/H]}{\Delta R}(dex/kpc)$from 4 to 22 kpc &-0.038 & -0.034 &  -0.036	&-0.032	&-0.038	&-0.036	&-0.038	\\
 $\frac{\Delta [el/H]}{\Delta R}(dex/kpc)$from 4 to 14 kpc &-0.054 & -0.047 &  -0.051	&-0.043	&-0.056	&-0.052	&-0.057	\\
 $\frac{\Delta [el/H]}{\Delta R}(dex/kpc)$from 16 to 22 kpc&-0.012 & -0.012 &  -0.012	&-0.012	&-0.011	&-0.012	&-0.011	 \\

\hline\hline						   
							   & Co	  & Sr	  & Y     & Zr	 & Ba	& Eu	& La   \\
\hline							   	           	       
 $\frac{\Delta [el/H]}{\Delta R}(dex/kpc)$from 4 to 22 kpc &-0.037&-0.020 &-0.020 &-0.022&-0.021&-0.030	&-0.021\\
 $\frac{\Delta [el/H]}{\Delta R}(dex/kpc)$from 4 to 14 kpc &-0.055&-0.016 &-0.016 &-0.019&-0.032&-0.036	&-0.032\\
 $\frac{\Delta [el/H]}{\Delta R}(dex/kpc)$from 16 to 22 kpc&-0.011&-0.010 &-0.010 &-0.013&-0.009&-0.013	&-0.008\\

\hline\hline

\end{tabular}
\end{minipage}
\end{table*}
\end{center}

\subsection{$\alpha$-elements (O-Mg-Si-S-Ca)}

We plot the results for these elements in Fig. \ref{F1} and \ref{F1b}.
There is a discrepancy between our predictions normalized to the solar abundances by
Asplund et al. (2005) and the mean abundance of these elements for Cepheids at
the solar galactocentric distance. The predictions of our model for these elements at the 
present time at 8 kpc are supersolar, whereas the mean abundances of Cepheids 
for Mg, Ca and O are subsolar and this difference is particularly  
evident for Mg.
This means that either our model predicts a too steep increase of 
metallicity in the last 4.5 Gyr or that the absolute abundances of Cepheids are 
underestimated. Some uncertainties in 
the absolute abundances could exist, but the slope of the abundance 
distributions should not be affected.
 If these subsolar abundances were real, then one might think that they are the effect
 of some additional infall episode, occurring in the last 4.5 Gyr.
However, the goal of this work is to reproduce the trend of
the gradients, so here this problem can be neglected and we can compare
the data with the model results  normalized to
 the mean abundances of the Cepheids at 8 kpc.
These results well reproduce the trend of the data for all the five elements.
Moreover, in the case of Si we note that the data show very little spread 
(as the small standard deviation values indicate)
and our model (the one normalized to the mean abundance at 8 kpc) perfectly 
lies over the mean value in each bin.
For S the values predicted by the model for small galactocentric distances 
are slightly low but inside the error bar of the data.
The trend for Ca is nicely followed by our model and the data for Ca show very little spread.
 On the other hand, the trend of Mg shows a shallower 
 slope toward the galactic center than the other $\alpha$-elements.
 This is probably due to the lack of Mg data for the stars located 
from 4 to 6.5 kpc, which determines the steep slope for the other $\alpha$-elements.
The results of the model, if we use the prescriptions for the halo gas density of model A by Chiappini
 et al. (2001) well reproduce the data up to 10 kpc but model B better reproduces the data
for larger galactocentric distances.
 For this reason in the next sections we will show only the results of model B. 
 In Table \ref{gradients}  we show the slopes of the gradients for all the studied elements, 
as predicted by model B at the present time.
 The  gradients become flatter towards the outermost disk regions, 
in agreement with the Cepheid data.
 Each element has a slightly different slope, due to the different production
 timescales and nucleosynthesis processes. In particular, $\alpha$-elements (O, Mg, Si, Ca etc..) 
generally have flatter slopes than the Fe-peak-elements. In addition, there are differences even among the
 $\alpha$-elements such as Si and Ca relative to Mg and O: the slightly steeper slope of Si and Ca is due
 to the fact that these elements are produced also by Type Ia SNe, whereas O and Mg are not. In general,
 elements produced on longer timescales have steeper gradients. 
This is confirmed by the observations not 
only of Cepheids but also of open clusters and HII regions. However, the predicted gradients 
for s- and r- process elements seem flatter  than all the others. The reason for this is that they 
are produced in very restricted stellar mass ranges producing an increase of their abundances at 
low metallicities until they reach a constant value for [Fe/H] $> -3.0$ (see Fig. \ref{Laresult}).
The variations between gradients are small but they may be statistically significant, 
in particular those derived from Cepheids: all the Cepheids have similar  
atmospheric parameters (atmospheric temperature, surface gravity); 
their relative abundances are much less affected than the absolute abundances
 by the effect
of using LTE models instead of recent NLTE, 3D models.
Therefore the gradients from Cepheids seem to be well established.

\begin{figure}
\begin{center}
\includegraphics[width=0.99\textwidth]{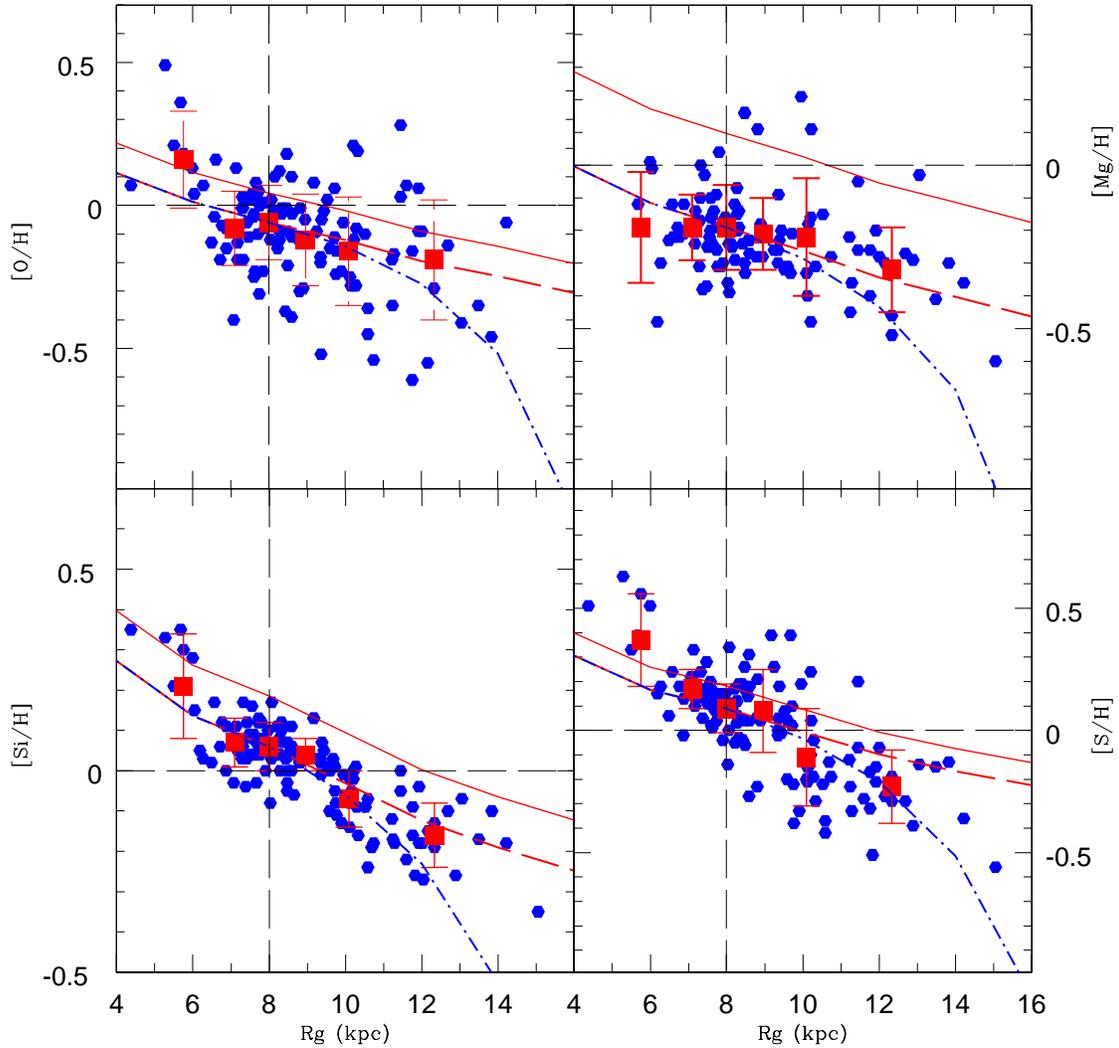}
\caption{Abundances for O, Mg, Si and S as functions of the galactocentric distance.
The blue dots are the data by 4AL, the red squares are the mean values inside each bin only for 
the data by 4AL and the error bars are the standard deviations (see Table \ref{meanAbunda}). 
The thin solid line is our model normalized to the observed solar abundances by Asplund et al. (2005),
 whereas the thick dashed line is normalized to the mean value at the bin centered 
in 8 kpc (the galactocentric distance of the Sun). The dash-dotted
line is the result of the model with the prescriptions for the halo gas density of model A by Chiappini 
et al. (2001) normalized to the mean value of the bin centered in 8 kpc (cfr. Sect.3).}
\label{F1}
\end{center}
\end{figure}

\begin{figure}
\begin{center}
\includegraphics[width=0.99\textwidth]{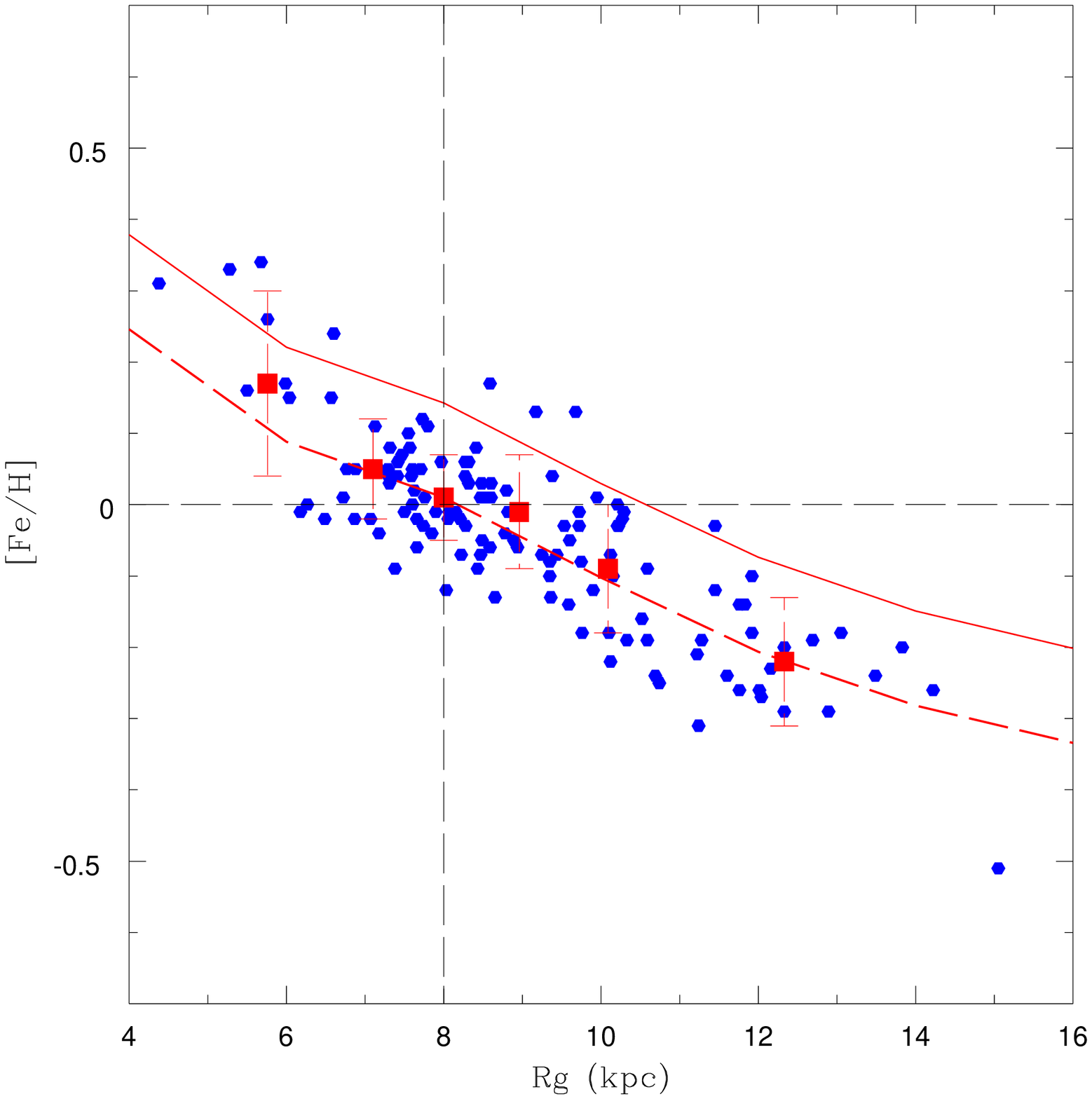}
\caption{The gradient of [Ca/H]. The models and the symbols are the same
as in Fig. \ref{F2b}.}
\label{F1b}
\end{center}
\end{figure}

\subsection{Iron peak elements (Sc-Ti-Co-V-Fe-Ni-Zn-Cu-Mn-Cr)}
The ten elements of the so-called iron peak are plotted in Figs. \ref{F2}, \ref{F2b}, \ref{F3} 
and \ref{F4} (for Cr).
The present time predictions of our model for
 iron peak abundances are super-solar at 8 kpc, as for the $\alpha$-elements.
On the other hand, the mean values for iron peak elements 
in Cepheids in the bin at 8 kpc are in general solar, except Zn,
which is super-solar and Sc and Co, which are sub-solar.
Nevertheless the model gives a prediction for the trends of the gradients for these elements
 which is very good, in particular in the cases of V, Fe, Ni, Mn and Cr, as shown 
by the results of the model normalized to the mean value of the bin centered at 8kpc.
A problem is present for Co, for which a too low abundance is predicted by the model
 at galactocentric distances  $>$ 12 kpc.

\begin{figure}
\begin{center}
\includegraphics[width=0.99\textwidth]{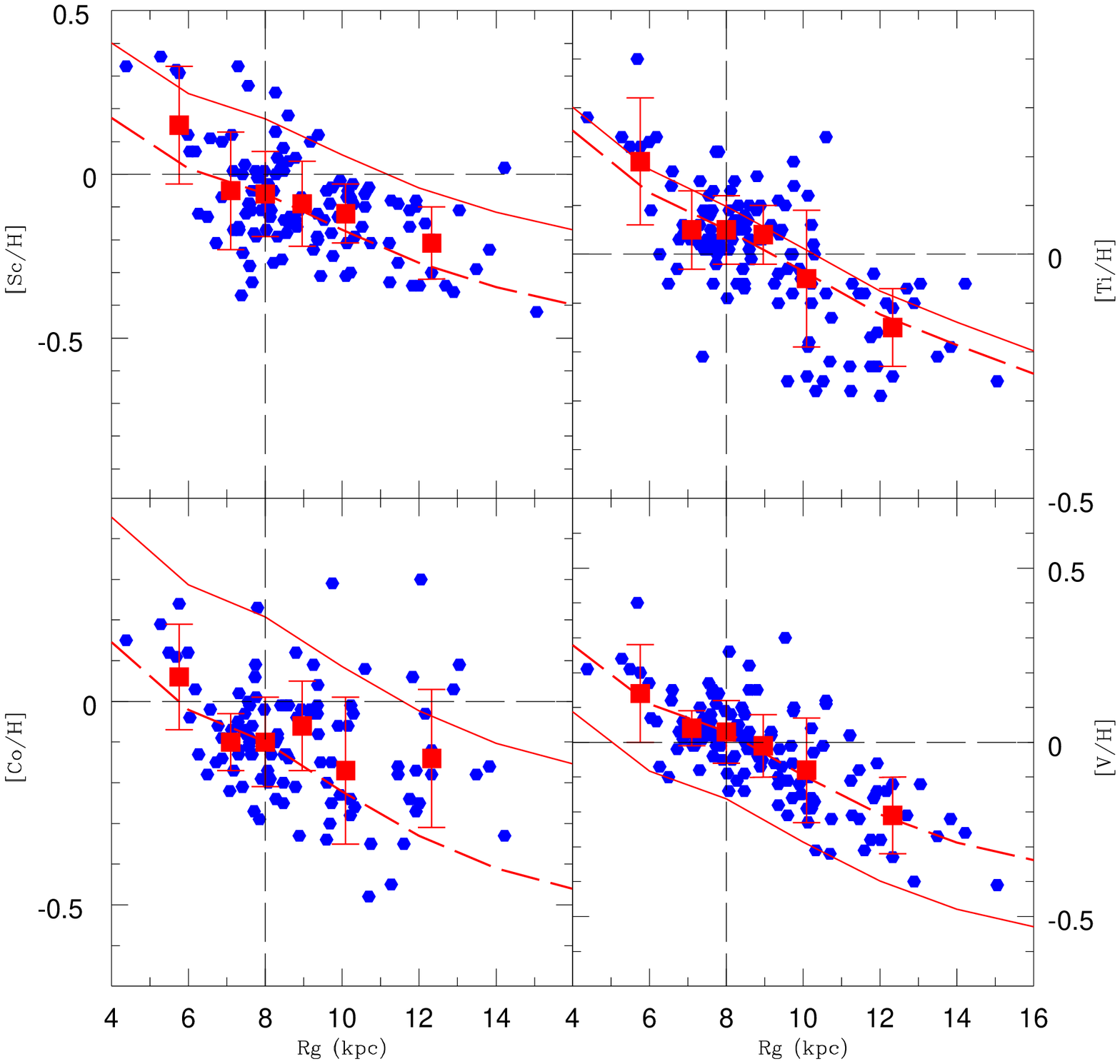}
\caption{Gradients for [Sc/H], [Ti/H], [Co/H] and [V/H].
 The models and the symbols are the same as in Fig. \ref{F1}.}
\label{F2}
\end{center}
\end{figure}

\begin{figure}
\begin{center}
\includegraphics[width=0.99\textwidth]{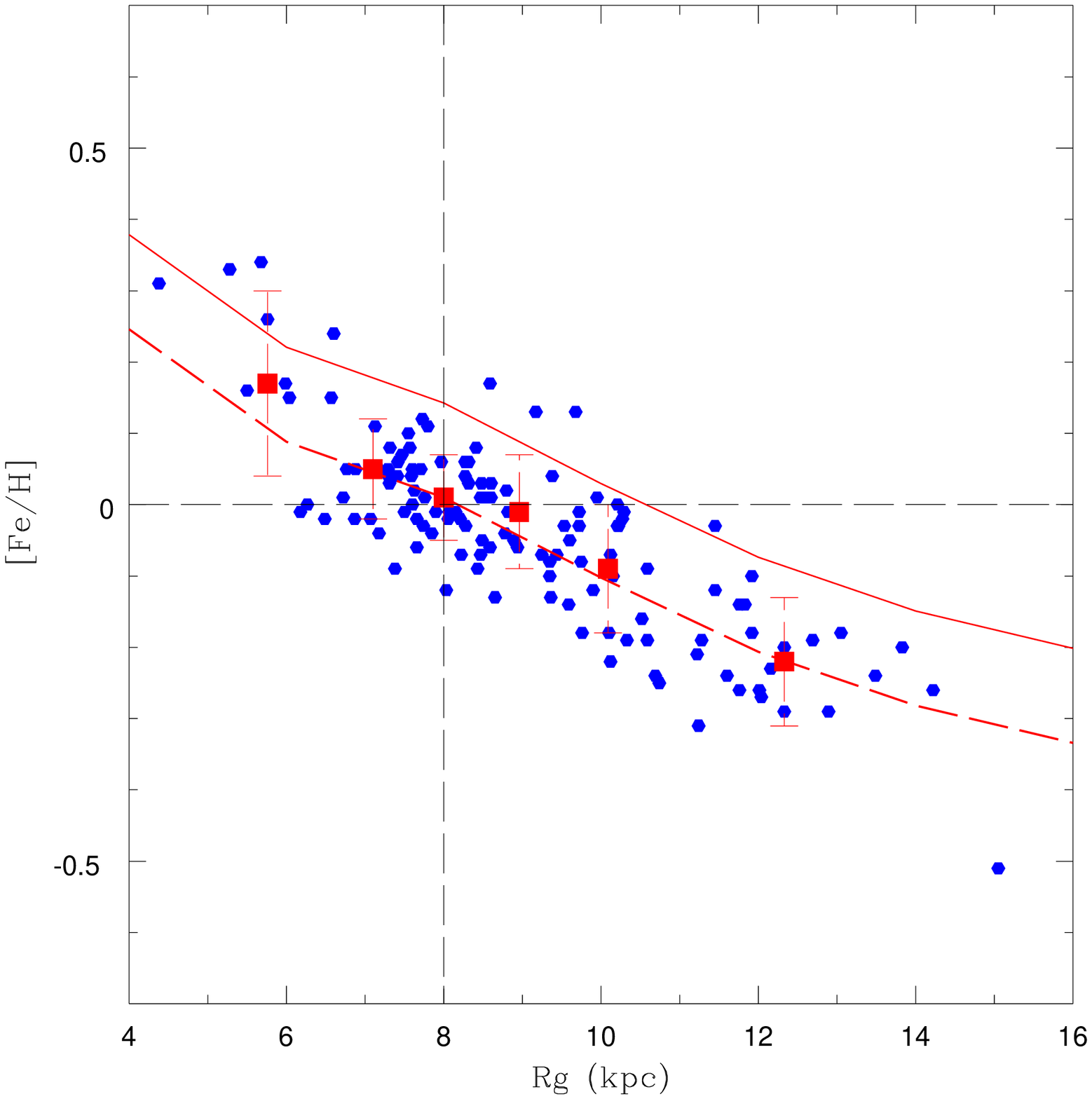}
\caption{The gradient of [Fe/H]. The models and the symbols are the same
as in Fig. \ref{F2b}.}
\label{F2b}
\end{center}
\end{figure}

\begin{figure}
\begin{center}
\includegraphics[width=0.99\textwidth]{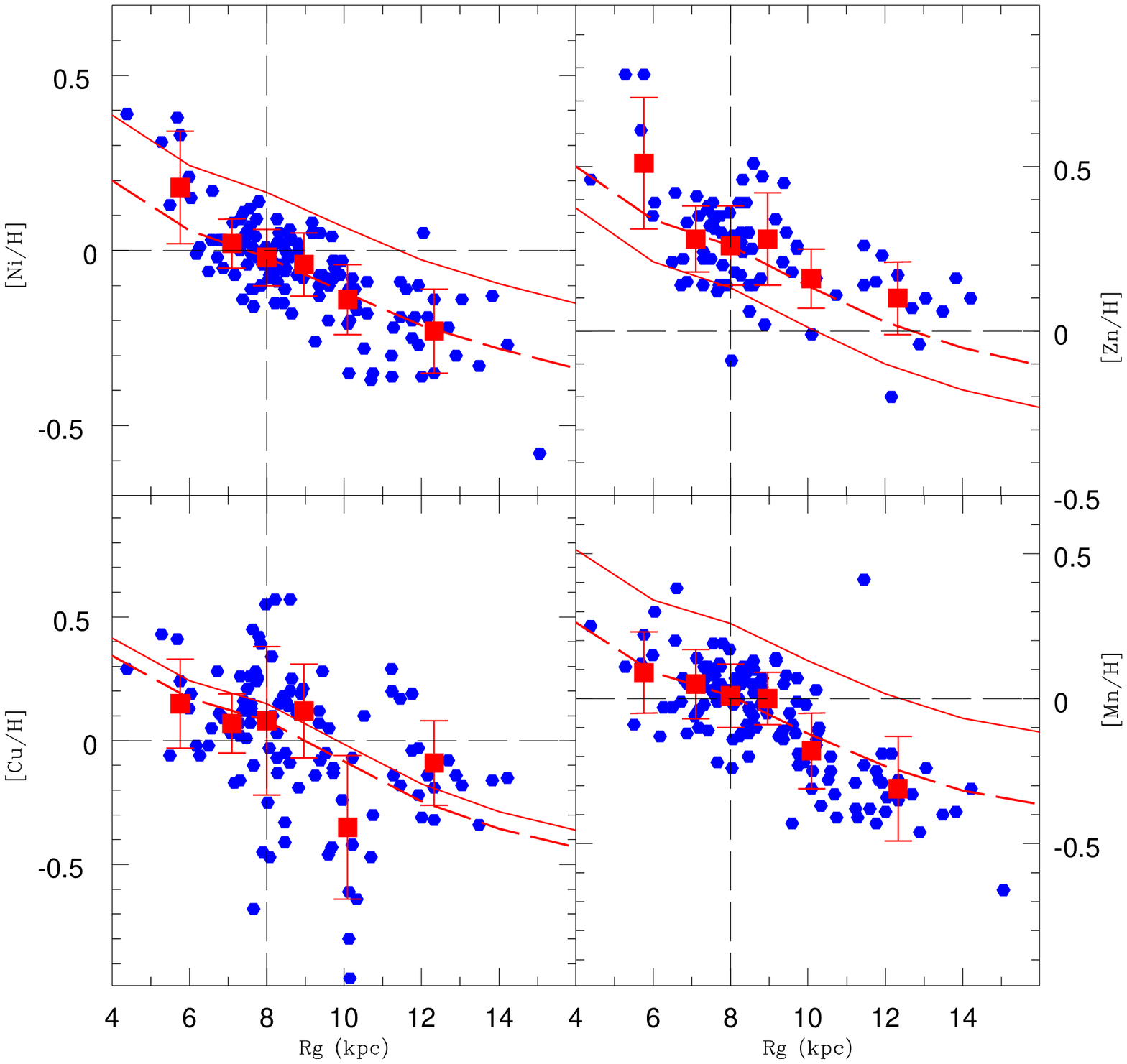}
\caption{Gradients for [Ni/H], [Zn/H], [Cu/H]and  [Mn/H].
 The models and the symbols are the same as in Fig. \ref{F1}.}
\label{F3}
\end{center}
\end{figure}

\subsection{Neutron capture elements (Sr-Y-Zr-Ba-La-Eu)}
As we have seen, neutron capture elements present a large spread 
at low metallicities. However, since the Cepheids are 
young metal-rich stars, this problem is not important.
As shown in Fig. \ref{F4b} and \ref{F4}, the  spread
in the data as a function of galactocentric distance is small. 
In the case of Eu our model well reproduces both the observed gradient and the mean value
for the Cepheid abundance at 8 kpc. 
On the other hand, the mean value of the La abundance in the data by 4AL
at the solar galactocentric distance is about a factor of 1.5 higher than the predicted abundance
by our model and the predictions for La show slightly steeper gradients than the  data.
The model well reproduces  the gradient for  Y and its mean value
for the Cepheids abundance at 8 kpc, as well as
 the gradient for Zr  whereas
 the present  predicted abundance at the solar distance for Zr is slightly higher.
In  Fig. \ref{F4b} and \ref{F4}, we show the predicted trend of the neutron capture
elements Ba and Sr. For these elements there are no data by 4AL;
 therefore, we just show our predictions,
which have to be confirmed by future observations.

\begin{figure}
\begin{center}
\includegraphics[width=0.99\textwidth]{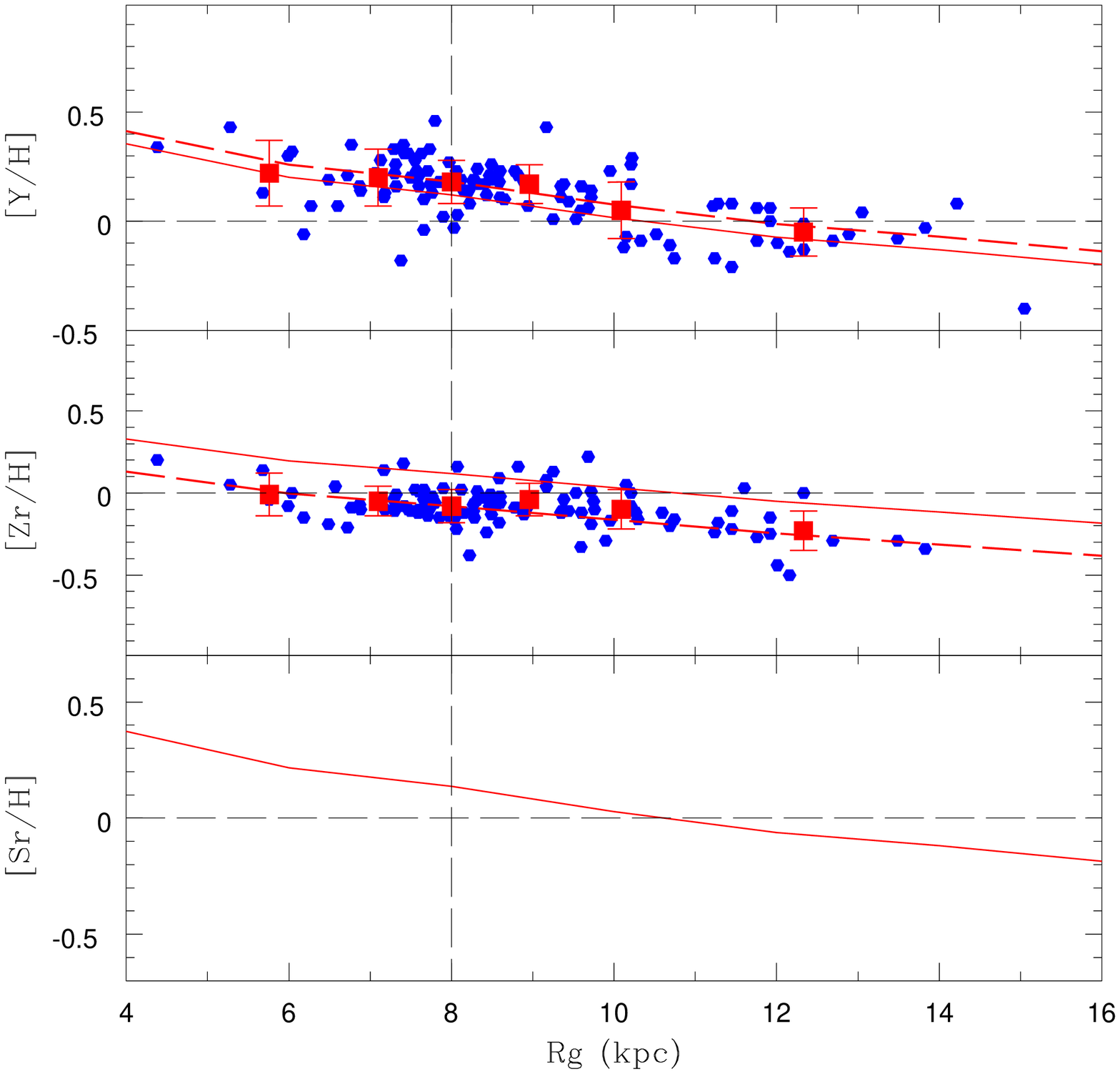}

\caption{Gradients for [Sr/H], [Y/H], and [Zr/H].
 The models and the symbols are the same as in Fig. \ref{F1}.
 Note that for Sr we show only the model predictions.}
\label{F4b}
\end{center}
\end{figure}

\begin{figure}
\begin{center}
\includegraphics[width=0.99\textwidth]{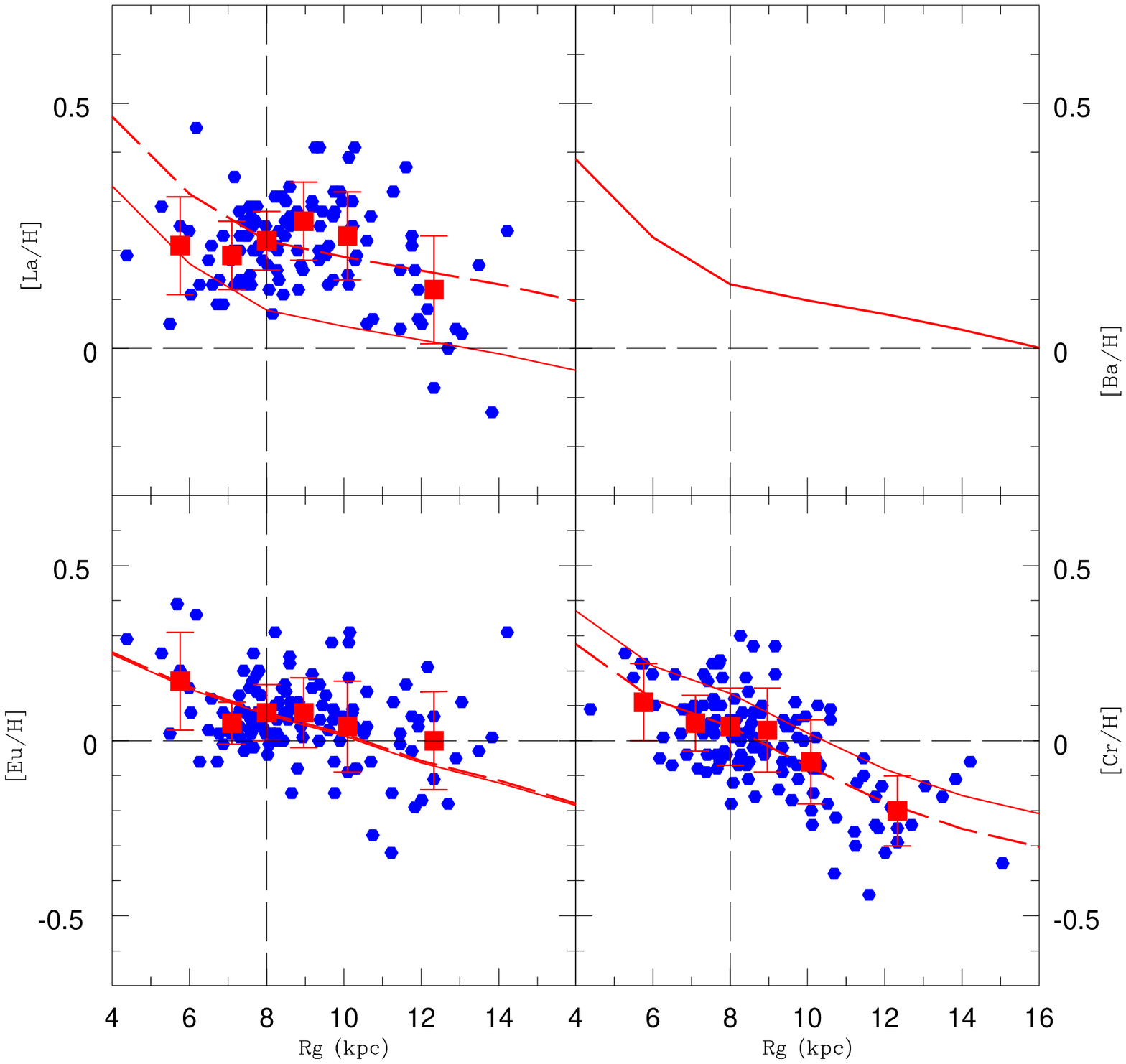}

\caption{Gradients for [La/H], [Ba/H], [Cr/H] and [Eu/H].
 The models and the symbols are the same as in Fig. \ref{F1}.
 Note that for Ba we show only the model predictions.}
\label{F4}
\end{center}
\end{figure}

\section[Comparison with other sets of data]{Predicted abundance gradients compared with other sets of data}

We compare the results of our models with different sets of observational data, as
described in Sect. \ref{data} Only the data by Yong et al. (2006) refer to
Cepheids. However, these data and 4AL data are not homogeneous because of
 the different way in which the abundances are derived. As a consequence of this, we apply 
the offsets calculated by Yong et al. (2006),
on the basis of a representative sample of stars analyzed and measured by both authors,
to compare the two sets of data.
We compare the Cepheids  and the Daflon \& Cunha (2004) data for OB stars
with the model predictions at the present time, normalized to the mean value at 8 kpc for the data
 by 4AL, whereas we compare observational data of red giants and open clusters
with the model predictions at the solar formation time,
 normalized at the  observed solar abundances 
by Asplund et al. (2005).  In Table \ref{gradients2}  we show the slopes 
of the gradients at the solar formation time for all the studied elements, 
as predicted by model B. 
The gradients generally tend to steepen with time (see Table \ref{gradients3}).
 This behavior is expected in the model taking into account the inside-out
formation of the disk, where the external disk regions are
still forming now and the abundance gradient is still building
up. In fact, at early epochs, the efficiency in the chemical
enrichment of the inner Galactic regions is low (owing to
the large amount of primordial infalling gas) leading to a
flat initial abundance gradient. Then, at late epochs, while
the SFR is still much higher in the central than in the
external regions, the infall of metal poor gas is stronger
in the outer than in the inner regions, thus steepening the
gradients. An important role in this scenario is also played by the 
gas density threshold. In fact, the gas density threshold, stopping the SFR,
tends to slow the chemical enrichment in the outskirts
which have a lower gas density compare to the central regions.

\begin{center}
\begin{table*}

\caption{Model results for   gradients at the solar formation time for each element. 
We show the gradients computed as a single slope, for all the range
 of galactocentric distance considered,  and as two slopes: 
from 4 to 14 kpc and from 16 to 22 kpc.}\label{gradients2}
\vspace{1.5cm}
\begin{minipage}{90mm}
\smallskip
\begin{tabular}{l|ccccccccc}
\hline\hline

\hline

 							   &     Fe &     O  &    Mg   &   Si   &     S  &    Ca  &   Cu  \\
\hline
 $\frac{\Delta [el/H]}{\Delta R}(dex/kpc)$from 4 to 22 kpc & -0.027 & -0.024 &	-0.026 & -0.026	& -0.026 & -0.026 & -0.038\\
 $\frac{\Delta [el/H]}{\Delta R}(dex/kpc)$from 4 to 14 kpc & -0.036 & -0.032 &  -0.033 & -0.035	& -0.035 & -0.035 & -0.052\\  
 $\frac{\Delta [el/H]}{\Delta R}(dex/kpc)$from 16 to 22 kpc& -0.016 & -0.019 &  -0.019 & -0.017	& -0.017 & -0.017 & -0.018\\

\hline\hline
							   &    Zn  &   Ni   &   Sc   &  Ti    & V	& Cr	 & Mn	\\
\hline							                   
 $\frac{\Delta [el/H]}{\Delta R}(dex/kpc)$from 4 to 22 kpc & -0.027 & -0.026 & -0.027 &	-0.026 & -0.027	& -0.027 & -0.027 \\	
 $\frac{\Delta [el/H]}{\Delta R}(dex/kpc)$from 4 to 14 kpc & -0.037 & -0.035 & -0.036 & -0.034 & -0.037	& -0.036 & -0.038 \\
 $\frac{\Delta [el/H]}{\Delta R}(dex/kpc)$from 16 to 22 kpc& -0.015 & -0.018 & -0.016 & -0.018 & -0.014	& -0.016 & -0.013 \\

\hline\hline						   
							   & Co	    & Sr     & Y      & Zr	 & Ba	& Eu	 & La   \\
\hline							   	           	       
 $\frac{\Delta [el/H]}{\Delta R}(dex/kpc)$from 4 to 22 kpc & -0.027 & -0.016 & -0.026 &	-0.016 & -0.020	& -0.020 & -0.022 \\
 $\frac{\Delta [el/H]}{\Delta R}(dex/kpc)$from 4 to 14 kpc & -0.037 & -0.020 & -0.033 & -0.020 & -0.016	& -0.016 & -0.019 \\	
 $\frac{\Delta [el/H]}{\Delta R}(dex/kpc)$from 16 to 22 kpc& -0.014 & -0.014 & -0.021 & -0.014 & -0.010	& -0.010 & -0.013 \\

\hline\hline

\end{tabular}
\end{minipage}
\end{table*}
\end{center}

\begin{center}
\begin{table*}

\caption{Model results for  gradients for each element at three different time:
4.5 Gyr, 9.5 Gyr (solar formation time) and 14 Gyr (present time).
We show the gradients computed as a single slope, for all the range
 of galactocentric distance considered.}\label{gradients3}
\vspace{1.5cm}
\begin{minipage}{90mm}
\smallskip
\begin{tabular}{l|ccccccccc}
\hline\hline

\hline

 	&     Fe &     O  &    Mg  &   Si   &     S  &    Ca  &   Cu  \\
\hline
4.5 Gyr & -0.012 & -0.014 & -0.014 & -0.013 & -0.013 & -0.013 &	-0.015\\ 
9.5 Gyr & -0.027 & -0.024 & -0.026 & -0.026 & -0.026 & -0.026 & -0.038\\
14 Gyr  & -0.036 & -0.028 & -0.031 & -0.033 & -0.034 & -0.034 & -0.050\\

\hline\hline
        &    Zn  &   Ni   &   Sc   &  Ti    & V	     &    Cr  & Mn	\\
\hline							                   
4.5 Gyr & -0.011 & -0.013 & -0.012 & -0.014 & -0.010 & -0.012 &	-0.009	\\
9.5 Gyr & -0.027 & -0.026 & -0.027 & -0.026 & -0.027 & -0.027 & -0.027	\\
14 Gyr  & -0.038 & -0.034 & -0.036 & -0.032 & -0.038 & -0.036 & -0.038	\\

\hline\hline						   
	& Co     & Sr     & Y      & Zr	    & Ba     & Eu	 & La   \\
\hline							   	           	       
4.5 Gyr & -0.010 & -0.007 & -0.015 & -0.007 & -0.005 & -0.006 &	-0.011\\
9.5 Gyr & -0.027 & -0.016 & -0.026 & -0.016 & -0.020 & -0.020 & -0.022\\
14 Gyr  & -0.037 & -0.021 & -0.030 & -0.021 & -0.032 & -0.032 & -0.031\\

\hline\hline

\end{tabular}
\end{minipage}
\end{table*}
\end{center}

\subsection{$\alpha$-elements (O-Mg-Si-S-Ca)}

In Fig. \ref{F5} and \ref{F5b},  we show the comparison for O, Mg, Si, S and Ca data with our model. 
Although the observations are from completely different types of 
astronomical objects (OB stars, red giants, open clusters and Cepheids),
 they  are in agreement with each other and with our model. 
Nevertheless, the data by Yong et al. (2006) and the data by Daflon \& Cunha (2004) 
show a larger spread than the data by 4AL, in particular for Ca.
The data by Carraro et al. (2004) for the open cluster Saurer 1 at 
the galactocentric distance of 18.7 kpc for all the considered elements, 
except Mg, are slightly above the predictions of our model.

\begin{figure}
\begin{center}
\includegraphics[width=0.99\textwidth]{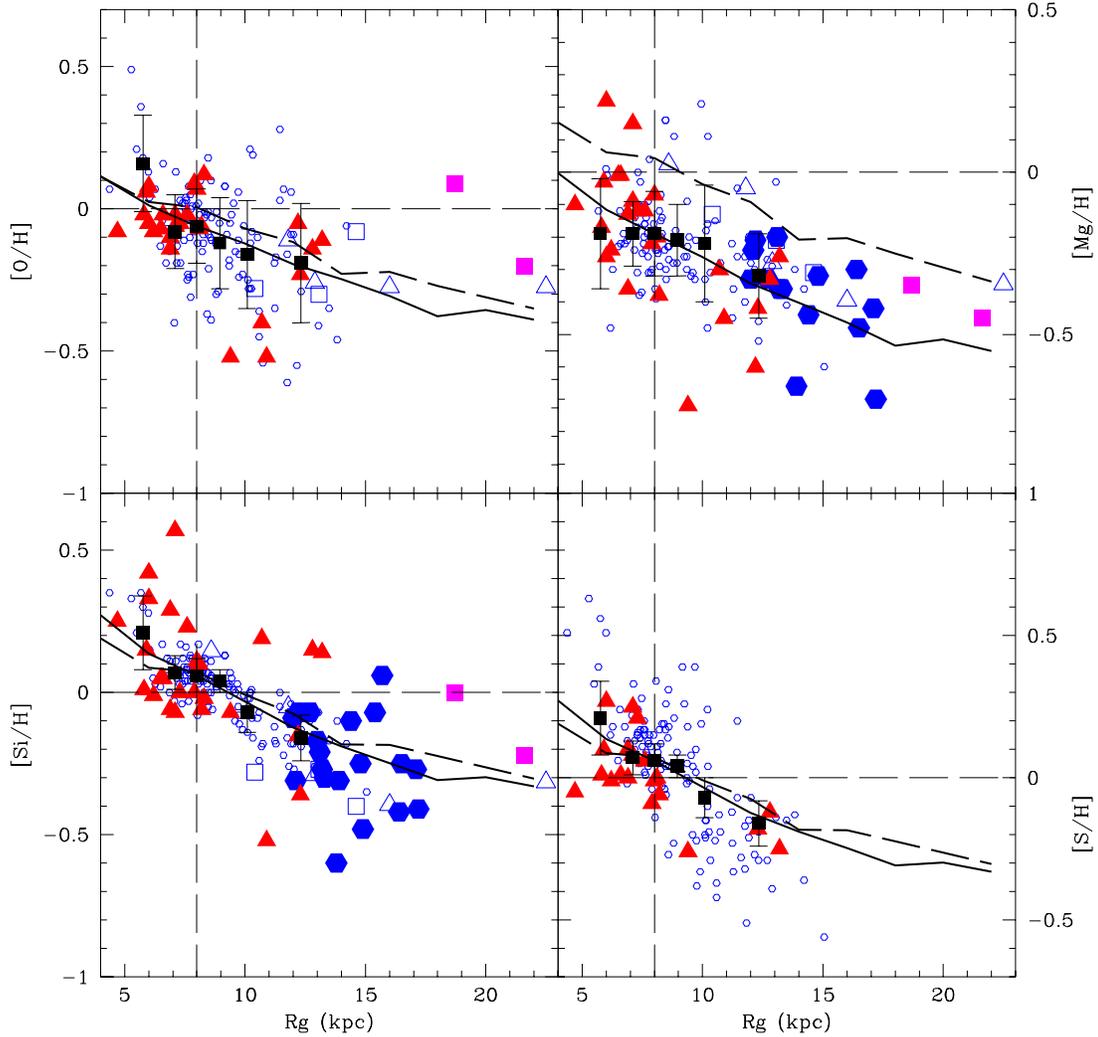}
\vspace{-1cm}
\caption{The gradients for O, Mg, Si and, S  compared with different sets of data.
The small  open circles are the data by 4AL, the black squares are the mean values inside each bin for 
the data by 4AL and the error bars are the standard deviations (see table \ref{meanAbunda}).
The red solid triangles are the data by Daflon \& Cunha (2004) (OB stars), the open blue 
squares are the data by Carney et al. (2005) (red giants),
the  blue solid hexagons are the data by Yong et al. (2006) (Cepheids), the blue open triangles are 
the data by Yong et al. (2005) (open clusters )and the magenta solid  squares are the
 data by Carraro et al. (2004) (open clusters).
The most distant value for Carraro et al. (2004) and Yong et al. (2005)
refers to the same object: the open cluster Berkeley 29.
The thin solid line is our model at the present time normalized to the mean value 
of the bin centered at 8 kpc for Cepheids stars by 4AL; the dashed line represents
the predictions of our model at the epoch of the formation of the solar system normalized
to the observed solar abundances by Asplund et al. (2005).
This prediction should be compared with the data for red giant stars and open clusters 
(Carraro et al. 2004; Carney et al. 2005; Yong et al. 2005).}
\label{F5}
\end{center}
\end{figure}

\begin{figure}
\begin{center}
\includegraphics[width=0.99\textwidth]{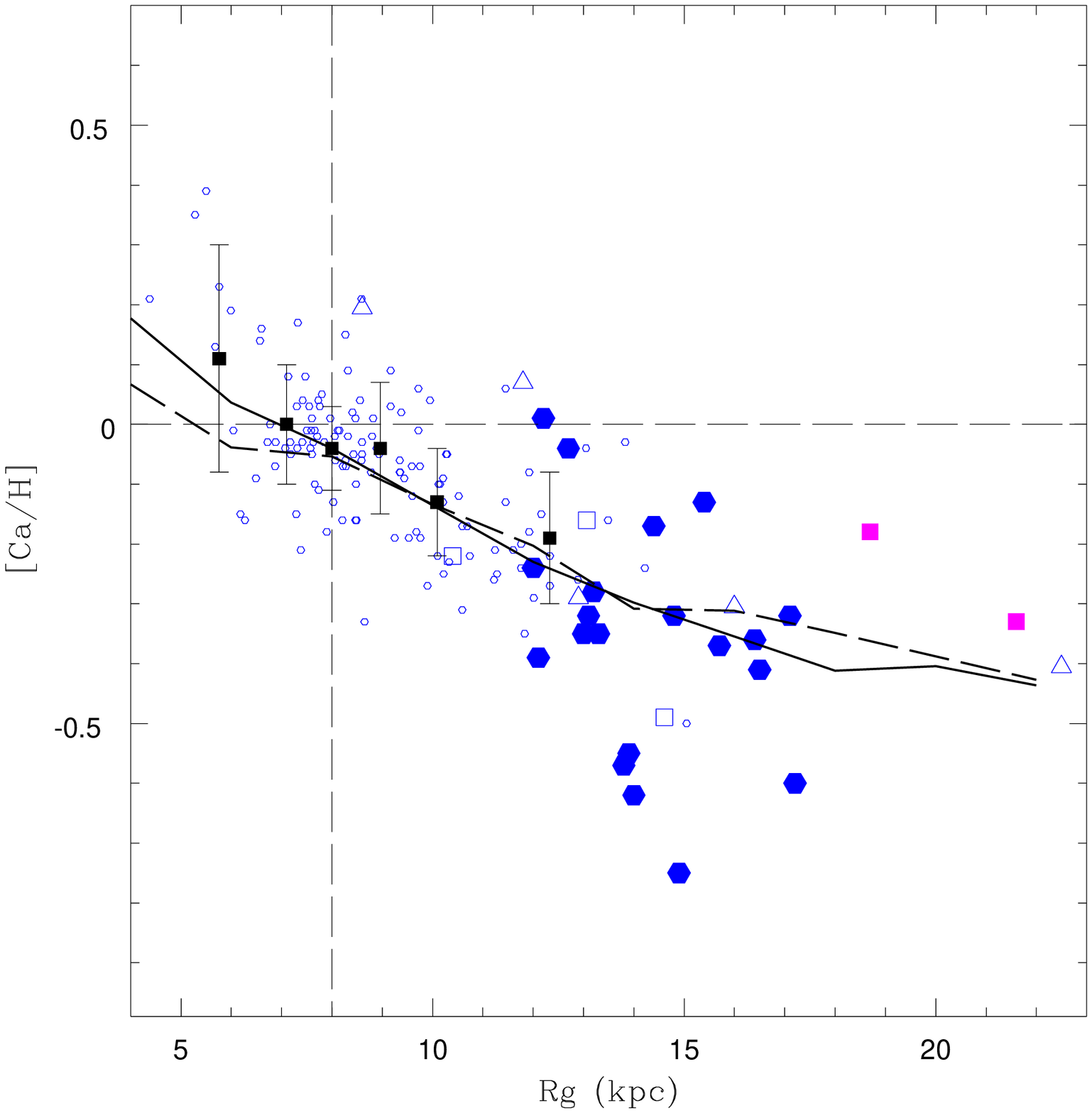}
\caption{The gradient of Ca. The models and the symbols are the same
as  in Fig. \ref{F5}.}
\label{F5b}
\end{center}
\end{figure}

\begin{figure}
\begin{center}
\includegraphics[width=0.99\textwidth]{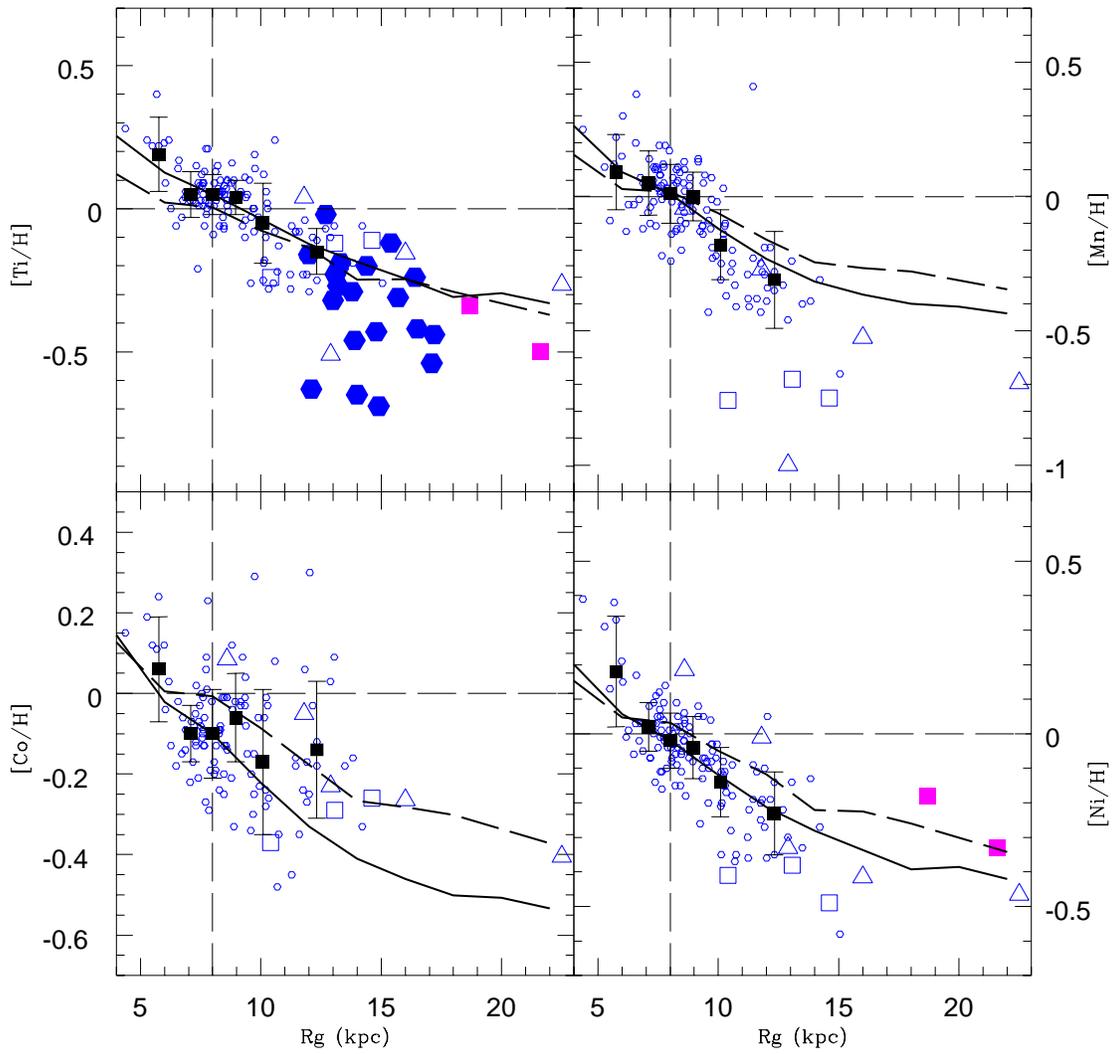}
\caption{Gradients for Ti, Mn, Co and Ni. The model and the symbols are the same
as  in Fig. \ref{F5}.}
\label{F6}
\end{center}
\end{figure}

\begin{figure}
\begin{center}
\includegraphics[width=0.99\textwidth]{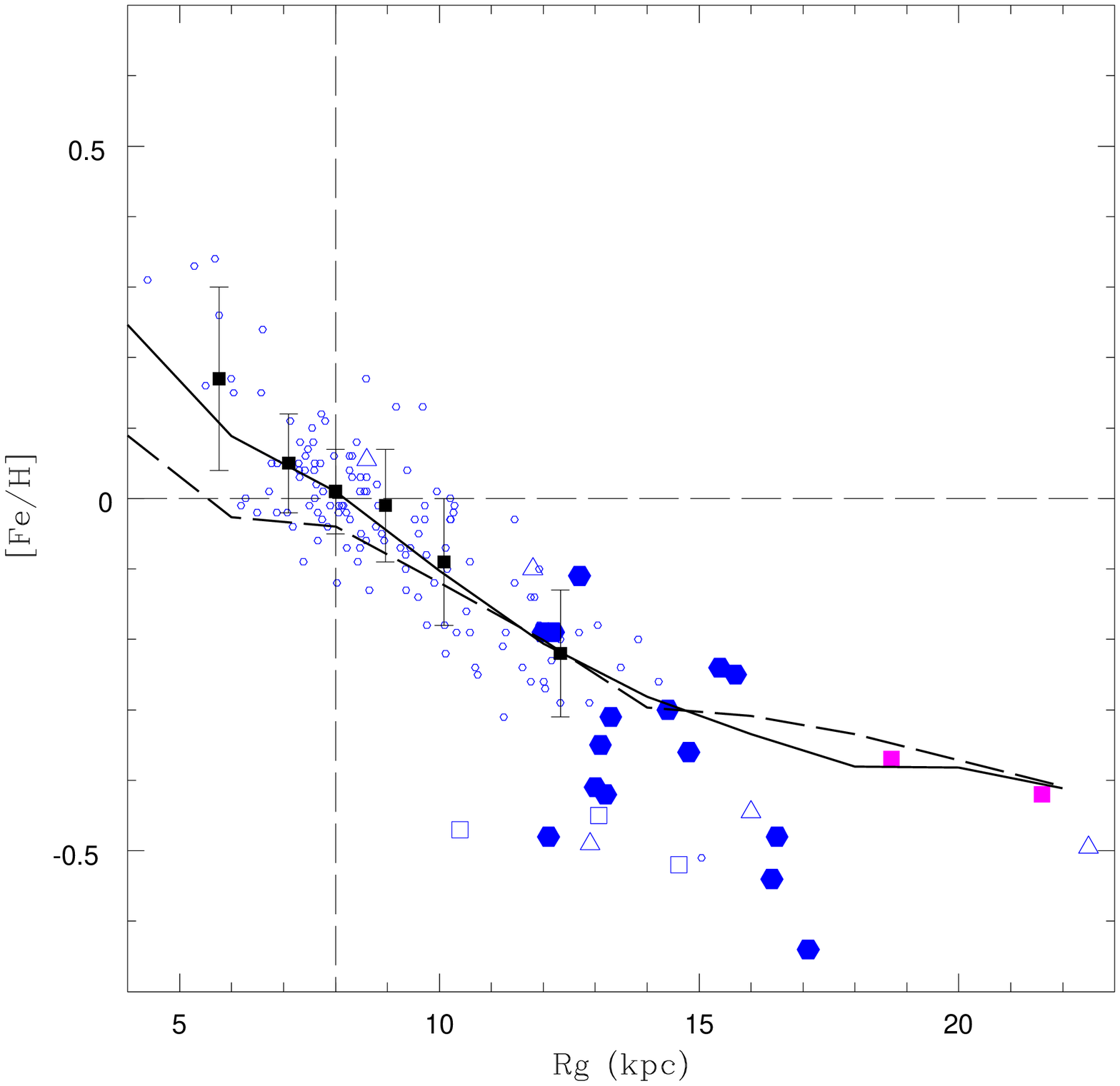}
\caption{The gradient of Fe. The models and the symbols are the same
as  in Fig. \ref{F5}.}
\label{F6b}
\end{center}
\end{figure}

\subsection{The iron peak elements (Ti-Mn-Co-Ni-Fe)}
We show the iron peak elements  in Fig. \ref{F6} and \ref{F6b}. 
The data by Yong et al. (2005)
seem to have a gradient in agreement with our model
if we take into account some possible offset in the data, as considered 
by Yong et al. (2006). In particular, the abundances of Mn in this data set are below
 our model predictions and the 4AL data.
In the data by Yong et al. (2005), the open cluster Berkeley 31,
 which is at about 13 kpc, shows abundances lower than those predicted by our model and 
the set of data by 4AL for all the iron peak elements, with the exception of Co.
On the other hand, the set of data by Carney et al. (2005) shows an almost flat
trend and again lower abundances for the iron peak elements
  than the abundances of the data by 4AL and those predicted by the model.
This is probably due to the fact that the data are from
 old  and evolved objects, as giant stars are, with a not well estimated age.
The data by Yong et al. (2006), which includes the abundances of Ti and Fe,
are in agreement with our model and the data by 4AL, even if they seem
to present  slightly steeper gradients. 
The open cluster abundances as measured by Carraro et al. (2004) are in agreement
with our model, in particular for Fe, whereas for Ti and Ni the model fits both open 
cluster abundances inside the error bars, which is about 0.2 dex.

\begin{figure}
\begin{center}
\includegraphics[ width=0.99\textwidth]{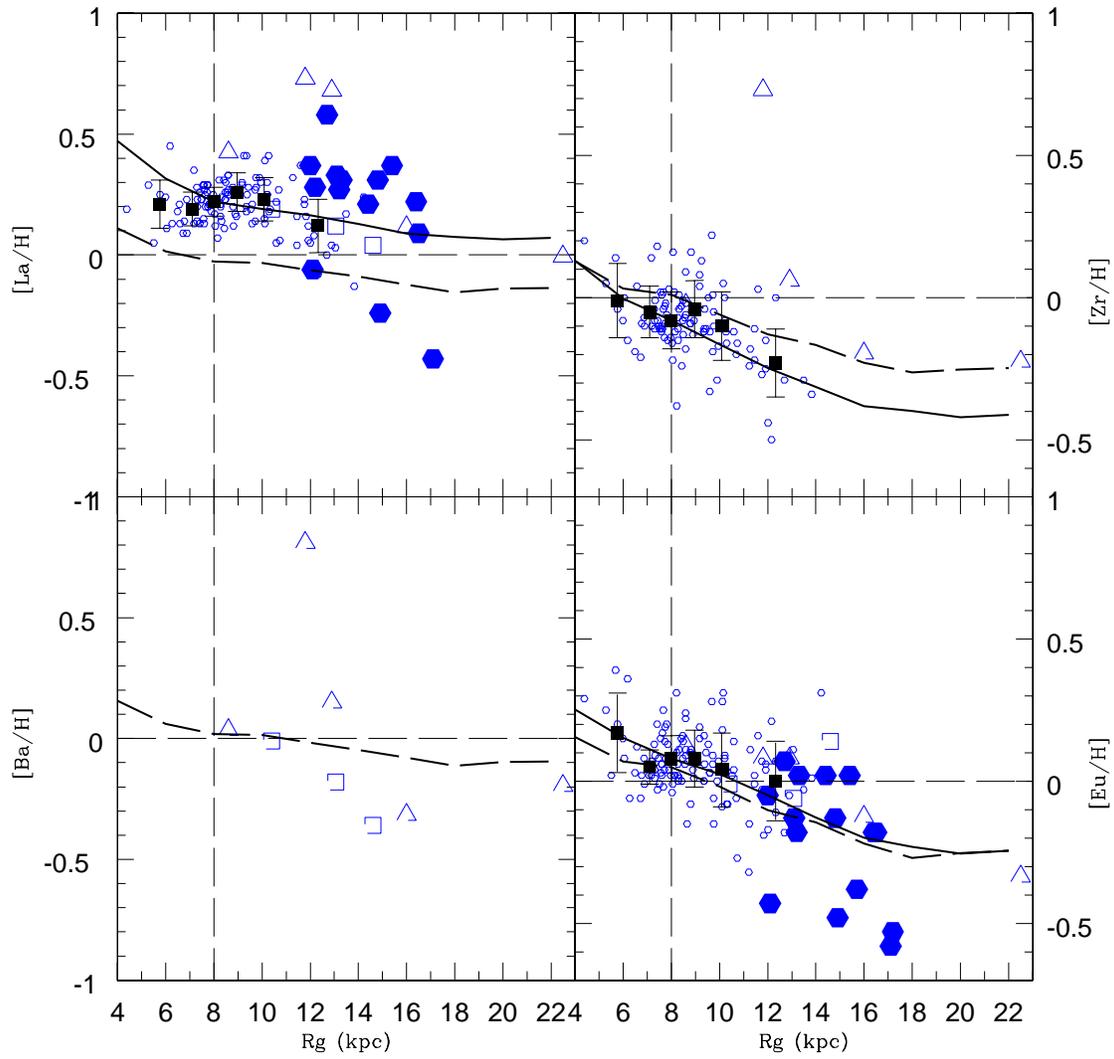}
\caption{Gradients for La, Zr Ba and Eu. The model and the symbols are the same
as  in Fig. \ref{F5}.}
\label{F7}
\end{center}
\end{figure}

\subsection{The neutron capture elements (Zr-Ba-La-Eu)}

We show the neutron capture  elements in Fig. \ref{F7}.
The data for Eu  are taken from the set of data by Yong et al. (2006, Cepheids),
Yong et al. (2005, open clusters) and Carney et al. (2005, red giants),
 and they are in agreement with our
model with the exception of a large spread in the data by Yong et al. (2006). 
Some problems arise for La. The trend of the gradients is similar for the sets of data
but the absolute values of the La abundances in the sets of data by Yong et al. (2006),
Yong et al. (2005) and Carney et al. (2005) are systematically lower od $sim$ 0.3 dex than
the ones of 4AL  and, without the offset, it is impossible  to make a comparison.
Therefore, we apply an offset of +0.3 dex to all
observational data (with the exception of the 4AL data) to better 
show all the sets of data. 
This can also be the consequence of the different way of calculating the abundances,
as explained in Yong et al. (2006). The most important results
are the slopes of the gradients rather than the absolute abundances.
With this offset applied to the data, the two open clusters (Berkeley 31 and NGC 2141),
 measured by Yong et al. (2005),
 still present  an abundance  of La larger than the one predicted by our model and the
 mean abundance 
of the data by 4AL; finally, the data of Yong et al. (2006) again have a large spread.
 Nevertheless, the comparison is acceptable and the abundance
of the most distant cluster is well fitted.

The results for  Zr gradient  at the solar time is 
in reasonable agreement with most of the open clusters data
measured by Yong et al. (2005). The absolute abundances of the the data 
are just slightly larger than the model results for three clusters, whereas  
Berkeley 31 and NGC 2141 present again an 
abundance of Zr larger than the one predicted by our model.
The results for Ba are similar to those for La. We have to
 apply an offset of +0.3 dex to the data by Carney et al. (2005),
for the reasons explained above. The results are also
 quite good with the exceptions of the two open clusters
mentioned before, which show a larger Ba abundance when compared 
to the results of our model.
For both these open clusters there is only one measured star 
and  it is possible that the stars chosen to be analyzed could be 
peculiar in terms of chemical abundances of s-process
elements and so they should not be considered  in deriving the gradients.

The data in the outer parts of the disk are
 still insufficient to completely constrain our
models. Moreover, the existing samples show scatter. Two factors can affect the abundance 
gradients in the outer parts: observational uncertainties in both abundances and distance, and 
the fact that the outer parts could reflect a more complex chemical evolution. Moreover,
there are some suggestions that the open clusters and giants in the outer part of the disk
could have been accreted. However, despite these uncertainties, our chemical 
evolution model, where the halo density is assumed constant with radius out
to $\sim$ 20 kpc, predicts abundance gradients in agreement with those measured
 in the outer disk.

\chapter [Inhomogeneous model for the Galactic halo]
{Inhomogeneous model for the Galactic halo: a possible explanation 
for the spread observed in s- and r-process elements}

\rightline{\emph{``There are mysteries which men can only guess at,}}
\rightline{\emph{ which age by age they may solve only in part.''}}
\rightline{\emph{by Bram Stoker}}
\vspace{1cm}

The neutron capture elements observed in low metallicity stars 
in the solar vicinity show a large spread,
compared to the small star to star scatter 
observed for the $\alpha$-elements.
In chapter 2 we have tried to explain this spread by means of different stellar yields
 for stars with the same mass, in the framework of a homogeneous model.
In this chapter  we show the results of a stochastic
chemical evolution model that we have developed in order to explain the
spread, following the suggestion of McWilliam et al. (1995).

\section{Observational data}
In this section we consider also the $\alpha-elements$ and not only the r-s process
elements. For the extremely  metal poor stars (-4$<$[Fe/H]$<$ -3), we adopted the recent
results from UVES Large Program "First Star'' (Cayrel  et al. 2004, Fran\c cois et al. 2007).
For the abundances in the remaining range of [Fe/H], we took published high quality data in the
literature from various sources: 
Beers et al. (1999), Burris (2000), Carney et al. (1997), Carretta et al. (2002), Cowan et al. (2005),
Edvardsson et al. (1993), Fulbright (2000, 2002), Gilroy et al. (1998), Gratton and Sneden (1988, 1994),
Honda et al. (2004), Ishimaru et al. (2004), Johnson (2002), Koch \& Edvardsson (2002), 
Mashonkina \& Gehren (2000, 2001), McWilliam et al. (1995), McWilliam \& Rich (1994), 
Nissen and Schuster (1997), Pompeia et al. (2003), Prochaska et al. (2000), Ryan et al. (1991, 1996), 
Stephens (1999)  and Stephens \& Boesgaard (2002).

\section{Inhomogeneous chemical evolution model for the Milky Way halo}

We model the chemical evolution of the halo of the Milky Way 
for the duration of 1 Gyr. We consider that the halo has formed by means of the assembly of 
many independent regions each with 	a typical volume of $10^{6}pc^{3}$.
Each region does not interact with  the
others. Inside each region the mixing is assumed to be instantaneous.
In each region we assume an infall episode with a timescale of 1Gyr and a threshold 
in the gas density for the star formation. 
We assume a timestep of 1 Myr.
When the threshold density is reached, the mass of gas which is
 transformed at each timestep in to stars, $M_{stars}^{new}$, 
is assumed to be proportional to $\rho_{gas}^{1.5}$.
We stop the star formation in each region when 
the total mass of the newly formed stars exceeds $M_{stars}^{new}$.
The mass of each star is assigned with a random function in the range between
0.1 and 80 $M_{\odot}$, weighted according to the IMF of Scalo (1986).
In this way, in each region, at each timestep, the $M_{stars}^{new}$ is the same but 
the total number and the masses of the stars are different.
At the end of its life, each star enriches the gas  with
 newly produced elements (see the next section), as a function of its mass and metallicity.
We calculate the cooling  timescale for SN bubbles in our environment at different times.
We found that the cooling timescale is always smaller than the typical timestep of our simulation,
so we decided to neglect this delay time  and the 
mixing in each simulated box is instantaneous.
The stellar lifetime is calculated as a function of the stellar mass, as described in the 
previous chapters. The model does not take in account the pollution produced by stars 
with mass $<3M_{\odot}$ because their lifetimes exceed the time considered in the model.

The existence of SNeIa is also taken into account, according to the prescriptions
 of Matteucci \& Greggio (1986)
In figure \ref{inomo1}, the SNeIa rate is compared to the SNeII rate.
Due to the threshold in the gas density that we impose, the star formation starts only after 250 Myr and 
at this stage the first SNeII start to explode. With a time delay of about 30Myr the first SNeIa
take place. In our model we consider the single degenerate scenario and  30Myr is 
the shortest timescale for a SNIa to explode, being the lifetime of the 
most massive progenitor leading to the formation  of  a C-O white dwarf.
We stop the star formation at 1 Gyr and the SNeII rate fall abruptly to zero, correctly,
being the longest lifetime of a SNII  30Myr, whereas the SNeIa continue to explode, being the lifetime
of the progenitors of a SNIa as long as 10 Gyr.
\begin{figure}
\begin{center}     
\includegraphics[width=0.99\textwidth]{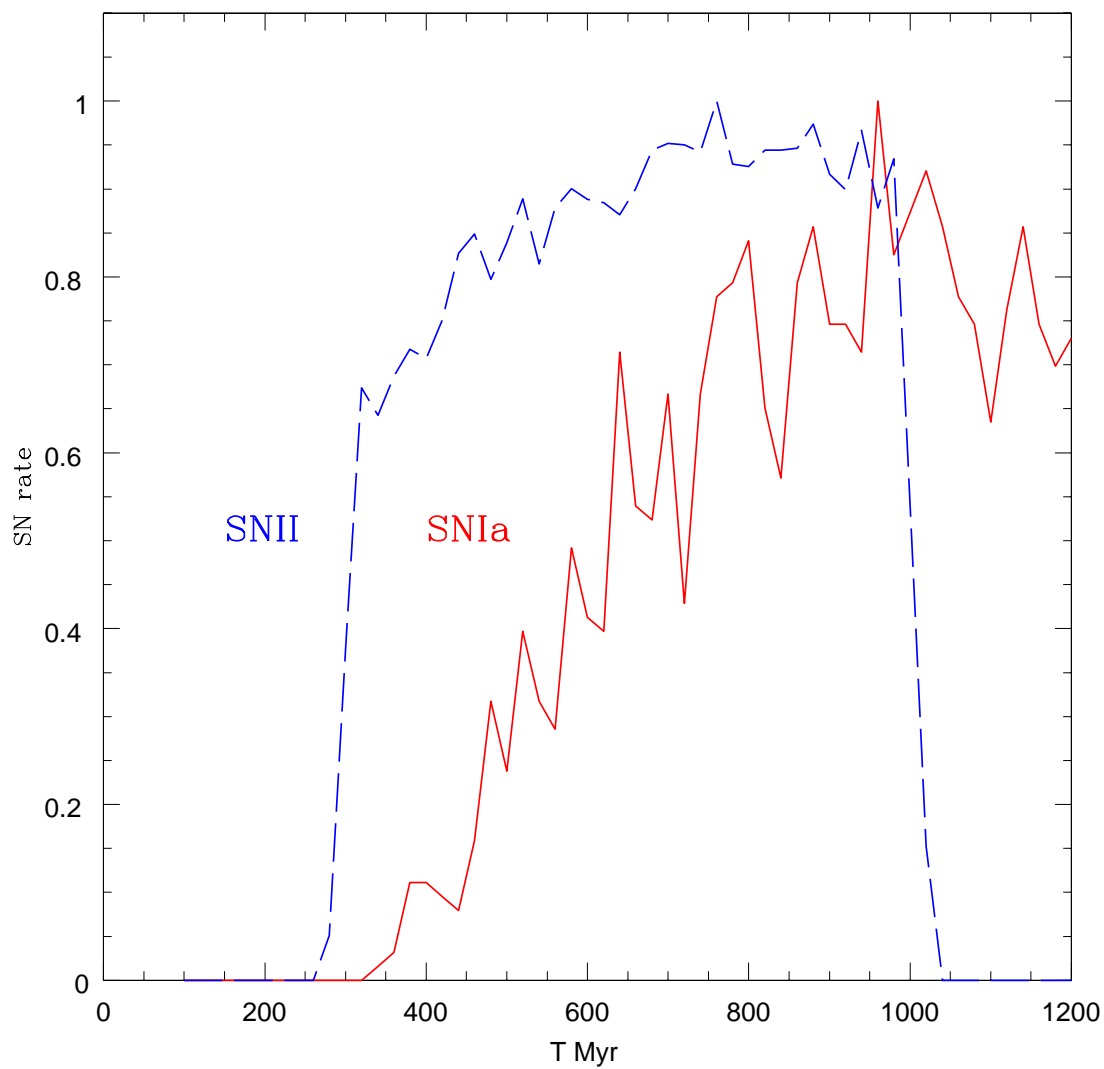} 
\caption{The SNIa rate (red line) and SNII rate (blue line)
in the halo. They are normalized to their maximum values.}
\label{inomo1}
\end{center}
\end{figure}

The model follows the chemical evolution of more than 20 elements
in each region.
The model  parameters of the chemical evolution (SFR, IMF, 
stellar lifetime, nucleosynthesis, gas threshold), are the same as those 
used in chapter 2 for the homogeneous model.
If the model is correct, our predictions will approximate the results of the homogeneous model
 as the number of stars increases. 
On the other hand, our model shows the spread that can be produced by different nucleosynthesis sites 
on the chemical enrichment at low metallicity, where the number of stars is low and
the random effects in the birth of stellar masses can be important.

\section{ Nucleosynthesis Prescriptions}
For the nucleosynthesis prescriptions of the Fe and the others elements 
(namely O, S, Si, Ca, Mg, Sc, Ti,
 V, Cr, Zn, Cu, Ni, Co and Mn), we adopted those suggested by Fran\c cois et al. (2004)
both for single stars and SNeIa, as we did in chapter 2.
We underline that the site of production of $\alpha$-elements and of Fe 
is the whole range of massive stars.
For the nucleosynthesis prescriptions of the r-process contribution
we used those of model 1  for Ba and Eu (see chapter 2.1), the results of chapter 2.2
for La and the results of the chapter 2.3 for Sr, Y and Zr.
These empirical yields have been chosen in order to reproduce 
the surface abundances for all these neutron capture elements 
in low metallicity stars 
as well as the Sr, Y, Zr, Ba, Eu and La solar abundances, taking in account the 
s-process contribution at high metallicities by means of the 
 yields of Busso et al. (2001) for lanthanum and barium
and  those  of Travaglio et al. (2004) for strontium, yttrium and
zirconium in the mass range 1.5-3$M_{\odot}$.
We have assumed that Sr, Y, Zr Ba and La are produced as r-process elements
by massive stars but only up to 30 $M_{\odot}$ 
 and  Eu is  also considered to be a purely r-process element produced
 in the  same range of masses.

\section{Results}

\subsection {The ratios of $\alpha$-elements and neutron capture elements to Fe}

We discuss here the results of our simulations compared to the observational data and to
 the prediction of the homogeneous model. 
We show the [Eu/Fe], [Ba/Fe], [La/Fe], [Sr/Fe], [Y/Fe] and [Zr/Fe] ratios 
as a function of [Fe/H] in the figures 
\ref{inomo2}, \ref{inomo3}, \ref{inomo4}, \ref{inomo5}, \ref{inomo6}, \ref{inomo7}, 
respectively; for  $\alpha$-elements we show [Si/Fe], [Ca/Fe] and [Mg/Fe] as a function of [Fe/H] 
in figures \ref{inomo8}, \ref{inomo9},  \ref{inomo10}, respectively.

\begin{figure}
\begin{center}
\includegraphics[width=0.99\textwidth]{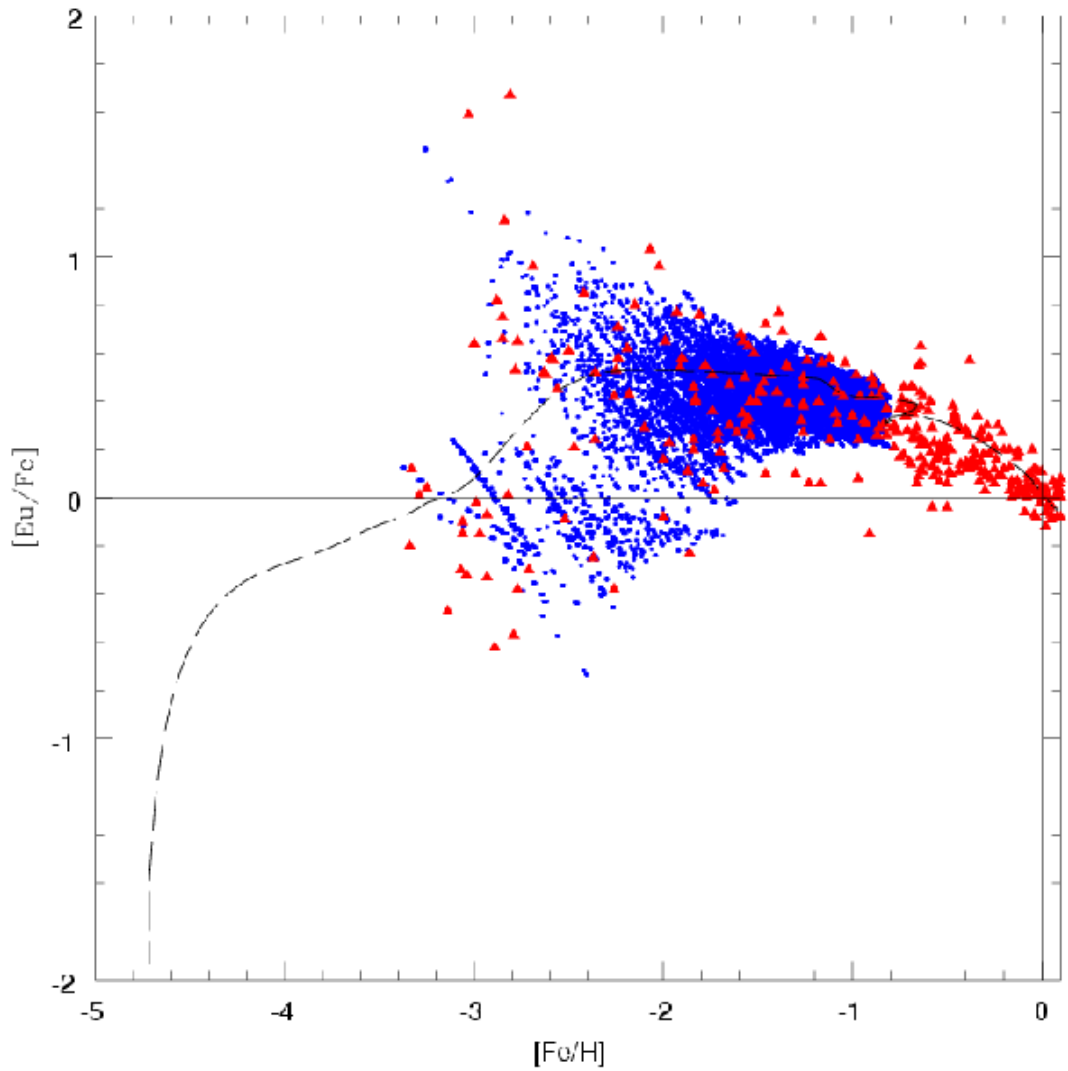}
\caption {[Eu/Fe] vs [Fe/H]. The abundances of simulated stars are indicated by the  blue dots,
the observational data by the  red triangles. 
The black line is the prediction of the homogeneous model.}\label{inomo2}
\end{center}
\end{figure}

\begin{figure}
\begin{center}
\includegraphics[width=0.99\textwidth]{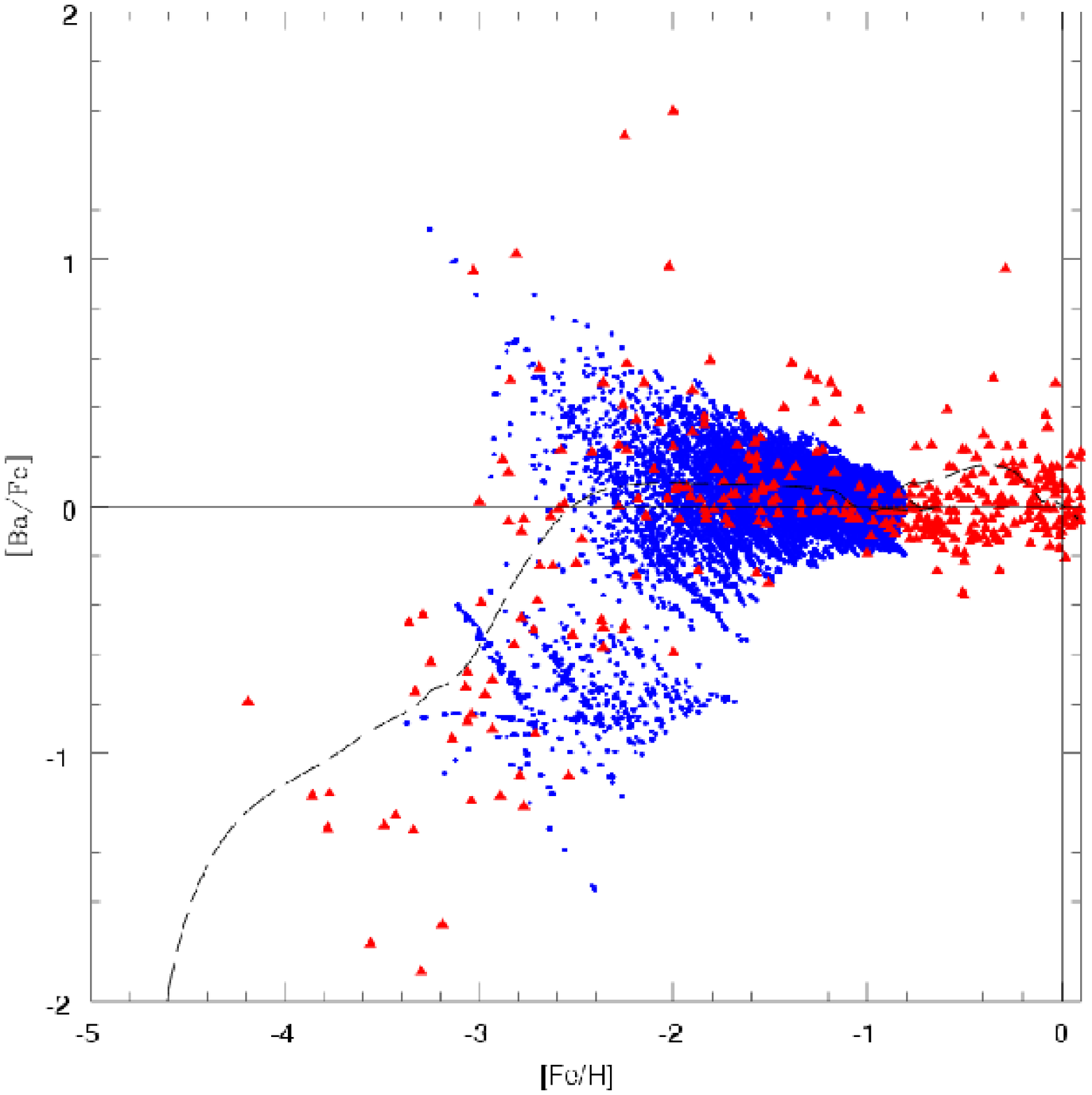}
\caption {As in figure 4.2 but for [Ba/Fe].}\label{inomo3}
\end{center}
\end{figure}

\begin{figure}
\begin{center}
\includegraphics[width=0.99\textwidth]{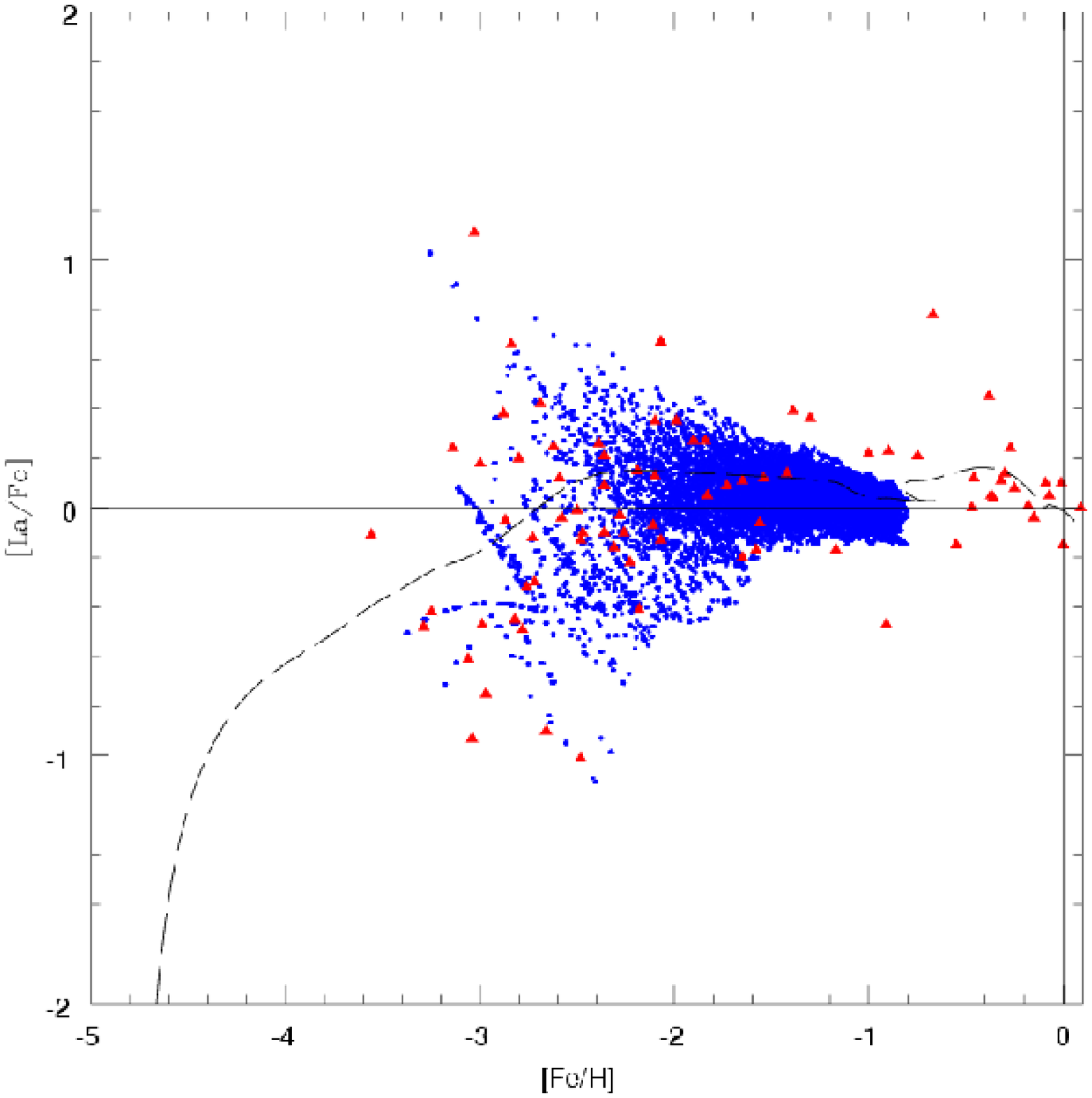}
\caption {As in figure 4.2 but for [La/Fe].}\label{inomo4}
\end{center}
\end{figure}

\begin{figure}
\begin{center}
\includegraphics[width=0.99\textwidth]{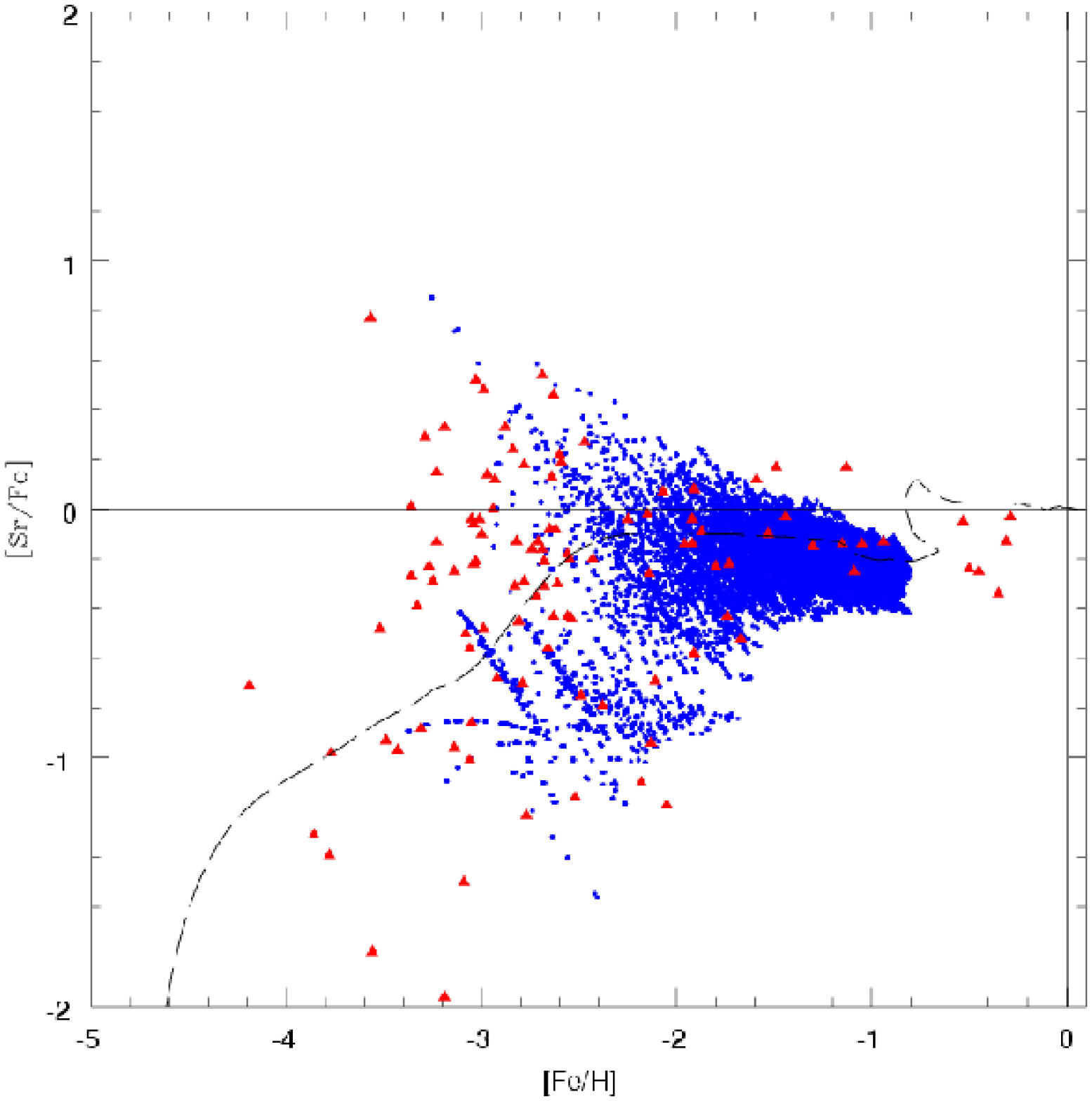}
\caption {As in figure 4.2 but for [Sr/Fe]. }\label{inomo5}
\end{center}
\end{figure}

\begin{figure}
\begin{center}
\includegraphics[width=0.99\textwidth]{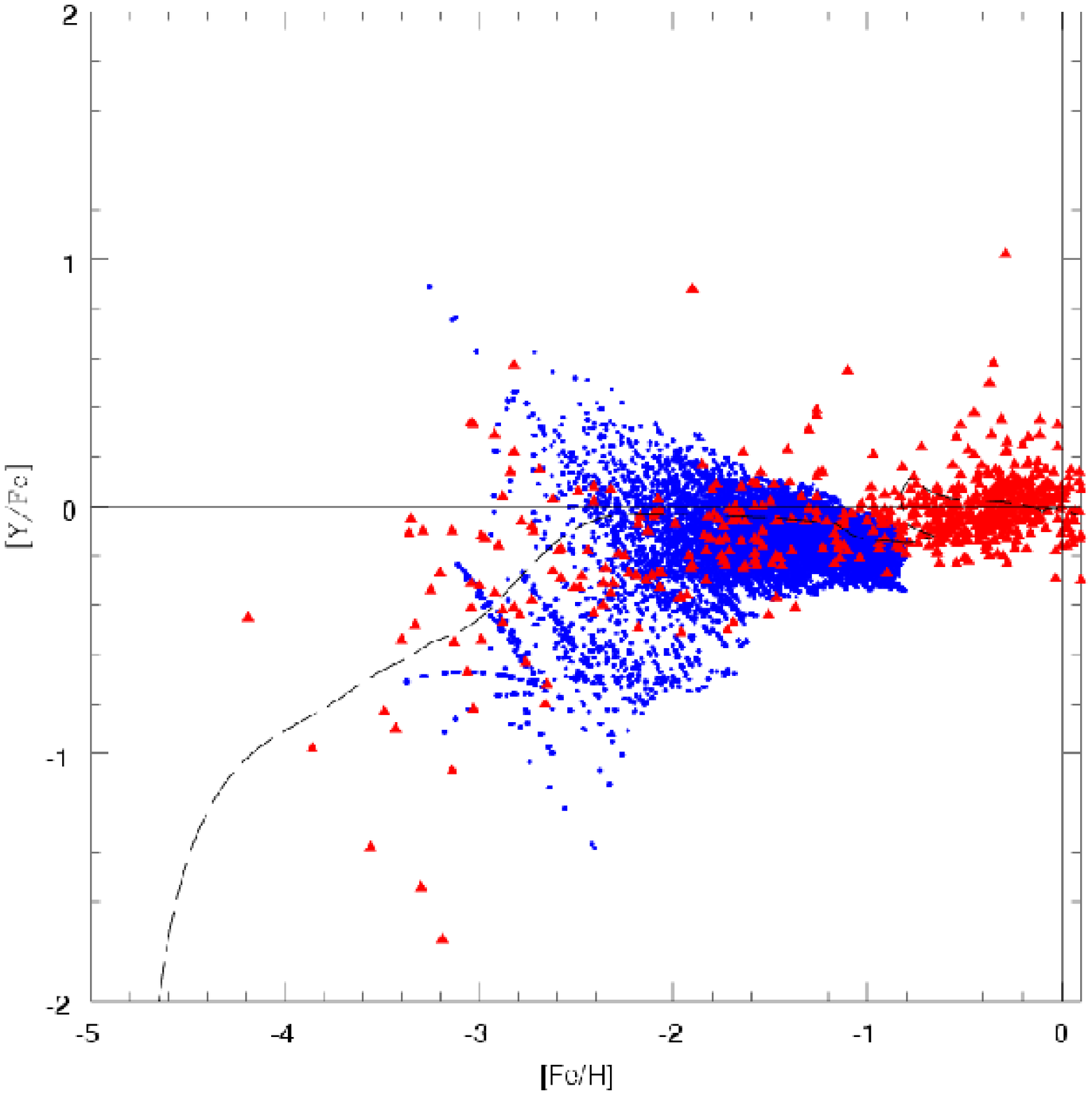}
\caption {As in figure 4.2 but for [Y/Fe].}\label{inomo6}
\end{center}
\end{figure}

\begin{figure}
\begin{center}
\includegraphics[width=0.99\textwidth]{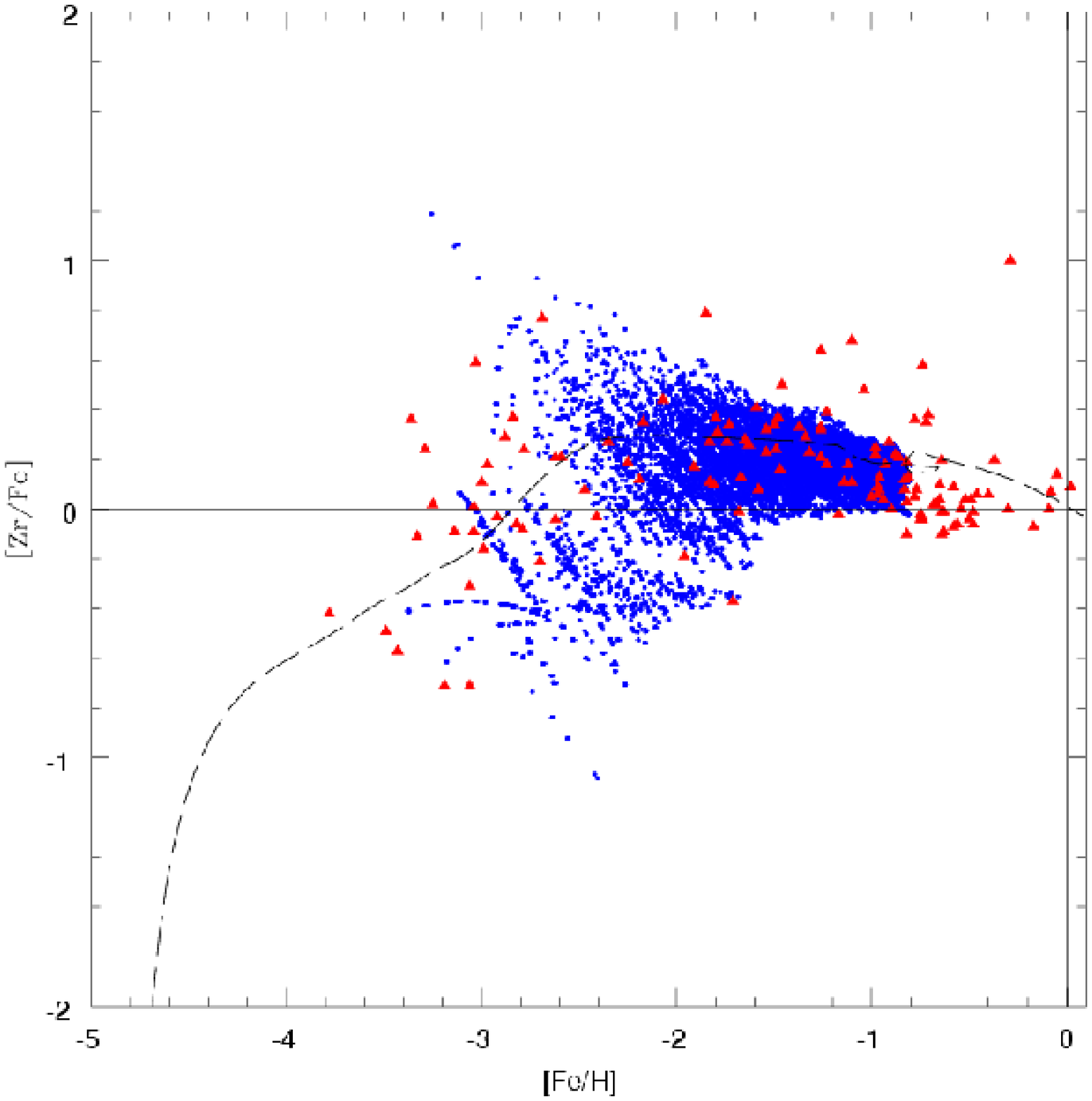}
\caption {As in figure 4.2 but for [Zr/Fe].}\label{inomo7}
\end{center}
\end{figure}

\begin{figure}
\begin{center}              
\includegraphics[width=0.99\textwidth]{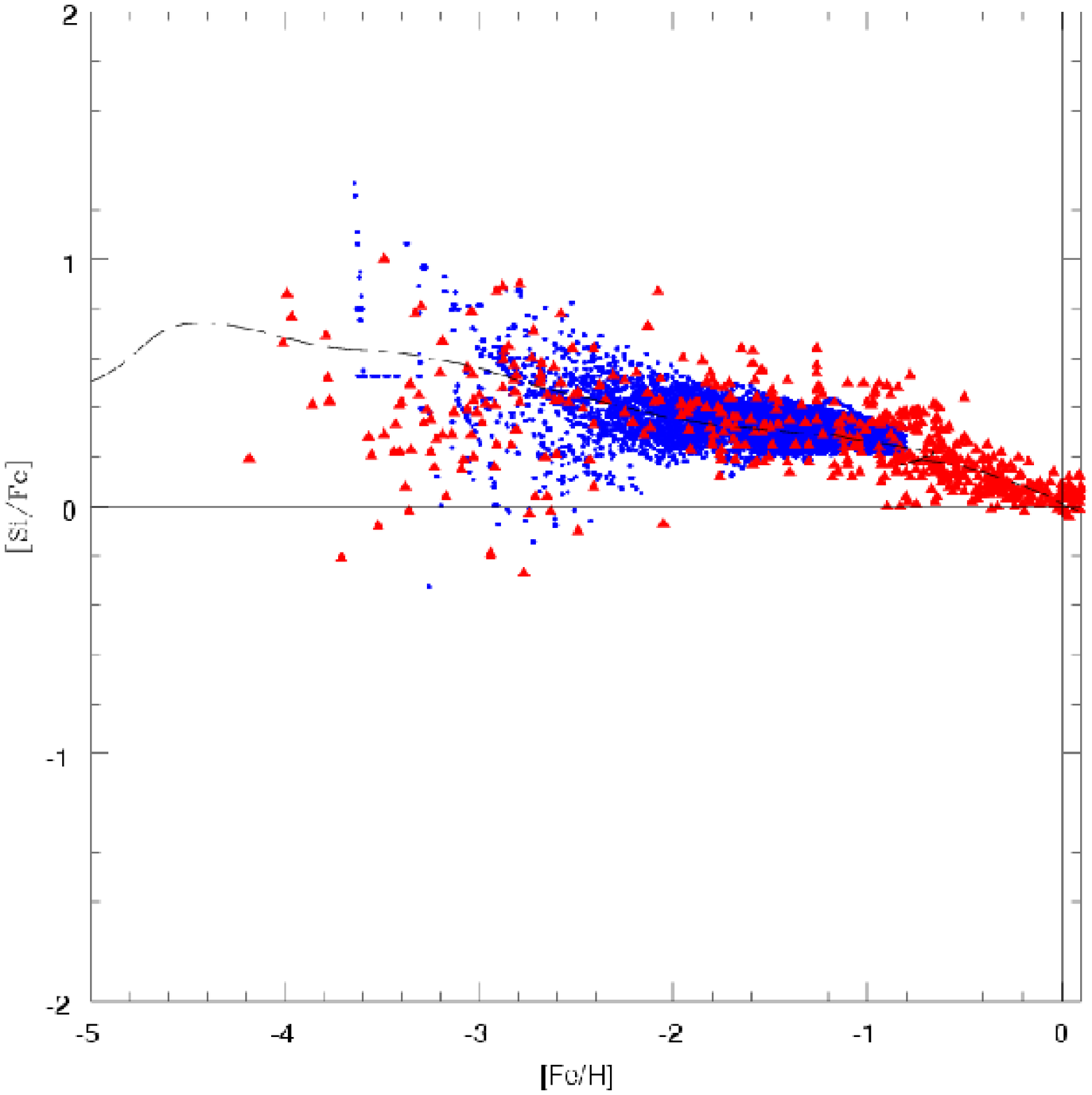} 
\caption{As in figure 4.2 but for [Si/Fe].}\label{inomo8}
\end{center}
\end{figure}

\begin{figure}
\begin{center}              
\includegraphics[width=0.99\textwidth]{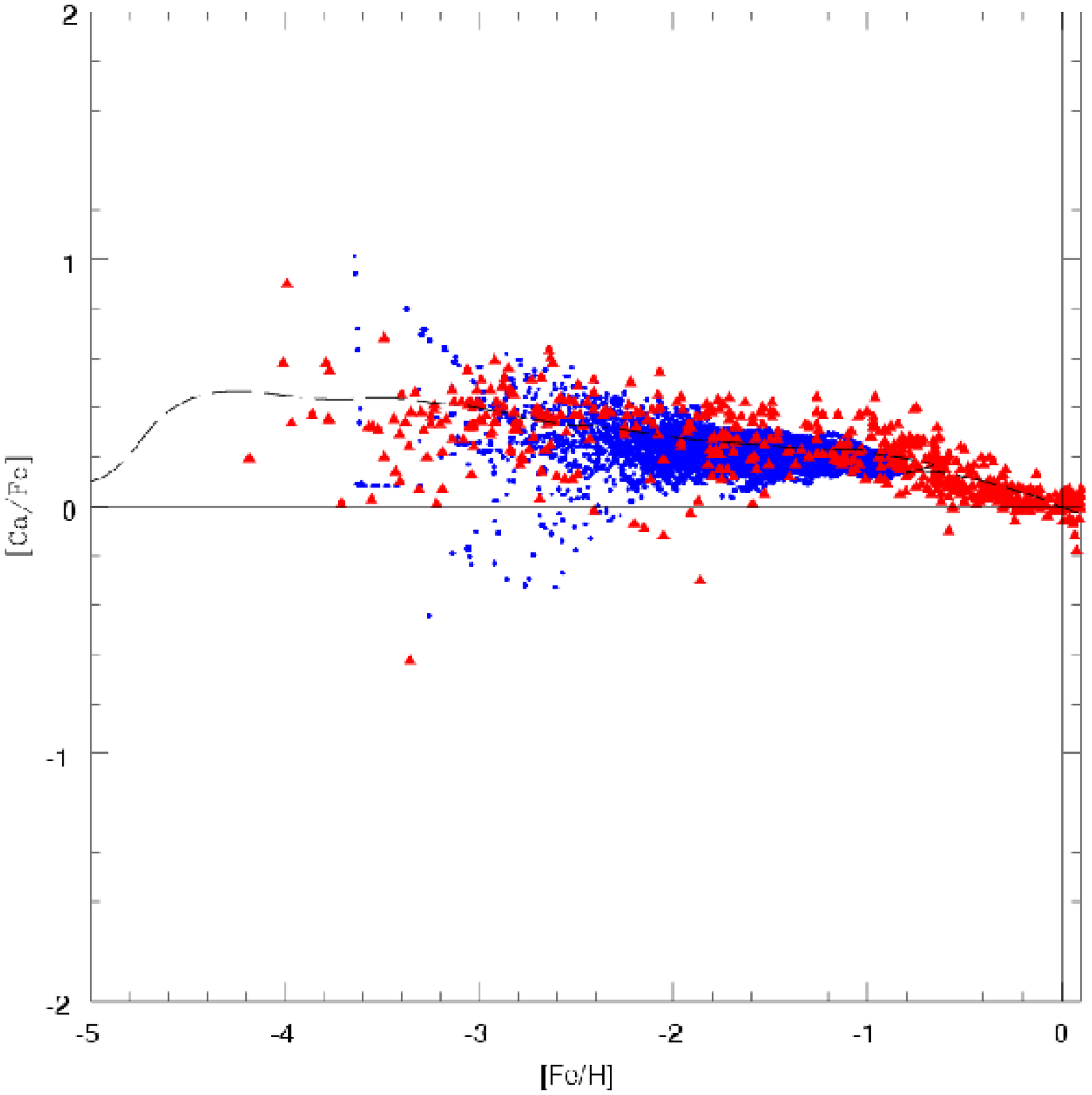} 
\caption{As in figure 4.2 but for [Ca/Fe].}\label{inomo9}
\end{center}
\end{figure}

\begin{figure}
\begin{center}              
\includegraphics[width=0.99\textwidth]{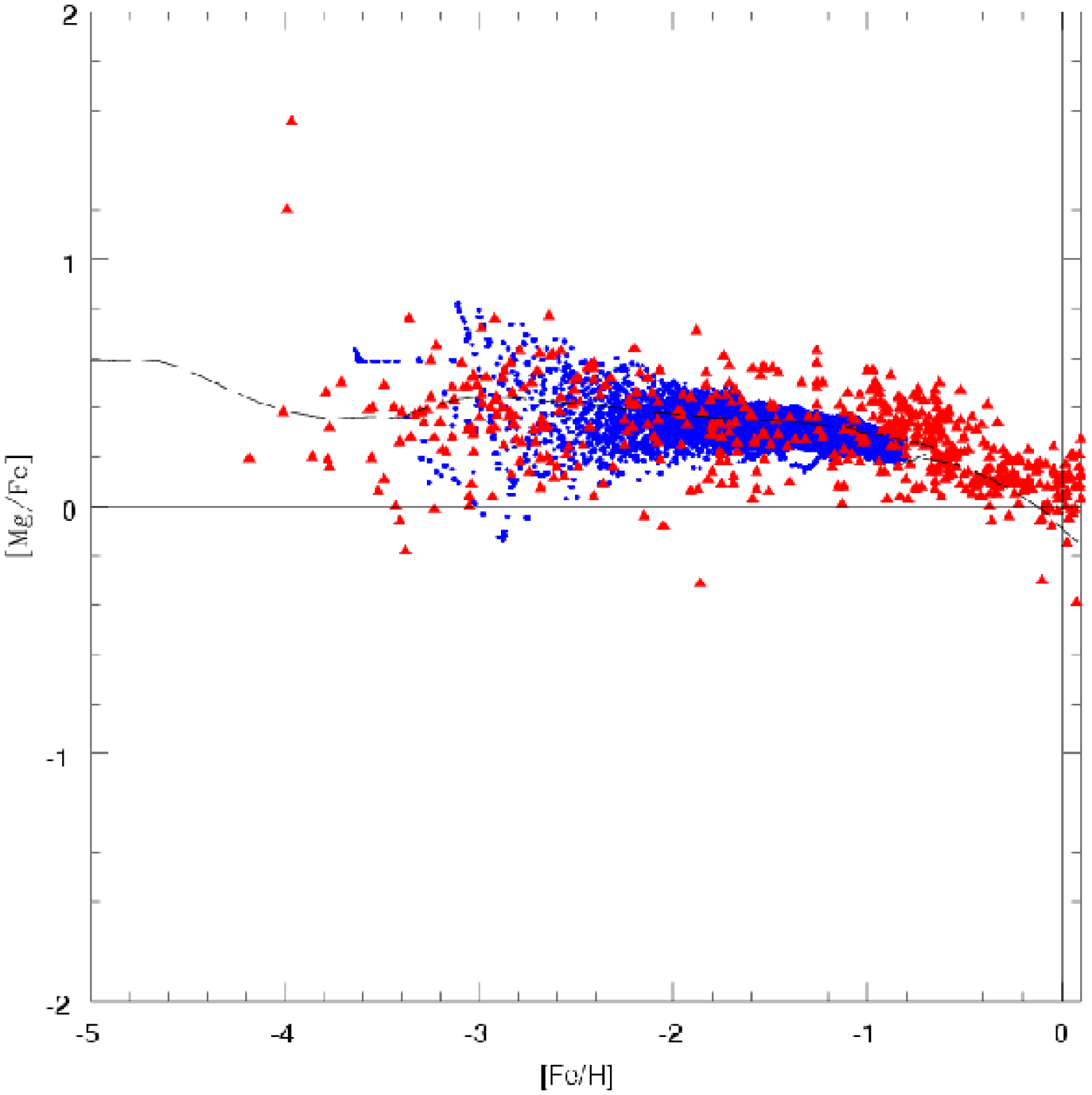} 
\caption{As in figure 4.2 but for [Mg/Fe].}\label{inomo10}
\end{center}
\end{figure}

Our aims is to explain how this model reproduces the large spread  in the  
abundances of  metal poor stars for the neutron capture elements
 and, at the same time, the small spread for the  $\alpha$-elements.
The chemical enrichment observed in metal poor stars is due to nucleosynthesis
in the massive stars.
The sites of production in massive stars for the
 $\alpha$-elements  and neutron capture elements are
 different (as described in the previous section):
the $\alpha$-elements and Fe are produced in the whole range of massive stars;
the neutron capture elements are produced only up to 30 $M_{\odot}$.
Therefore, in regions with many stars less massive than 30 $M_{\odot}$, the
 ratio of neutron capture elements over Fe is
high. The opposite happens in regions where most of the stars are more massive
than 30 $M_{\odot}$. This fact produces, in our inhomogeneous model,
 a large spread for the rates of neutron capture elements to Fe,
 but not for the ratio of the  $\alpha$-elements to Fe, since 
the  $\alpha$-elements and Fe are produced in the same range.

In the figures \ref{inomo11}, \ref{inomo12}, we show the relative frequency 
of stars at a given [El/Fe] ratio  for different enrichment phases.
To calculate these frequencies, we have used only the stars which still
exist nowadays in the halo, those with a mass $< 0.8M_{\odot}$.
 The different enrichment phases: $[Fe/H]< -3$, $-3<[Fe/H]<-2$ and $-2<[Fe/H]<-1$
are given in the panels from top to bottom for Si and Eu.
In these figures the black lines are the predictions of the model and the red lines are
the observational data.
\begin{figure}
\begin{center}              
\includegraphics[width=0.99\textwidth]{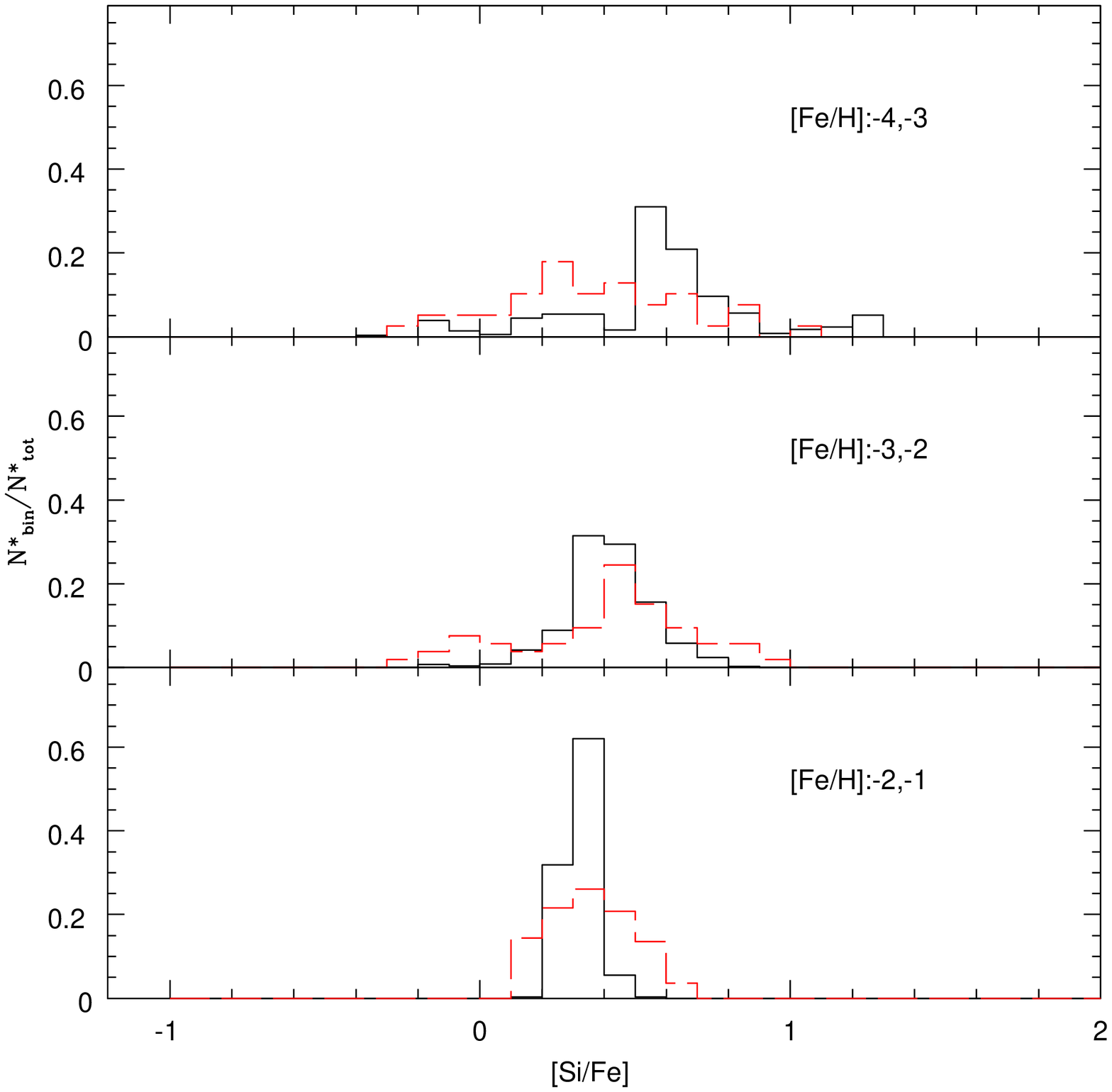} 
\caption{The relative frequency of stars at a given [Si/Fe] ratio for different enrichment phase.
In black line the predictions of the model, in red line the observational data}
\label{inomo11}
\end{center}
\end{figure}
\begin{figure}
\begin{center}              
\includegraphics[width=0.99\textwidth]{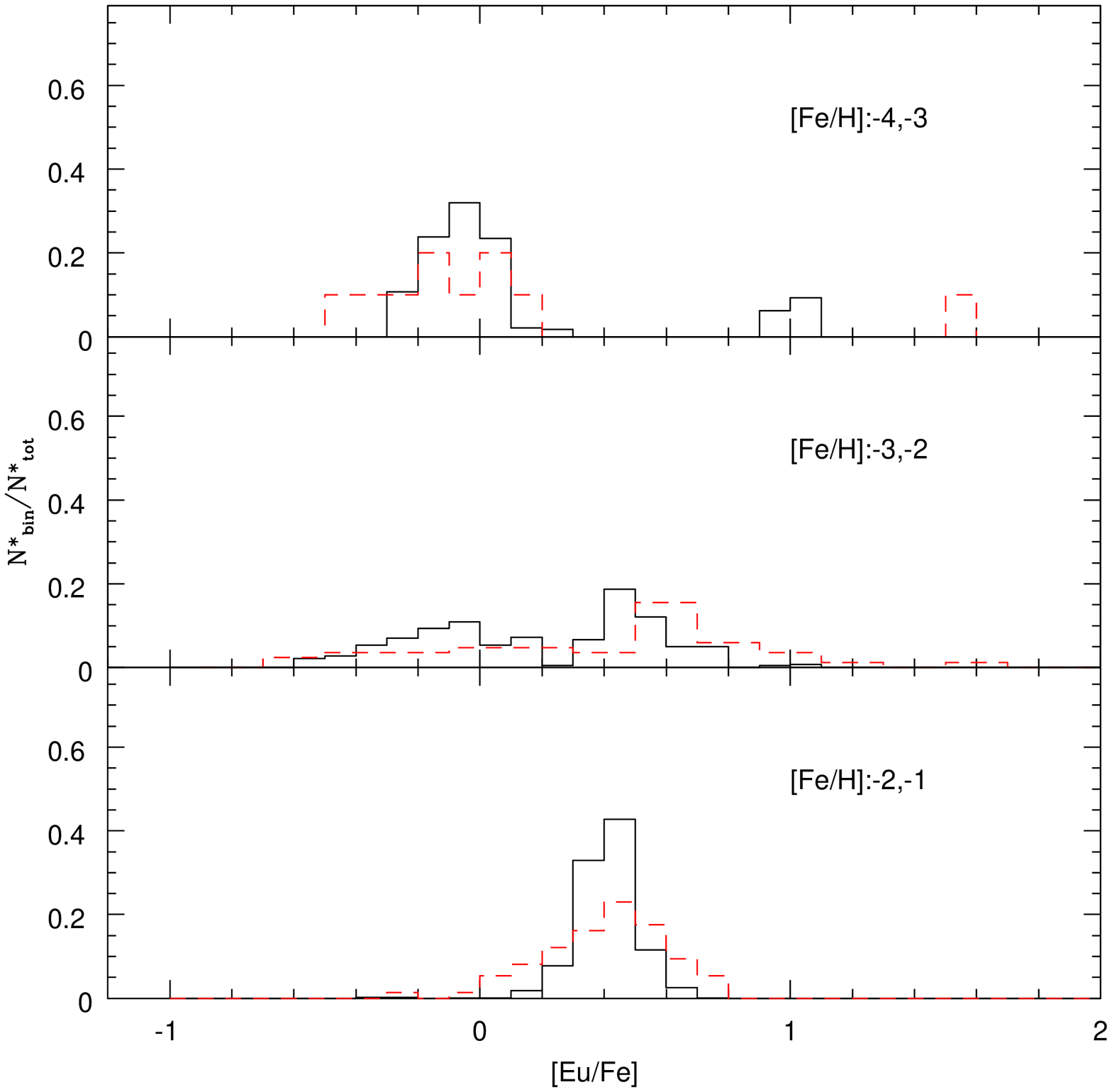} 
\caption{The relative frequency of stars at a given [Eu/Fe] ratio for different enrichment phase.
In black line the predictions of the model, in red line the observational data}
\label{inomo12}
\end{center}
\end{figure}
At intermediate metallicity, the model predicts for the Si already a quite narrow distribution,
 whereas for Eu the distribution of the stars is broad,	in agreement with the observations.
At higher metallicities both elements have predicted distributions
 narrower than the observed ones and this could be explained by the observational errors.
In the low metallicity range some discrepancies can be noticed. 
For instance, the predicted [Si/Fe] distribution is slightly shifted toward higher metallicity ratios
compared to observations. Therefore a more careful analysis of nucleosynthesis yields
is required (for instance using metallicity dependent Si yields).
For what concerns Eu, the results at low [Fe/H] cannot predict the r-process rich stars,
being the highest ratio of [Eu/Fe] predicted by the model $\sim 0.4 dex$ lower than the 
observed one in these stars

The model does not reproduce well the total number of stars present in the range $-4<[Fe/H]<-3$:
the number of simulated stars is too low. This is due to a too fast rise 
of [Fe/H].
Moreover, the model predicts that $\sim$ 5\% of the simulated stars are metal free.
This is the fraction of stars formed in the first 
5 Myr, in which no SN has yet exploded and enriched the ISM.
We underline that the star formation can be slowed down by a different infall rate 
which would modulate a different star formation and this can solve
these two problems. The problem of the metal free stars can be also solved by adopting
a different IMF. The results of many  works  (see 
 Larson 1998, Abel, Bryan  \& Norman 2000, Hernandez \& Ferrara 2001,
 Nakamura \& Umemura 2001, 
Mackey, Bromm, \& Hernquist 2003) predict that the 
star formed in a metal free gas  must be massive stars.
So the long living  stars which are low mass stars, start to born when the ISM has been already 
enriched by these massive and metal free stars, the so-called Population III.
The effect of the Pop III on the global chemical evolution should be negligible  (see Matteucci \& 
Pipino 2005, Matteucci \& Calura 2005, Ballero et al. 2006),
 being the total amount of recycled mass  small. 

Nevertheless, we have decided, preliminary, to use the same parameters and conditions of the homogeneous 
model, to test the validity of the inhomogeneous one. It is clear from the figures
that this new model for a high number of star formation events well approximate the homogeneous model,
which is shown in the figures by the black solid line.

\subsection{The ratios [Ba/Eu] and [Ba/Y]}

Our model  predicts a spread in the abundances ratios if the ratios of the yields of 
the same elements are different as a function of the stellar mass.
This is not the case for what concern the yields of neutron capture elements
considered in our nucleosynthesis prescriptions. 
In figure \ref{inomo13} we show the ratio [Ba/Eu] versus [Fe/H] . 
As expected the results of the model show a  small spread,
 as for the plot for [$\alpha$/Fe], being the  
production ratio between Ba and Eu almost constant as a function of the stellar mass.
\begin{figure}
\begin{center}              
\includegraphics[width=0.99\textwidth]{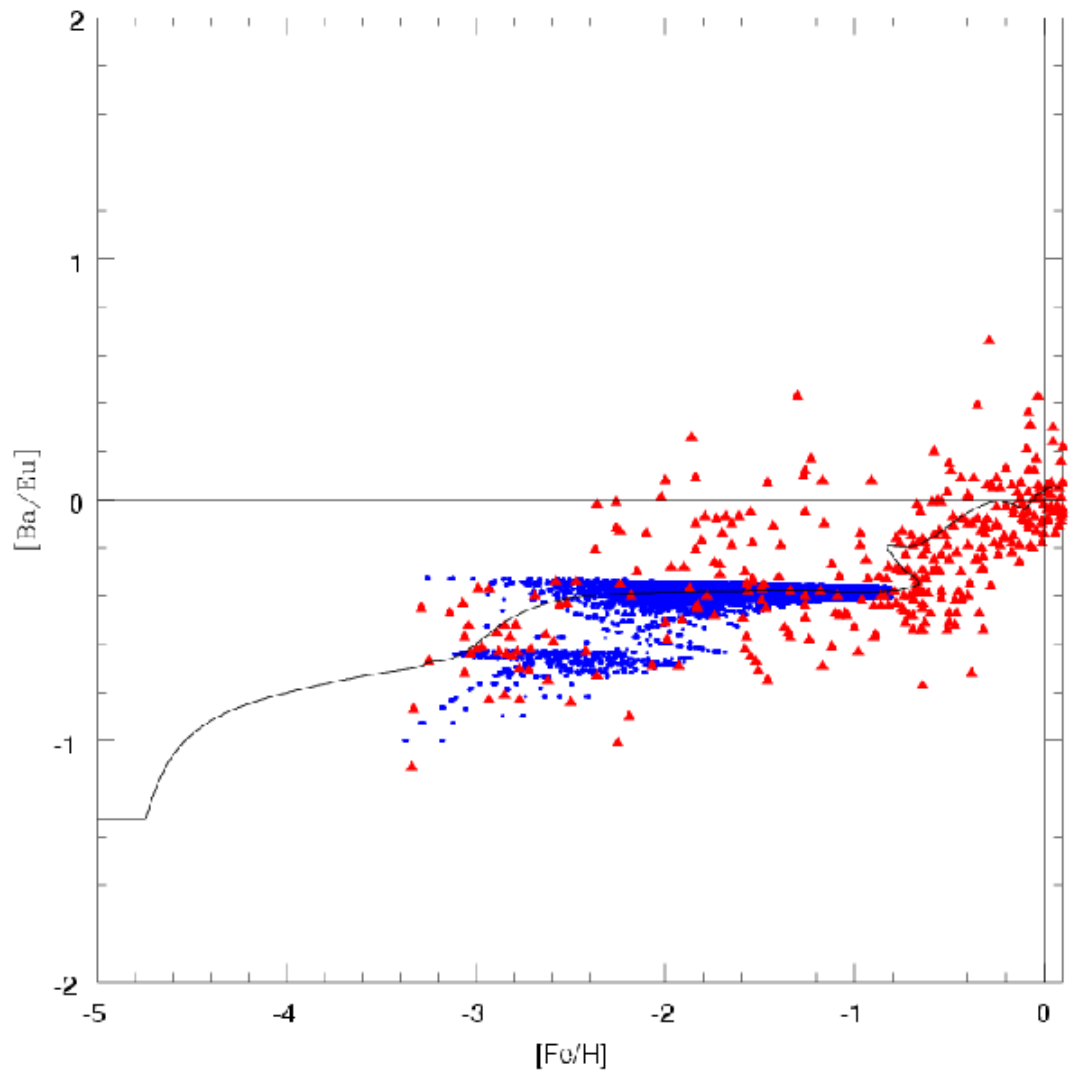} 
\caption{[Ba/Eu] vs [Fe/H]. The abundances of simulated stars in blue dots,
observational data in red triangles. The black line is the prediction of the homogeneous
model}\label{inomo13}
\end{center}
\end{figure}
On the other hand, the observational data show a spread. 
We observe that most of the data seem to have a higher Ba abundance
than that predicted by our model. 
Taking into account this fact, the observational spread
 could be explained either by an earlier 
production of Ba by s-process in intermediate stars from 3 to 8$M_{\odot}$ 
(that we do not take in account in our nucleosynthesis),
 or by a self enrichment of the observed stars due to dredge-up or to a
binary system with an AGB mass transfer 
 to the presently observed companion star (see Aoki et al. 2006).

In the figure \ref{inomo14}, we show the ratio [Ba/Y] versus [Fe/H]. 
The results of the model relative to this ratio is 
not satisfactory. The observational data show a very large scatter 
at [Fe/H] $\sim$ -3 that is not be predicted by our model.
Contrarily to the [Ba/Eu] ratio, for which the available data
show a moderate spread at [Fe/H] $\sim$ -3, [Ba/Y] ratio shows
a large spread. 
A possible way to explain this spread could be that the r-process
yields that we use in our model, are indicative of the mean contribution 
by r-process to the abundances of these elements. We recall
from the introduction of this work, that these two elements
are in different peaks both for what concern s-process 
(not so important at this stage),
 and for what concern r-process.
It could be that, as introduced by Otsuki et al. (2003),
the r-process is not unique but consists of different 
contributions and what we use are only the mean values.
Massive stars may produce r-process with different patterns,
probably as functions of the multiple factors which
influence r-process. 
In our inhomogeneous model we use only one pattern for all 
the neutron capture elements. This could be the reason 
why we are not able to well reproduce the spread 
for the ratios of neutron capture elements. 
Moreover, this problem should be more visible
when we compare neutron capture elements belonging to
different peaks as in the case of [Ba/Y].

\begin{figure}
\begin{center}              
\includegraphics[width=0.99\textwidth]{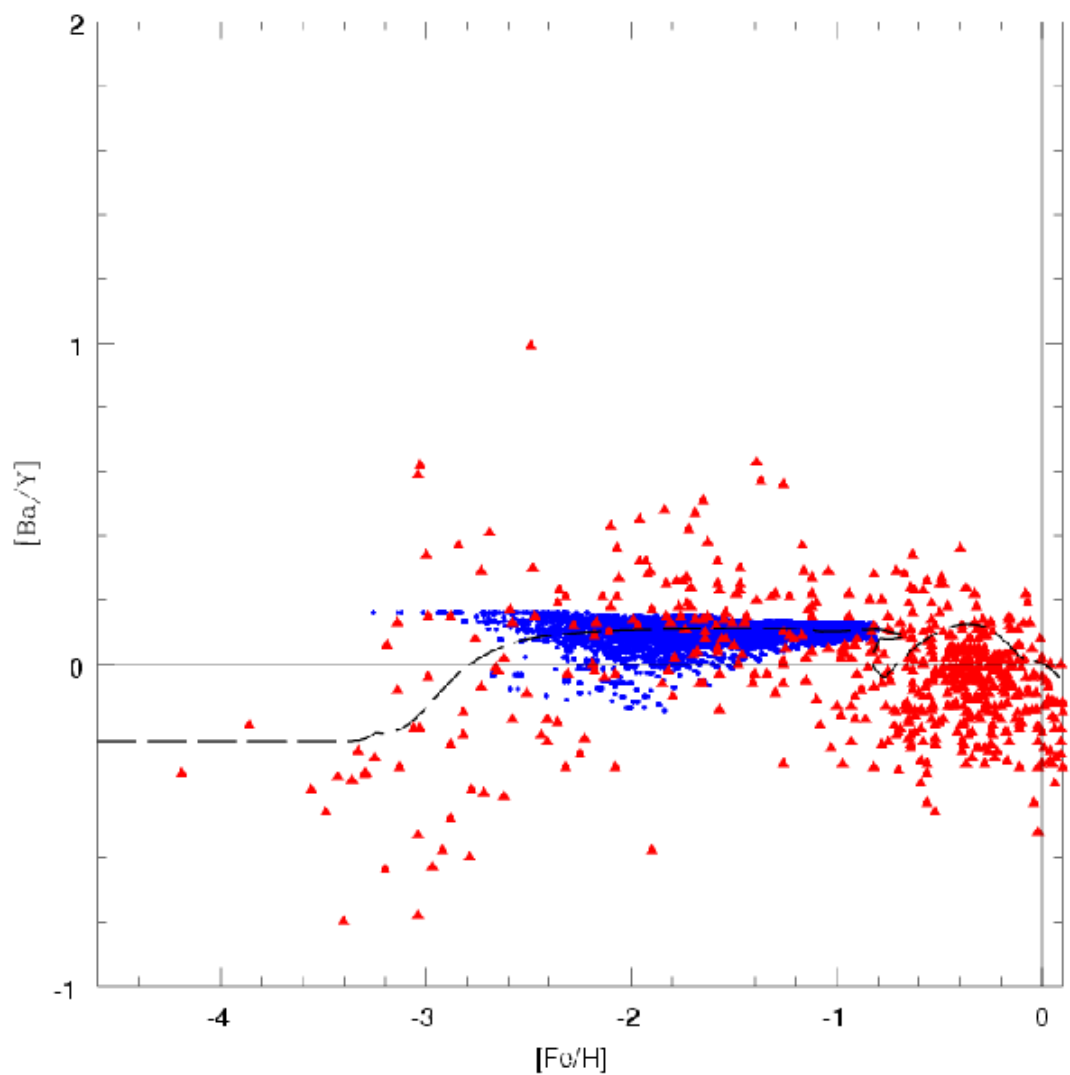} 
\caption{[Ba/Y] vs [Fe/H]. The abundances of simulated stars in blue dots,
observational data in red triangles. The black line is the prediction of the homogeneous
model}\label{inomo14}
\end{center}
\end{figure}

\chapter [Chemical evolution of Ba and Eu in Local dSph galaxies] 
{Chemical evolution of barium and europium in Local dwarf spheroidal galaxies}

\rightline{\emph{``"From a drop of water," said the writer, "a logician could infer the possibility }}
\rightline{\emph{of an Atlantic or a Niagara without having seen or heard of one or the other. }}
\rightline{\emph{So all life is a great chain, the nature of which is known whenever we are }}
\rightline{\emph{shown a single link of it.'' by Sir Arthur Conan Doyle}}

\vspace{1cm}

In this chapter we present the results obtained in collaboration 
with Gustavo Lanfranchi (Department of Astronomy, University 
of S\~ao Paulo). The success of LM03 and LM04  
models (see chapter 1) in reproducing several 
observational constraints allows us to use them as tools to test the
theories about the sites of production and the processes responsible for
the synthesis of Ba and Eu in dSph galaxies. By adopting the
nucleosynthesis prescriptions for these elements which are able to
reproduce the most recent observed data for our Galaxy, as shown in chapter 2,
 and comparing the predictions of the models with observational data,
it is possible to verify if the assumptions made regarding the 
nucleosynthesis of Ba and Eu can also fit the data of local dSph 
galaxies.

\section{Observational Data}

Recently, red giant stars of dSph galaxies have been the subject 
of several works with the aim of determining with high-resolution
spectroscopy the abundance of several chemical elements
including heavy elements such as barium and europium
(Bonifacio et al. 2000; Shetrone, Cot\'e $\&$ Sargent 2001; 
Shetrone et al. 2003; Venn et al. 2004; Sadakane et al. 2004; 
Fulbright, Rich $\&$ Castro, 2004; Geisler et al. 2005). 
From these observations we gathered the data from the galaxies
that were analyzed in LM03 
and LM04 and for which there are abundance determinations for 
both Ba and Eu. They are Carina, Draco, Sculptor, Ursa Minor and Sagittarius.
Despite of the relative small number of data points, it is 
possible to compare the observed abundance ratios 
with the model predictions. We choose to compare the observed
ratios [Ba/Fe], [Eu/Fe] and [Ba/Eu] with the ones predicted by 
the models, since these ratios can provide some clues not only
to the nucleosynthesis of Ba and Eu, but also to all s-process
and r-process elements. 

In order to properly compare different 
data from different authors with the predictions of the models we adopted 
the abundance values of Shetrone, Cot\'e $\&$ Sargent (2001), 
Shetrone et al. (2003), and  Sadakane et al. (2004) updated by
Venn et al. (2004). Venn et al. (2004) homogenized the atomic data 
for spectral lines of Ba and Eu providing data with improved quality which allow a
consistent comparison between data from different sources. Otherwise,
the effect of combining these different data would be seen as a larger spread 
in the abundances and possibly in the abundance ratios of 0.1 to 0.2 dex
(see Venn et al. 2004). In the case of Bonifacio et al. (2000) data, Eu 
is obtained using hyper-fine splitting (HFS) (see their Table 5), but Ba 
is not. The authors claimed that the Ba abundances obtained with HFS would 
exhibit no significant difference since the line observed (Ba II 6496.9) 
is a strong line which is not affected by this correction (Bonifacio 
private communication, see also Shetrone et al. 2003).

Some of the observed stars, however, exhibit anomalous values of
[Ba/H] or [Eu/H], and for this reason were excluded from the sample. 
Two stars in Ursa Minor, K and 199 (in Shetrone, Cot\'e $\&$ 
Sargent 2001), exhibit heavy-element abundance ratios enhanced
relative to those typical for other dSph stars: the Ursa Minor K star
has an abundance pattern dominated by the s-process and
was classified as a Carbon star while Ursa Minor 199 is 
dominated by r-process (see also Sadakane et al. 2004). In Sculptor,
there are also two stars with enhanced heavy-element abundance:
Sc982 (Geisler et al. 2005) and Sculptor H-400 (Shetrone et al. 2003).
While Shetrone et al. (2003) claimed that the r-process dominated 
abundance could be attributed to inhomogeneous mixing of the SNe II
yields, Geisler et al. (2005) classified Sc982 as a heavy element star
which could have been enriched by an other star, which is now dead.
Either way, all these stars do not exhibit an abundance pattern
characterized only by the nucleosynthesis process occurring inside
the star, but also one which was contaminated by external factors. 
The maintenance of these stars in the sample could lead to
an erroneous comparison with the model predictions and, as 
a consequence, to a misleading interpretation and to wrong conclusions
regarding the processes and the site of production of the heavy
elements analyzed. Therefore, we excluded these stars from our sample, 
whereas all the other stars were considered and included in 
the comparisons with the models predictions.

%%%%%%%%%%%%%%%%%%%%%%%%%%%%%%%%%%%%%%%%%%%%%%%%%%%%%%%%%%%%%%%%%%%
\begin{table*}
%\begin{flushleft} 
\begin{center}\scriptsize  
\caption{Models for dSph galaxies. $M_{tot}^{initial}$ 
is the baryonic initial mass of the galaxy, $\nu$ is the star-formation 
efficiency, $w_i$ is the wind efficiency, and $n$, $t$ and $d$ 
are the number, time of occurrence and duration of the SF 
episodes, respectively.}\label{lanfa1}
\vspace{1.5cm}
\begin{tabular}{lccccccc}  
\hline\hline\noalign{\smallskip}  
galaxy &$M_{tot}^{initial} (M_{\odot})$ &$\nu(Gyr^{-1})$ &$w_i$
&n &t($Gyr$) &d($Gyr$) &$IMF$\\    
\noalign{\smallskip}  
\hline
Sculptor &$5*10^{8}$ &0.05-0.5 &11-15 &1 &0 &7 &Salpeter\\
Draco  &$5*10^{8}$ &0.005-0.1 &6-10 &1 &6 &4 &Salpeter\\
Ursa Minor &$5*10^{8}$ &0.05-0.5 &8-12 &1 &0 &3 &Salpeter\\
Carina &$5*10^{8}$ &0.02-0.4 &7-11 &2 &6/10 &3/3 &Salpeter\\
Sagittarius &$5*10^{8}$ &1.0-5.0 &9-13 &1 &0 &13 &Salpeter\\
\hline\hline
\end{tabular}
\end{center}
%\end{flushleft}
\end{table*} 
%%%%%%%%%%%%%%%%%%%%%%%%%%%%%%%%%%%%%%%%%%%%%%%%%%%%%%%%%%%%%%%%

\section{Chemical evolution models for the Local dSph galaxies} 

We use in this work the same chemical evolution model for dSph galaxies
galaxies as described in LM03 and LM04. The model is able to reproduce
the [$\alpha$/Fe] ratios, the present gas mass and the inferred total mass
of five dSph galaxies of the Local Group, namely Carina, Draco, Sculptor, 
Sagittarius and Ursa Minor, and also the stellar metallicity 
distribution of Carina (Koch et al. 2004).  The scenario representing 
these galaxies is characterized by one long episode (two episodes
in the case of Carina) of star formation (SF) with very low efficiencies
(except in the case of Sagittarius) - $\nu$ = 0.001 to 0.5 Gyr $^{-1}$ -
and by the occurrence of very intense galactic winds - $w_i$ = 6-13. 
The model allows one to follow in detail the evolution of the 
abundances of several chemical elements, starting from 
the matter reprocessed by the stars and restored into the 
ISM by stellar winds and type II and Ia supernova explosions.

The main features of the model are:

\begin{itemize}

\item
one zone with instantaneous and complete mixing of gas inside
this zone;

\item
no instantaneous recycling approximation, i.e. the stellar 
lifetimes are taken into account;

\item
the evolution of several chemical elements (H, D, He, C, N, O, 
Mg, Si, S, Ca, Fe, Ba and Eu) is followed in detail;

\end{itemize}

In the scenario adopted in the previous works, the dSph galaxies 
form through
a continuous and fast infall of pristine gas until a mass of
$\sim 10^8 M_{\odot}$ is accumulated.  One crucial 
feature in the evolution of these galaxies is the occurrence 
of galactic winds, which develop when the thermal 
energy of the gas equates its binding energy (Matteucci $\&$
Tornamb\'e 1987). This quantity is strongly influenced by 
assumptions concerning the presence and distribution 
of dark matter (Matteucci 1992). A diffuse ($R_e/R_d$=0.1, 
where $R_e$ is the effective radius of the galaxy and $R_d$ is 
the radius of the dark matter core) but massive 
($M_{dark}/M_{Lum}=10$) dark halo has been assumed for each galaxy. 

In table \ref{lanfa1} we summarize the adopted parameters for the models
of dSph galaxies.

\section{Results}
 
\subsection{Europium}
%%%%%%%%%%%%%%%%%%%%%%%%%%%%%%%%%%%%%%%%%%%%%%%%%%%%%%%%%%%%%%%%%%%
\begin{figure}
\begin{center}
\includegraphics[width=0.99\textwidth]{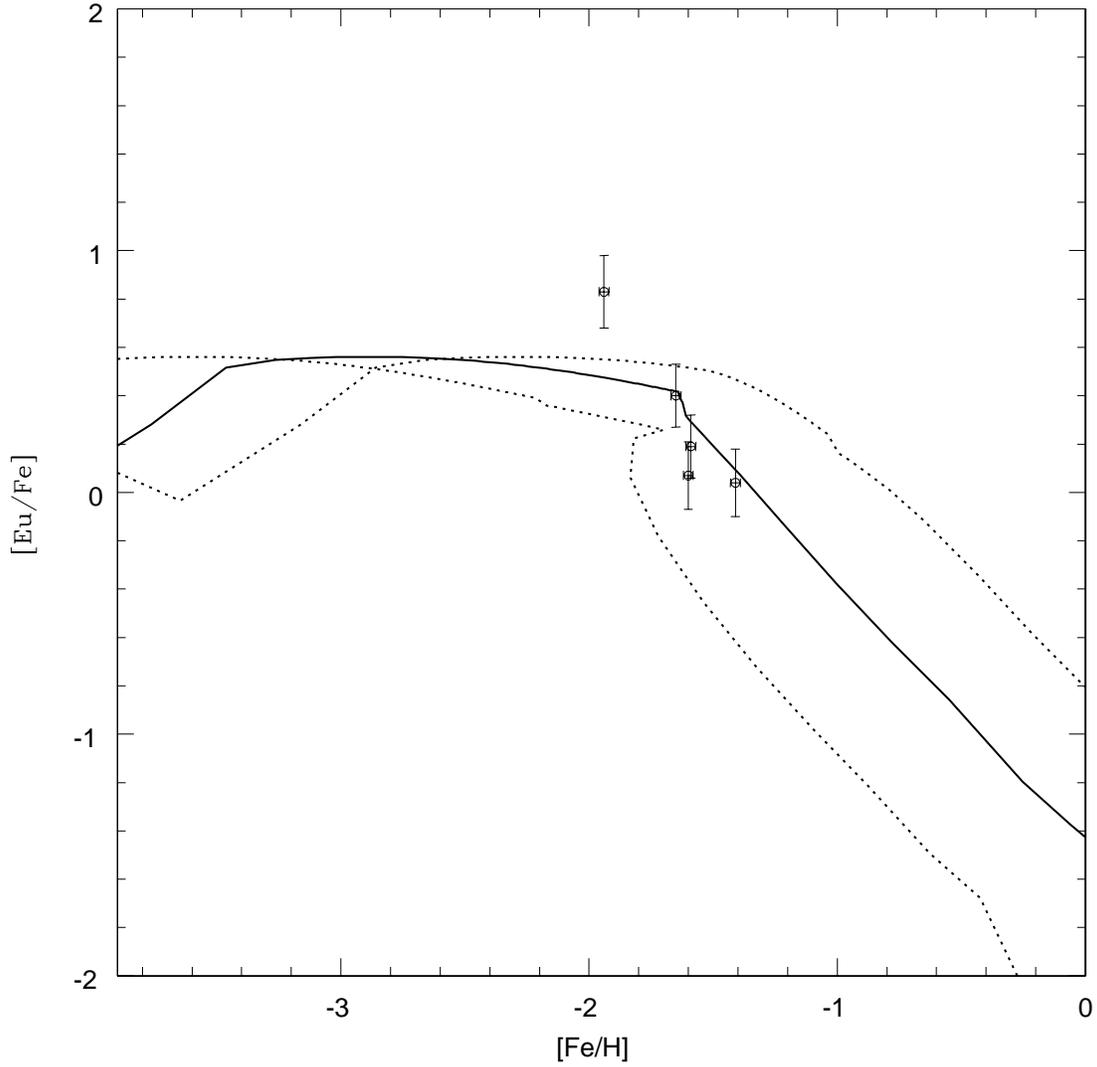}

\caption{[Eu/Fe] vs. [Fe/H] observed in Carina dSph 
galaxy compared to the predictions of the chemical evolution
model for Carina. The solid line represents the best model 
($\nu = 0.1\;Gyr^{-1}$, w$_i$ = 7) and the dotted
lines the lower ($\nu = 0.02\;Gyr^{-1}$) and upper 
($\nu = 0.4\;Gyr^{-1}$) limits for the SF efficiency.}
\label{lf1}
\end{center}
\end{figure}
%%%%%%%%%%%%%%%%%%%%%%%%%%%%%%%%%%%%%%%%%%%%%%%%%%%%%%%%%%%%%%%%%%%

The [Eu/Fe] ratio as a function of [Fe/H] observed in the four 
Local Group dSph galaxies is compared with the model predictions 
in the Figures \ref{lf1} to \ref{lf4} (Carina, Draco, Sculptor and Ursa Minor,
respectively). The predicted behaviour seen in the plots is the same for 
all galaxies: [Eu/Fe] is almost constant with supra-solar values
($\sim 0.5$ dex) until [Fe/H] $\sim -1.7$ dex (depending on the 
galaxy). Above this metallicity, the [Eu/Fe] values start decreasing
fast in Sculptor and Carina (there are no points at these metallicities
for Draco and Ursa Minor) similar to what is observed in the 
case of the [$\alpha$/Fe] ratio. This behaviour is consistent with the production of
Eu by r-process taking place in massive stars with $M > 10 M_{\odot}$.
Stars in this mass range have short lifetimes
and enrich
the ISM at early stages of galactic evolution
giving rise to high values of [Eu/Fe], since the production of Fe
in these stars is lower than in type Ia SNe occurring later. 
When the SNe Ia begin to occur,
the Fe abundance increases and, consequently, the [Eu/Fe] ratio
decreases, as one can see also in the data.
%%%%%%%%%%%%%%%%%%%%%%%%%%%%%%%%%%%%%%%%%%%%%%%%%%%%%%%%%%%%%%%%%%%
\begin{figure}
\begin{center}
\includegraphics[width=0.99\textwidth]{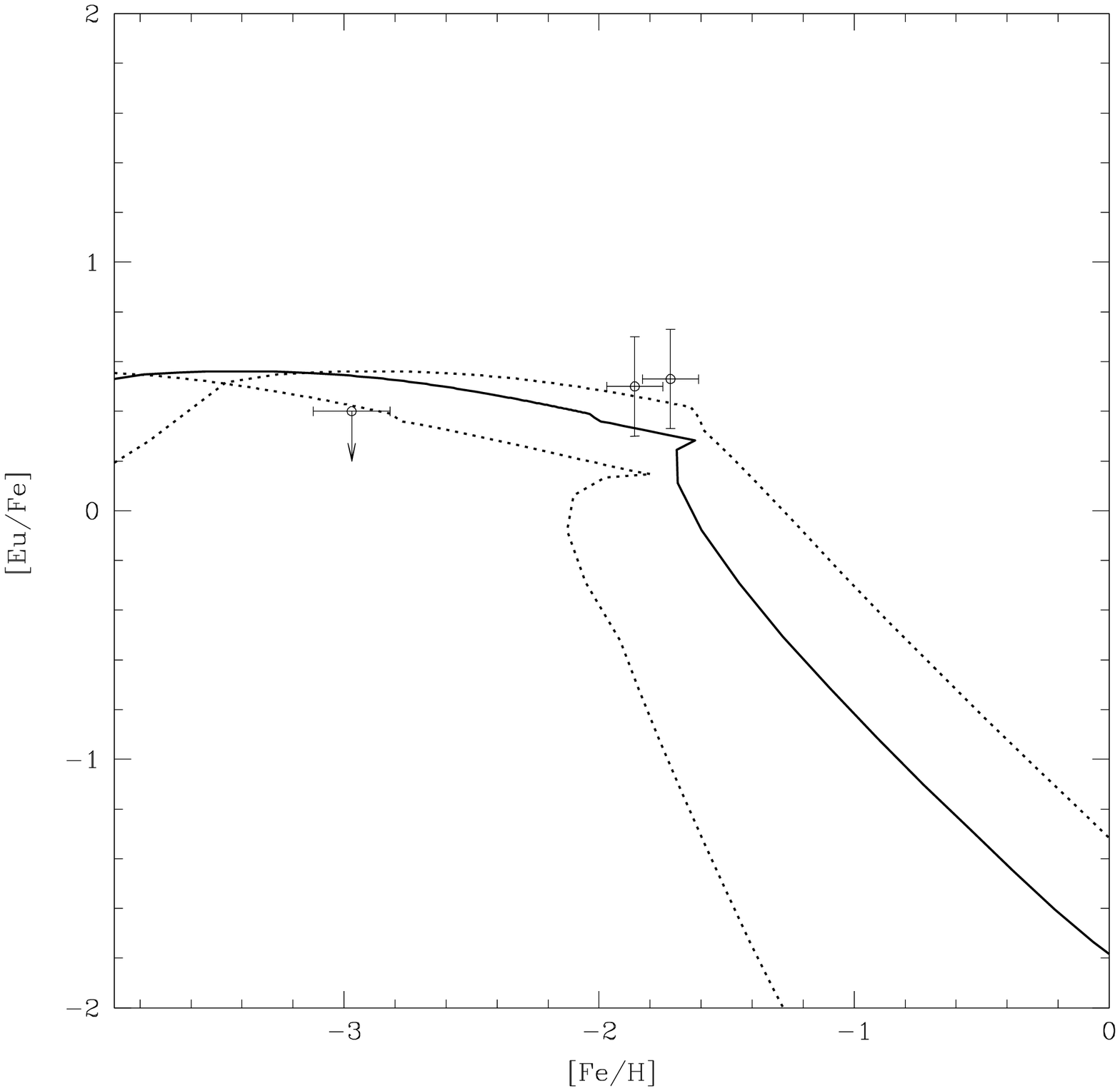}

\caption{[Eu/Fe] vs. [Fe/H] observed in Draco dSph 
galaxy compared to the predictions of the chemical evolution
model for Draco. The solid line represents the best model 
($\nu = 0.03\;Gyr^{-1}$, w$_i$ = 6) and the dotted
lines the lower ($\nu = 0.005\;Gyr^{-1}$) and upper 
($\nu = 0.1\;Gyr^{-1}$) limits for the SF efficiency.}
\label{lf2}
\end{center}
\end{figure}

%%%%%%%%%%%%%%%%%%%%%%%%%%%%%%%%%%%%%%%%%%%%%%%%%%%%%%%%%%%%%%%%%%%

%%%%%%%%%%%%%%%%%%%%%%%%%%%%%%%%%%%%%%%%%%%%%%%%%%%%%%%%%%%%%%%%%%%
\begin{figure}

\begin{center}
\includegraphics[width=0.99\textwidth]{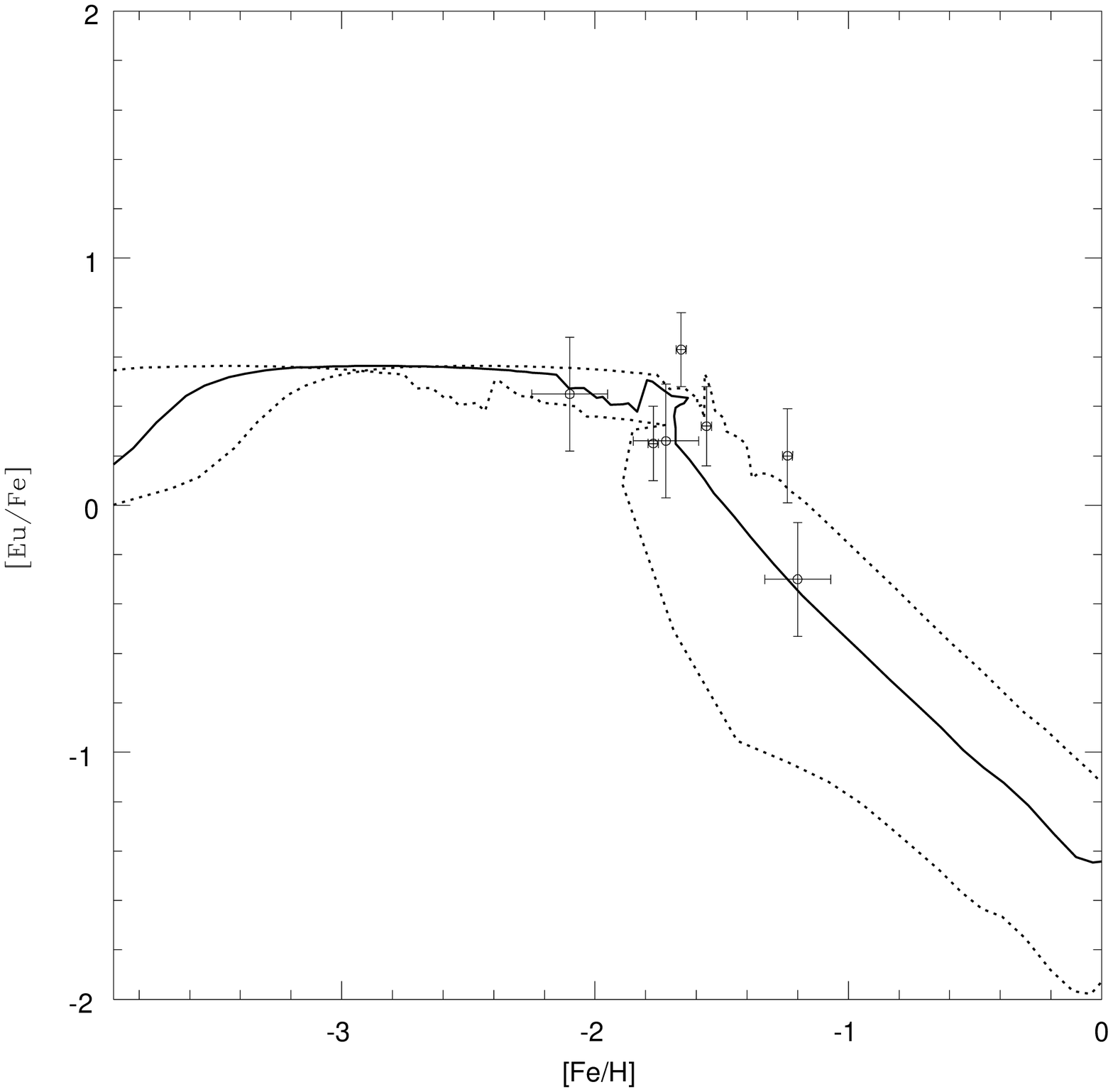}
\caption{[Eu/Fe] vs. [Fe/H] observed in Sculptor dSph 
galaxy compared to the predictions of the chemical evolution
model for Sculptor. The solid line represents the best model 
($\nu = 0.2\;Gyr^{-1}$, w$_i$ = 13) and the dotted
lines the lower ($\nu = 0.05\;Gyr^{-1}$) and upper 
($\nu = 0.5\;Gyr^{-1}$) limits for the SF efficiency.}
\label{lf3}
\end{center}
\end{figure}
%%%%%%%%%%%%%%%%%%%%%%%%%%%%%%%%%%%%%%%%%%%%%%%%%%%%%%%%%%%%%%%%%%%
The predicted [Eu/Fe] ratios in all four dSph galaxies  well reproduce 
the observed trend: an almost constant value 
at low metallicities, and an abrupt decrease starting at [Fe/H] 
$>$ -1.7 dex. In the model this decrease is caused not only by
the nucleosynthesis prescriptions and stellar lifetimes, but also 
by the effect of a very intense galactic wind on the star 
formation rate and, consequently,
on the production of the elements involved. 
In fact, since the wind is very efficient, a large fraction of the gas reservoir
is swept from the galaxy. At this point, the SF is almost halted
and the production of Eu goes down to negligible values. The injection
of Fe in the ISM, on the other hand, continues due to the large
lifetimes of the stars responsible for its production. The main 
result is an abrupt decrease in the [Eu/Fe] ratios, larger than the
one that one would expect only from the nucleosynthesis point of view if there
was no such intense wind. The abrupt decrease follows the 
trend of the data very well, especially in the case of Sculptor and
Carina. For these two galaxies there are stars observed with
metallicities higher than the one corresponding to the time when the wind
develops ([Fe/H]  $>$ -1.7 dex), and which are characterized by lower 
values of
[Eu/Fe], in agreement with our predictions. The observed
stars of the other two dSph galaxies, Draco and Ursa Minor, exhibit
[Fe/H] values which place them before the occurrence of SNe Ia, 
so it is not possible to verify if the abrupt decrease in the 
[Eu/Fe] occurs also in these objects. Only observations of more
stars will confirm the trend. It should be said again that
the same phenomena explain very well the [$\alpha$/Fe] ratios 
and the final total mass and present day gas mass observed in these 
galaxies (LM03, LM04). 
%%%%%%%%%%%%%%%%%%%%%%%%%%%%%%%%%%%%%%%%%%%%%%%%%%%%%%%%%%%%%%%%%%%
\begin{figure}

\begin{center}
\includegraphics[width=0.99\textwidth]{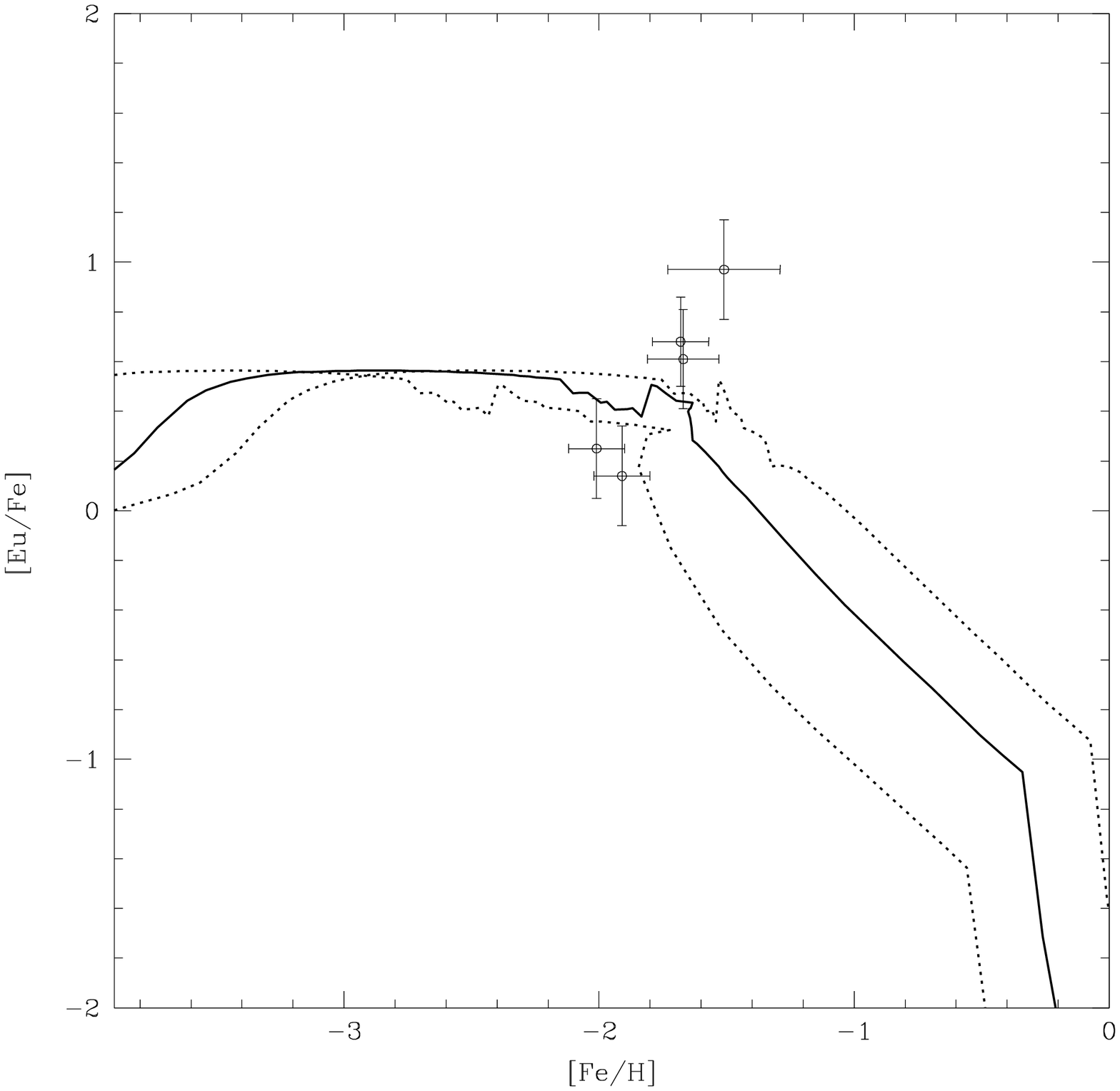}

\caption{[Eu/Fe] vs. [Fe/H] observed in Ursa Minor dSph 
galaxy compared to the predictions of the chemical evolution
model for Ursa Minor. The solid line represents the best model 
($\nu = 0.2\;Gyr^{-1}$, w$_i$ = 10) and the dotted
lines the lower ($\nu = 0.05\;Gyr^{-1}$) and upper 
($\nu = 0.5\;Gyr^{-1}$) limits for the SF efficiency.}
\label{lf4}
\end{center}
\end{figure}
%%%%%%%%%%%%%%%%%%%%%%%%%%%%%%%%%%%%%%%%%%%%%%%%%%%%%%%%%%%%%%%%%%%

The small differences in the SF and wind efficiencies do not affect 
strongly the
predictions of the models. As one can see in Table 1, the range of 
values for the SF efficiency is practically the same for Carina, Sculptor
and Ursa Minor ($\nu$ = 0.02-0.4, 0.05-0.5, 0.05-0.5 $Gyr^{-1}$, 
respectively), whereas Draco observational constraints are reproduced by
a model with lower values of $\nu$, $\nu$ = 0.005-0.1 $Gyr^{-1}$.
These values reflect in very similar curves for the first three  
galaxies and a curve for Draco with only a small difference, 
namely a [Eu/Fe] ratio which starts decreasing very slowly at metallicities
lower ([Fe/H] $\sim$ -2.0 dex) than in the other 
three galaxies. However, the abrupt decrease starts at a similar
point. The same similarity can be seen in the values of the wind efficiency:
Carina - w$_i$ = 7-11, Draco - w$_i$ = 6-10, Sculptor - w$_i$ = 11-15 -
and Ursa Minor - w$_i$ = 8-12. Only Sculptor is characterized by
a wind efficiency a little bit higher, but this fact does not influence
the pattern of the abundances significantly. They all exhibit an
intense decrease in the [Eu/Fe] ratio after the wind develops. The small
differences in the ranges of values for w$_i$ are related more directly
to the gas mass and total mass observed.

What should be highlighted is that the nucleosynthesis prescriptions
adopted here allow the models to reproduce very well the data, 
supporting the 
assumption that Eu, also in dSph galaxies, is a pure r-process 
element synthesized in massive stars in the range $M = 10-30 
M_{\odot}$, as it is in the Milky Way (see chapter 2 and Cescutti et al. 2006).
 Besides that, the low SF efficiencies
and the high wind efficiencies are required also to explain
the [Eu/Fe] observed pattern, especially the abrupt decrease of 
the data in some dSph galaxies.

\subsection{Barium}

%%%%%%%%%%%%%%%%%%%%%%%%%%%%%%%%%%%%%%%%%%%%%%%%%%%%%%%%%%%%%%%%%%%
\begin{figure}
\begin{center}
\includegraphics[width=0.99\textwidth]{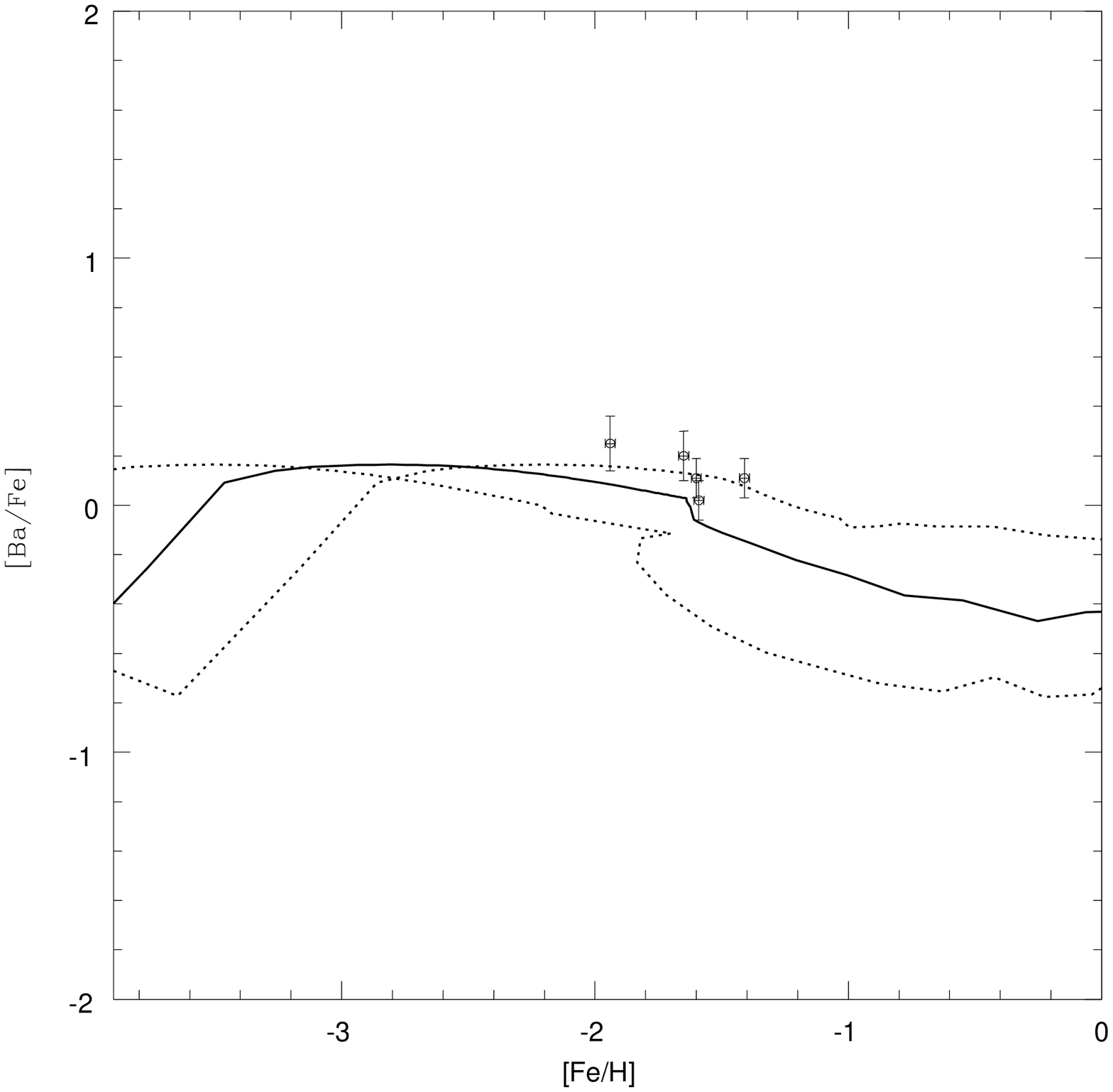}

\caption{[Ba/Fe] vs. [Fe/H] observed in Carina dSph 
galaxy compared to the predictions of the chemical evolution
model for Carina. The solid line represents the best model 
($\nu = 0.1\;Gyr^{-1}$, w$_i$ = 7) and the dotted
lines the lower ($\nu = 0.02\;Gyr^{-1}$) and upper 
($\nu = 0.4\;Gyr^{-1}$) limits for the SF efficiency.}\label{lf5}
\end{center}
\end{figure}
%%%%%%%%%%%%%%%%%%%%%%%%%%%%%%%%%%%%%%%%%%%%%%%%%%%%%%%%%%%%%%%%%%%
The evolution of [Ba/Fe] as a function of [Fe/H] predicted by the models
and compared to the observed data in four 
Local Group dSph galaxies is shown in the Figures \ref{lf5} to \ref{lf8}
(Carina, Draco, Sculptor and Ursa Minor,
respectively). One can easily notice that the predicted curves
exhibit a similar behaviour in all four galaxies: the predicted [Ba/Fe]
ratio increases fast at very low metallicities ([Fe/H] $<$ -3.5 dex),
then remains almost constant, close to the solar value, at low-intermediate 
metallicities (-3.5$<$ [Fe/H] $<$ -1.7 dex) and then starts decreasing soon 
after the occurrence of the galactic wind at relatively high
metallicities ([Fe/H] $>$ -1.7 dex). 
In this case, the decrease is not so intense as it is in the 
case of [Eu/Fe], due to the differences in the nucleosynthesis 
of Ba and Eu. 

%%%%%%%%%%%%%%%%%%%%%%%%%%%%%%%%%%%%%%%%%%%%%%%%%%%%%%%%%%%%%%%%%%%
\begin{figure}
\begin{center}
\includegraphics[width=0.99\textwidth]{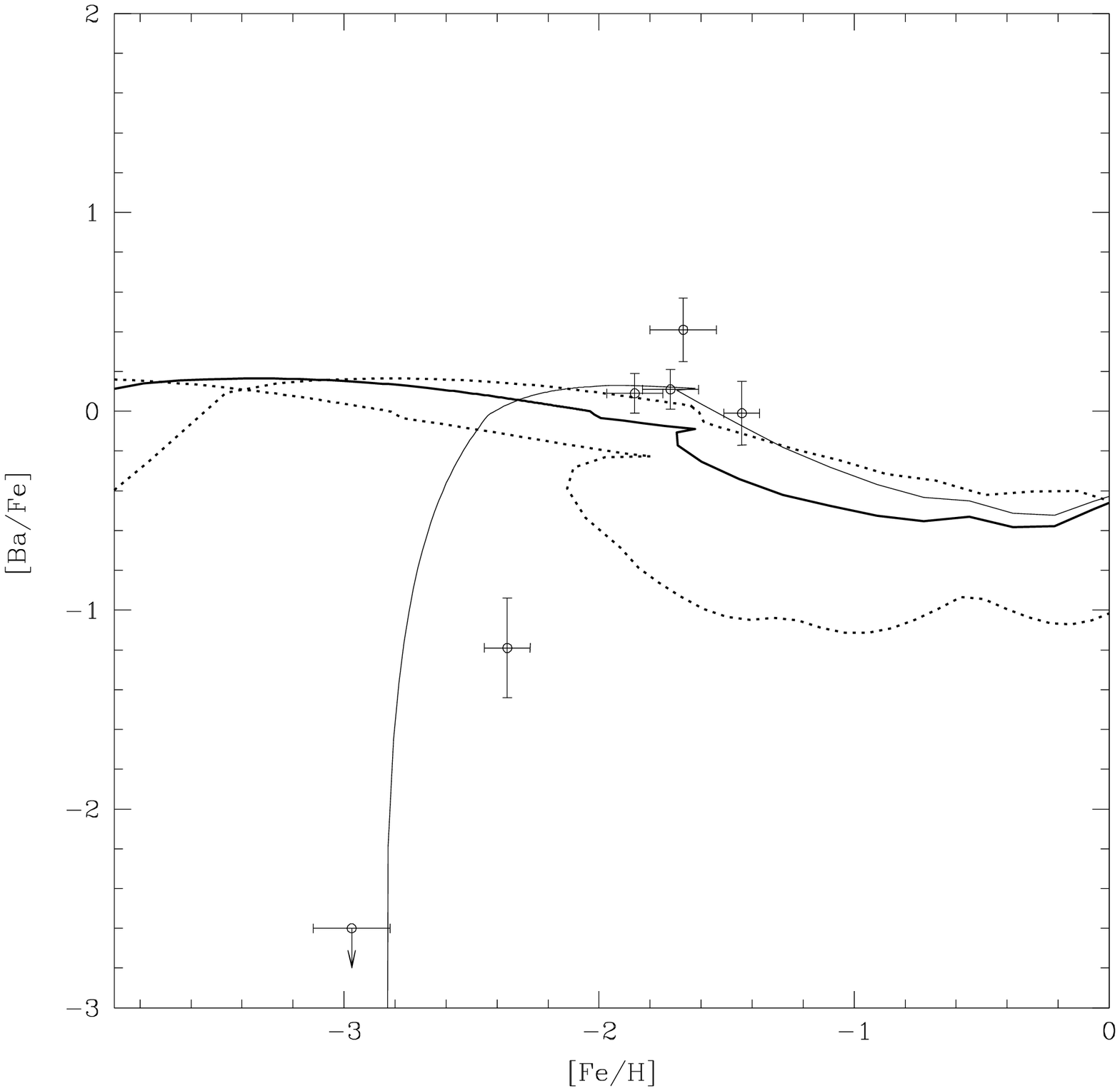}
\caption{[Ba/Fe] vs. [Fe/H] observed in Draco dSph 
galaxy compared to the predictions of the chemical evolution
model for Draco. The solid line represents the best model 
($\nu = 0.03\;Gyr^{-1}$, w$_i$ = 6) and the dotted
lines the lower ($\nu = 0.005\;Gyr^{-1}$) and upper 
($\nu = 0.1\;Gyr^{-1}$) limits for the SF efficiency. The thin line 
represents the best model without Ba production in massive stars.}
\label{lf6}
\end{center}
\end{figure}
%%%%%%%%%%%%%%%%%%%%%%%%%%%%%%%%%%%%%%%%%%%%%%%%%%%%%%%%%%%%%%%%%%%

The predicted shape of the [Ba/Fe] vs. [Fe/H] relation in dSph galaxies 
can be associated 
to the two different Ba 
contributions, from stars in different mass ranges (high masses - 
10 to 30 $M_{\odot}$ - and low masses - 1 to 3 $M_{\odot}$).
In the low metallicity
portion of the plot 
the production of Ba is dominated by the r-process taking place
in massive stars which have lifetimes in the range from 6 to 25 Myr.
Therefore,  the [Ba/Fe] ratio increases fast 
reaching values above solar already at [Fe/H] $\sim$ -3.5 dex and 
stays almost constant up to [Fe/H]=-1.7 dex.
It is worth noting that the massive star 
contribution is more  clearly seen when the SF efficiency is low. 
In this regime, in fact, the stars are formed slowly and the 
difference between the contribution of stars of different 
masses is more evident, since the increase of the metallicity and 
the evolution of the galaxy proceed at a low speed. On the other hand,
when the SF 
efficiency is higher (like in the Milky Way), the early contribution of 
massive stars is 
more difficult to distinguish, because of the much
faster increase in metallicity. 

At low-intermediate metallicities (-3.5 $<$ [Fe/H] $<$ -1.7 dex),
the production of Ba is still the one by r-process taking place
in massive stars, in particular in those with masses around the lower 
limit for 
the r-process Ba producers ($\sim 10M_{\odot}$).

The contribution to s-process Ba enrichment from low mass stars (LMS), with  
lifetimes from 400Myr to 10 Gyr,  affects significantly the predicted 
[Ba/Fe] ratio only after the onset of the wind, consequently 
only after the occurrence of the first SNe Ia. At this stage, the 
[Ba/Fe] starts to decrease rapidly, since the first SNe Ia are
injecting large amounts of Fe into the ISM. Together with the
enrichment of Fe, the SNe Ia release also large quantities of energy
in the ISM which gives rise to a galactic wind. As the galactic
wind starts, the SFR goes down to very low values and the production
of Ba is limited only to the LMS, especially those at the low mass
end. The injection of Ba in the ISM at this stage is, however,
not so effective due to the galactic wind which removes
a large fraction of the material freshly released in the hot 
medium (Ferrara $\&$ Tolstoy 1999, Recchi et al. 2001, 2004). 
The effect of the Ba production in LMS is particularly 
important to slow down the abrupt decrease
in [Ba/Fe] after the occurrence of the galactic wind. If this 
production is not taken in account, the [Ba/Fe] values after the
onset of the wind would go down faster to very low values. 

%%%%%%%%%%%%%%%%%%%%%%%%%%%%%%%%%%%%%%%%%%%%%%%%%%%%%%%%%%%%%%%%%%%
\begin{figure}
\begin{center}
\includegraphics[width=0.99\textwidth]{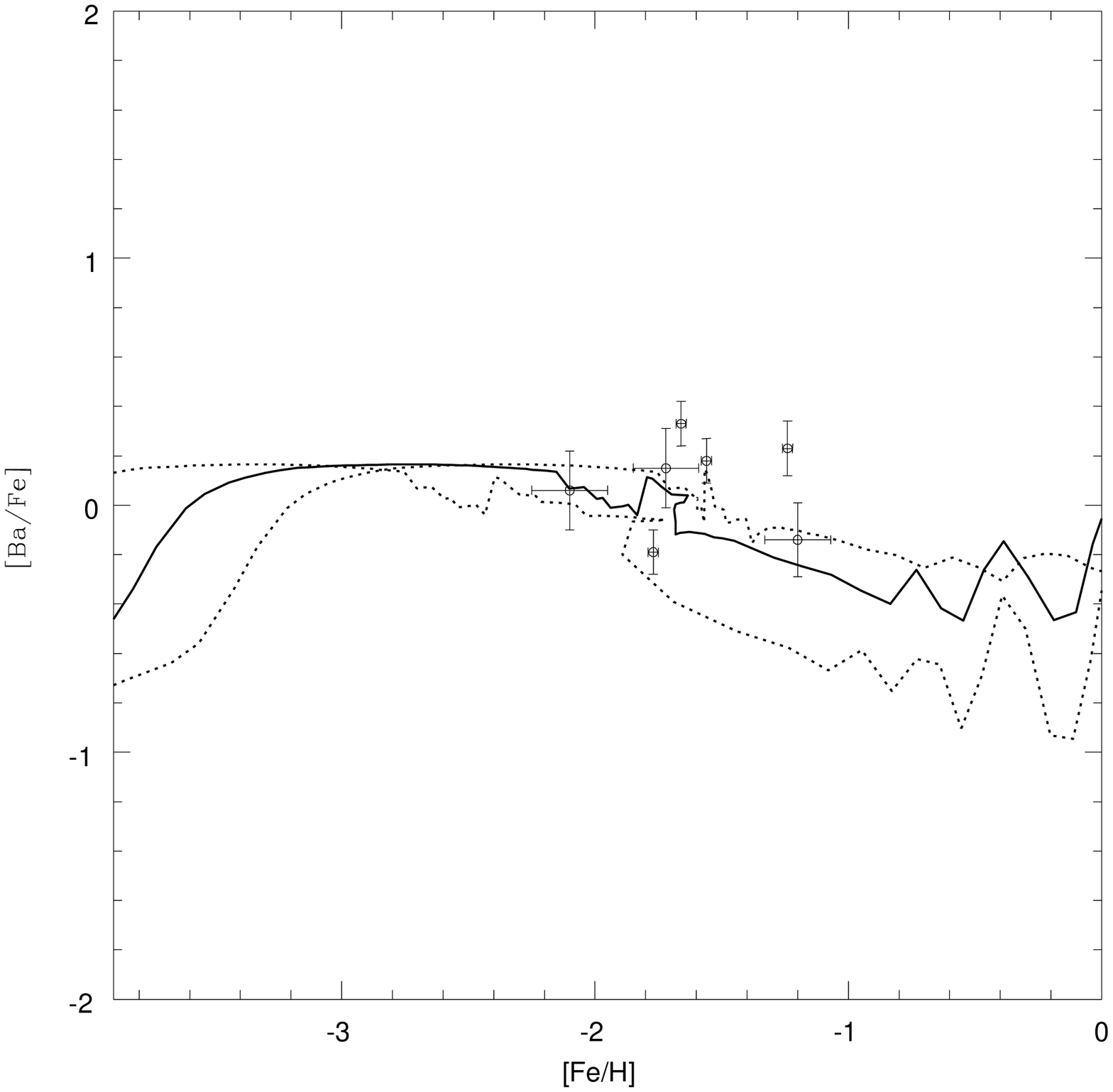}

\caption{[Ba/Fe] vs. [Fe/H] observed in Sculptor dSph 
galaxy compared to the predictions of the chemical evolution
model for Sculptor. The solid line represents the best model 
($\nu = 0.2\;Gyr^{-1}$, w$_i$ = 13) and the dotted
lines the lower ($\nu = 0.05\;Gyr^{-1}$) and upper 
($\nu = 0.5\;Gyr^{-1}$) limits for the SF efficiency.}
\label{lf7}
\end{center}
\end{figure}
%%%%%%%%%%%%%%%%%%%%%%%%%%%%%%%%%%%%%%%%%%%%%%%%%%%%%%%%%%%%%%%%%%%
One can see in the Figures \ref{lf5} to \ref{lf8} that the observational 
trends at high metallicities are very well reproduced by the model 
predictions 
supporting the assumptions made regarding the 
nucleosynthesis of Ba. As already 
mentioned the contribution from LMS to the enrichment
of Ba becomes important starting from  intermediate to high 
metallicities ([Fe/H] $>$ -1.9 dex), depending on the SF efficiency
adopted. In this 
metallicity range, the data of all four galaxies are very well 
reproduced, including the stars with low values of [Ba/Fe]
which should have formed soon after the onset of the 
galactic wind.
 
On the other hand, at low metallicities ([Fe/H] $<$ -2.4 dex)
only the observational trend of Carina and Sculptor are well 
fitted by the model predictions. In Ursa Minor and Draco there 
are a few stars
which exhibits a very low [Ba/Fe] 
($\sim$ -1.2 dex) at low [Fe/H] (Figures \ref{lf6} and \ref{lf8}). These
points are well below the predicted curves and close to the values
of the Milky Way stars at similar metallicities, which are 
reproduced by a chemical evolution model with the same 
nucleosynthesis prescriptions adopted here  but with
a higher SF efficiency. In general, it seems like if the data for the solar 
neighborhood show values of [Ba/Fe] 
lower than in the dSph galaxies at the same metallicity, although this fact should 
be confirmed by more data. For the $\alpha$-elements is the opposite, dSph
stars show lower [$\alpha$/Fe] ratios than Galactic stars at the same 
metallicity (Shetrone \& al. 2001; Tolstoy et al. 2003).
LM03 and LM04 suggested that the difference in the behaviour of 
$\alpha$-elements in the Milky Way and dSph galaxies should be ascribed to their 
different SF histories. In particular, the lower [$\alpha$/Fe] 
ratios in dSph galaxies
are due to their low star formation efficiency which produces a slow 
increase of the [Fe/H] with the consequence of having the Fe restored 
by type Ia SNe, and
therefore a decrease of the [$\alpha$/Fe] ratios, 
at lower [Fe/H] values than in the Milky Way. This effect has been 
described in Matteucci (2001) and is a consequence of the time-delay 
model applied to systems with different star formation histories.

Therefore, in the light of what is said above, 
can we explain also the differences between 
the predicted [Ba/Fe] in 
dSph galaxies and in the Milky Way? Again, the SF efficiency is the major 
responsible parameter for this difference. In the Milky Way 
model the SF efficiency is much larger (10 - 100 times) than 
the ones adopted for the dSph galaxies of the sample analyzed here.
In the low efficiency regime, the contribution from 
LMS appears at lower metallicities than in the high SF regime, exactly 
for the same reason discussed for the [$\alpha$/Fe] ratios. As a 
consequence, we predict a longer plateau for the [Ba/Fe] ratio in dSph galaxies 
than in the solar neighborhood and starting at lower metallicities.
This prediction should in the future be confirmed or rejected by more data 
at low metallicities in dSph galaxies.

%%%%%%%%%%%%%%%%%%%%%%%%%%%%%%%%%%%%%%%%%%%%%%%%%%%%%%%%%%%%%%%%%%%
\begin{figure}

\begin{center}
\includegraphics[width=0.99\textwidth]{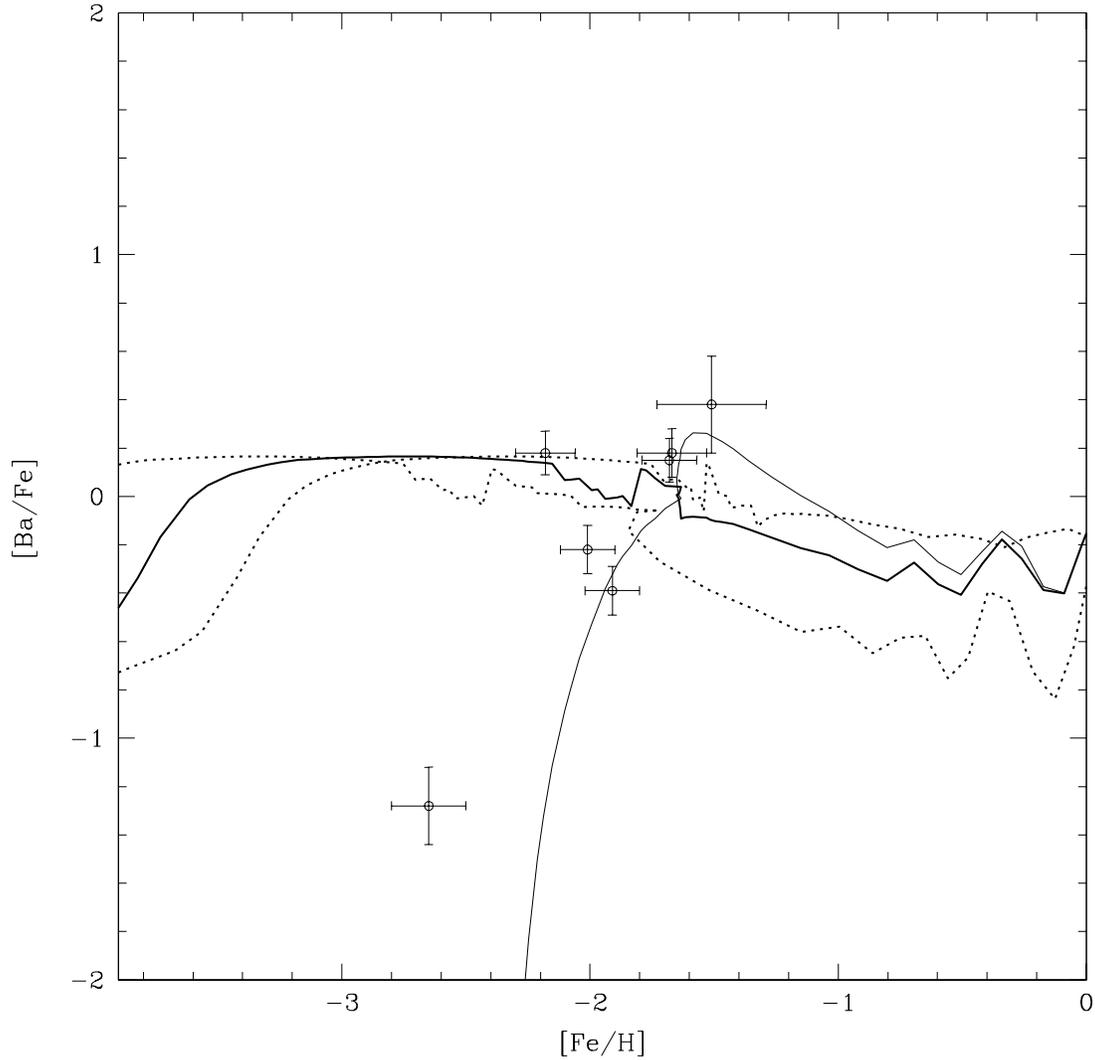}

\caption{[Ba/Fe] vs. [Fe/H] observed in Ursa Minor dSph 
galaxy compared to the predictions of the chemical evolution
model for Ursa Minor. The solid line represents the best model 
($\nu = 0.2\;Gyr^{-1}$, w$_i$ = 10) and the dotted
lines the lower ($\nu = 0.05\;Gyr^{-1}$) and upper 
($\nu = 0.5\;Gyr^{-1}$) limits for the SF efficiency. The thin line 
represents the best model without Ba production in massive stars.}
\label{lf8}
\end{center}
\end{figure}

\begin{figure}
\begin{center}
\includegraphics[width=0.99\textwidth]{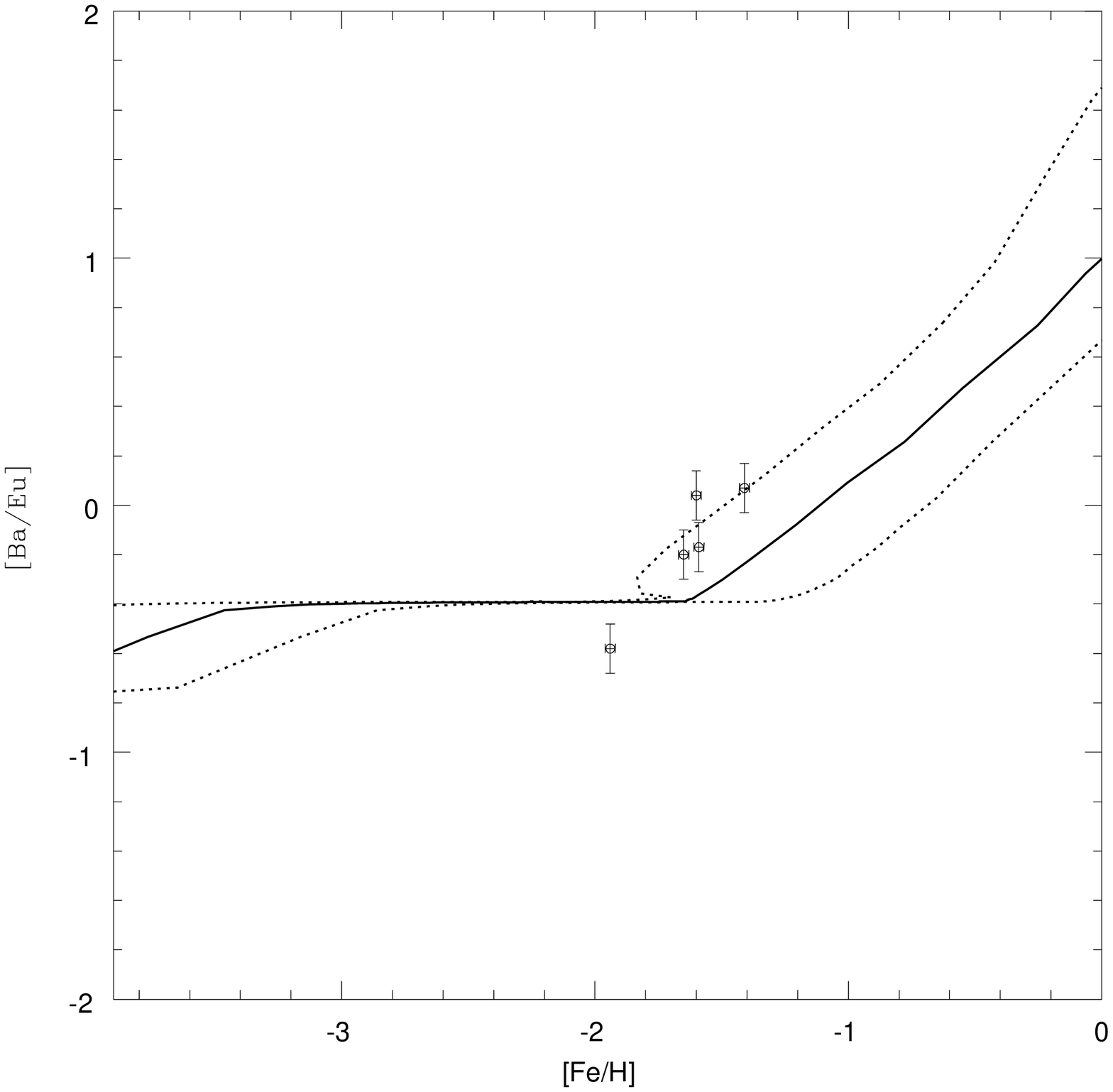}

\caption{[Ba/Eu] vs. [Fe/H] observed in Carina dSph 
galaxy compared to the predictions of the chemical evolution
model for Carina. The solid line represents the best model 
($\nu = 0.1\;Gyr^{-1}$, w$_i$ = 7) and the dotted
lines the lower ($\nu = 0.02\;Gyr^{-1}$) and upper 
($\nu = 0.4\;Gyr^{-1}$) limits for the SF efficiency.}
\label{lf9}
\end{center}
\end{figure}
%%%%%%%%%%%%%%%%%%%%%%%%%%%%%%%%%%%%%%%%%%%%%%%%%%%%%%%%%%%%%%%%%%%

Since there are no observed stars at  
low [Fe/H] in Carina and Sculptor while there are three stars 
(one with an upper limit) with very 
low [Ba/Fe] in Draco and Ursa Minor, one could argue that the Ba production 
from massive stars is not necessary.
To better see the effect of the r-process Ba production from massive stars,
we computed models suppressing this contribution. 
In such a case, the 
predictions of the models lie below all the observed data and are not 
capable of fitting the stars with low [Ba/Fe]. 
%If, on the other hand, 
%besides suppressing the contribution from r-processed Ba synthesized in 
%massive stars one expands the production of s-processed Ba to stars 
%with $M = 3 - 4 M_{\odot}$, with an yield $X_{Ba}$ = 0.5 x $10^{-6}$ 
%in this mass range,
%the models predict a trend similar to that observed 
%(see the thin lines in  Figures \ref{lf6} and
%8). The assumption  of Ba production 
%by s-process in stars with $M = 3 - 4 M_{\odot}$ is justified 
%by the fact that this production is predicted by models of stellar 
%evolution (Gallino et al. 1998, Busso et al. 2001), even
%though there are no such yields available in the literature. 
%Besides that, Travaglio et al. (1999) 
%suggested that the dominant production of Ba comes from stars with  
%$2 - 4 M_{\odot}$.
Nevertheless, one can see from the thin lines in Figures \ref{lf6} and \ref{lf8} that 
the increase of the [Ba/Fe] ratio occurs at
metallicities similar to those of  
the stars with low [Ba/Fe]. Besides that, this
model also reproduces the high values of [Ba/Fe] at high
metallicities and the [Ba/Eu] observed (see next section). In that sense, 
the observed low values of [Ba/Fe], if confirmed by more observations,
could be explained by a model with Ba produced only by s-process contribution from LMS. 
Only more observations of stars with [Fe/H] $<$ -2.0
 in dSph galaxies could solve this problem.

%%%%%%%%%%%%%%%%%%%%%%%%%%%%%%%%%%%%%%%%%%%%%%%%%%%%%%%%%%%%%%%%%%%
\begin{figure}

\begin{center}
\includegraphics[width=0.99\textwidth]{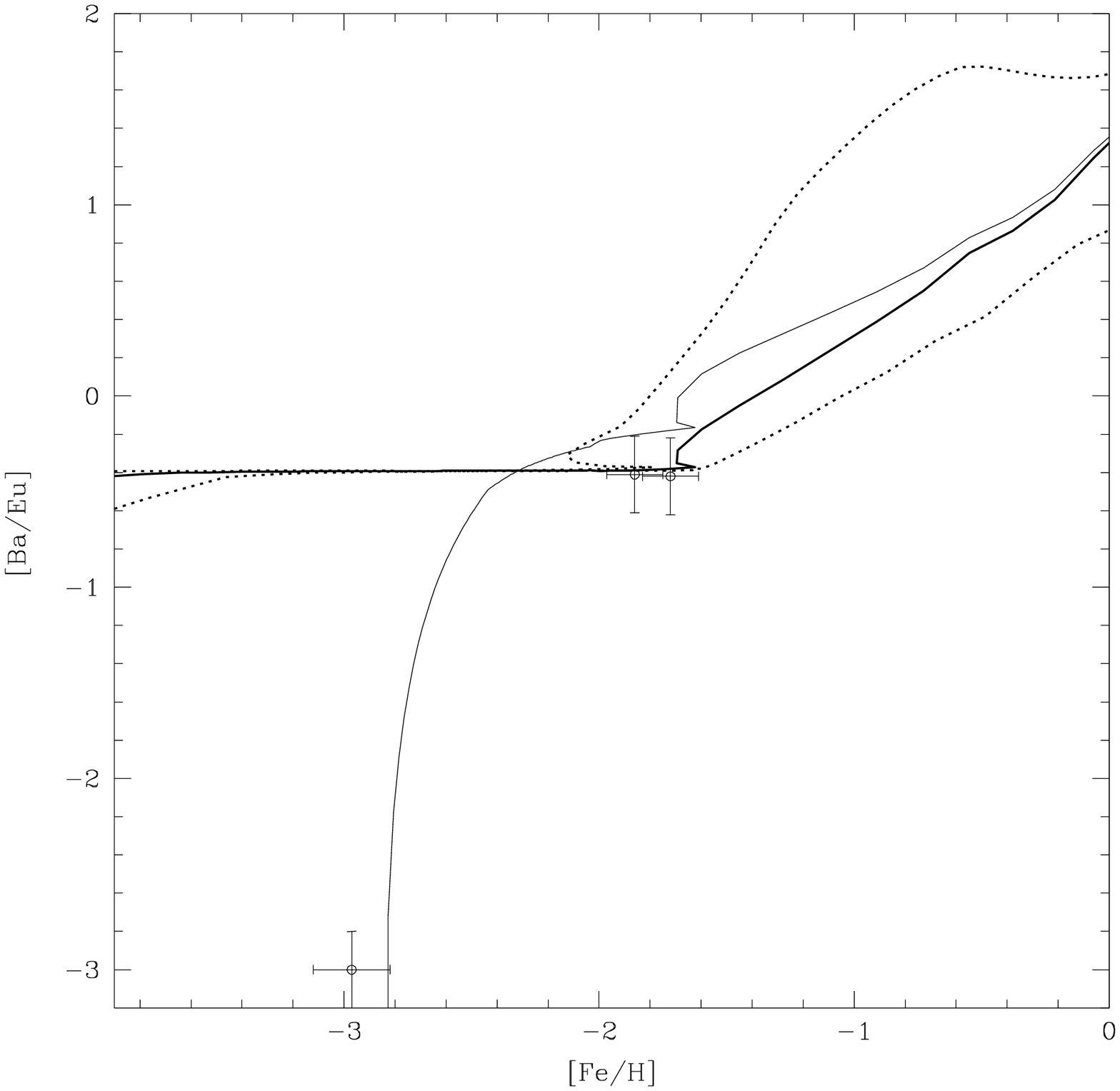}

\caption{[Ba/Eu] vs. [Fe/H] observed in Draco dSph 
galaxy compared to the predictions of the chemical evolution
model for Draco. The solid line represents the best model 
($\nu = 0.03\;Gyr^{-1}$, w$_i$ = 6) and the dotted
lines the lower ($\nu = 0.005\;Gyr^{-1}$) and upper 
($\nu = 0.1\;Gyr^{-1}$) limits for the SF efficiency. The thin line 
represents the best model without Ba production in massive stars.}
\label{lf10}
\end{center}
\end{figure}
%%%%%%%%%%%%%%%%%%%%%%%%%%%%%%%%%%%%%%%%%%%%%%%%%%%%%%%%%%%%%%%%%%%

%Also in the case of [Ba/Fe], the minor differences in the range of
%values adopted for the SF and wind efficiencies among the galaxies
%analyzed do not affect significantly the results. This is due to 
%the fact that the adopted values for all four galaxies are very similar,
%and so are the predicted curves. The most important point is
%that the assumed nucleosynthesis prescriptions produce a good fit to the
%observations, confirming the proposed scenario for the evolution of 
%the dSph galaxies.

\subsection{The ratio [Ba/Eu]} 

The comparison between the observed [Ba/Eu] as a function of [Fe/H] 
and the predicted curves for the four dSph galaxies is shown
in Figures \ref{lf9} to \ref{lf12}. The models predict a similar pattern for all four
galaxies: an almost constant sub-solar value at low metallicities
([Fe/H] $<$ -1.7 dex) and, after that, a strong increase. This 
pattern is explained again by the adopted nucleosynthesis
and by the effect of the galactic wind on the SFR and, consequently,
on the production of Ba and Eu. At the early stages of evolution, the
high mass stars provide the major contribution to the enrichment of the
ISM medium. Since Ba and Eu are both produced by the r-process taking place
in massive stars,
they both are injected in the ISM when the gas metallicity is still low. 
The difference
is that Eu is considered to be a pure r-process element, while the
fraction of Ba that is produced by the r-process is low and its 
bulk originates instead from LMS.
This fact translates into the sub-solar pattern observed in the predicted
curves: more Eu than Ba is injected in the ISM at low metallicities, at an
almost constant rate. When the LMS start to die and the first SNe Ia 
start exploding the scenario changes significantly. The LMS
inject a considerable amount of Ba into the ISM causing an increase in the
[Ba/Eu] ratio. Besides that, the energy
released by the SNe Ia contributes to the onset of the galactic wind. 
Since the wind is very intense, it removes from the galaxy a large 
fraction of the gas reservoir which feeds the SF. Consequently, 
the SFR drops down considerably and also the  production of Eu 
by massive stars, because the number of new formed stars is almost 
negligible. Barium, on the other hand,
continues to be produced and injected in the ISM by the LMS
(s-process). This fact induces the increase of [Ba/Eu] to be even more
intense, as one can see in the predicted curves (Figures \ref{lf9} to \ref{lf12}).

%%%%%%%%%%%%%%%%%%%%%%%%%%%%%%%%%%%%%%%%%%%%%%%%%%%%%%%%%%%%%%%%%%%
\begin{figure}

\begin{center}
\includegraphics[width=0.99\textwidth]{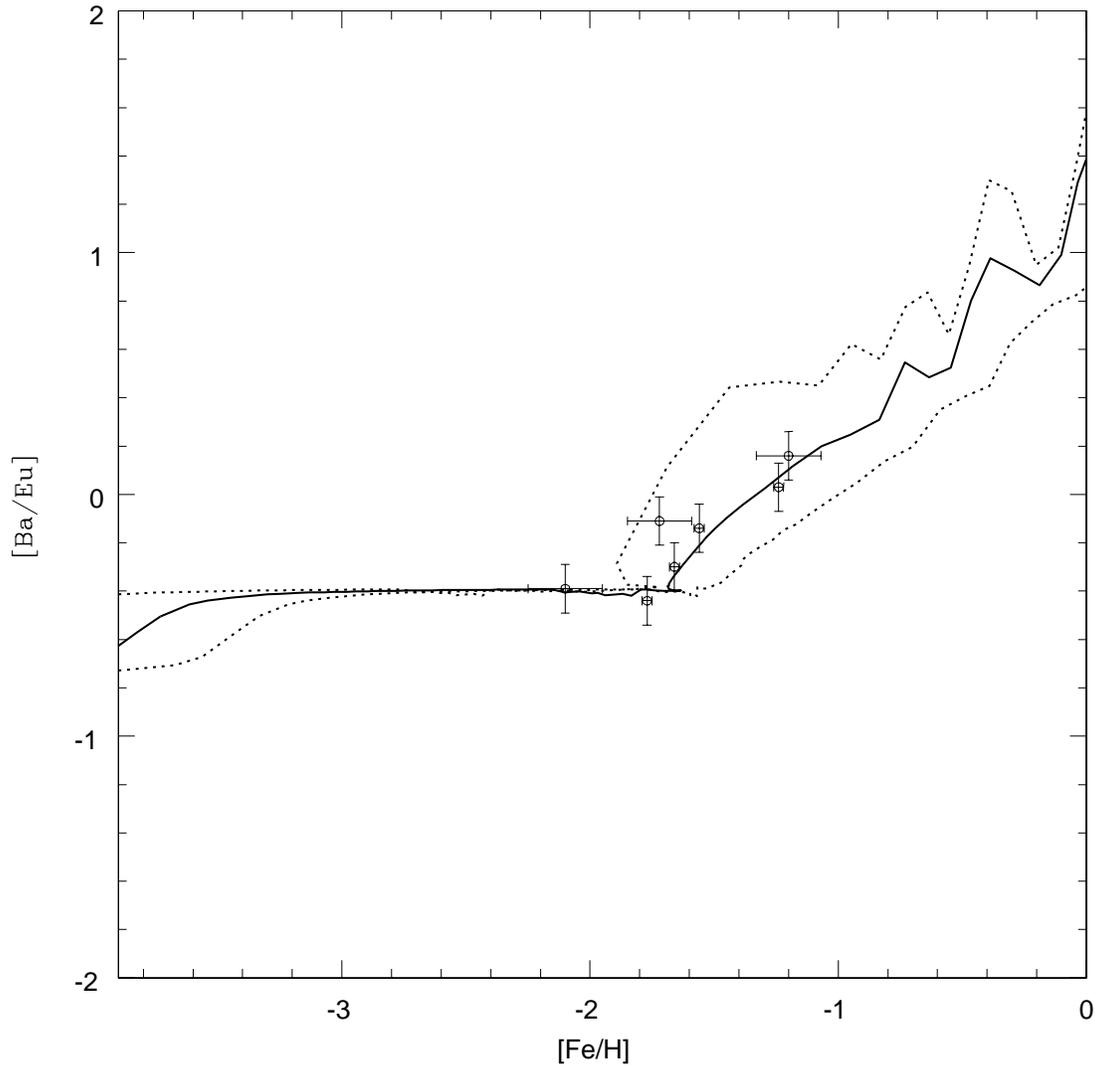}

\caption{[Ba/Eu] vs. [Fe/H] observed in Sculptor dSph 
galaxy compared to the predictions of the chemical evolution
model for Sculptor. The solid line represents the best model 
($\nu = 0.2\;Gyr^{-1}$, w$_i$ = 13) and the dotted
lines the lower ($\nu = 0.05\;Gyr^{-1}$) and upper 
($\nu = 0.5\;Gyr^{-1}$) limits for the SF efficiency.}
\label{lf11}
\end{center}

\end{figure}
%%%%%%%%%%%%%%%%%%%%%%%%%%%%%%%%%%%%%%%%%%%%%%%%%%%%%%%%%%%%%%%%%%%
The observed trend is very well reproduced by the predicted curves
in all four galaxies, especially in the case of Carina and Sculptor
(Figures \ref{lf9}  and \ref{lf11}, respectively).
The abundance pattern of these two galaxies not only exhibits the 
"plateau" at low metallicities, but also the sudden observed increase of
[Ba/Eu] after the onset of the wind, suggesting that the adopted
nucleosynthesis prescriptions for both Ba and Eu are appropriate
and that the scenario described by the chemical evolution models
is suitable to explain the evolution of these galaxies. In the case
of Draco and Ursa Minor (Figures \ref{lf10} and \ref{lf12}, respectively), there are
no stars with metallicities larger than [Fe/H] $\sim$ -1.7 dex, the one
characteristic for the onset of the galactic wind. Therefore, one 
cannot verify if this scenario (after the occurrence of the wind)
holds also for these systems. However, the "plateau" is very well 
reproduced, even though there is some dispersion in the data, 
especially in the case of Ursa Minor.

%%%%%%%%%%%%%%%%%%%%%%%%%%%%%%%%%%%%%%%%%%%%%%%%%%%%%%%%%%%%%%%%%%%
\begin{figure}

\begin{center}
\includegraphics[width=0.99\textwidth]{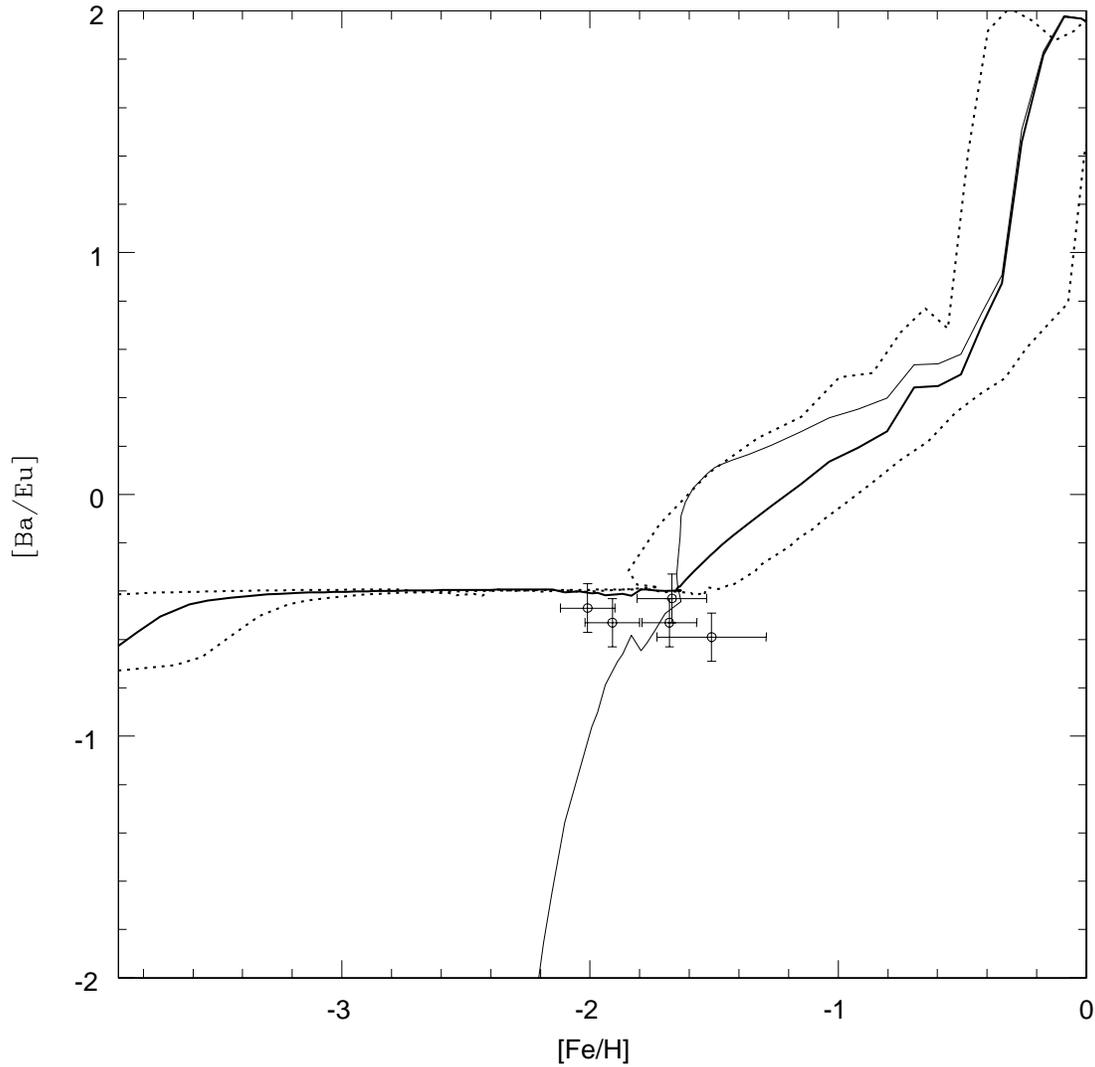}

\caption{[Ba/Eu] vs. [Fe/H] observed in Ursa Minor dSph 
galaxy compared to the predictions of the chemical evolution
model for Ursa Minor. The solid line represents the best model 
($\nu = 0.2\;Gyr^{-1}$, w$_i$ = 10) and the dotted
lines the lower ($\nu = 0.05\;Gyr^{-1}$) and upper 
($\nu = 0.5\;Gyr^{-1}$) limits for the SF efficiency. The thin line 
represents the best model without Ba production in massive stars.}
\label{lf12}
\end{center}
\end{figure}
%%%%%%%%%%%%%%%%%%%%%%%%%%%%%%%%%%%%%%%%%%%%%%%%%%%%%%%%%%%%%%%%%%%
It is important to stress that the predicted [Ba/Eu] reproduces
all the observed trends, and that no star, given the uncertainties
(with the exception of Draco 
119, which exhibits a very uncertain value due to the limits on 
Ba and Eu abundances),
lies outside the predictions, as it was the case for the two stars 
with very low [Ba/Fe]. This fact suggests strongly 
that the outsider stars must be examined separately.

\subsection{The Sagittarius dSph galaxy}  

In this section, we present the predictions for [Ba/Fe], [Eu/Fe] and 
[Ba/Eu] as functions of [Fe/H] in Sagittarius dSph galaxy. Even
though there are only two stars (Bonifacio et al. 2000)
observed with Ba and Eu, it is 
interesting to compare the predictions of the models to the data and to
analyze how these ratios would behave in this dSph galaxy. As
mentioned in LM04, the Sagittarius dSph galaxy exhibits chemical
properties which distinguish this galaxy from the other Local Group dSph 
galaxies. In particular, the SF efficiency (required to reproduce the
observed [$\alpha$/Fe] ratios) and the predicted metallicity 
distribution of this galaxy differ a lot from the other dSph 
galaxies analyzed - Draco, Carina, Ursa Minor, Sextan and Sculptor - 
being more similar to the values assumed for the Milky Way disc. The 
required SF 
efficiency is much higher ($\nu$ = 1 - 5 Gyr$^{-1}$
compared to $\nu$ = 0.01 - 0.5 Gyr$^{-1}$) and the stellar metallicity
distribution exhibits a peak at higher metallicities ([Fe/H] $\sim$ -0.6 
dex) than the other dSph galaxies ([Fe/H] $\sim$ -1.6 dex) and 
close to the one from the solar neighborhood.  As a consequence, one 
would expect also [Ba/Fe], [Eu/Fe] in Sagittarius to be different 
from the patterns observed in the other four dSph and more similar to 
those observed in the metal-poor stars of the
Milky Way. 

In order to predict the evolution of Ba and Eu as functions of Fe, we 
made use of the Sagittarius dSph model as described in LM04, without 
any changes in the most important parameters, such as SF efficiency 
and wind efficiency, and with the same nucleosynthesis prescriptions 
adopted for the other dSph galaxies. This procedure is justified 
by the fact that no modifications were required for the LM04 models 
of the other galaxies (Carina, Draco, Sculptor and Ursa Minor) to
fit the observed [Ba/Fe], [Eu/Fe] and [Ba/Eu].

%%%%%%%%%%%%%%%%%%%%%%%%%%%%%%%%%%%%%%%%%%%%%%%%%%%%%%%%%%%%%%%%%%%
\begin{figure}
\begin{center}
\includegraphics[width=0.99\textwidth]{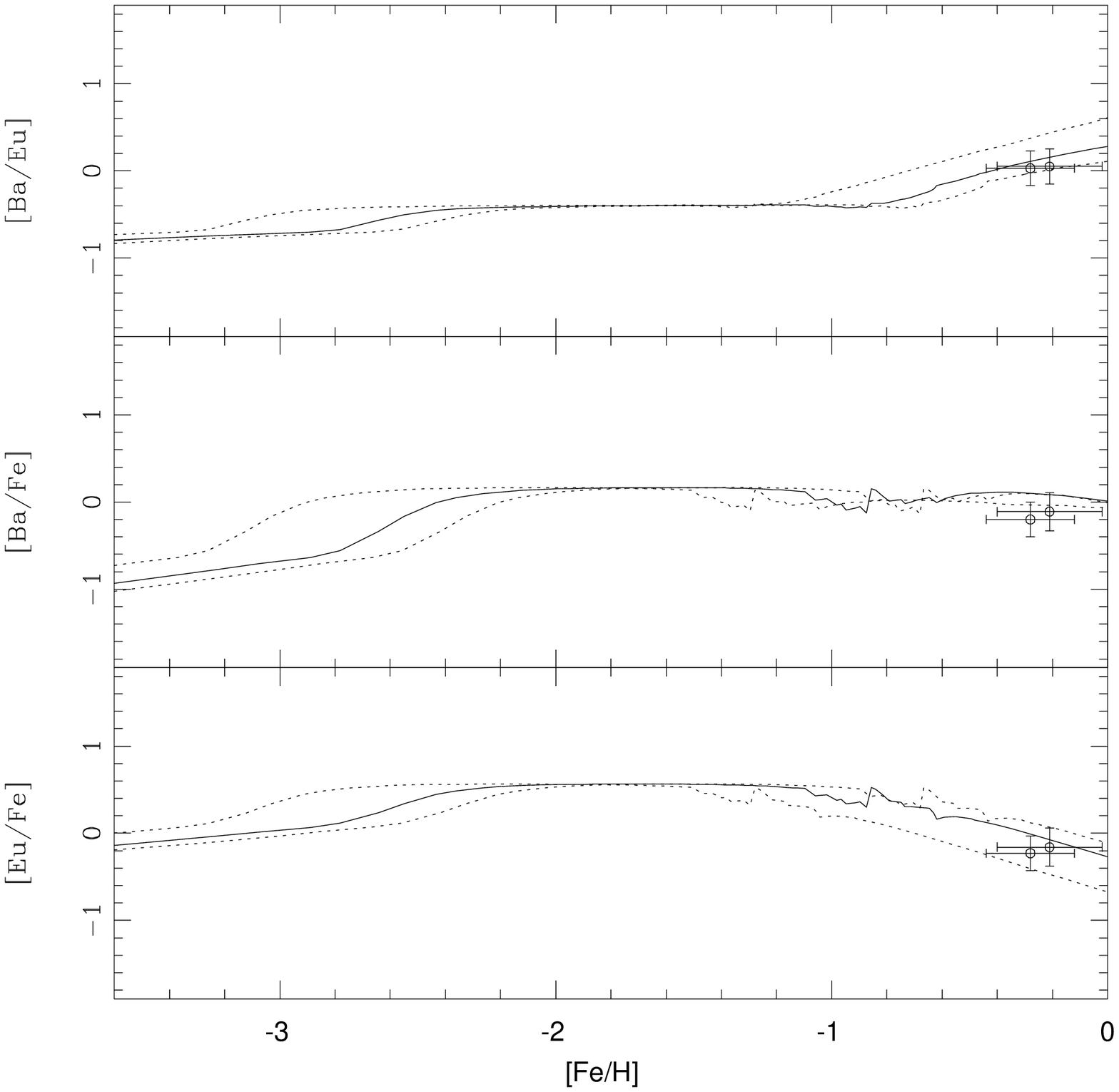}
\caption{The predicted evolution of Ba and Eu as function of [Fe/H] for
Sagittarius dSph galaxy compared with the data. The solid line represents
the best model  ($\nu =
3\;Gyr^{-1}$, w$_i$ = 9) and the dotted lines the lower ($\nu =
1\;Gyr^{-1}$) and upper  ($\nu = 5\;Gyr^{-1}$) limits for the SF
efficiency. } \label{lf13}
\end{center}
\end{figure}
%%%%%%%%%%%%%%%%%%%%%%%%%%%%%%%%%%%%%%%%%%%%%%%%%%%%%%%%%%%%%%%%%%%

In Figure \ref{lf13}, the predictions for Sagittarius dSph galaxy model for 
[Ba/Fe], [Eu/Fe] and [Ba/Eu] are shown in comparison with the data. 
As one can clearly see, all three predicted ratios reproduce very well the 
data and exhibit significant differences (in particular [Ba/Fe])
when compared to the predictions (and observations) for 
the other dSph galaxies. The decrease in the [Eu/Fe] at relatively 
high metallicities ([Fe/H] $\sim$ -1.7 dex) observed in the four 
dSph galaxies, and attributed to the effect of the galactic wind on 
the SFR, is less intense in the case of Sagittarius. Moreover, one cannot 
see the high values of [Ba/Fe] at low metallicities ([Fe/H] 
$<$ -3.0 dex ), which were explained as an effect of the low SF 
efficiency.  Also, the predicted [Ba/Eu] ratios do not show the 
almost constant  "plateau" observed at low and intermediate metallicities 
in the other dSph  galaxies. All these differences can be also 
found when one compares the pattern of these ratios in dSph galaxies 
with those in the metal-poor stars of the Milky Way. The differences 
between the predictions of Sagittarius and the other dSph and
the similarities with the Milky Way can be attributed to the high
values of the SF efficiency adopted for Sagittarius when compared 
to the other dSph galaxies. These high values, in fact, are more 
similar to the values generally adopted for the solar neighborhood 
(Chiappini, Matteucci $\&$ Gratton, 1997). 
Consequently, one could suggest that the chemical evolution of 
Sagittarius follows roughly that of the Milky Way disc at the solar 
neighborhood in contrast to the other dSph Galaxies,
which exhibit a much slower chemical evolution.

\chapter [Comparison between  the MW and Sculptor]
{Comparison between  the evolution of Ba, La, Eu and Y 
in the Milky Way and in the dwarf spheroidal galaxy Sculptor}

\rightline{\emph{``Passion is like the lightning, it is beautiful, }}
\rightline{\emph{and it links the earth to heaven, but alas it blinds!''}}
\rightline{\emph{by H. Rider Haggard}}

\vspace{1cm}

In this chapter we use the results of the previous chapters, in particular
chapter 2 and chapter 5, to compare the results of the Milky Way
to those of the dwarf spheroidal galaxies. In particular, we choose Sculptor as typical dwarf spheroidal
galaxy. We also include Y and La in our discussion, besides Eu and Ba discussed in the
previous chapter.
We show how in these two different systems, the Milky
Way and Sculptor, the results of the chemical evolution
 predicts different abundance ratios and these results are
confirmed by the observational data.
We concentrate on the ratios of neutron capture elements Y-Ba-La-Eu, which are
the neutron capture elements measured in Sculptor.

\section{Results}

\subsection{Ratios of neutron capture elements over iron}

In Fig. \ref{chap6-1} we plot [Y/Fe]  vs [Fe/H]. The data for Sculptor are
in a small range of [Fe/H] and present a quite large scatter. Nevertheless,
Sculptor has a mean ratio [Y/Fe] below the mean ratio of the Milky Way 
in the same range of [Fe/H], and this confirms the result of our model
for Sculptor. The model predicts a strong decrease of the ratio [Y/Fe] due to the galactic wind
which takes place for [Fe/H] $>$ -2.

\begin{figure}
\begin{center}
\includegraphics[width=0.99\textwidth]{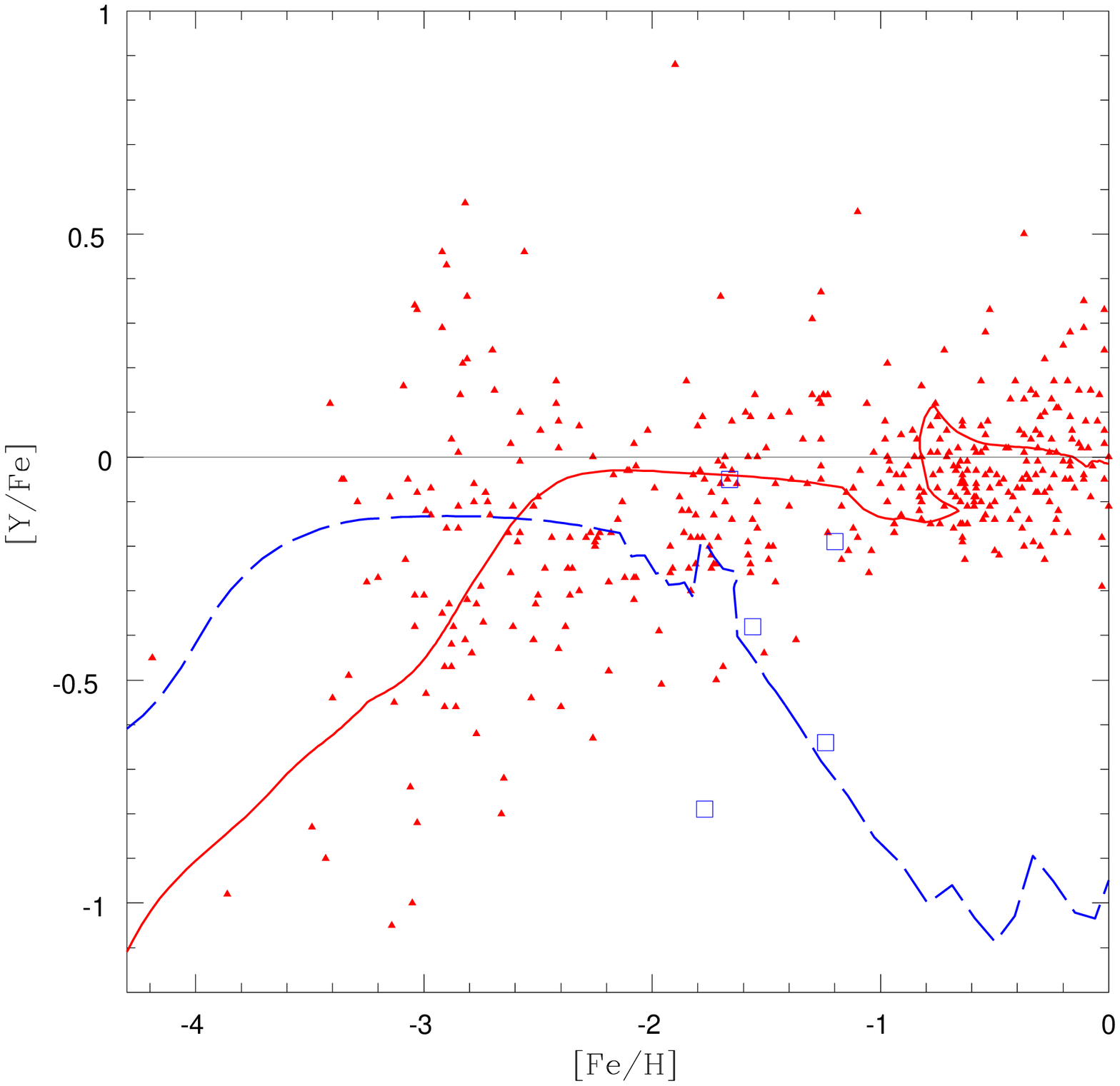}
\caption{[Y/Fe]  vs [Fe/H]: the red dots are the observational data for the 
for the Milky Way, whereas the blue open squares are the data for Sculptor. 
The results of the model for the Milky Way are plotted in red solid line
 and the results for Sculptor in blue dashed line.}
\label{chap6-1}
\end{center}
\end{figure}

The steepness of the decrease depends on the s-process contribution to each
neutron capture elements. The s-process enrichment due to low-mass stars is 
delayed with respect to the galactic wind time for Sculptor;
 as a result, the elements with a contribution by the s-process 
show a smaller decrease of their ratios relative to iron.

In fact, the predictions of the model for the ratio of [Y/Fe], 
and yttrium has a contribution of s-process , show a smaller decrease 
than the results for the ratio of europium over iron (see Fig. \ref{chap6-2}).

\begin{figure}
\begin{center}
\includegraphics[width=0.99\textwidth]{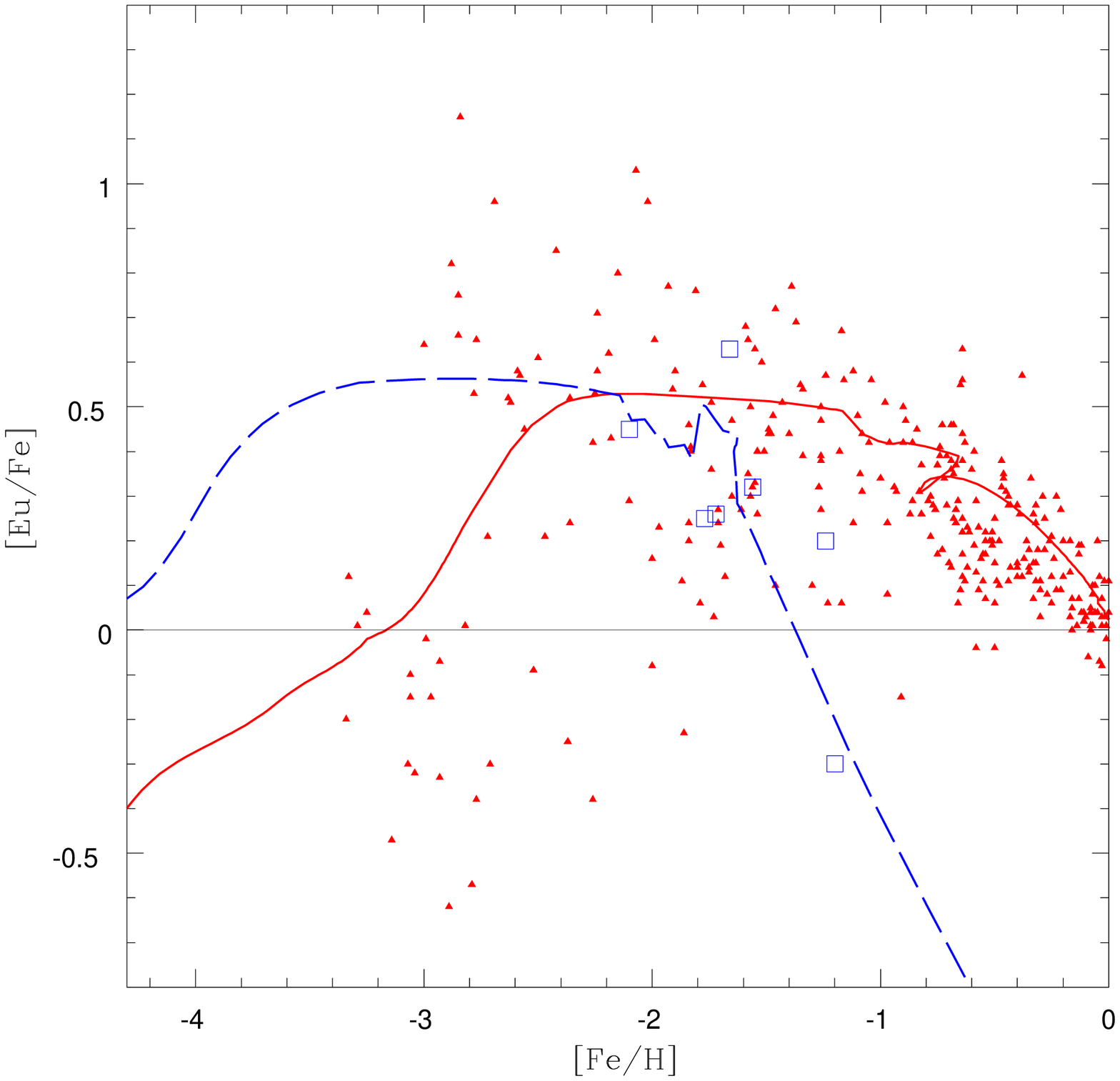}
\caption{[Eu/Fe] vs [Fe/H]: the red dots are the observational data 
for the Milky Way, whereas the blue open squares are the data for Sculptor. 
The results of the model for the Milky Way are plotted in red solid line
 and the results for Sculptor in blue dashed line.}
\label{chap6-2}
\end{center}
\end{figure}

 Europium  has a negligible production by s-process and 
the model predicts a very steep decrease for its ratio over iron in Sculptor
and this well fits  the observational data for this galaxy  whereas the data 
for the Milky Way do not show this so steep decrease at this stage.

For barium and lanthanum the results are similar to those of yttrium;
 this is in agreement with the data of Sculptor (see Fig. \ref{chap6-3},  \ref{chap6-4}),
even if  in these cases there are not significant differences
with the data of the Milky Way.

\begin{figure}
\begin{center}
\includegraphics[width=0.99\textwidth]{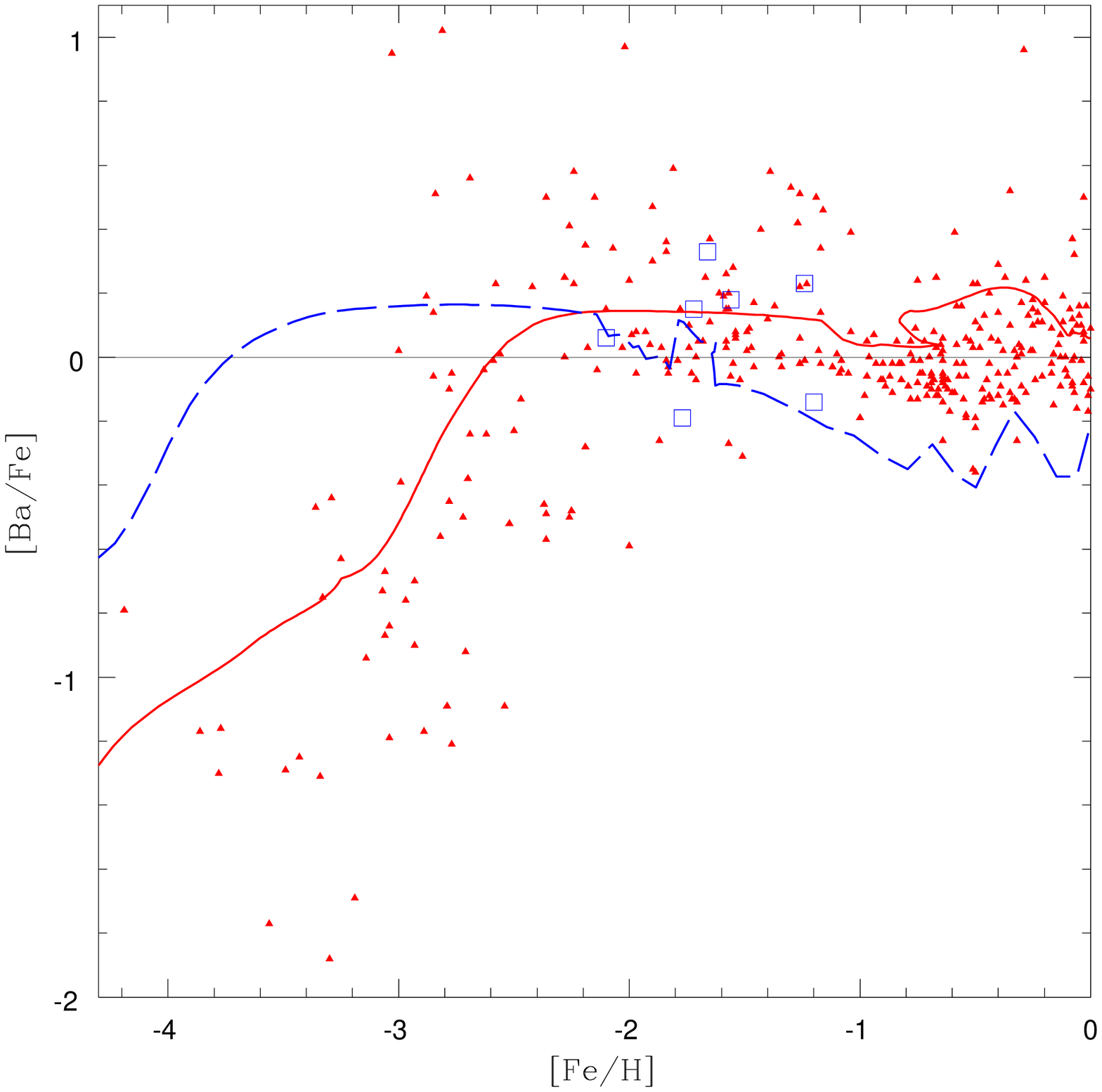}
\caption{[Ba/Fe] vs [Fe/H]: the red dots are the observational data
for the Milky Way, whereas the blue open squares are the data for Sculptor. 
The results of the model for the Milky Way are plotted in red solid line
 and the results for Sculptor in blue dashed line.}
\label{chap6-3}
\end{center}
\end{figure}

\begin{figure}
\begin{center}
\includegraphics[width=0.99\textwidth]{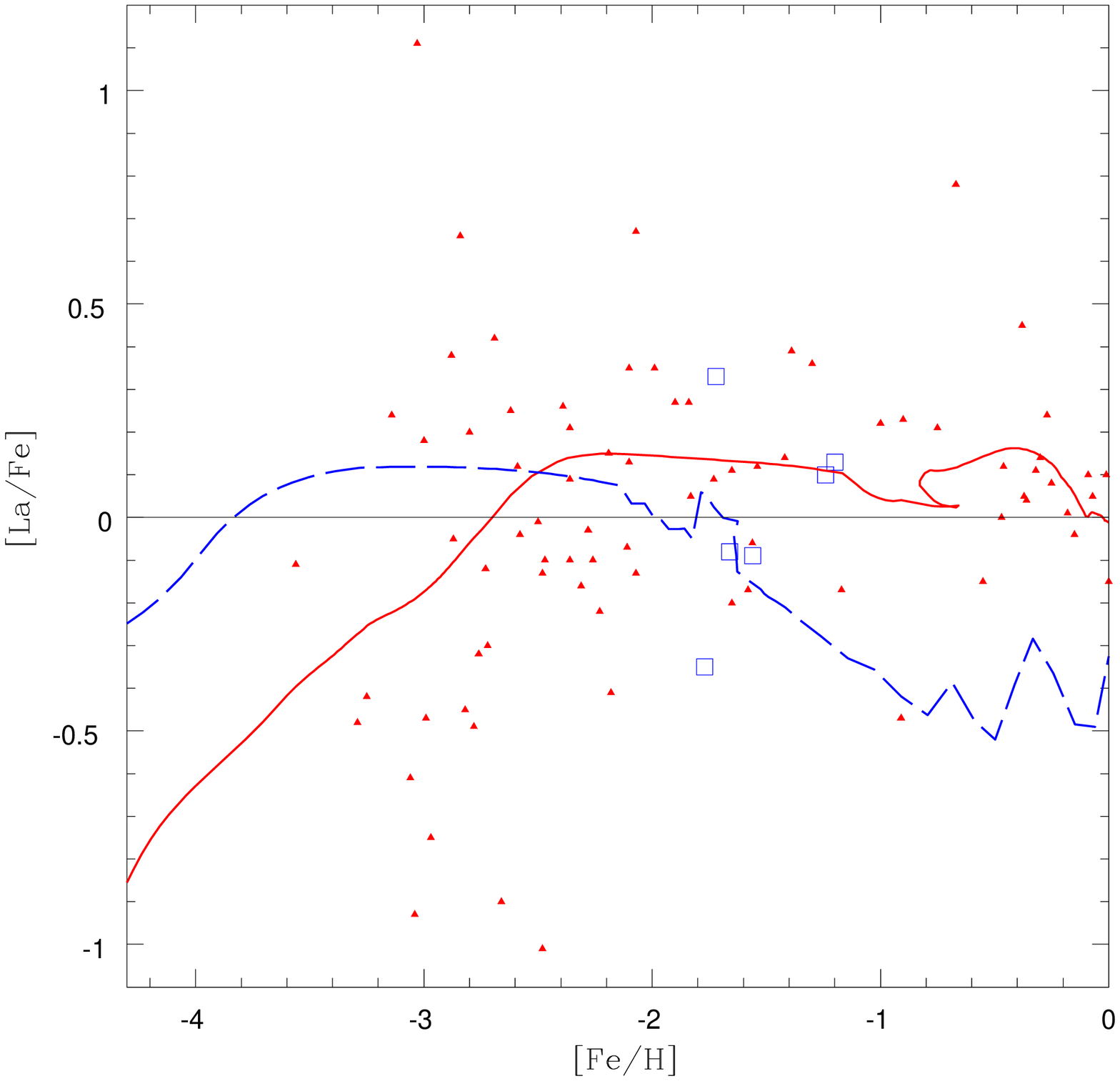}
\caption{[La/Fe] vs [Fe/H]: the red dots are the observational data  
for the Milky Way, whereas the blue open squares are the data for Sculptor. 
The results of the model for the Milky Way are plotted in red solid line
 and the results for Sculptor in blue dashed line.}
\label{chap6-4}
\end{center}
\end{figure}

The two different models, which share the same nucleosynthesis, give different
results for the two systems due to the different conditions in which the chemical
evolution take place, as  the presence of a galactic wind in Sculptor,
 the different star formation history and the different efficiency for star formation
(for others details see chapter 2 and 5).

It is worth noting that the main difference in the [Y, Ba, La, Eu/Fe] in Sculptor 
(and in general in the dSph galaxies) and in the Milky Way is due to the time delay
model, namely the different timescales with which different elements are produced, 
in conjunction with different star formation histories. In the figures \ref{chap6-1},
\ref{chap6-2}, \ref{chap6-3} and \ref{chap6-4}, we see that the abundance ratios in Sculptor
are higher than in the Milky Way at very low metallicities ([Fe/H] $<$ -3.0 dex).
This is due to the fact that [Fe/H] grows at a different rate in Sculptor and in the Milky Way.
In particular, in Sculptor the SFR is lower than  in the Milky Way, thus a given [Fe/H]
corresponds to different cosmic epochs in the two systems. 
%Therefore, [Fe/H]=-3.0 dex,
%for example corresponds in Sculptor to a cosmic time where the r-process contribution
%have already well enriched the ISM, contrary to what happens in the Milky Way.

\subsection{Neutron capture elements ratios: 
[Y/Eu] - [Ba/Eu] - [La/Eu] - [Ba/Y] - [La/Y]}

In Fig.\ref{chap6-5} we show the ratio [Ba/Eu]. The trends of the data
in the two systems are quite different. In the Milky Way, the ratio  is almost flat up to 
[Fe/H]$\sim$ -1 at a value of about [Ba/Eu]$\sim$ -0.5, then it increases up to the solar value,
in agreement with our model and it is due to s-process enrichment by low mass stars.
In the dSph galaxy Sculptor the enrichment by s-process has
a different timescale. When the low mass stars 
start to die and enrich the interstellar medium, the [Fe/H] in Sculptor
is only $\sim$  -2 dex. So the ratio [Ba/Eu] in our models starts to increase 
at lower metallicities in Sculptor than in the Milky Way and this is what is observed in the 
data. This is a remarkable result and we give here for the first time this
kind of interpretation.

\begin{figure}
\begin{center}
\includegraphics[width=0.99\textwidth]{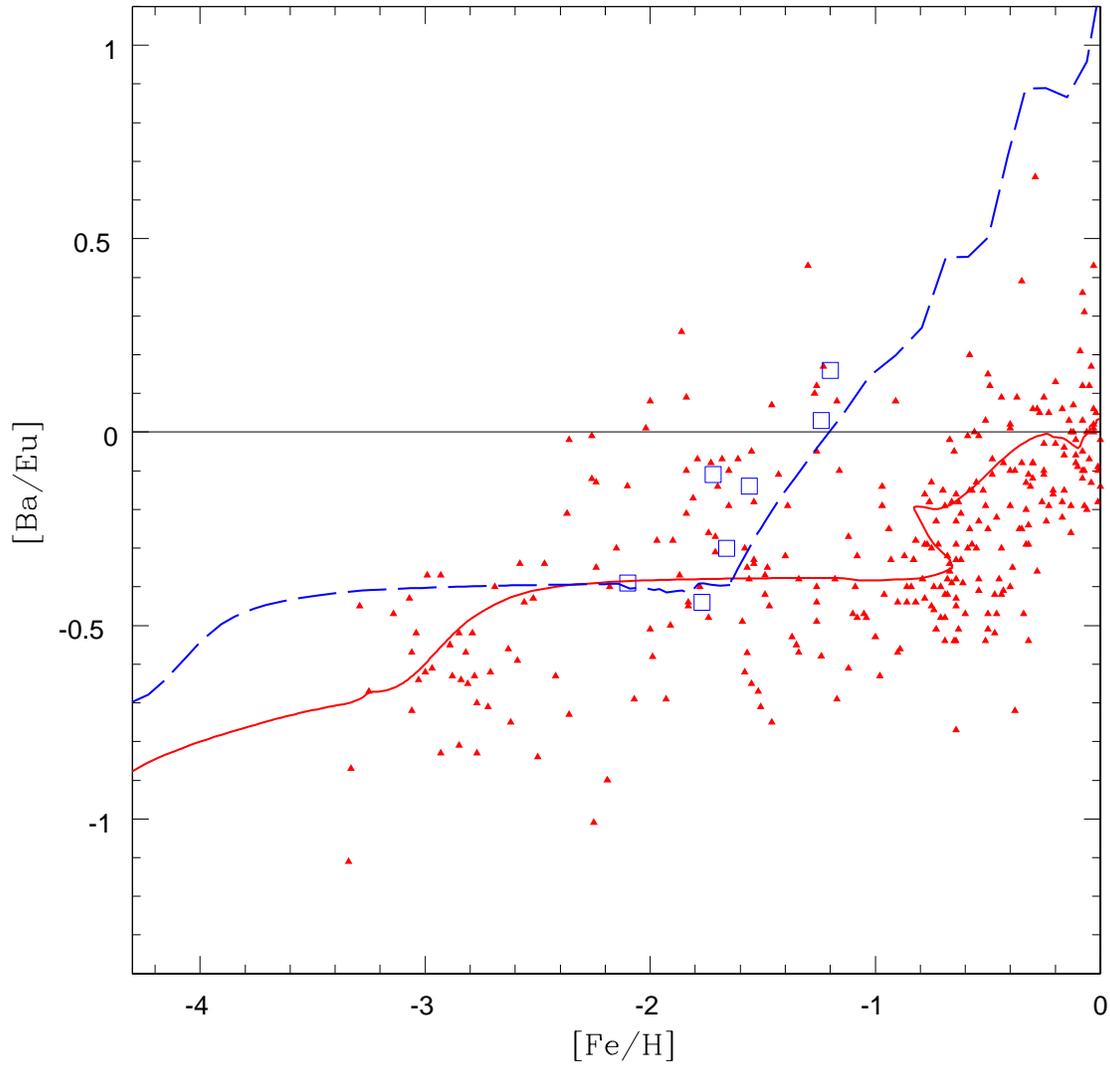}
\caption{[Ba/Eu] vs [Fe/H]: the red dots are the observational data 
for the Milky Way, whereas the blue open squares are the data for Sculptor. 
The results of the model for the Milky Way are plotted in red solid line
 and the results for Sculptor in blue dashed line.}
\label{chap6-5}
\end{center}
\end{figure}

For lanthanum we obtain the results shown in Fig. \ref{chap6-6}. As expected, 
the models give results very similar to those of Ba, being the nucleosynthesis similar.
The observational data for the Milky Way are few, but the trend resembles the one of the barium.
The trend of the data for Sculptor is the same as for Ba and the model reproduces this trend.

\begin{figure}
\begin{center}
\includegraphics[width=0.99\textwidth]{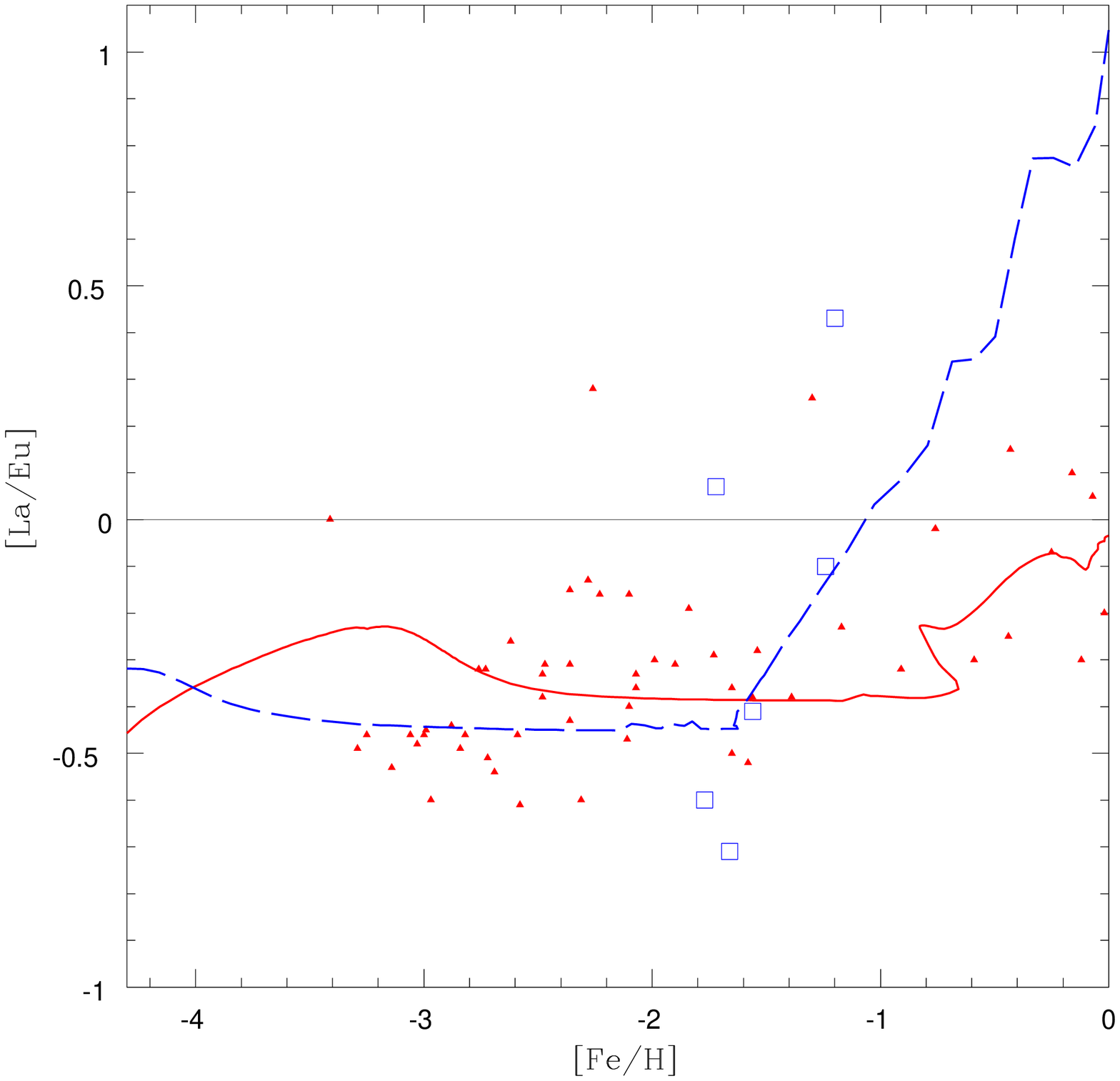}
\caption{ [La/Eu] vs [Fe/H]: the red dots are the observational data  
for the Milky Way, whereas the blue open squares are the data for Sculptor. 
The results of the model for the Milky Way are plotted in red solid line
 and the results for Sculptor in blue dashed line.}
\label{chap6-6}
\end{center}
\end{figure}

In the case of yttrium, the results of the model for Sculptor do not show
a large difference compared to the results for the Milky Way (see Fig. \ref{chap6-7}).
However, the two observational points for Sculptor with the highest [Fe/H] show a large difference,
so it is hard to say which is the real observational trend
and if we take the mean value of the two points the results are quite good.

\begin{figure}
\begin{center}
\includegraphics[width=0.99\textwidth]{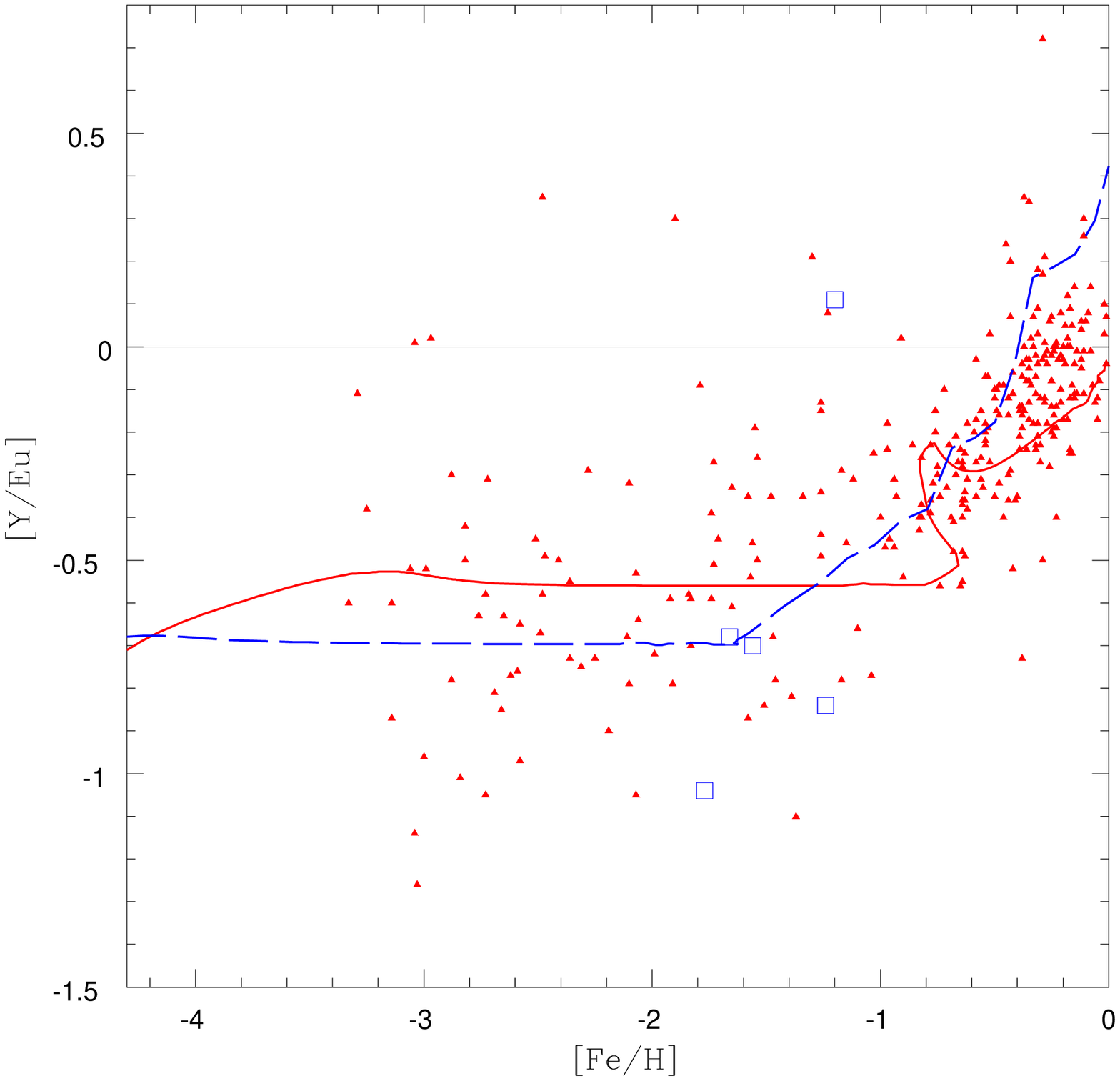}
\caption{ [Y/Eu] vs [Fe/H]: the red dots are the observational data  
for the Milky Way, whereas the blue open squares are the data for Sculptor. 
The results of the model for the Milky Way are plotted in red solid line
 and the results for Sculptor in blue dashed line.}
\label{chap6-7}
\end{center}
\end{figure}

In Fig. \ref{chap6-8}, \ref{chap6-9} we show 
the ratios [Ba/Y] and [La/Y]. Concerning the Milky Way data, the model
fits well the observational data for the ratios [Ba/Y] and for [La/Y], even if
the data for [La/Y] are few. % Yttrium  has a larger 
%contribution by low mass stars than lanthanum and barium and because of this,
%coupled with the galactic wind, 
We note that the results for the [Ba/Y] and for [La/Y] ratios
for Sculptor increase for [Fe/H]$>$ -2. The observational data for Sculptor show 
substantially the same behaviors, even if we find again that  the two observational data 
with the highest [Fe/H] show discrepancy and the model passes about the mean
 value of the two data.

\begin{figure}
\begin{center}
\includegraphics[width=0.99\textwidth]{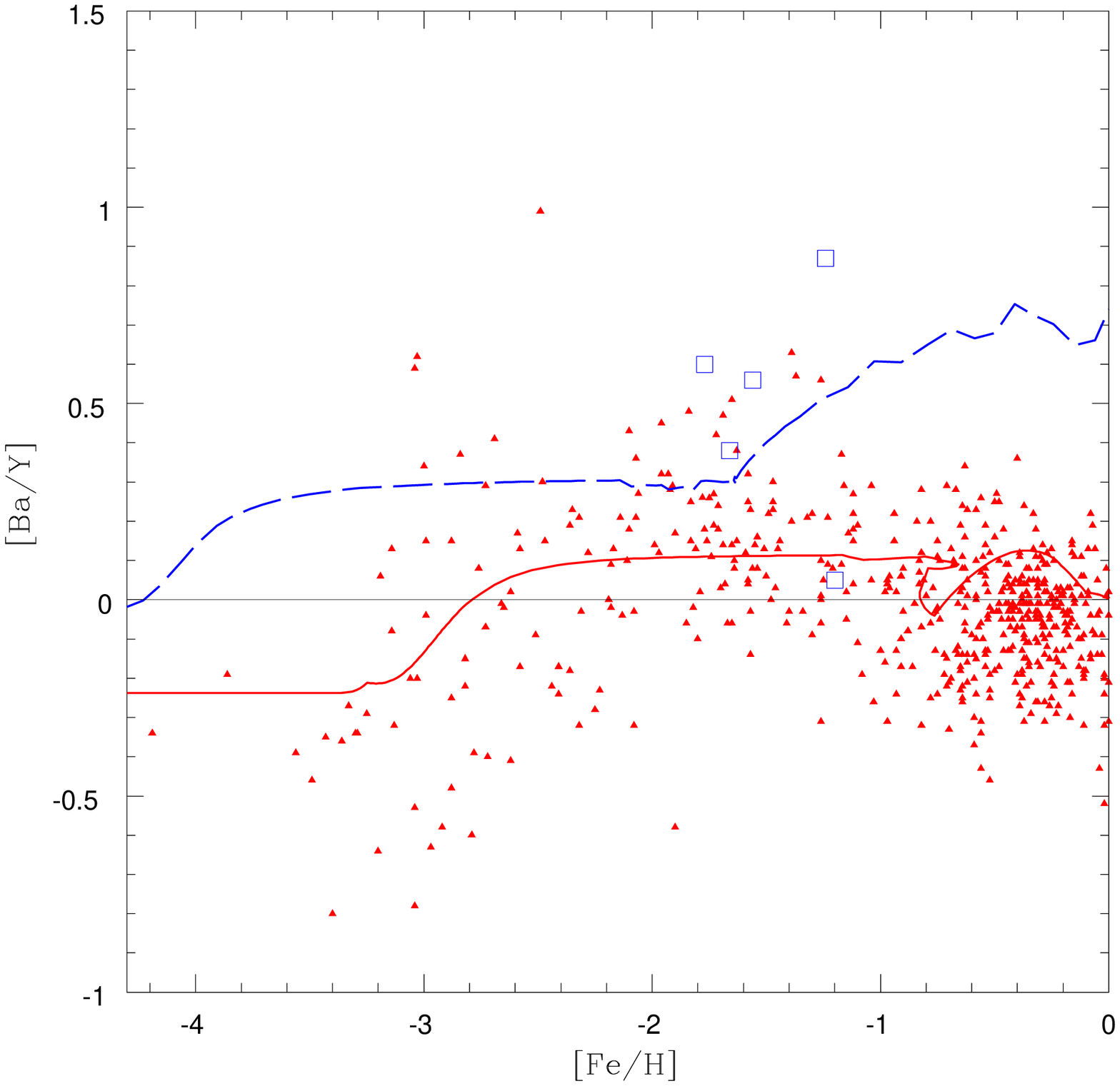}
\caption{[Ba/Y] vs [Fe/H]: the red dots are the observational data   
for the Milky Way, whereas the blue open squares are the data for Sculptor;
The results of the model for the Milky Way are plotted in red solid line
 and the results for Sculptor in blue dashed line.}
\label{chap6-8}
\end{center}
\end{figure}

\begin{figure}
\begin{center}
\includegraphics[width=0.99\textwidth]{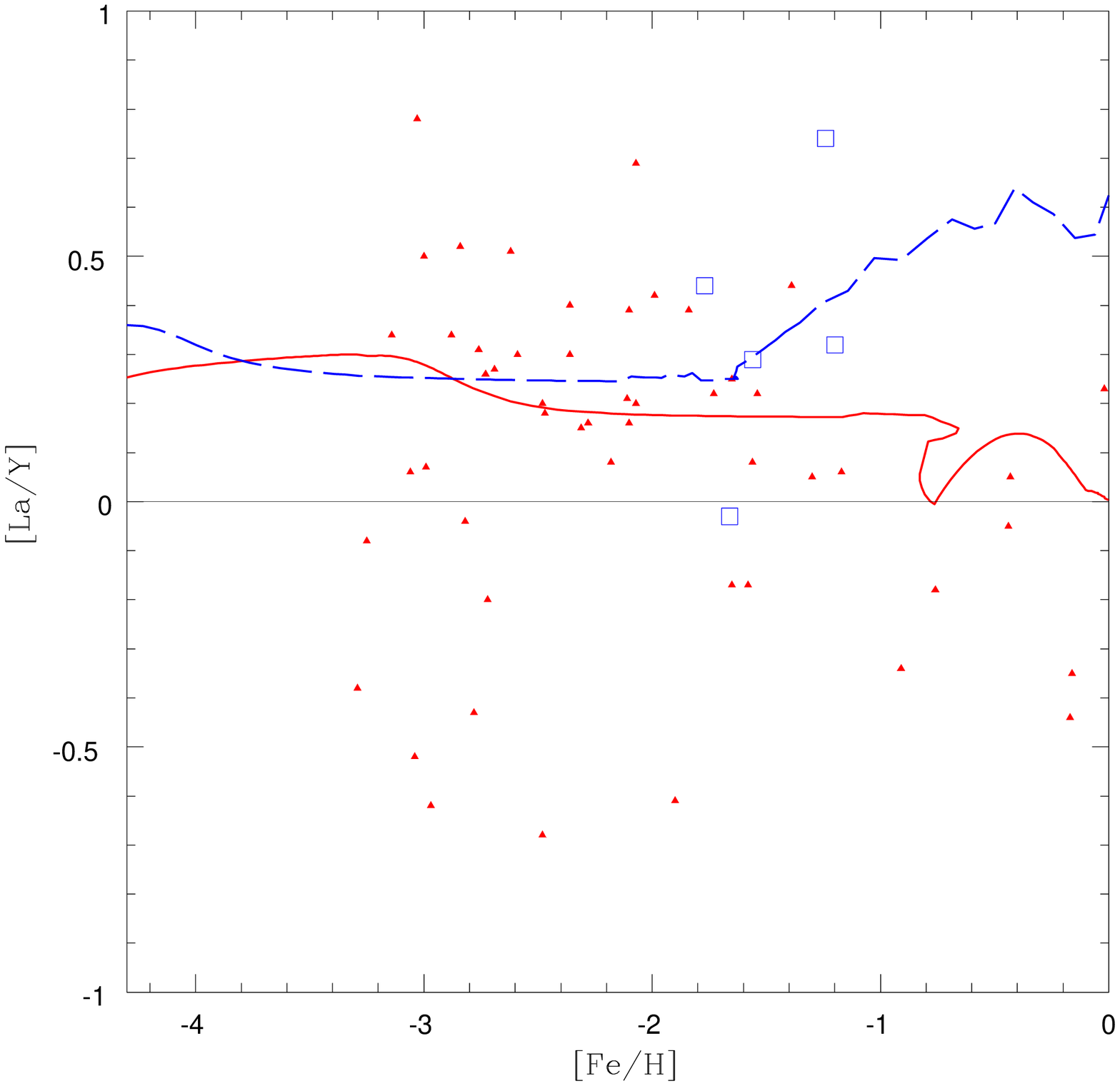}
\caption{[Ba/Y] vs [Fe/H]: the red dots are the observational data   
for the Milky Way, whereas the blue open squares are the data for Sculptor;
The results of the model for the Milky Way are plotted in red solid line
 and the results for Sculptor in blue dashed line.}
\label{chap6-9}
\end{center}
\end{figure}

Therefore, we conclude that the chemical evolution models, both the one for the Milky Way and the one
for Sculptor, well fit the observational data. Moreover, they are able to explain 
why the abundance ratios for neutron capture elements in the two systems
are different for the same [Fe/H], in terms of different SFR history, 
star formation efficiencies and galactic winds.

\chapter{Conclusions}

\rightline{\emph{``... it is possible to make even Venus herself vanish from the firmament}}
\rightline{\emph{ by a scrutiny too sustained, too concentrated, or too direct.''}}
\rightline{\emph{ by Edgar Allan Poe }}

\vspace{1cm}

\section{Chemical evolution in the solar vicinity}

In the first part of this work, the main goal 
was to follow the evolution of neutron capture elements by 
means of a chemical evolution model reproducing the abundance trends 
for other elements.
We used the Chiappini et al. (1997) model in its latest version 
as described in Chiappini et al. (2003) and Fran\c cois et al. (2004).
We have used empirical yields for stars with mass 
$>8M_{\odot}$, producing r-process elements.
For the r-process elements there are not solid theoretical yields, 
since the mechanism involved in their production, 
the so-called r-process, it is still not well understood.
We conclude that Ba, La, Sr, Y and Zr need two components: an s-process main component 
originating in low mass stars (1-3 $M_{\odot}$) plus an r-component originating in stars in 
the range 10-30$M_{\odot}$. This range is different from the one suggested by Travaglio 
et al. (1999) and Travaglio et al. (2004) and
it has been obtained by requiring the best fit of the new and accurate data.
For Eu we estimate that it is mainly produced by an r-process and that stars in 
the same mass range, 10-30$M_{\odot}$, should be considered as the progenitors of this 
element.

The nearly constant value of the ratio [Ba/Eu] produced in massive stars
by the r-process can be used to estimate  the fraction of barium in the solar abundance
produced  by  the slow process. We have obtained in this
way a fraction that is slightly different from the previous results: 60\% instead of 80\%.
The yields that we derived and even the fact that the ratio of the r- process production  
of Eu and Ba seems to be be nearly constant could be very useful in studies involving nucleosynthesis 
in stellar models and even in nuclear physics studies.
The open question remains the observed, and probably real, large spread in the [el/Fe]
ratios for the neutron capture elements at [Fe/H]. Our homogeneous model can in fact reproduce
the mean trend of the the data but not the spread.

\section{Abundance gradients in the Milky Way}

The aim of this chapter was to compare new observational data 
on the radial gradients for  19 chemical elements (including s- and r-process elements)
with the predictions of our chemical evolution model for the Milky Way.
This model has been tested on the properties of the solar vicinity and contains
a set of yields which best fits the abundances and abundance ratios in the solar vicinity,
 as shown in Fran\c cois et al. (2004).

The bulk of observational data comes from the abundances derived in 
a large number of Cepheids observed by
 4AL. For the first time with these data it has been possible to verify the predictions for many heavy
elements with statistical validity.

The comparison between model predictions and observational data showed that our model well reproduces
the gradients of almost all the elements that we analyzed. 
Since abundance gradients can impose strong constraints  both on the mechanism
of galaxy formation, in particular of the galactic disk, and on the nucleosynthesis 
prescriptions, we can conclude that:

\begin{itemize}

\item The model for the Milky Way disk formation, assuming an inside-out building-up of the disk, 
 as suggested originally by Matteucci \& Fran\c cois (1989), can be considered successful;
 for almost all the considered elements, we find a good fit
to the observational data ranging from 5 to 17 kpc. In particular, the model assuming
a constant total surface mass density for the halo best fits the data of 
Cepheids. At large galactocentric distances the halo mass distribution influences 
the abundance gradients (see Chiappini et al. 2001).

\item 
In our chemical evolution model we adopt a threshold in the gas density for star
 formation in the disk of $7M_{\odot}pc^{-2}$, whereas for 
the halo phase we have several options with and without a threshold. 
The threshold in the halo, when considered, is $4M_{\odot}pc^{-2}$.
We also assume a constant surface mass density for the halo or variable with 
galactocentric distance. 
This is important for the gradients at very large galactocentric distances, 
where the enrichment from the halo predominates over the enrichment occurring 
in the thin disk, thus influencing the abundances at such large distances.
We conclude that to reproduce the flat gradients suggested by the abundance measurements
at large galactocentric distances, we need to assume a constant density distribution
and a threshold in the star formation during the halo phase.
However, there are still many uncertainties in the data at very large
galactocentric distances and only more data will allow us to draw firm 
conclusions on this important point.

\item By means of the assumed nucleosynthesis prescriptions, we have 
successfully reproduced 
the abundance gradients  of each specific element, 
as well as  the [el/Fe] vs [Fe/H] relations
in the solar neighborhood,  for both  
neutron capture elements and  for $\alpha$- and
iron peak elements.

\end {itemize}

\section{Inhomogeneous model for the Galactic halo}

In this chapter we tried to solve the problem of the spread at low
metallicity of the [el/Fe] ratios for neutron capture elements.
We developed a new model for the chemical evolution of the halo 
in the Milky Way. We showed in this model that a random birth of stellar masses, coupled with 
the different mass ranges  responsible for the production of
 $\alpha$-elements and neutron capture elements, respectively,
can explain the large spread in the abundances of 
metal poor stars for neutron capture elements and the smaller
spread for $\alpha$-elements. 
We used for this model the same parameters of the homogeneous one.
We adopted this point of view because these parameters have been already constrained
to give good results compared to the observational data at higher metallicities.
In fact, toward high metallicities ([Fe/H]$>$-2.0 dex) the model naturally gives 
results compatible with the homogeneous model.
However, this set of parameters is a starting point and the model 
still needs a better investigation of the parameter space;
in particular, for what concern the early galactic stages, the model predicts
a too fast increase of the [Fe/H] and produces a too small number of stars with a
metallicities [Fe/H]$<$ -3. Moreover this new model generates 
a too large number of metal free stars.
To avoid these problems a different and slower star formation history can be 
used, in order to provide a smoother increase of the metallicities.
To solve the problem of the metal free stars, a top-heavy IMF can be
applied for the very first period. 
In fact, many theoretical models for the metal free stars predict the 
existence of a top-heavy IMF (see Larson 1998, Abel, Bryan  \& Norman 2000, 
Hernandez \& Ferrara 2001, Nakamura \& Umemura 2001, Mackey, Bromm, \& Hernquist 2003).

\section{Dwarf spheroidal galaxies}

In the last part of this work, we have applied the nucleosynthesis prescriptions for 
the neutron capture elements, derived in chapter 2 for the Milky Way, 
to five dSph galaxies, in order to verify 
the assumptions regarding the production of these elements.
We have implemented these new nucleosynthesis prescriptions
 in a chemical evolution model, which was able to reproduce
several observational constraints of the Local Group dSph galaxies 
(such as [$\alpha$/Fe], present day gas mass, estimated final total 
mass, metallicity distribution). The model galaxies are characterized by 
the SF prescriptions, such as the number and the duration of the SF episodes and
also by the wind efficiency. The prescriptions for the SF history are taken from 
the suggestion arising from their color-magnitude diagrams.
In the chapter 5, we have compared the results of the chemical evolution of Ba and Eu 
in 5 dSph galaxies (Draco, Sculptor, Carina, Ursa minor and Sagittarius);
in chapter 6 we have  compared the chemical evolution of the dSph Sculptor
with the one the Milky Way and in this case we have shown also the results for Y and La.
The main conclusions can be summarized as follows: 

\begin{itemize}

\item The observed [Eu/Fe], [Ba/Fe] and [Ba/Eu] ratios are very well reproduced by the 
models for Draco, Carina, Ursa Minor, Sculptor and Sagittarius with the same nucleosynthesis
 that we have used in the Milky Way.
There is the exception of two stars (one in Draco and another in Ursa Minor) which exhibit very low
values of [Ba/Fe] at low metallicities ([Fe/H] $<$ -2.4 dex);
these two stars could be explained by a model in which Ba is produced
only by s-process occuring in stars with masses in the range 
1 - 3 $M_{\odot}$, but they could be also anomalous stars.

\item The evolution of [Y/Fe], [La/Fe], [Y/Eu], [La/Eu], [Ba/Y] and [La/Y] ratios,
which we have followed only for the dSph Sculptor,
confirms the goodness of the prescriptions also for these elements when
we compare the results of the model to the observational data.

\item The comparison between Sculptor and the Milky Way 
permits us to show the different predicted and  observed values 
in these two systems. 
Although more data for the dSph galaxies are necessary,
we can interpret these different ratios at a given metallicity
as due to the much less efficient star formation that we need to adopt
to reproduce the abundances of dSph galaxies, compared to the Milky Way.
In the star formation regime adopted for the dSph galaxies, in fact, 
the metallicity increases more slowly and the 
different contributions for the enrichment of neutron capture elements in
 the ISM appear at lower metallicities than in the Milky Way. 

\item The dSph galaxies cannot be 
the building blocks of the Milky Way galaxy, being their
chemical evolution and chemical composition very different.
This cannot rule out that clumps of gas, progenitors of dSph galaxies
in a isolate evolution,  merged to our Galaxy at a very early 
stage;  on the other hand, interactions between the Milky Way and dSph galaxies
do exist as shown by the galaxy Sagittarius, which is gradually being disrupted
 by the tidal interaction with our Galaxy, as suggested also by the same chemical 
signature of  Sagittarius and the globular cluster Palomar 12, which is 
nowadays in the halo of the Galaxy (Sbordone et al. 2006).

\item We underline that only more data, possibly extended toward
 larger metallicities, and a robust
statistical basis can confirm definitively our results.  

\end{itemize}

\backmatter % FONDAMENTALE !

\chapter{Acknowledgements}

\vspace{1cm}

First of all, I wish to thank Francesca Matteucci, who led me during these three
years. All her experience, all her suggestions and
criticisms made me grow up as a scientist but also as a person.
My gratitude goes also to Patrick Fran\c cois, who provided new and unpublished data
as well as many unvaluable suggestions and for inviting me at the
Observatoire de Paris.
I wish to warmly thank  Gustavo Lanfranchi for the fruitful collaboration
on the chemical evolution of dwarf spheroidal galaxies.
I also  wish  to thank Cristina Chappini, Francesco Calura, Antonio Pipino 
and Simone Recchi for all the interesting and stimulating discussions.
Finally, I whish to thank professor Jim Truran for his enlightening 
suggestions.

\listoffigures

\listoftables


\begin{thebibliography}{1000}

\bibitem{Abel}
Abel T., Bryan G.L., Norman M.L.,
2000, ApJ, 540, 39

\bibitem{Anders}
 Anders E., Grevesse N.,
1989, GeCoA, 53, 197

%\bibitem[\protect\citeauthoryear{Akerman et al.}{2004}]{b01}
%  Akerman C. J., Carigi L., Nissen P. E., Pettini M., Asplund M.,
%  2004, A\&A, 414, 931

\bibitem[\protect\citeauthoryear{Alib\'es et al.}{2001}]{b02}
  Alib\'es A., Labay J., Canal R.,
  2001, A\&A, 370, 1103

\bibitem{Andrievsky1}
  Andrievsky S.M., Bersier D., Kovtyukh V.V. et al
  2002, A\&A 384, 140


\bibitem{Andrievsky2}
  Andrievsky S.M., Kovtyukh V.V., Luck R.E. et al
  2002, A\&A 381, 32

\bibitem{Andrievsky3}
  Andrievsky S.M., Kovtyukh V.V., Luck R.E. et al
  2002, A\&A 392, 491

\bibitem{Andrievsky4}
  Andrievsky S.M., Luck R.E., Martin P. et al
  2004, A\&A 413, 159


%\bibitem[\protect\citeauthoryear{Allende Prieto et al.}{2002}]{b1} 
%  Allende Prieto C., Lambert D.L., Asplund M., 
%  2001, ApJ, 556, L63

\bibitem{Argast D}
Argast D., Samland M., Gerhard O.E., Thielemann, F.-K.,
2000, A\&A, 356, 873

\bibitem[\protect\citeauthoryear{Argast et al.}{2004}]{b114}
  Argast D., Samland M., Thielemann F.-K., Qian Y.-Z.,
  2004, A\&A, 416, 997


\bibitem{ Arimoto N.} Arimoto N., Yoshii Y., 1987, A\&A, 173, 23 


\bibitem[\protect\citeauthoryear{Arlandini et al.}{1999}]{b112} 
 Arlandini C., K\"appeler F.,  Wisshak K. et al.,
 1999, ApJ, 525, 886


\bibitem[\protect\citeauthoryear{Asplund et al.}{2005}]{b41}
  Asplund M., Grevesse N., Sauval A. J.,
  2005, ASPC, 336, 25A	


\bibitem{Ballero S.K.}
Ballero S.K., Matteucci F., Chiappini C.,
2006, NewA, 11, 306


\bibitem{Baraffe}
Baraffe  I., El Eid M.F., Prantzos N.,
1992, A\&A, 258, 357


\bibitem[\protect\citeauthoryear{Beers et al.}{1992}]{b2} 
  Beers T.C., Preston G.W., Shectman S.A.,
  1992, A.J., 103, 1987

\bibitem[\protect\citeauthoryear{Beers et al.}{1999}]{b3}
  Beers T.C., Rossi S., Norris J.E., Ryan S.G., Shefler T.,
  1999, A.J., 117, 981


\bibitem{Boisser}
  Boissier S., Prantzos N., 
  1999, MNRAS, 307, 857

\bibitem{Bonifacio1}
Bonifacio P., Hill V., Molaro P., Pasquini L., Di Marcantonio P.,
Santin P., 2000, A\&A, 359, 663 

\bibitem{Bonifacio2}
Bonifacio P., Sbordone L., Marconi G., Pasquini L., Hill V., 2004, 
A\&A, 414, 503

\bibitem[\protect\citeauthoryear{Bono et al.}{2005}]{b44}
  Bono G., Marconi M., Cassisi S. et al.
  2005, ApJ, 621, 966


\bibitem[\protect\citeauthoryear{Burris et al.}{2000}]{b500}
  Burris D.L., Pilachowski C.A., Armandroff T.E. et al.,
  2000, ApJ, 544, 302



\bibitem[\protect\citeauthoryear{Busso et al.}{2001}]{b4}        
  Busso M., Gallino R., Lambert D.L., Travaglio, C., Smith V.V.,
  2001, ApJ, 557, 802


\bibitem[\protect\citeauthoryear{Busso et al.}{1999}]{b5} 
  Busso M., Gallino R., Wasserburg G.J.,
  1999, ARA\&A ,37, 239

\bibitem[\protect\citeauthoryear{Cappellaro et al.}{1999}]{b6} 
  Cappellaro E., Evans R., Turatto M.,
  1999, A\&A, 351, 459

%\bibitem[\protect\citeauthoryear{Carretta et al.}{2002}]{b7} 
%  Carretta E., Gratton R.G., Cohen J.G.,
%  2002, AJ, 124, 481


\bibitem[\protect\citeauthoryear{Carney et al.}{1997}]{b8} 
  Carney B.W., Wright J.S., Sneden C. et al.,
  1997, A.J., 114, 363


\bibitem{Carigi}
Carigi L., Hernandez X., Gilmore G., 2002, MNRAS, 334, 117 



\bibitem{Carney et al.}
  Carney B.W., Yong D., Teixera de Almeida M.L., Seitzer P.,
  2005, AJ, 130, 1111

\bibitem[\protect\citeauthoryear{Carraro et al.}{2004}]{b92}
  Carraro G., Bresolin  F., Villanova S. et al.,
  2004, AJ, 128, 1676


\bibitem{Carraro}
Carraro G., Chiosi C., Girardi L., Lia C., 2001, MNRAS, 327, 69


\bibitem[\protect\citeauthoryear{Cayrel et al.}{2004}]{b34} 
 Cayrel R., Depagne E., Spite M. et al., 2004, A\&A, 416, 1117 


%\bibitem{Cayrel2}
%Cayrel R., Hill V., Beers T.C., Barbuy B., Spite M., Spite F., Plez B., Andersen J.,
%Bonifacio P., François P., Molaro P., Nordström B., Primas F.,
%2001, Nature, 409, 691

\bibitem[\protect\citeauthoryear{Cescutti et al.}{2006}]{b100} 
  Cescutti G., Fran\c cois P., Matteucci F., Cayrel R., Spite M.,
  2006, A\&A, 448, 557

\bibitem{Cescutti2} Cescutti G.,  Matteucci F., Fran\c cois P., Chiappini C., 2006,
 astro-ph/0609813   	   


\bibitem{Chang et al.}
  Chang R. X., Hou J. L., Shu C. G., Fu C.Q.,
  1999, A\&AS, 141, 491 			



\bibitem[\protect\citeauthoryear{Chiappini et al.}{1997}]{b9} 
  Chiappini C., Matteucci F., Gratton R.G.,
  1997, ApJ, 477, 765


\bibitem[\protect\citeauthoryear{Chiappini et al.}{2003}]{b130} 
  Chiappini C., Matteucci F., Meynet G.,
   2003b, A\&A, 410, 257	

\bibitem{Chiappini et al.}
Chiappini C., Matteucci F., Padoan P., 2000, ApJ, 528, 711



\bibitem[\protect\citeauthoryear{Chiappini et al.}{2001}]{b120} 
  Chiappini C., Matteucci F., Romano D.,
  2001, ApJ, 554, 1044

\bibitem[\protect\citeauthoryear{Chiappini et al.}{2003}]{b93} 
  Chiappini C., Romano D., Matteucci F., 
  2003a, MNRAS, 339, 63


\bibitem[\protect\citeauthoryear{Clayton \& Rassbach}{1967}]{b10} 
  Clayton D.D., Rassbach, M.E., 
  1967, ApJ, 168, 69


\bibitem[\protect\citeauthoryear{Cowan et al.}{2005}]{b141} 
  Cowan J.J., Sneden C., Beers T.C. et al.
  2005, ApJ, 627, 238

\bibitem{Cowan2}
Cowan J.J., Sneden C., Burles S., 
Ivans I.I., Beers T.C., Truran J.W., 
Lawler J.E., Primas F., Fuller G.M., 
Pfeiffer B., Kratz K.-L.,  2002, ApJ, 572, 861



\bibitem[\protect\citeauthoryear{Daflon \& Cunha}{2004}]{b143}
  Daflon, S., Cunha K.,
  2004, ApJ, 617, 1115


\bibitem{Dolphin} 
Dolphin A.E., MNRAS, 2002, 332, 91

\bibitem{Dopita}
Dopita M.A., Ryder S.D., 1994, ApJ, 430, 163

\bibitem[\protect\citeauthoryear{Edvardsson et al.}{1993}]{b11}
  Edvardsson B., Andersen J., Gustafsson B. et al.,
 1993, A\&A, 275, 101

\bibitem[\protect\citeauthoryear{Feast \& Walker}{1987}]{b144}
  Feast M. W., Walker A. R.,
  1987, ARA\&A, 25, 345


\bibitem[\protect\citeauthoryear{Fran\c cois et al.}{2005}]{b113}
  Fran\c cois P.,  Depagne E., Hill V. et al., 
  2007, in preparation

\bibitem[\protect\citeauthoryear{Fran\c cois et al.}{2004}]{b110}
  Fran\c cois P., Matteucci F., Cayrel R. et al.,
  2004, A\&A, 421, 613

\bibitem{Freiburghaus C.}
Freiburghaus C., Rembges J.-F., Rauscher T., Kolbe E., 
Thielemann F.-K., Kratz K.-L., Pfeiffer B., Cowan J.J.,
1999, ApJ, 516, 381


\bibitem[\protect\citeauthoryear{Freiburghaus et al.}{1999}]{b111}
  Freiburghaus C., Rosswog S., Thielemann F.-K.,
  1999, ApJ, 525L, 121


\bibitem[\protect\citeauthoryear{Fulbright}{2000}]{b12}
Fulbright J.P.,
2000, AJ, 120, 1841

\bibitem{Fulbright2}
Fulbright J.P.,
2002, AJ, 123, 404

\bibitem{Fulbright3}
Fulbright, J.P., Rich R.M., Castro S., 2004, ApJ, 612, 447

\bibitem{Gallino}
Gallino R., Arlandini C., Busso M., 
Lugaro M., Travaglio C., Straniero O., 
Chieffi A., Limongi M., 1998, ApJ, 497, 388

\bibitem{Geisler}
Geisler D., Smith V.V., Wallerstein G., 
Gonzalez G., Charbonnel C., 2005, AJ, 129, 1428



\bibitem[\protect\citeauthoryear{Gilroy et al.}{1988}]{b13} 
  Gilroy K.K., Sneden C., Pilachowski C.A., Cowan J.J.,
  1988, ApJ, 327, 298

\bibitem[\protect\citeauthoryear{Gratton \& Sneden }{1988}]{b14} 
  Gratton R.G., Sneden C.,
  1988, A\&A, 204, 193

\bibitem[\protect\citeauthoryear{Gratton \& Sneden }{1994}]{b15} 
  Gratton R.G., Sneden C.,
  1994, A\&A, 287, 927

\bibitem{Greggio1}
Greggio, L., Renzini, A., 1983a, A\&A, 118, 217


\bibitem{Greggio2}
Greggio, L., Renzini, A., 1983b, in. Frascati Workshop on First Stellar
Generations, Vulcano, Italy, Societa Astronomica Italiana,
vol. 54, no. 1, p. 311-319   



\bibitem[\protect\citeauthoryear{Grevesse \& Sauval}{1998}]{b17} 
  Grevesse N., Sauval A.J.,
  1998, Space Sci. Rev., 85, 161

\bibitem{Hartwick}
Hartwick F. 1976, ApJ, 209,418

%\bibitem[\protect\citeauthoryear{Henry et al}{2000}]{200}	
%  Henry R. B. C., Edmunds M. G., K\"oppen J.
%  2000, ApJ, ,541, 660

\bibitem{Hernandez}
Hernandez X., Ferrara A.,
2001, MNRAS, 324, 484



\bibitem{Hernandez2}
Hernandez X., Gilmore G., Valls-Gabaud D., 2000, MNRAS, 317, 831


\bibitem{Hill}
Hill V., et al.,  2002, A\&A, 387, 560
 
\bibitem[\protect\citeauthoryear{Honda et al.}{2004}]{b170} 
  Honda S., Aoki W., Kajino T. et al.,
  2004, ApJ, 607, 474

\bibitem[\protect\citeauthoryear{Hou et al.}{2000}]{b210} 
  Hou J. L., Prantzos N., Boissier S.,
  2000, A\&A, 362, 921

\bibitem{Iben1}
Iben, I. Jr., Renzini A., 1982a, ApJ,  259,  79

\bibitem{Iben2}
Iben, I. Jr., Renzini A., 1982b, ApJ, 263, 23

\bibitem{Ikuta}
Ikuta C., Arimoto N.,
1999, PASJ, 51, 459


\bibitem[\protect\citeauthoryear{Ishimaru \& Wanajo}{1999}]{b171} 
  Ishimaru Y., Wanajo S.,
  1999, ApJ, 511L, 33


\bibitem[\protect\citeauthoryear{Ishimaru et al.}{2004}]{b172} 
  Ishimaru Y., Wanajo S., Aoki, W., Ryan, S.G.,
  2004, ApJ, 600L, 47 


\bibitem{Iwamoto}
Iwamoto, K., Brachwitz F., Nomoto K.,
 Kishimoto N., Umeda H., Hix W.R., Thielemann F.-K.,
1999, ApJS, 125, 439



\bibitem{Johnson1}
Johnson J.A.,
2002, ApJS, 139, 219


\bibitem{Johnson2} 
Johnson J.A., Bolte M., 
2001, ApJ, 554, 888

\bibitem[\protect\citeauthoryear{Johnson \& Bolte}{2002}]{b97} 
Johnson J.A., Bolte M.,
2002, ApJ, 579, 616




\bibitem[\protect\citeauthoryear{K\"appeler et al.}{1989}]{b18} 
  K\"appeler F., Beer H., Wisshak K.,
  1989, Rep.Prog.Phys., 52, 945

\bibitem[\protect\citeauthoryear{Kennicutt}{1989}]{b19} 
  Kennicutt R.C.Jr., 
  1989, ApJ, 344, 685


\bibitem[\protect\citeauthoryear{Kennicutt}{1998}]{b20} 
  Kennicutt R.C.Jr., 
  1998, ApJ, 498, 541


\bibitem[\protect\citeauthoryear{Koch \& Edvardsson}{2002}]{b190} 
  Koch A., Edvardsson B.,
  2002, A\&A,381, 500

\bibitem{Koch}
Koch  A., Grebel E.K., Harbeck D., Wilkinson M., Kleyna J., Gilimore G.,
Wyse R.F. G., Evans W.,
2004,ANS, 325, 44	

\bibitem{kroupa}
Kroupa P., Tout C.A., Gilmore G. 1993, MNRAS, 262, 545

\bibitem{Lamb}
Lamb S.A., Howard W.M., Truran J.W., Iben I.Jr.,
1977, ApJ, 217, 213

\bibitem{Lanfranchi1}
Lanfranchi G., Matteucci F., 2003, MNRAS, 345, 71
 
\bibitem{Lanfranchi2}
Lanfranchi G., Matteucci F., 2004, MNRAS, 351, 1338


\bibitem{Langer}
Langer N., Henkel C., 1995, Space Sci. Rev., 74, 343


\bibitem{Larson}	
Larson R.B.,
1998, MNRAS, 301, 569

%\bibitem[\protect\citeauthoryear{Liang et al.}{2001}]{b250} 
%  Liang, Y. C., Zhao G., Shi J. R.,
%  2001 A\&A, 374, 936	


\bibitem[\protect\citeauthoryear{Limongi \& Chieffi}{2003}]{b91} 
  Limongi  M., Chieffi A., 2003, ApJ, 592, 404



\bibitem[\protect\citeauthoryear{Luck et al.}{2003}]{b260} 
  Luck R.E., Gieren W.P., Andrievsky S.M. et al. 
  2003, A\&A, 401, 939

\bibitem{Mackey}
Mackey J., Bromm V., Hernquist L.,
2003, ApJ, 586, 1


\bibitem{MacLow}
MacLow M., Ferrara A.,  1999, ApJ, 513, 142

\bibitem{Mader}
Maeder A., \& Meynet G., 1989, A\&A, 210, 155 


%\bibitem[\protect\citeauthoryear{Madau et al.}{1998}]{b21} 
%  Madau P., della Valle M., Panagia N.,
%  1998, MNRAS, 297, 17


\bibitem[\protect\citeauthoryear{Martin \& Kennicutt}{2001}]{b22} 
Martin C.L., Kennicutt R.C.Jr., 
2001, ApJ, 555, 301


\bibitem[\protect\citeauthoryear{Mashonkina \& Geheren}{2000}]{b252} 
  Mashonkina L., Gehren T.,
  2000, A\&A, 364, 249

\bibitem[\protect\citeauthoryear{Mashonkina \& Geheren}{2001}]{b251} 
  Mashonkina L., Gehren T.,
  2001, A\&A , 376, 232


\bibitem[\protect\citeauthoryear{Mathews et al}{1992}]{b220}   
  Mathews G.J., Bazan G., Cowan J. J.,
  1992, ApJ, 391, 719

\bibitem{Mathews}
  Mathews G.J., Cowan J. J.,
  1990, Nature, 345, 491



\bibitem{Matteucci0}
Matteucci F., 1986, ApJ, 305, 81


\bibitem{Matteucci1}
Matteucci F., 1992, ApJ, 397, 32


\bibitem{Matteucci2}
Matteucci F., 1996, FCPh, 17, 283


\bibitem{Matteucci3} 
Matteucci F., Brocato E., 1990, ApJ, 365, 539

\bibitem{Matteucci4} 
Matteucci F., Calura F.,  2005, MNRAS, 360, 447

\bibitem{Matteucci5}
Matteucci F., Chiosi C., 1983, A\&A, 123, 121

\bibitem[\protect\citeauthoryear{Matteucci5 \&Fran\c cois}{1989}]{b23} 
Matteucci F., Fran\c cois P.,
1989, MNRAS, 239, 885


\bibitem[\protect\citeauthoryear{Matteucci6 \& Greggio}{1986}]{b24} 
Matteucci F., Greggio L.,
1986, A\&A, 154, 279

%\bibitem[\protect\citeauthoryear{Matteucci et al.}{1993}]{b25} 
%  Matteucci M.F., Raiteri C.M., Busso M., Gallino R., Gratton R.G.,
%  1993, A\&A , 272, 421

\bibitem{Matteucci7} Matteucci F., Pipino A.,
2005, MNRAS, 357, 489

\bibitem{Matteucci8}
Matteucci F., Recchi S., 2001, ApJ, 558, 351


\bibitem{Matteucci9}
Matteucci F., Romano D., Molaro P.,
1999 A\&A, 341, 45

\bibitem{Matteucci99} 
Matteucci F., Tornamb\'{e} A., 1987, A\&A, 185, 51


\bibitem[\protect\citeauthoryear{Mazzali \& Chugai}{1995}]{b250} 
  Mazzali, P. A., Chugai, N. N.,
  1995, A\&A, 303, 118

\bibitem[\protect\citeauthoryear{McWilliam et al.}{1995}]{b26} 
  McWilliam A., Preston G.W., Sneden C., Searle L.,
  1995, AJ, 109, 2757

\bibitem[\protect\citeauthoryear{McWilliam \& Rich}{1994}]{b325} 	
  McWilliam  A., Rich R. M.,
  1994, ApJS, 91, 749

\bibitem{Mcwilliam3}
McWilliam A., Searle L.,
1999,Ap\&SS, 265, 133

\bibitem{Meynet}
Meynet G., Maeder A., 2002, A\&A, 390, 561.


\bibitem{Nakamura} Nakamura F., Umemura M.,
2001, ApJ, 548, 19

%\bibitem[\protect\citeauthoryear{Nissen et al.}{2002}]{b27} 
%  Nissen P.E, Primas F., Asplund M., Lambert D.L.,
%  2002, A\&A, 390, 235

\bibitem[\protect\citeauthoryear{Nissen \& Schuster}{1997}]{b28}
  Nissen P.E., Schuster W.J., 
  1997, A\&A, 326, 751

%\bibitem[\protect\citeauthoryear{Norris et al.}{2001}]{b29}
%  Norris J.E., Ryan S.G., Beers T.C.,
%  2001, ApJ, 286, 644

\bibitem{} 
Nomoto K., Hashimoto M., Tsujimoto T., Thielemann F.-K., Kishimoto
 N., Kubo Y., Nakasato N., 1997, Nucl. Phys. A, 616, 79

\bibitem{Otsuki}
Otsuki, K., Mathews G.J., Kajino T.,
2003, NewA, 8, 767


\bibitem[\protect\citeauthoryear{Pagel \& Tautvaisienne}{1995}]{b30} 
Pagel B.E.J., Tautvaisiene G.,
1995, MNRAS, 276, 505

\bibitem[\protect\citeauthoryear{Pagel \& Tautvaisienne}{1997}]{b300}
Pagel B.E.J., Tautvaisiene G.,
1997, MNRAS, 288, 108

\bibitem{Peters}
Peters J.G.,
1968, ApJ, 154, 225

\bibitem[\protect\citeauthoryear{Pompeia et al.}{2003}]{b345}
  Pompeia L., Barbuy B., Grenon M.,
  2003, ApJ, 592, 1173

\bibitem{Portinari}
Portinari L., Chiosi C. 1999, A\&A, 350, 827


\bibitem[\protect\citeauthoryear{Prantzos \& Boisser}{2000}]{b42}
   Prantzos N., Boisser S., 
  2000, MNRAS, 313, 338

\bibitem{Prochaska}
Prochaska  J.X., Naumov S.O., Carney B.W., McWilliam A., Wolfe A.M.,
2000, AJ, 120, 2513


%\bibitem[\protect\citeauthoryear{Qian \& Wasserburg}{2001}]{b304}
%  Qian Y.Z., Wasserburg G.J.,
%  2001, ApJ, 559, 925

%\bibitem[\protect\citeauthoryear{Qian \& Wasserburg}{2003}]{b301}
%  Qian Y.Z., Wasserburg G.J.,
%  2003, ApJ, 588, 1099



\bibitem[\protect\citeauthoryear{Raiteri et al.}{1992}]{b302}
  Raiteri C.M., Gallino R., Busso M.,
  1992, ApJ, 387, 263

%\bibitem[\protect\citeauthoryear{Raiteri et al.}{1993}]{b303}
%  Raiteri C.M., Gallino R., Busso M.,
%  1993, ApJ, 419, 207

\bibitem{Raiteri4}
 Raiteri C.M., Villata M., Gallino R., Busso M., Cravanzola, A.,
1999, ApJ, 518, L91

\bibitem[\protect\citeauthoryear{Rauscher et al.}{2002}]{b31} 
  Rauscher T., Heger A., Hoffmann R.D., Woosley S.E.,
  2002, ApJ, 576, 323

\bibitem{Recchi1}
Recchi S., Matteucci F., D'Ercole A., 2001, MNRAS, 322, 800 

%\bibitem{}
%Recchi S., Matteucci F., D'Ercole A., Tosi M., 2002, A\&A, 384, 799 

\bibitem{Recchi2}
Recchi S., Matteucci F., D'Ercole A., Tosi M., 2004, A\&A, 426, 37

\bibitem[\protect\citeauthoryear{Reid}{1993}]{b32} 
  Reid M.J.,
  1993, ARAA, 31, 345

%\bibitem[\protect\citeauthoryear{Ryan et al.}{1991}]{b33}
%  Ryan S., Norris J.E., Bessel M.S.,
%  1991, AJ, 102, 303


\bibitem[\protect\citeauthoryear{Ryan et al.}{1996}]{b330}
  Ryan S., Norris J.E., Beers  T.C.,
  1996, ApJ, 471, 254


\bibitem{Sadakane}
Sadakane K., Arimoto N., Ikuta C., Aoki W.,
Jablonka P., Tajitsu A., 2004, PASJ, 56, 1041


\bibitem{Salpeter} 
Salpeter E.E., 1955, ApJ, 121, 161

\bibitem{Sbordone} 
Sbordone L., Bonifacio P., Buonanno R., Marconi G., Monaco L., Zaggia S.,
2006, astro-ph 0612125


\bibitem[\protect\citeauthoryear{Scalo}{1986}]{b37}
Scalo J.M.,
1986, FCPh, 11, 1

\bibitem{Scalo2}
Scalo J.M., 1998 in ``The Stellar Initial Mass Function''
 ASP Conf. Ser. Vol. 142 p.201


\bibitem{Schmidt}
Schmidt M., 1959, ApJ, 129, 243
%1963, ApJ, 137, 758



\bibitem{Shetrone1}
Shetrone M., 2004, astro/ph 0411030
%, to appear in L. Pasquini \& S. Randich,
%eds. Chemical Abundances and Mixing 
%in Stars in the Milky Way Galaxy and its Satellites, ESO 
%Astrophysics Symposia, Springer-Verlag Press

\bibitem{Shetrone2}
Shetrone M., C\^ot\'e P., Sargent W.L.W., 
2001, ApJ, 548, 59

\bibitem{Shetrone3}
Shetrone M., Venn K.A., Tolstoy E., Primas F., 
2003, AJ, 125, 684


\bibitem{Siess}
Siess L., Livio M., Lattanzio J., 2002, ApJ, 570, 329

\bibitem{Smecker-Hane}
Smecker-Hane T., McWilliam A., 1999, in Hubeny I. et al.,
eds. Spectrophotometric Dating of Stars and Galaxies, ASP 
Conference Proceedings, Vol. 192, p.150 

\bibitem{Sneden1}
Sneden C., Cowan J.J., Ivans I.I., Fuller G.M., Burles S., Beers T.C., Lawler J.E.,
2000, ApJ, 533, 139

\bibitem{Sneden2}
Sneden C., Cowan J.J., Burris D.L.,  Truran J.W., 
1998, ApJ, 496, 235

\bibitem{Sneden3}
Sneden C., Cowan J.J., Lavier J.E., Burles S., Beers T.C., Fuller G.M.,
2002, ApJ, 566, L25

\bibitem{Sneden4}
Sneden C., Cowan J.J., Lawler J.E., Ivans I.I., Burles S., 
Beers T.C., Primas F., Hill V., Truran J.W.,
Fuller G.M., Pfeiffer B., Kratz K.-L.,
2003, ApJ, 591, 936


\bibitem[\protect\citeauthoryear{Stephens}{1999}]{b35}
  Stephens A., 
  1999, AJ, 117, 1771

\bibitem{Stephens2}
Stephens A., Boesgaard A.M.,
2002, AJ, 123, 1647

\bibitem{Thielemann} 
Thielemann F.K., Nomoto K., Hashimoto M., 1996, ApJ, 
460, 408

\bibitem{Tinsley1}
Tinsley B.M., 1979, ApJ, 229, 1046 

\bibitem{Tinsley2}
Tinsley B.M., 1980, FCPh, 5, 287


\bibitem{Tolstoy}
Tolstoy E., Venn K.A., Shetrone M., Primas F., Hill V.,
Kaufer A., Szeifert T., 2003, AJ, 125, 707 

\bibitem{Tosi}
Tosi M., 1988, A\&A, 197, 338



\bibitem[\protect\citeauthoryear{Travaglio et al.}{2001}]{b38} 
  Travaglio C., Galli D., Burkert A.,
  2001, ApJ, 547, 217

\bibitem[\protect\citeauthoryear{Travaglio et al.}{1999}]{b36} 
  Travaglio C., Galli D., Gallino R. et al.,
  1999, ApJ, 521, 691


\bibitem[\protect\citeauthoryear{Travaglio et al.}{2004}]{b38} 
Travaglio C., Gallino R., Arnone E., Cowan J., Jordan F., Sneden C.,
2004, ApJ, 601 864



\bibitem{Truran}
Truran J.W., 1981, A\&A, 97, 391


\bibitem[\protect\citeauthoryear{Tsujimoto et al.}{1999}]{b386} 
  Tsujimoto T., Shigeyama T., Yoshii Y.,
  1999, ApJ, 519L, 63  


\bibitem{Umeda}
Umeda H., Nomoto K. 2002, ApJ, 565, 385



\bibitem[\protect\citeauthoryear{van den Hoek \& Groenewegen}{1997}]{b365}
  van den Hoek L.B.,  Groenewegen M.A.T.
  1997, A\&A Suppl., 123, 305


\bibitem{Venn}
Venn K. A., Irwin M., Shetrone M.D., Tout, C.A., Hill V., Tolstoy E.,
2004, AJ, 128, 1177


\bibitem{Ventura}
Ventura P., D'Antona F., Mazzitelli I., 2002, A\&A, 393, 21



\bibitem{Wanajo}
Wanajo S., Tamamura M., Itoh N., Nomoto K., 
Ishimaru Y., Beers T.C., Nozawa S.,
2003, ApJ, 593, 968


\bibitem{Whelan}
Whelan J., Iben I. Jr., 1973, ApJ, 186, 1007


\bibitem{Wheeler}
Wheeler J. C., Sneden C., Truran J.W.Jr., 1989, ARA\&A, 27, 279


%\bibitem[\protect\citeauthoryear{Wheeler et al.}{1998}]{b382} 
%  Wheeler J.C., Cowan J.J., Hillebrandt W.,
%  1998, ApJ, 493L, 101


\bibitem[\protect\citeauthoryear{Woolf et al.}{1995}]{b501} 
  Woolf  V.M., Tomkin J., Lambert D.L.,
  1995, ApJ, 453, 660


\bibitem[\protect\citeauthoryear{Woosley \& Hoffmann}{1992}]{b383} 
  Woosley S.E., Hoffmann, R.D.,
  1992, ApJ, 395, 202

\bibitem[\protect\citeauthoryear{Woosley \& Weaver.}{1995}]{b384} 
  Woosley S.E., Weaver, T.A.,
  1995, ApJ, 101, 181

\bibitem[\protect\citeauthoryear{Woosley et al.}{1994}]{b385} 
  Woosley S.E., Wilson J.R., Mathews G.J., Hoffman R.D., Meyer B.S.,
  1994, ApJ, 433, 229


\bibitem[\protect\citeauthoryear{Yong et al.}{2005}]{b370}
  Yong D., Carney B.W., Teixera de Almeida M.L,
  2005, AJ, 130, 597

\bibitem[\protect\citeauthoryear{Yong et al.}{2006}]{b371}
  Yong D., Carney B.W., Teixera de Almeida M.L, Pohl B.L,
  2006, AJ, 131, 2256



\end{thebibliography}
\end{document}